\documentclass[prb,twocolumn,superscriptaddress,showpacs,floatfix,eqsecnum]{revtex4-2}

\usepackage{amsmath,amssymb,bm}
\usepackage{graphicx}
\usepackage{xcolor}

\pdfoutput=1

\begin{document}

\title{
Three-body Fermi-liquid corrections for Andreev transport through quantum dots
}

\author{Akira Oguri}
\affiliation{
Department of Physics, Osaka City University, Sumiyoshi-ku, Osaka, 558-8585, Japan
}
\affiliation{ 
NITEP, Osaka Metropolitan University, Sumiyoshi-ku, Osaka, 558-8585, Japan
}

 \author{Masashi Hashimoto}
 \affiliation{
 Department of Physics, Osaka City University, Sumiyoshi-ku, 
 Osaka, 558-8585, Japan}

\author{Yoshimichi Teratani}
\affiliation{
Department of Physics, Osaka City University, Sumiyoshi-ku, 
Osaka, 558-8585, Japan}
\affiliation{ 
NITEP, Osaka Metropolitan University, Sumiyoshi-ku, Osaka, 558-8585, Japan
}

\date{\today}

\begin{abstract}
We study  crossed Andreev reflection occurring in quantum dots connected to 
one superconducting lead and two normal leads at low temperatures $T$.  
Specifically, we derive an exact formula for the conductance 
up to order $T^2$ in the large superconducting gap limit,   
which is expressed in terms of the transmission probabilities of Cooper pairs 
 and interacting Bogoliubov quasiparticles. 
Our formulation is based on the latest version of Fermi-liquid theory 
for the Anderson impurity model, 
which has clarified the quasiparticle energy shifts of order 
$\omega^2$ and $T^2$---namely, 
corrections of the same order as those arising from the finite lifetime of 
quasiparticles---can be expressed exactly in terms of three-body correlations 
of impurity electrons. 
We also demonstrate how the three-body contributions evolve and affect 
the Cooper-pair tunneling as the Andreev level moves away from the Fermi level, 
using the numerical renormalization group approach.  
The results show that the Cooper-pair contribution to the $T^2$ terms of 
the local and nonlocal conductances 
becomes comparable to the Bogoliubov-quasiparticle contribution  
in the parameter region, in which superconducting proximity effects 
dominate over the Kondo effect.
\end{abstract}

\maketitle

\section{Introduction}

Andreev reflection in quantum dots (QD) connected 
to multiterminal networks consisting of normal (N) and superconducting (SC) leads  
is an active field of current research
\cite{PhysRevB.89.045422,HLee2013,Bordoloi2022,Golubev_2007}.
In particular,  crossed Andreev reflection (CAR) is one of the most interesting processes 
arising from Cooper-pair tunneling 
in which an incident electron entering from a normal lead forms 
a Cooper pair with another electron from 
a different normal lead to tunnel into the superconducting leads, 
leaving behind a hole in the normal lead where the second electron came from.
The CAR is also related, 
via time-reversal symmetry, to Cooper-pair splitting processes, in which 
a Cooper pair emitted from the superconducting lead    
 splits into two entangled electrons that propagate into different normal leads 
\cite{Hofstetter_2009,Schindele_2012,Das_2012,PhysRevB.90.235412,PhysRevLett.114.096602,Borzenets2016,Tan2020,PhysRevB.85.035419,PhysRevB.98.241414,PhysRevB.99.115127,PhysRevLett.120.087701,PhysRevB.99.075429,Ranni_2021}.  

\begin{figure}[b]

\includegraphics[width=0.9\linewidth]{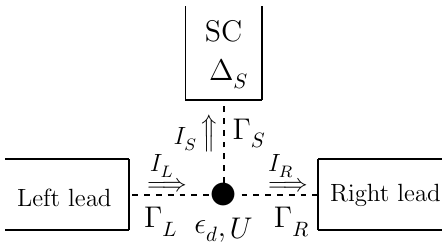}

\caption{
Anderson impurity ({\large $\bullet$}) 
coupled to one superconducting (SC) lead and two normal leads on the left and right. 
 $\Delta_{S}$ is the SC gap, and    
 $\epsilon_d^{}$ and $U$ are the level position and Coulomb interaction 
of the impurity electrons. 
$\Gamma_L^{}$, $\Gamma_R^{}$, and $\Gamma_S^{}$ 
represent the tunneling rates between the impurity site and 
 the left ($L$),  right ($R$), and  SC leads,  respectively.  
}
\label{fig:system}

\end{figure}

Moreover, the Coulomb interaction between electrons in quantum dots induces 
an interesting crossover between the Kondo singlet state and the superconducting 
singlet state 
\cite{RozhkovArovas1999,ClerkAmbegaokar2000,Vecino_2003,Yoshioka_2000,Choi2004,SianoEgger2004,Oguri_2004,Bauer_2007,YoshihideTanaka_2007,Karrasch2008,Florens2009,Hecht_2008,YoichTanaka_2007,Yamada_2011,Buizert_2007,Deacon_2010,Deacon_2010Rapid,Governale_2008,Futterer2009,Eldridge_2010,Michalek_2013,Michalek_2015,Weymann2015,Oguri_2013,PhysRevB.87.115409,Domanski_2017,Wrzeinfmmode_2017,PhysRevLett.129.207701,Zitko2013,Zonda2017,Zonda2023,PhysRevB.99.045120,PhysRevB.101.205422}.
Contributions from CAR can be probed through the nonlocal conductance 
measured from the current flowing from the quantum dot to the normal drain electrode  
under a bias voltage applied to the source electrode 
 (e.g., $I_R^{}$ shown in Fig.\ \ref{fig:system}). 
However, the current flowing between two normal leads 
also contains contributions from single-electron tunneling,  
 in which an incident electron from the source electrode 
is transmitted directly to the drain electrode through the quantum dot. 
It is therefore essential to clarify the conditions under which Cooper-pair tunneling 
dominates over the single-electron tunneling or becomes comparable to it,   
including the competing effects of electron correlations and quantum interference  
\cite{Hashimoto2024,Tanaka_2008}.
In the present work, 
we demonstrate that the contributions from these different tunneling processes 
can, in principle, be separated by observing both the local and nonlocal conductances 
associated with the current  $I_L^{}$ and $I_R^{}$ shown in Fig.\ \ref{fig:system},     
and also show that three-body correlations between electrons in the quantum dot  
play a significant role in the conductance behavior at low but finite temperatures.

The low-energy properties of Kondo systems in quantum dots 
have been extensively studied through highly sensitive transport measurements 
\cite{Goldhaber-Gordon1998nature,Goldhaber-Goldon1998PRL,Cronenwett-Oosterkamp-Kouwenhoven1998,vanderWiel2000,KondoCloud2020,GrobisGoldhaber-Gordon,ScottNatelson,Heiblum,Delattre2009,KobayashiKondoShot,Ferrier2016,Hata2021,Costi2022, Svilansexperimentthermopower}. 
Transport properties have also been theoretically investigated for electrical currents 
\cite{Hershfield1,Izumida2001,AO2001,Sela-Malecki2009,Aligia2012,Aligia2014},  
shot noise \cite{Hershfield2,GogolinKomnikPRL,Sela2006,Golub,OguriSakanoFujii2011}, 
and thermal conductivity \cite{CostiZlatic2010, Costimagthermopower}. 
The underlying physics of low-lying energy states that exhibit universal behavior 
can be explained in terms of Fermi-liquid theory  
\cite{NozieresFermiLiquid,YamadaYosida2,YamadaYosida4,ShibaKorringa,Yoshimori}.    
In particular, recent developments in Fermi-liquid theory 
have revealed the fact that three-body correlations between 
impurity electrons play an essential role in the next-to-leading-order terms 
of the transport coefficients 
at finite frequencies $\omega$, temperatures $T$, and bias voltages $eV$, 
when particle-hole symmetry, 
time-reversal symmetry, or both are broken
\cite{Mora_etal_2009,Mora2009,MoraMocaVonDelftZarand,FMvDM2018,AO2017_I,AO2017_II,AO2017_III,KarkiMora2018, KarkiKiselev, MocaMora2018, Teratani2020PRL, Oguri2022, Tsutsumi2021,Tsutsumi2023, teratani2024thermoelectric,MotoyamaUinf2025}.    
This follows from rigorous proofs showing that the quasiparticle energy shifts of order 
$\omega^2$,  $T^2$, and $(eV)^2$ ---that is, 
corrections of the same order as those arising from the finite lifetime of 
quasiparticles--- can be expressed in terms of three-body correlations 
of impurity electrons.

Recent experiments on nonlinear current and thermocurrent 
through a quantum dot in the Kondo regime 
have demonstrated magnetic-field universality 
in quantitative agreement with theoretical predictions 
that include three-body correlations \cite{Hata2021,Costi2022}.
Therefore, it is natural to expect analogous low-energy Fermi-liquid effects 
in quantum dots connected to superconductors.
Crossed Andreev reflection in the quantum dot connected 
to multiterminal networks, as illustrated in Fig.\ \ref{fig:system},
has been studied intensively, focusing mainly on  
the effect of Coulomb interactions $U$ on the nonlocal conductance  
\cite{Futterer2009,Eldridge_2010,Michalek_2013,Michalek_2015,Weymann2015}.
However, the universal low-energy Fermi-liquid behavior 
arising from three-body correlations has not been investigated,   
except for the seminal work by Moca {\it et al.\ }\cite{MocaMora2018}, 
which clarified the behavior of nonlinear current noise in a two-terminal S-QD-N junction  
near the singlet-doublet quantum phase transition at $T=0$,  
in the regime in which the superconducting gap $\Delta_S^{}$ 
is much larger than the tunneling coupling $\Gamma$ 
between the QD and a single normal lead, 
based on a mapping onto the $U\to\infty$ Anderson model.

The purpose of this paper is to present a comprehensive analysis 
of Andreev transport through a strongly correlated quantum state 
at finite temperatures $T\neq 0$, 
with explicit inclusion of three-body correlations.
To this end, we focus on a quantum-dot system 
connected to two normal leads and one superconducting lead, 
which is described by the Anderson impurity model. 
We derive a formula for the linear conductance up to order $T^2$ 
based on the latest  version of Fermi-liquid theory in the large superconducting-gap limit 
and show that the conductance is determined by the transmission probabilities  
of Cooper pairs, 
$\mathcal{T}_{\mathrm{CP}}^{}(T)$, 
and Bogoliubov quasiparticles,  
$\mathcal{T}_{\mathrm{BG}}^{}(T)$.  
Their low-temperature expansions are given by  
\begin{align*}
\mathcal{T}_{\mathrm{CP}}^{} (T) \, =  & \  
\frac{1}{4}\, \sin^2 \theta\, 
\sin^2 \bigl(\delta_{\uparrow} + \delta_{\downarrow})   
\, + 
  c_{T}^{\mathrm{CP}} \bigl(\pi T\bigr)^2 \,+ \, 
O(T^4) \,, 
\\
\mathcal{T}_{\mathrm{BG}}^{} (T)  \, =&  \  
\frac{1}{2}\,\sum_{\sigma=\uparrow,\downarrow} \sin^2 \delta_{\sigma}^{}
\,-c_{T}^{\mathrm{BG}} \, \bigl(\pi T\bigr)^2 \,+\,O(T^4)  \,.  
\end{align*}
The first terms on the right-hand side 
correspond to the ground-state values, where 
$\delta_{\sigma}^{}$ denotes the phase shift of interacting Bogoliubov quasiparticles 
and $\theta$ is the Bogoliubov rotation angle \cite{Hashimoto2024}.
The coefficients 
$c_{T}^{\mathrm{CP}}$ and $c_{T}^{\mathrm{BG}}$  
of the $T^2$ terms are expressed in terms of the linear susceptibilities 
$\chi_{\sigma_1\sigma_2}^{}$
and the three-body correlation functions 
$\chi^{[3]}_{\sigma_1\sigma_2\sigma_3}$ 
of the Bogoliubov quasiparticles.
We show that the Cooper-pair contribution $c_{T}^{\mathrm{CP}}$ can be 
expressed in the following form:
\begin{align*}
c_{T}^{\mathrm{CP}}
 =&  \,  
- \frac{\pi^2}{3} \, 
\frac{1}{4}\,
\sin^2 \theta \, 
 \Biggl[\,
-\,\cos (2\delta_{\uparrow}^{}+2\delta_{\downarrow}^{})\,
\left( \, \chi_{\uparrow\uparrow}^{}
- \chi_{\downarrow\downarrow}^{} \right)^2
\\
& +\, 
\frac{\sin (2 \delta_{\uparrow}^{} + 2 \delta_{\downarrow}^{}) 
}{2\pi}
\left(\,
\chi_{\uparrow\uparrow\uparrow}^{[3]}\,  
  +   
\chi_{\uparrow\downarrow\downarrow}^{[3]}
+
\chi_{\downarrow\downarrow\downarrow}^{[3]}\,  
 +   
\chi_{\downarrow\uparrow\uparrow}^{[3]}
\,\right)
\,\Biggr] .
 \rule{0cm}{0.5cm}
\end{align*}
By contrast, the Bogoliubov-quasiparticle contribution 
$c_{T}^{\mathrm{BG}}$ 
is identical to that for  
normal electrons tunneling through an N/QD/N junction, studied previously  
\cite{MoraMocaVonDelftZarand,FMvDM2018,AO2017_I,AO2017_II,AO2017_III}. 
In particular, at zero magnetic field,  the Cooper-pair contribution 
 $c_{T}^{\mathrm{CP}}$   
 is determined solely by the three-body correlations. 

We also calculate these coefficients, together with the three-body correlation functions, 
using Wilson's numerical renormalization group (NRG) at zero magnetic field.
The results show that  
the Cooper-pair contribution $c_{T}^{\mathrm{CP}}$ 
becomes comparable to $c_{T}^{\mathrm{BG}}$
in the parameter region 
where superconducting pair correlation is enhanced in the quantum dot,
whereas $c_{T}^{\mathrm{BG}}$ dominates the $T^2$ term in the Kondo regime. 
These contributions from Cooper pairs and Bogoliubov quasiparticles,  
$\mathcal{T}_{\mathrm{CP}}^{}(T)$ and 
$\mathcal{T}_{\mathrm{BG}}^{}(T)$,   
can be experimentally determined by measuring 
both the nonlocal conductance $\partial I_{R}^{}/\partial V_{L}^{}$ 
and the local conductance $\partial I_{L}^{}/\partial V_{L}^{}$ 
under a finite bias voltage $V_{L}^{}$ applied to the source electrode.

This paper is organized as follows.
In Sec.\ \ref{sec:Formulation}, 
we introduce an Anderson impurity model for quantum dots 
 connected to normal and superconducting leads,    
and reformulate the Hamiltonian and correlation functions 
in terms of interacting Bogoliubov quasiparticles.   
Section \ref{sec:multi_terminal_formulation} is devoted to   
the Fermi-liquid description of interacting Bogoliubov quasiparticles,  
 in which three-body correlations play an essential role.
An exact conductance formula valid up to order $T^2$  
is also presented, while detailed derivations are deferred to the final part of this paper.  
In Sec.\ \ref{CARZeroMag},  
we present NRG results that demonstrate  
 how the three-body correlations affect 
the next-to-leading-order terms in crossed Andreev reflection.  
In Sec.\ \ref{sec:3body_correction_derivation}, 
we provide the derivation of the conductance formula to order $T^2$, 
leaving the precise analysis of the superconducting collision integrals 
required for the complete proof to the Appendix.
A summary and discussion are  given in Sec.\ \ref{summary}.

 \section{Fermi-liquid description for Bogoliubov quasiparticles
}
\label{sec:Formulation}

In this section, we begin with an Anderson impurity model  
describing a single quantum dot (QD) connected to 
one superconducting (SC) lead and two normal (N) leads,
and then transform it into a system of interacting Bogoliubov 
quasiparticles, the total number of which is conserved.
We also introduce three-body correlation functions of the Bogoliubov quasiparticles, 
which determine the next-to-leading-order terms of 
the transport coefficients in the low-energy Fermi-liquid regime.

\subsection{Anderson impurity model for the CAR}

We consider the Anderson impurity model 
for the quantum dot system as shown in Fig.\ \ref{fig:system}, 
\begin{align}
&   \   H \   = \, 
H_d^{0}
\,+\, 
H_d^{U}
\,+\, 
H_{\text{N}}^{}  \,+\,H_{\text{TN}}^{} 
\,+\,H_{\text{S}}^{} +H_{\text{TS}}^{}\,,
\label{eq:total_H_single}
\\
&H_d^{0}
 \,=\, 
\xi_d^{}  \bigl( n_d^{} -1 \bigr) 
- b \, \bigl(
n_{d\uparrow}^{} - n_{d\downarrow}^{} 
 \bigr) ,
\label{eq:H_dot_0}
\\
& H_d^{U} =\, 
\frac{U}{2}\bigl( n_d^{} -1 \bigr)^2 ,
\label{eq:H_dot_U}
 \\
&  
H_{\text{N}}^{} 
 \,=\, 
\sum_{j=L,R}\sum_{\sigma} 
\int_{-D}^{D}  \! d\epsilon \,\epsilon\,
 c^\dagger_{\epsilon j \sigma}c^{ }_{\epsilon j \sigma},
\label{eq:H_N}
\\
  &  
H_{\text{TN}}^{} =  
-
\! 
\sum_{j=L,R} 
v_{j}^{}
\sum_{\sigma} 
\int_{-D}^{D}  \! d\epsilon \,\sqrt{\rho_c^{}}\,
\Bigl( c^\dagger_{\epsilon j\sigma}d^{}_\sigma 
+ \mathrm{H.c.}
\Bigr),
\label{eq:H_TN}
\\
&H_{\text{TS}} = \, 
-\, v_{S}^{}
 \sum_{\sigma} \! 
 \int_{-D_S^{}}^{D_S^{}}  \!\!  d\epsilon \,\sqrt{\rho_S^{}}\, 
 \Bigl( s^\dagger_{\epsilon\sigma} d^{}_\sigma 
 + \mathrm{H.c.}
 \Bigr),
\\
& H_{\text{S}} \,= \,  
\sum_{\sigma} \!
\int_{-D_S^{}}^{D_S^{}}  \!\! d\epsilon \,\epsilon\,
s^\dagger_{\epsilon\sigma} s^{}_{\epsilon\sigma} 
  + 
\int_{-D_S^{}}^{D_S^{}}  \!\! d\epsilon 
\left( \Delta_S^{} \,s^\dagger_{\epsilon\uparrow} 
s^\dagger_{\epsilon\downarrow} 
+  \mathrm{H.c.} \right). 
\end{align}
Here, $H_d^{0}$ and $H_d^{U}$, 
correspond to the Hamiltonian for electrons in the QD.
The parameter $\xi_d^{} \equiv \epsilon_d^{} +{U}/{2}$ is defined in terms of  
the discrete energy level $\epsilon_d^{}$ 
and the Coulomb interaction $U$. 
The Zeeman energy arising from a magnetic field $B$ applied to the QD is given by  
 $b\equiv \mu_B^{} B$, with $\mu_B^{}$ the Bohr magneton. 
The operator
 $d^\dagger_\sigma$ creates 
 an electron with spin $\sigma$ ($=\uparrow,\downarrow$), 
and  $n_d^{} \equiv n_{d\uparrow}^{}+n_{d\downarrow}^{}$ 
is the total number operator, with 
$n_{d\sigma}^{}\equiv d^\dagger_\sigma d^{ }_\sigma$.

  $H_{\text{N}}$ describes 
the conduction electrons in the normal leads, 
for which the density of states is assumed to be a constant  
  $\rho_c^{}=1/(2D)$, with $D$ the half-width of the bands. 
The Fermi level is set at the center 
 of the conduction bands, $E_F^{} \equiv 0$.
The operator  $c^\dagger_{\epsilon j\sigma}$ 
 creates a conduction electron 
with spin $\sigma$ and energy $\epsilon$ in the  
left ($j =L$) or right ($j=R$) lead.  
The operators satisfy the anti-commutation relation,
$\{ c^{}_{\epsilon j\sigma}, 
c^{\dagger}_{\epsilon'j'\sigma'}
\} = \delta_{jj'} \,\delta_{\sigma\sigma'}   
\delta(\epsilon-\epsilon')$,
 which is normalized by the Dirac delta function.  
$H_{\text{TN}}^{}$ describes 
the tunnel coupling between the QD and the normal leads.
The level broadening of the discrete energy level 
 in the QD is given by $\Gamma_N \equiv \Gamma_L + \Gamma_R$, 
with $\Gamma_{j}^{} \equiv \pi \rho_c^{} v_{j}^2$.

 $H_{\text{S}}^{}$ and $H_{\text{TS}}^{}$ 
describe the superconducting lead and its tunnel coupling to the QD, 
characterized by an $s$-wave superconducting gap  
 $\Delta_{S}^{} \equiv \left| \Delta_{S}^{}\right|\,e^{i\phi_S^{}}$. 
The operator  
$s^\dagger_{\epsilon\sigma}$ 
creates an electron with spin $\sigma$ and energy $\epsilon$ in the SC lead, 
with $D_S^{}$ the half-band width and $\rho_S^{} =1/(2D_S^{})$.
The coupling strength between the QD and the SC lead is 
given by $\Gamma_S \equiv \pi \rho_S^{} v_{S}^2$, 
and it serves as one of the key parameters characterizing 
the superconducting proximity effect in this system.

The current operators, $\widehat{I}_{R,\sigma}^{}$, flowing from 
the QD to the right normal lead,  
and $\widehat{I}_{L,\sigma}^{}$, flowing  
from the left normal lead to the QD,  
are given by  
\begin{align}
\widehat{I}_{R,\sigma}^{}
 \,= & \   i\,
e v_{R}^{}  \int_{-D}^{D} \!\! d\epsilon\,
\sqrt{\rho_c^{}}\, 
\Bigl(
c^{\dagger}_{\epsilon,R,\sigma} d^{}_{\sigma} 
-d^{\dagger}_{\sigma} c_{\epsilon,R,\sigma}^{}\Bigr)  
\label{eq:IR_def}
,\\
\widehat{I}_{L,\sigma}^{} \,= &\  - i\,
e v_{L}^{}  \int_{-D}^{D} \!\! d\epsilon\,
\sqrt{\rho_c^{}}\, 
\Bigl(
c^{\dagger}_{\epsilon,L,\sigma} d^{}_{\sigma} 
-d^{\dagger}_{\sigma} c_{\epsilon,L,\sigma}^{}\Bigr)  ,
\label{eq:IL_def}
\end{align}
 for each spin component $\sigma$. 
We consider the situation in which a bias voltage $V_{j}^{}$ is applied  
to the normal leads ($j=L, R$) via   
their chemical potentials, such that $\mu_{j}^{} \equiv  eV_{j}^{}$   
and $V \equiv V_{L}^{}-V_{R}^{}$.
In addition,  the chemical potential $\mu_S^{}$ of the SC lead  
 is set at the center of the conduction band, i.e., $\mu_S^{} \equiv 0$.

\subsection{Effective Hamiltonian for large-gap limit}
\label{subsec:Large_gap_limit}

In this paper, we investigate transport properties at finite temperatures 
 much lower than the superconducting energy gap. 
Specifically, we consider the large-gap limit, 
$\left| \Delta_{S}^{}\right| \to \infty$, 
in which the influence of Cooper-pair tunneling on the QD and normal leads 
can be described by an effective Hamiltonian 
$\mathcal{H}_\mathrm{eff}^{}$ \cite{Oguri_2004,YoichTanaka_2007},
\begin{align}
\mathcal{H}_\mathrm{eff}^{} 
  \,\equiv & \      
\Delta_d^{} \,d^{\dagger}_{\uparrow} d^{\dagger}_{\downarrow}
+ \Delta_d^{*} \, d^{}_{\downarrow} d^{}_{\uparrow} 
+ H_\mathrm{dot}^{} 
+ H_{\text{N}}^{}  +H_{\text{TN}}^{} , 
\label{eq:Heff_single}
\\  
 \Delta_d^{} \, \equiv & \ \Gamma_{S}^{}\,e^{i\phi_S^{}} 
\,, \qquad 
H_\mathrm{dot}^{} \,\equiv \,H_d^{0} + H_d^{U} 
\,. 
\rule{0cm}{0.5cm}
\end{align}
Here,  the SC part of the Hamiltonian,  
$H_{\text{S}} +H_{\text{TS}}$,  
is replaced by the local pair potential 
$\Delta_d^{}$ induced at the impurity site. 
Specifically, the pair creation and annihilation terms,  
$d^{\dagger}_{\uparrow} d^{\dagger}_{\downarrow}$ 
and $d^{}_{\downarrow} d^{}_{\uparrow}$, 
govern the Cooper-pair tunneling between the SC lead 
and the QD, with coupling strength $\Gamma_{S}^{}$. 
Namely, in the large-gap limit, 
the current $I_{S}^{}$ flowing from the QD into the SC lead,
 illustrated in Fig.\ \ref{fig:system}, is induced solely by the Cooper-pair tunneling, 
since the single-electron tunneling 
(or tunneling of unpaired Bogoliubov quasiparticles) 
between the QD and the SC lead, which occurs 
at finite $\left|\Delta_{S}^{}\right|$, is suppressed.    
In the following, we will choose the gauge of the electron operators 
 such that it absorbs the Josephson phase of $\Delta_d^{}$,
 i.e., $\phi_S^{}=0$.

This effective Hamiltonian possesses a global U(1) symmetry in the Nambu 
pseudo-spin space, which becomes explicit upon performing    
the Bogoliubov transformation, 
\begin{align}
 \begin{pmatrix}
  \gamma_{d\uparrow}^{} \cr
  \gamma_{d\downarrow}^{\dagger} \rule{0cm}{0.6cm}\cr
 \end{pmatrix}
=  
 \bm{\mathcal{U}}^\dagger
 \begin{pmatrix}
  d_{\uparrow}^{} \cr
  d_{\downarrow}^{\dagger} \rule{0cm}{0.6cm}\cr
 \end{pmatrix}
,
\quad 
 \begin{pmatrix}
  \gamma_{\epsilon, j, \uparrow}^{} \cr
  -\gamma_{-\epsilon,j, \downarrow}^{\dagger} 
\rule{0cm}{0.6cm}\cr
 \end{pmatrix}
 =   
 \bm{\mathcal{U}}^\dagger
 \begin{pmatrix}
  c_{\epsilon, j, \uparrow}^{} \cr
  -c_{-\epsilon,j, \downarrow}^{\dagger} 
\rule{0cm}{0.6cm}\cr
 \end{pmatrix}
.
\label{eq:Bogoliubov_tans_single}
\end{align}
The matrix $\bm{\mathcal{U}}$ can be expressed in the following form:    
\begin{align}
& 
\bm{\mathcal{U}}
\,\equiv \, 
\begin{pmatrix} 
\cos \frac{\theta}{2} & 
\, -
\sin \frac{\theta}{2} \\
\sin \frac{\theta}{2} & 
\rule{0cm}{0.5cm} 
\quad   
\cos \frac{\theta}{2}
\end{pmatrix} 
\,,
\label{eq:Bogoliubov_rotation}
\\
& 
\cos \frac{\theta}{2} =  
\sqrt{\frac{1}{2}\left(1+\frac{\xi_{d}}{E^{}_A}\right)} , 
\quad 
\sin \frac{\theta}{2} =  
\sqrt{\frac{1}{2}\left(1-\frac{\xi_{d}}{E^{}_A}\right)} .
\rule{0cm}{0.9cm}
\nonumber 
\end{align}
Here, $E_A^{} \equiv \sqrt{\xi_d^2 + \Gamma_S^2}$, 
$\sin \theta = \frac{\Gamma_S^{}}{E^{}_A}$, 
and $\cos \theta = \frac{\xi_d^{}}{E^{}_A}$.   
This transformation maps   
the effective Hamiltonian $\mathcal{H}_\mathrm{eff}^{}$    
onto a system of interacting Bogoliubov quasiparticles, which is 
described by a standard Anderson impurity model 
 with a conserved charge, 
\begin{align}
&
\!\!\!\! 
\mathcal{H}_\mathrm{eff}^{} 
=\,   E^{}_A
\left(\sum_\sigma \gamma^\dagger_{d\sigma}\gamma^{}_{d\sigma}-1\right)
- b\, \Bigl(
\gamma^\dagger_{d\uparrow}\gamma^{}_{d\uparrow}
-\gamma^\dagger_{d\downarrow}\gamma^{}_{d\downarrow}
 \Bigr)
\nonumber \\
& 
\!\!\!\! 
+ \frac{U}{2}\left(\sum_\sigma \gamma^\dagger_{d\sigma}
\gamma^{}_{d\sigma} - 1 \right)^2 
 + 
\sum_{j=L,R} \sum_{\sigma} 
\int_{-D}^{D}  \! d\epsilon \,\epsilon\,  
 \gamma^\dagger_{\epsilon,j,\sigma}
 \gamma^{ }_{\epsilon,j,\sigma} 
\nonumber \\
&
\!\!\!\! 
+ 
\sum_{j=L,R} 
v_{j}^{}
\sum_{\sigma} 
\int_{-D}^{D}  \! d\epsilon \,
\sqrt{\rho_c^{}}\,
\left(
\gamma^\dagger_{\epsilon,j,\sigma}
\gamma^{ }_{d\sigma}
+ \mathrm{H.c.} 
\right)\, . 
\label{eq:SingleDot_new}
\end{align}
This representation explicitly shows that 
the total number of Bogoliubov quasiparticles 
$\mathcal{N}_{\gamma}^{}$ is conserved, 
reflecting the global $U(1)$ symmetry along the principal axis 
of the Nambu pseudo-spin space,  
\begin{align}
&\mathcal{N}_{\gamma}^{} 
\equiv   
\sum_\sigma  
\gamma^{\dagger}_{d\sigma}\gamma^{}_{d\sigma} 
+ 
\sum_{j=L,R} 
\sum_\sigma  
\int_{-D}^{D}  \! d\epsilon \,
\gamma^{\dagger}_{\epsilon,j,\sigma} \gamma^{}_{\epsilon,j,\sigma} 
\,.  
\label{eq:Bogolon_total_charge}
\end{align}
In particular, the occupation number $q_{d\sigma}^{}$ of Bogoliubov quasiparticles 
in the QD plays a central role in the ground-state properties,
\begin{align}
   q_{d\sigma}^{} \,  \equiv \,  
\gamma^\dagger_{d\sigma}\gamma^{}_{d\sigma} \,, 
\qquad \quad 
  q_d^{} \, & \equiv \, 
 q_{d\uparrow}^{} 
   + 
 q_{d\downarrow}^{} \, .
\label{eq:def_Q}
 \end{align}

\begin{figure}[t]
\includegraphics[width=\linewidth]{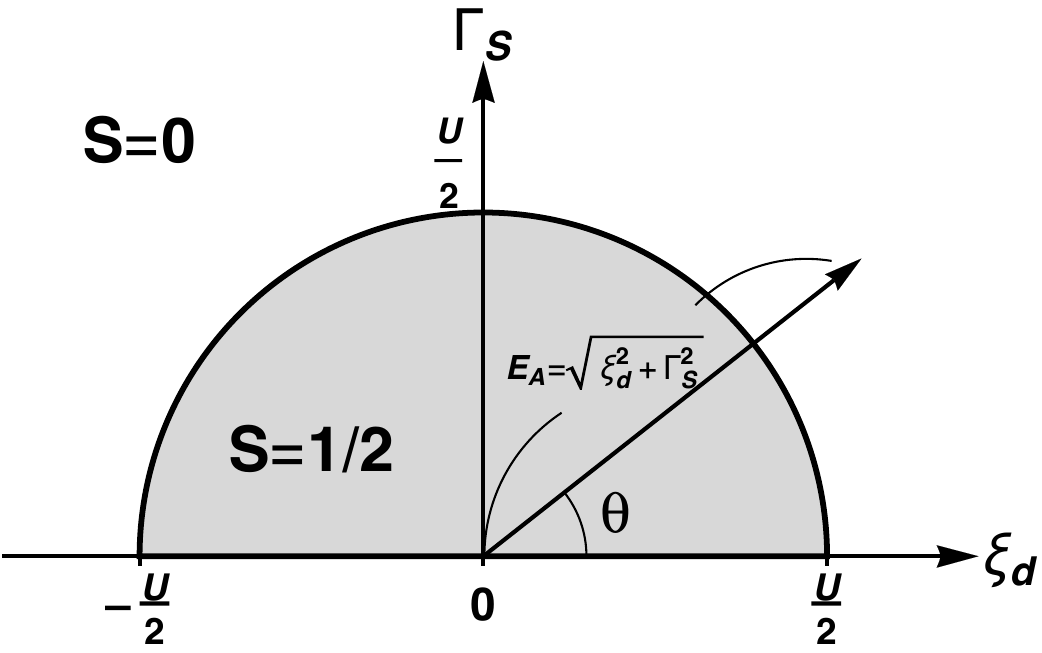}
\caption{Parameter space of $\mathcal{H}_\mathrm{eff}^{}$ 
at zero magnetic field $b=0$, defined in Eq.\ \eqref{eq:Heff_single}.
Here,  $\theta \equiv  \cot^{-1}(\xi_d^{}/\Gamma_S^{})$ 
is the Bogoliubov rotation angle, with $\xi_d^{} \equiv \epsilon_d^{} +U/2$.    
The semicircle corresponds to the line along which 
the energy of the Andreev level, $E_A^{} \equiv \sqrt{\xi_d^2+\Gamma_S^2}$, 
coincides with $U/2$.
In the atomic limit $\Gamma_N ^{} =0$, 
the ground state is a magnetic spin doublet ($S=1/2$) inside the semicircle, 
which is eventually screened by conduction electrons to form the Kondo singlet 
when the tunnel coupling $\Gamma_N^{}$ is switched on,  
whereas outside the semicircle the ground state 
is a spin singlet ($S=0$) because of Cooper pairing. 
}
\label{fig:SingleDotPhase}
\end{figure}

Figure \ref{fig:SingleDotPhase} illustrates the parameter space  of 
  $\mathcal{H}_\mathrm{eff}^{}$ at zero magnetic field,  $b=0$. 
 The polar angle $\theta =  \cot^{-1}(\xi_d^{}/\Gamma_S^{})$  
corresponds to the Bogoliubov rotation angle.    
In the atomic limit $\Gamma_N^{}\to 0$, 
  the impurity level is occupied 
by a single Bogoliubov quasiparticle,  $\langle q_{d}^{}\rangle = 1.0$,
inside the semicircle $E_A^{} < U/2$, 
whereas it becomes empty, $\langle q_{d}^{}\rangle = 0.0$, 
outside the semicircle  $E_A^{} > U/2$. 
As the coupling $\Gamma_N^{}$ is switched on, 
conduction electrons from the normal leads screen   
the local moment,  and the Kondo singlet becomes 
the ground state inside the semicircle. 
Valence-fluctuation behavior of the Bogoliubov quasiparticles
also appears near the boundary $E_A^{} \simeq U/2$. 
In particular, it manifests itself 
as a crossover between the Kondo singlet and the superconductivity-induced  singlet 
at $\theta \simeq \pi/2$.
Correspondingly, crossed Andreev scattering is enhanced at $T=0$ 
in the angular range $\pi/4 \lesssim \theta \lesssim 3\pi/4$ 
 in the vicinity just outside the semicircle for $E_A^{} \gtrsim U/2$ 
\cite{Hashimoto2024}.

\subsection{Phase shift of interacting Bogoliubov quasiparticles}
\label{subsec:def_renormalized_parameters}

We use the following retarded Green's function 
to describe the transport properties of the QD   
connected to the SC lead:
\begin{align}
 &   
\!\!\!
\bm{G}^r_{dd} (\omega) 
\equiv  -i\int_0^{\infty} \! dt \, e^{i\left(\omega + i0^+ \right)t} 
\nonumber \\
 & \qquad \qquad \ \ 
\times 
 \begin{pmatrix}
\left\langle\left\{d^{}_{\uparrow}(t),\,d^\dagger_{\uparrow} \right\} 
\right\rangle_{V}^{} & 
\left\langle\left\{d^{}_{\uparrow}(t),\,d^{}_{\downarrow} \right\} 
\right\rangle_{V}^{}\\
\left\langle\left\{d^\dagger_{\downarrow}(t),\,d^\dagger_{\uparrow} \right\} 
\right\rangle_{V}^{} & 
\left\langle\left\{d^\dagger_{\downarrow}(t),\,d^{}_{\downarrow} \right\} 
\right\rangle_{V}^{}
\rule{0cm}{0.6cm}
\end{pmatrix}.
\label{eq:impurity_Nambu_Green's_function}
\end{align}
Here, the average $\langle \cdots\rangle_{V}^{}$ 
is defined with respect to the nonequilibrium steady state, 
constructed at finite bias voltage $eV$ and temperature $T$, 
using the Keldysh formalism \cite{Hershfield1,Caroli_1971,Keldysh}.
This matrix, $\bm{G}^r_{dd}$, can be 
diagonalized by the Bogoliubov transformation,  
\begin{align}
&
\!\!\!\! 
\widetilde{\bm{G}}^r_{}(\omega)
\,  \equiv \,
\bm{\mathcal{U}}^\dagger\,  
\bm{G}^r_{dd}(\omega)\ 
\bm{\mathcal{U}}^{}  
\nonumber \\ 
 & 
\!\!\!\! 
 = 
 \begin{pmatrix}
\widetilde{G}^r_{1}(\omega) & 0 \\
0 & \widetilde{G}^r_{2}(\omega)
\rule{0cm}{0.4cm}
\end{pmatrix} 
 = 
 \begin{pmatrix}
G^r_{\gamma,\uparrow}(\omega) & 0 \\
0 & - G^a_{\gamma,\downarrow}(-\omega)
\rule{0cm}{0.4cm}
\end{pmatrix} .
\rule{0cm}{0.9cm}
 \label{eq:GreenBogoliubovTrans}
\end{align}
Here, $G^r_{\gamma,\sigma}$ and  $G^a_{\gamma,\sigma}$ 
correspond to the retarded and advanced Green's functions, respectively,  
for Bogoliubov quasiparticles in the QD,  
 \begin{align}
 G^r_{\gamma,\sigma}(\omega)
\equiv& \,   -i\int_0^{\infty} \!  dt \, e^{i\left(\omega + i0^+\right)t} 
\left\langle\left\{\gamma^{}_{d\sigma}(t),\, 
\gamma^\dagger_{d\sigma} \right\} \right\rangle_{V}^{}  
\nonumber 
\\
=&  
\ \frac{1}{\omega\,- E_{A,\sigma}^{} +i\Gamma_{N}
-\Sigma_{\gamma,\sigma}^{r}(\omega) 
}, 
\label{eq:retarded_G_gamma}
 \end{align}
and   $ G^a_{\gamma,\sigma}(\omega) 
= \left\{G^r_{\gamma,\sigma}(\omega)\right\}^*$.
Here, $E_{A,\uparrow}^{} \equiv E_{A}^{}- b$ and  
$E_{A,\downarrow}^{} \equiv E_{A}^{}+ b$,   
and 
 $\Sigma_{\gamma,\sigma}^{r}(\omega)$  represents 
the self-energy correction arising from the interaction 
Hamiltonian $H_d^{U}$  defined in Eq.\ \eqref{eq:H_dot_U}, 
which can alternatively be expressed in the form  
$H_d^{U}=\frac{U}{2} \bigl( q_{d}^{} -1 \bigr)^2$.  
In the noninteracting limit $H_d^{U} \to 0$, 
Eq.\ \eqref{eq:retarded_G_gamma} describes 
an Andreev resonance level located  
at $\omega=E_{A,\sigma}^{}$ with a width $\Gamma_{N}$.

 The phase shift $\delta_\sigma$ of the Bogoliubov quasiparticles, 
defined by  
$G_{\gamma,\sigma}^r(0) = -\left| G_{\gamma,\sigma}(0)\right| 
e ^{i \delta_{\sigma}^{}}$  
at $T=eV=0$, plays a central role in the ground-state properties. 
In particular,  the Friedel sum rule holds 
for the average occupation number $\left \langle q_{d\sigma}^{} \right\rangle$
of the Bogoliubov quasiparticles in the QD, namely,  
 \begin{align}
\left \langle q_{d\sigma}^{} \right\rangle
\, \xrightarrow{\,T \to 0 \,} \,\frac{\delta_\sigma}{\pi}.
\label{eq:Friedel_Q}
\end{align}
Here, $\langle  \mathcal{O} \rangle
= \mathrm{Tr} 
\bigl[\,e^{-\beta \mathcal{H}_\mathrm{eff}^{}} \mathcal{O}\,\bigr]
/\mathrm{Tr}\,e^{- \beta \mathcal{H}_\mathrm{eff}^{}}$ 
denotes the equilibrium average, with $\beta \equiv1/T$.
The phase shift reaches $\delta_{\sigma}^{} =\pi/2$ at $E_A^{}=0$ 
and decreases as $E_A^{}$ increases along the radial direction
in the  $\xi_d^{}$--$\Gamma_S^{}$ plane  
shown in Fig.\ \ref{fig:SingleDotPhase},   
while it remains independent of the polar angle $\theta$.

The ground-state properties of electrons, such as 
the occupation number  $\left\langle n_{d} \right \rangle$ 
 and the superconducting pair correlation function 
$\bigl \langle d^\dagger_{\uparrow}\,d^\dagger_{\downarrow}
 +d_{\downarrow}\,d_{\uparrow} \bigr \rangle$, 
can be derived from $\langle q_{d}^{} \rangle$ as follows: 
\begin{align}
   \left\langle n_{d}^{} \right \rangle - 1 
\ =& \, \Bigl(\langle q_{d}^{} \rangle-1\Bigr)\,\cos \theta \, , 
 \label{eq:SingleDotM}
\\
   \left \langle d^\dagger_{\uparrow}\,d^\dagger_{\downarrow}
 +d_{\downarrow}\,d_{\uparrow} \right \rangle
\ =&  \, \Bigl(\langle q_{d}^{} \rangle -1\Bigr) \,\sin \theta \,.
 \label{eq:SingleDotK}
\end{align}
These two averages correspond, respectively, to the projections of a vector 
of magnitude $\langle q_{d}^{} \rangle-1$, oriented along the principal axis,  
onto the $z$ and $x$ axes of the Nambu pseudo-spin space.
Similarly, the local magnetization $m_d^{}$, which can be induced in the QD  
 at finite magnetic fields, can be expressed as 
\begin{align}
  m_d^{} \, \equiv \ &  \  \left \langle n_{d\uparrow}^{} \right \rangle 
- \left \langle n_{d\downarrow}^{} \right \rangle
\ = \, 
\left \langle q_{d\uparrow}^{} \right\rangle -
\left \langle q_{d\downarrow}^{} \right\rangle \,.
 \label{eq:SingleDotm}
\end{align}

\subsection{Susceptibilities and three-body correlations}

The low-energy properties of 
the effective Hamiltonian $\mathcal{H}_\mathrm{eff}^{}$     
can be described within the framework of Fermi-liquid theory 
using a set of essential parameters: 
the occupation numbers 
$\bigl \langle q_{d\sigma}^{} \bigr\rangle$,
the linear susceptibilities $\chi_{\sigma_1\sigma_2}^{}$, 
and the three-body correlation functions $\chi^{[3]}_{\sigma_1\sigma_2\sigma_3}$, 
defined by  
\begin{align}
   \langle q_{d\sigma}^{} \rangle 
  &= \,\frac{\partial\Omega_\mathrm{eff}^{}}{\partial E_{A,\sigma}^{}}, 
  \\
  \chi_{\sigma_1\sigma_2}^{} 
  &\equiv \,
-\frac{\partial^2\Omega_{\mathrm{eff}}^{}}
{\partial E_{A,\sigma_1}^{}\partial E_{A,\sigma_2}^{}}
  \ = \int_0^\beta \!\! d\tau\,\left\langle \delta q_{d\sigma_1}^{}(\tau)\,
\delta q_{d\sigma_2}^{}\right \rangle,
  \\
  \chi^{[3]}_{\sigma_1\sigma_2\sigma_3}
  &\equiv \,
 - \frac{\partial^3\Omega_{\mathrm{eff}}^{}}
{\partial E_{A,\sigma_1}^{} 
 \partial E_{A,\sigma_2}^{} \partial E_{A,\sigma_3}^{}}
  \nonumber \\
  & 
\!\!\!\!\!\!\!\!\! 
\!\!\!\!\!\!\!\!
= \,
-\int_0^\beta \!\! d\tau \!\! \int_0^\beta \!\! d\tau' 
\left 
\langle T_\tau \delta q_{d\sigma_1}^{}(\tau)\, \delta q_{d\sigma_2}^{}(\tau')\, 
\delta q_{d\sigma_3}^{}\right\rangle.
\end{align}
Here,  
 $\Omega_\mathrm{eff}^{}\equiv
-\frac{1}{\beta}\mathrm{ln}
\bigl[\,\mathrm{Tr}\,e^{-\beta \mathcal{H}_\mathrm{eff}^{}}\bigr]$ 
is the free energy defined with respect to $\mathcal{H}_\mathrm{eff}^{}$. 
The operator 
$\delta q_{d\sigma}^{}(\tau) \equiv e^{\tau \mathcal{H}_\mathrm{eff}^{}} 
\delta q_{d\sigma}^{}\,e^{-\tau \mathcal{H}_\mathrm{eff}^{}}$  
represents the fluctuation 
$\delta q_{d\sigma}^{}\equiv q_{d\sigma}^{} - \langle q_{d\sigma}^{}\rangle$ 
of local Bogoliubov quasiparticles, and  
 $T_\tau$ is the imaginary-time ordering operator. 
Specifically, 
the ground-state values of these correlation functions, namely,   
  $\langle q_{d\sigma}^{}\rangle$, 
  $\chi_{\sigma_1\sigma_2}^{}$, and 
  $\chi^{[3]}_{\sigma_1\sigma_2\sigma_3}$ defined at $T=0$, 
completely determine the transport properties 
in the Fermi-liquid regime up to the next-to-leading order.

One of the typical Fermi-liquid corrections arising from many-body scattering 
appears in the $T$-linear specific  heat  of impurity electrons    
$\mathcal{C}_\mathrm{imp}^{\mathrm{heat}}$, 
which can be expressed in terms of  
the diagonal components of the linear susceptibility $\chi_{\sigma\sigma}^{}$,   
\begin{align}
\mathcal{C}_\mathrm{imp}^{\mathrm{heat}} \,=\,   
\gamma_\mathrm{imp}^{}  \, T \,,   
\qquad 
\gamma_\mathrm{imp}^{} 
\,\equiv \, \frac{\pi^2}{3}\sum_\sigma \chi_{\sigma\sigma}^{} \,.    
\end{align}
This formula was derived by Yamada and Yosida 
using the Ward identity \cite{YamadaYosida2,ShibaKorringa,Yoshimori}.  
In particular, at zero magnetic field $b=0$, where 
the two diagonal components coincide, i.e., 
 $\chi_{\uparrow\uparrow}^{}=\chi_{\downarrow\downarrow}^{}$,   
 the $T$-linear specific heat can be written as
 $\mathcal{C}_\mathrm{imp}^{\mathrm{heat}}
=\frac{\pi^2}{6}\frac{T}{T^{*}}$,  
where the Fermi-liquid energy scale $T^{*}$ is defined by   
\begin{align}
T^{*}  
\equiv \, 
\frac{1}{4\chi_{\uparrow\uparrow}^{}}
\,.
\label{eq:T*_def}
\end{align}
Correspondingly,  the off-diagonal susceptibility 
$\chi_{\uparrow\downarrow}^{}$ can be naturally scaled by $T^*$,    
which leads to the definition of the Wilson ratio $R$:
\begin{align}
R -1 \,\equiv\, 
-\frac{\chi_{\uparrow\downarrow}^{}}{\chi_{\uparrow\uparrow}^{}}
\ = \,   - 4T^* \chi_{\uparrow\downarrow}^{}
\,.
\label{eq:Wilson_ratio_def}
\end{align}

The three-body correlations 
$\chi^{[3]}_{\sigma_1\sigma_2\sigma_3}$
arise when the particle-hole symmetry, the time-reversal symmetry or both are broken,   
and they contribute to next-to-leading-order transport properties,   
such as  the $T^2$  and $(eV)^2$ terms in the conductance. 
Physically, the three-body correlations give rise to energy shifts 
of order $\omega^2$, $T^2$, and  $(eV)^2$ in the low-lying excitations, 
corresponding to
 Fermi-liquid corrections of the same order 
 as those arising from the finite lifetime of quasiparticles.

\section{Linear response theory for CAR}
\label{sec:multi_terminal_formulation}

In this section, 
we show how the contributions of crossed Andreev reflection  
to the nonlocal conductance of the multi-terminal network 
can be described within the framework of 
the Fermi-liquid theory for interacting Bogoliubov quasiparticles.
Specifically, we consider the linear-response current 
$I_{j}^{}\equiv
\sum_{\sigma}
\langle \widehat{I}_{j,\sigma}^{} \rangle_{V}^{}$ for $j=L,R$, 
induced by small bias voltages $V_L$ and $V_R$ 
 applied to the left and right leads, respectively,   
\begin{align}
  I_{R}^{} \,  =&  \ \ g_{RL}^{} V_L^{}  \,-\,  g_{RR}^{} V_R^{}  \,,
\\
  I_{L}^{} \,  =&   - g_{LR}^{} V_R^{}  \,+\,  g_{LL}^{} V_L^{}   \,,    
\rule{0cm}{0.5cm} 
\end{align}
where the current directions are specified as illustrated in Fig.\ \ref{fig:system}.

\subsection{General formula for $T^2$ conductance}
\label{subsec:General_formula_result}

Here  we highlight one of the central results of this work,
leaving the derivations to Sec.\ \ref{sec:3body_correction_derivation}.  
The low-temperature expansion of the linear conductance 
 $g_{jj'}^{}=\partial I_j{^{}}/\partial V_{j'}^{}$  
 can be calculated exactly up to terms of order $T^2$, 
\begin{align}
g_{RL}^{} =&  \  
g_{LR}^{}=  
 \frac{2e^2}{h} \, \frac{4\Gamma_{R}^{}\Gamma_{L}^{}}
{\Gamma_{N}^{2}}
\Bigl[\,
\mathcal{T}_{\mathrm{BG}}^{} (T) 
\,-\, 2\mathcal{T}_{\mathrm{CP}}^{} (T) 
\,\Bigr], 
\label{eq:gRL}
\\
g_{RR}^{}\, =& \  
 \frac{2e^2}{h} 
\left[\,
\frac{4\Gamma_{R}^{}\Gamma_{L}^{}}{\Gamma_{N}^2}\   
\mathcal{T}_{\mathrm{BG}}^{} (T) 
\,+\, 
\frac{4\Gamma_{R}^{2}}{\Gamma_{N}^2}\   
2\mathcal{T}_{\mathrm{CP}}^{} (T) 
\,\right],
\label{eq:gRR}
 \\
g_{LL}^{}\, =& \  
 \frac{2e^2}{h}  
\left[\,
\frac{4\Gamma_{R}^{}\Gamma_{L}^{}}{\Gamma_{N}^2}\   
\mathcal{T}_{\mathrm{BG}}^{} (T) 
\,+\, 
\frac{4\Gamma_{L}^{2}}{\Gamma_{N}^2}\   
2\mathcal{T}_{\mathrm{CP}}^{} (T) 
\,\right]. 
\label{eq:gLL}
\end{align}
Specifically, the transmission probability for Bogoliubov quasiparticles, 
  $\mathcal{T}_{\mathrm{BG}}^{}(T)$, 
and that for Cooper pairs, 
$\mathcal{T}_{\mathrm{CP}}^{}(T)$, 
can be expanded up to order $T^2$ as follows:
\begin{align}
\mathcal{T}_{\mathrm{BG}}^{} (T)  \, =&  \  
\frac{1}{2}\,\sum_{\sigma} \sin^2 \delta_{\sigma}^{}
\,-c_{T}^{\mathrm{BG}} \, \bigl(\pi T\bigr)^2 \,+\,\cdots \,,  
\label{eq:Transmission_BG_def}
\\
\mathcal{T}_{\mathrm{CP}}^{} (T) \, =  & \  
\frac{1}{4}\, \sin^2 \theta\, 
\sin^2 \bigl(\delta_{\uparrow}+ \delta_{\downarrow})   
\, + 
  c_{T}^{\mathrm{CP}} \bigl(\pi T\bigr)^2 \,+ \,\cdots \,.
\label{eq:Transmission_CP_def}
\end{align}

At $T=0$, the Cooper-pair contribution is determined by 
the amplitude of the off-diagonal 
components of $\bm{G}_{dd}^{r}(\omega)$ at $\omega=0$,
whereas the diagonal components correspond to 
the single-electron contribution 
$\mathcal{T}_{\mathrm{ET}}^{}(T)$,   
\begin{align}
\!\!\!   
\mathcal{T}_{\mathrm{CP}}^{}(0)  =  & \ 
\frac{\Gamma_N^{2}}{2} 
\left[\  \Bigl | \bigl\{\bm{G}_{dd}^{r}(0)\bigl\}_{12}^{}\Bigr|^2 + \Bigl | \bigl\{\bm{G}_{dd}^{r}(0)\bigl\}_{21}^{}\Bigr|^2 \ \right] , 
\label{eq:T_CP_single}
\\
\!\!\!   
\mathcal{T}_{\mathrm{ET}}^{}(0) = & \ 
\frac{\Gamma_N^{2}}{2} 
\left[\ 
\Bigl | \bigl\{\bm{G}_{dd}^{r}(0)\bigl\}_{11}^{}\Bigr|^2 
+\Bigl | \bigl\{\bm{G}_{dd}^{r}(0)\bigl\}_{22}^{}\Bigr|^2
\ \right] .
\label{eq:T_ET_single}
\end{align}
These probabilities also satisfy the optical theorem \cite{Hashimoto2024},
which clarifies the relation between their contributions, 
\begin{align}
\mathcal{T}_{\mathrm{ET}}^{} (0) +
\mathcal{T}_{\mathrm{CP}}^{} (0) \,=\, 
\mathcal{T}_{\mathrm{BG}}^{} (0) \,, 
\end{align}
and are bounded as   
 $0 \leq \mathcal{T}_{\mathrm{BG}}^{}(0) \leq 1$ 
and $0 \leq \mathcal{T}_{\mathrm{CP}}^{}(0) \leq 1/4$. 

While the zero-temperature transport is determined by  
 the phase shifts $\delta_{\sigma}^{}$ 
and the Bogoliubov angle $\theta$, 
 the $T^2$ coefficients 
$c_{T}^{\mathrm{BG}}$ and 
$c_{T}^{\mathrm{CP}}$, appearing in Eqs.\ \eqref{eq:Transmission_BG_def} and 
\eqref{eq:Transmission_CP_def},   
additionally depend on the susceptibilities 
$\chi_{\sigma_1\sigma_2}^{}$
and the three-body correlation functions 
$\chi^{[3]}_{\sigma_1\sigma_2\sigma_3}$ 
of interacting Bogoliubov quasiparticles, 
\begin{align}
&c_{T}^{\mathrm{BG}}
\, \equiv \, 
\frac{\pi^2}{3} 
\,\frac{1}{2}\,
\sum_{\sigma=\uparrow,\downarrow}
\,\left[
\,-\,\cos 2 \delta_{\sigma}^{}
\Bigl(
 \, \chi_{\sigma\sigma}^{2}
+  2
\, \chi_{\uparrow\downarrow}^{2}\,
\Bigr)
\right.
\nonumber \\
& 
\qquad \qquad \qquad \qquad \   
\left.
\,+\, 
\frac{\sin 2 \delta_{\sigma}^{}}{2\pi} 
\left(\,
\chi_{\sigma\sigma\sigma}^{[3]}
  +  
\chi_{\sigma \overline{\sigma}\overline{\sigma}}^{[3]}
\,\right)
\right], 
\label{eq:cT_BG_def}
\\
&c_{T}^{\mathrm{CP}}
\, \equiv  \, 
- \frac{\pi^2}{3} \, 
\frac{1}{4}\,
\sin^2 \theta \, 
 \Biggl[\,
-\,\cos (2\delta_{\uparrow}^{}+2\delta_{\downarrow}^{})\,
\left( \, \chi_{\uparrow\uparrow}^{}
- \chi_{\downarrow\downarrow}^{} \right)^2
\nonumber 
\rule{0cm}{0.8cm}
\\
& 
\qquad 
+\,
\frac{\sin (2 \delta_{\uparrow}^{} + 2 \delta_{\downarrow}^{}) 
}{2\pi}
\left(\,
\chi_{\uparrow\uparrow\uparrow}^{[3]}\,  
  +   
\chi_{\uparrow\downarrow\downarrow}^{[3]}
+
\chi_{\downarrow\downarrow\downarrow}^{[3]}\,  
 +   
\chi_{\downarrow\uparrow\uparrow}^{[3]}
\,\right)
\,\Biggr] .
\label{eq:cT_CP_def}
\rule{0cm}{0.6cm}
\end{align}
Here, $\overline{\sigma}$ represents 
 an opposite spin component of $\sigma$, i.e., 
$\overline{\uparrow}=\downarrow$ and
$\overline{\downarrow}=\uparrow$.


The Bogoliubov-quasiparticle contribution $\mathcal{T}_{\mathrm{BG}}^{} (T)$ 
 is fully determined by the effective Hamiltonian 
$\mathcal{H}_\mathrm{eff}^{}$ in Eq.\ \eqref{eq:SingleDot_new}  
and is independent of the angle $\theta$.  
Its expression, given in Eqs.\ 
\eqref{eq:Transmission_BG_def} and \eqref{eq:cT_BG_def},
coincides with that for normal electrons tunneling through an N/QD/N junction 
\cite{MoraMocaVonDelftZarand,FMvDM2018,AO2017_I,AO2017_II,AO2017_III}.   
In contrast, the Cooper-pair contribution $\mathcal{T}_{\mathrm{CP}}^{}(T)$ 
depends on the rotation angle $\theta$ through the superconducting coherence factor  
$\sin^2 \theta$. Specifically,  both the leading term 
$\frac{1}{4}\, \sin^2 \theta\, 
\sin^2 \bigl(\delta_{\uparrow}+ \delta_{\downarrow})$  and  
the $T^2$ coefficient $c_{T}^{\mathrm{CP}}$ are proportional to $\sin^2 \theta$.
Therefore, the ratio $\mathcal{T}_{\mathrm{CP}}^{}(T)/\sin^2 \theta$ 
depends only on the radial coordinate in the parameter space shown  
in Fig.\ \ref{fig:SingleDotPhase}.   
Note that  
$\mathcal{T}_{\mathrm{BG}}^{} (T)$ 
and $\mathcal{T}_{\mathrm{CP}}^{} (T)$ 
depend on the tunneling couplings only through the total 
hybridization width 
$\Gamma_{N}^{}=\Gamma_{L}^{}+\Gamma_{R}^{}$.
Thus,  the $L$--$R$ tunnel asymmetry, 
$(\Gamma_{L}^{}-\Gamma_{R}^{})/\Gamma_{N}^{}$, 
influences the linear current $I_{j}^{}$ 
solely via the prefactors 
$\Gamma_{R}^{}\Gamma_{L}^{}/\Gamma_{N}^{2}$, 
$\Gamma_{R}^{2}/\Gamma_{N}^{2}$, and  
$\Gamma_{L}^{2}/\Gamma_{N}^{2}$ 
that appear explicitly in Eqs.\ \eqref{eq:gRL}--\eqref{eq:gLL}.

In the nonlocal conductance $g_{RL}^{}$,  
the Cooper-pair contribution  $\mathcal{T}_{\mathrm{CP}}^{}(T)$ 
enters with the opposite sign to its contribution to the local conductance $g_{LL}^{}$, 
since in the CAR process the electrons from the right (drain) normal lead 
 flow in opposite directions to form a Cooper pair. 
The current $I_{S}^{}$ flowing from the QD into the SC lead 
(see Fig.\ \ref{fig:system})  
is determined solely by Cooper-pair tunneling processes, 
\begin{align}
I_{S}^{} 
\equiv  & \ I_{L}^{} -  I_{R}^{}
 \ =  
\frac{4e^2}{h} \, 
4\mathcal{T}_{\mathrm{CP}}^{} (T) \,  V_\mathrm{SN}^{} , 
\label{eq:Js_general}
  \\
V_\mathrm{SN}^{}  \equiv  & \ 
\frac{\Gamma_{L}^{} V_L^{}+\Gamma_{R}^{} V_R^{}}{\Gamma_{L}^{}+\Gamma_{R}^{}} ,
\rule{0cm}{0.6cm}
\end{align}
where $V_\mathrm{SN}^{}$ can be regarded as 
a symmetrized bias voltage between the SC and the normal leads.  
The corresponding conductance,  
$g_\mathrm{SN}^{} \equiv dI_{S}^{}/dV_\mathrm{SN}^{}$, 
is equivalent to that investigated previously 
for an N/QD/SC junction \cite{YoichTanaka_2007},  
and reaches its maximum value of $4e^2 /h$ 
at $T=0$ when  $4\mathcal{T}_{\mathrm{CP}}^{}(0)= 1$.  
Similarly, the average of currents $I_{L}^{}$ and  $I_{R}^{}$ 
 flowing between the QD and the normal leads can be expressed as  
\begin{align}
\frac{I_{L}^{} + I_{R}^{}}{2} 
\,  = &  \ \,   
 \frac{2e^2}{h} \ \frac{4\Gamma_{R}^{}\Gamma_{L}^{}}
{(\Gamma_{R}^{}+\Gamma_{L}^{})^2}\, 
\mathcal{T}_{\mathrm{BG}}^{}(T)
\, \left( V_L^{} - V_R^{}\right)
\nonumber \\
&   
+ \, \frac{2e^2}{h} \, 
 \frac{\Gamma_{L}^{}-\Gamma_{R}^{}}
{\Gamma_{L}^{}+\Gamma_{R}^{}}\  
4 \mathcal{T}_{\mathrm{CP}}^{}(T)
\,
 V_\mathrm{SN}^{}
\rule{0cm}{0.6cm} \,. 
\end{align}
When the tunnel couplings are symmetric, i.e., $\Gamma_{L}^{}=\Gamma_{R}^{}$,
the Cooper-pair contribution from $\mathcal{T}_{\mathrm{CP}}^{}(T)$ 
to this average current vanishes, since the currents associated with Andreev scattering 
of electrons from the left and right normal leads cancel each other out.

\subsection{Conductance at zero magnetic field}

Here, we present simplified form of 
the conductance formulas applicable at zero magnetic field.   
In the absence of a magnetic field ($b=0$), 
time-reversal symmetry reduces 
the number of independent components of 
the correlation functions: 
$\delta_{\uparrow}^{}=\delta_{\downarrow}^{}$, 
$\chi_{\uparrow\uparrow}^{}=\chi_{\downarrow\downarrow}^{}$, 
$\chi_{\uparrow\uparrow\uparrow}^{[3]}
=\chi_{\downarrow\downarrow\downarrow}^{[3]}$, and
$\chi_{\uparrow\downarrow\downarrow}^{[3]}
=\chi_{\downarrow\uparrow\uparrow}^{[3]}$. 
Among these,  the two independent linear susceptibilities, 
$\chi_{\uparrow\uparrow}^{}$ and 
$\chi_{\uparrow\downarrow}^{}$, 
can be expressed in terms of the Fermi-liquid energy scale  
$T^*$ and the Wilson ratio $R$, defined 
in Eqs.\ \eqref{eq:T*_def} and \eqref{eq:Wilson_ratio_def}, respectively.
Therefore, the transport probabilities 
$\mathcal{T}_{\mathrm{BG}}^{} (T)$ and  
$\mathcal{T}_{\mathrm{CP}}^{} (T)$ simplify 
 and take the following forms: 
\begin{align}
\mathcal{T}_{\mathrm{BG}}^{} (T)  
\, \xrightarrow{\,b\to 0\,}&  \  
 \sin^2 \delta
\,-\,C_{T}^{\mathrm{BG}} 
\, \left(\frac{\pi T}{T^*}\right)^2 
\,+\,\cdots \,,  
\label{eq:Transmission_BG_def_b=0} 
\\
\mathcal{T}_{\mathrm{CP}}^{} (T) 
\, \xrightarrow{\,b\to 0\,}&  \  
\frac{1}{4}\, \sin^2 \theta\, 
\sin^2 2 \delta   
\, +\,   C_{T}^{\mathrm{CP}} 
\, \left(\frac{\pi T}{T^*}\right)^2 
 \,+ \,\cdots \,.
\label{eq:Transmission_CP_def_b=0} 
\end{align}
Here, the dimensionless coefficients 
$C_{T}^{\mathrm{BG}}$ and $C_{T}^{\mathrm{CP}}$ 
of the $T^2$ terms are defined by 
\begin{align}
C_{T}^{\mathrm{BG}}
= & \ 
\frac{\pi^2}{48} 
\,\left[\,
-\cos 2 \delta\, \Bigl\{ 1 +  2 (R-1)^{2} \Bigr\} 
+ \,\Theta_{T}^{\mathrm{BG}}
\,\right], 
\label{eq:cT_BG_def_b=0}
\\
\Theta_{T}^{\mathrm{BG}}  \equiv & \  \,  
\frac{\sin 2 \delta }{2\pi \chi_{\uparrow\uparrow}^{2}}
\left(
\chi_{\uparrow\uparrow\uparrow}^{[3]} 
  +   
\chi_{\uparrow\downarrow\downarrow}^{[3]}
\right) , 
\label{eq:Theta_BG_def_b=0}
\end{align}
and 
\begin{align}
& 
\!\!\!\!\!\!\!\!\!\!\!\!\!\!\!\!
\!\!\!\!\!\!\!\!\!\!\!\!\!\!\!\!
\!\!\!\!\!\!\!\!\!\!\!\!\!\!\!\!
\!\!\!\!
C_{T}^{\mathrm{CP}}
=\, 
\frac{\pi^2}{48} \, 
\Theta_{T}^{\mathrm{CP}}
\, \sin^2 \theta \, ,  
\label{eq:cT_CP_def_b=0}
\\
& 
\!\!\!\!\!\!\!\!\!\!\!\!\!\!\!\!
\!\!\!\!\!\!\!\!\!\!\!\!\!\!\!\!
\!\!\!\!\!\!\!\!\!\!\!\!\!\!\!\!
\!\!\!\!
\Theta_{T}^{\mathrm{CP}} \equiv \,   
- \,\frac{\sin 4 \delta}{4\pi \chi_{\uparrow\uparrow}^{2}}
\left(
\chi_{\uparrow\uparrow\uparrow}^{[3]}
  +   
\chi_{\uparrow\downarrow\downarrow}^{[3]}
\right) \,.
\label{eq:Theta_CP_def_b=0}
\end{align}
Note that, at zero magnetic field,  the Cooper-pair contribution 
$C_{T}^{\mathrm{CP}}$ is determined solely by the dimensionless 
three-body correlation, 
 $\Theta_{T}^{\mathrm{CP}}$, which is proportional to the sum 
$\chi_{\uparrow\uparrow\uparrow}^{[3]}+ 
\chi_{\uparrow\downarrow\downarrow}^{[3]}$.

\subsubsection{Three-body correlations for $H_d^{U}=0$}

\label{subsubsec:CAR_U=0}

In the noninteracting case,  $H_d^{U}=0$, 
all off-diagonal components of the linear susceptibilities and three-body 
correlations vanish. 
However, 
the diagonal components, $\chi_{\sigma\sigma}$ and 
 $\chi_{\sigma\sigma\sigma}^{[3]}$, remain finite,  
reflecting the Pauli exclusion principle acting on Bogoliubov quasiparticles 
with the same spin $\sigma$. 
These diagonal components can be derived from derivatives of the 
noninteracting phase shift  
$\delta_{0}^{}=\cot^{-1} (E_{A}^{}/\Gamma_N^{})$ 
with respect to $E_{A}^{}$,
\begin{align}
\chi_{\uparrow\uparrow}^{0} 
\,\equiv & \ 
\frac{\sin^2 \delta_{0}^{} }{\pi\Gamma_N^{}}
\ = \,  \frac{1}{\pi}\frac{\Gamma_N^{}}{E_{A}^{2}+\Gamma_N^{2}}\,,  
\\
\Theta_{T}^{\mathrm{BG}}
\, \xrightarrow{\,H_d^{U}\to 0\,}& \ 
\frac{\sin 2\delta_{0}^{}}{2\pi \left\{ \chi_{\uparrow\uparrow}^0 \right\}^2} 
 \frac{\partial \chi_{\uparrow\uparrow}^0}{\partial E_{A}^{}}
\ =\,
\frac{-2E_{A}^{2}}
{E_A^{2}+\Gamma_N^{2}}\,, 
\\
\Theta_{T}^{\mathrm{CP}}  
\, \xrightarrow{\,H_d^{U}\to 0\,} & \  
- \frac{\sin 4\delta_{0}^{}}{4\pi \left\{ \chi_{\uparrow\uparrow}^0 \right\}^2} 
 \frac{\partial \chi_{\uparrow\uparrow}^0}{\partial E_{A}^{}}
\ =\, 
 \frac{2 E_{A}^{2}
\,\Bigl(E_{A}^{2} - \Gamma_N^2 \Bigr) }
{\Big(E_{A}^{2} + \Gamma_N^2 \Bigr)^2}  \,.
\end{align}

\begin{figure}[t]
\leavevmode 
\begin{minipage}[t]{1\linewidth}
 \includegraphics[width=0.75\linewidth]{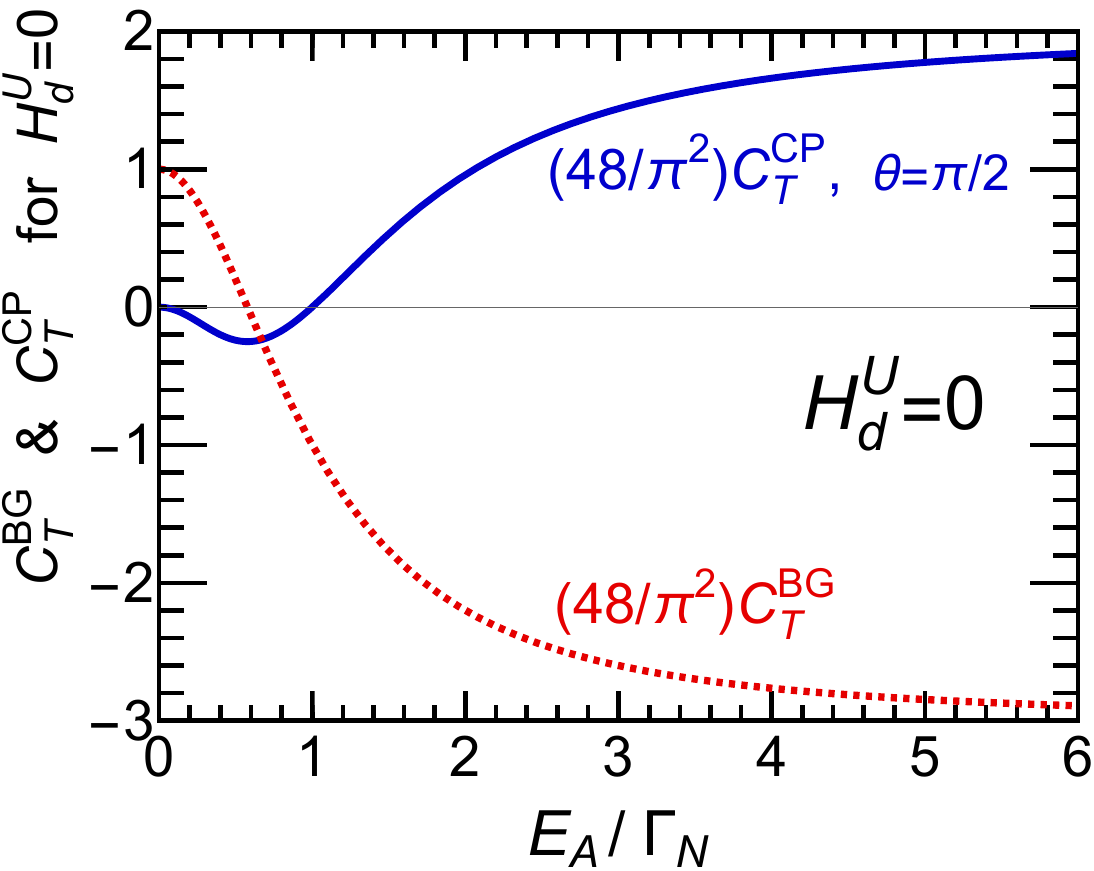}
\rule{0.006\linewidth}{0cm}
\caption{
Coefficients $C_{T}^{\mathrm{BG}}$ and $C_{T}^{\mathrm{CP}}$ 
for noninteracting electrons ($H_d^{U}=0$),  
 plotted as functions of  $E_{A}^{}/\Gamma_{N}^{}$ 
for the Bogoliubov angle $\theta=\pi/2$.}
\label{fig:Cooper_T2_U0}
\end{minipage}
\end{figure}

Figure \ref{fig:Cooper_T2_U0} demonstrates 
 the behavior of 
 $C_{T}^{\mathrm{BG}}$ and $C_{T}^{\mathrm{CP}}$ 
for $\theta=\pi/2$ 
in the noninteracting case $\mathcal{H}_{d}^{U}=0$, 
for which the Fermi-liquid energy scale and the Wilson ratio 
are given by 
 $T_0^{*} = 1/(4\chi_{\uparrow\uparrow}^{0})$ 
and $R-1 \to 0$, respectively. 
The Cooper-pair contribution $C_{T}^{\mathrm{CP}}$, 
which is solely determined by the three-body correlation 
$\Theta_{T}^{\mathrm{CP}}$,  
 takes negative values in the range $0<E_{A}^{}<\Gamma_{N}^{}$, 
as the phase shift  $\delta_{0}^{}$ varies from ${\pi}/{2}$ to ${\pi}/{4}$, 
reflecting the sign of $\sin 4\delta_{0}^{}$. 
In contrast, the Bogoliubov-quasiparticle contribution 
$C_{T}^{\mathrm{BG}}$ decreases monotonically as $E_{A}^{}$ increases.

\subsubsection{NRG approach to interacting Bogoliubov quasiparticles}

In order to clarify how the Coulomb interaction affects the coefficients
$C_{T}^{\mathrm{CP}}$ and $C_{T}^{\mathrm{BG}}$,  
we calculate the correlation functions of the Bogoliubov quasiparticles,  
$\delta_{\sigma}^{}$,   $\chi_{\sigma_1\sigma_2}^{}$, 
and $\chi^{[3]}_{\sigma_1\sigma_2\sigma_3}$ 
as functions of $E_A^{}/U$ and $U/(\pi \Gamma_{N}^{})$,  
by applying the NRG approach \cite{KWW1,KWW2,HewsonOguriMeyer} 
to the effective Hamiltonian $\mathcal{H}_\mathrm{eff}^{}$. 
We choose the discretization parameter as  
$\Lambda=2.0$, $\Gamma_{N}/D=1/(100\pi)$, 
and typically retain 4000 low-lying excited states. 
Note that the dependence of the conductance on the Bogoliubov angle  $\theta$ 
is entirely determined by the superconducting coherence factor   
 $\sin^2 \theta$, as discussed above.

\begin{figure}[t]
\hspace{0.015\linewidth}
\includegraphics[width=0.453\linewidth]{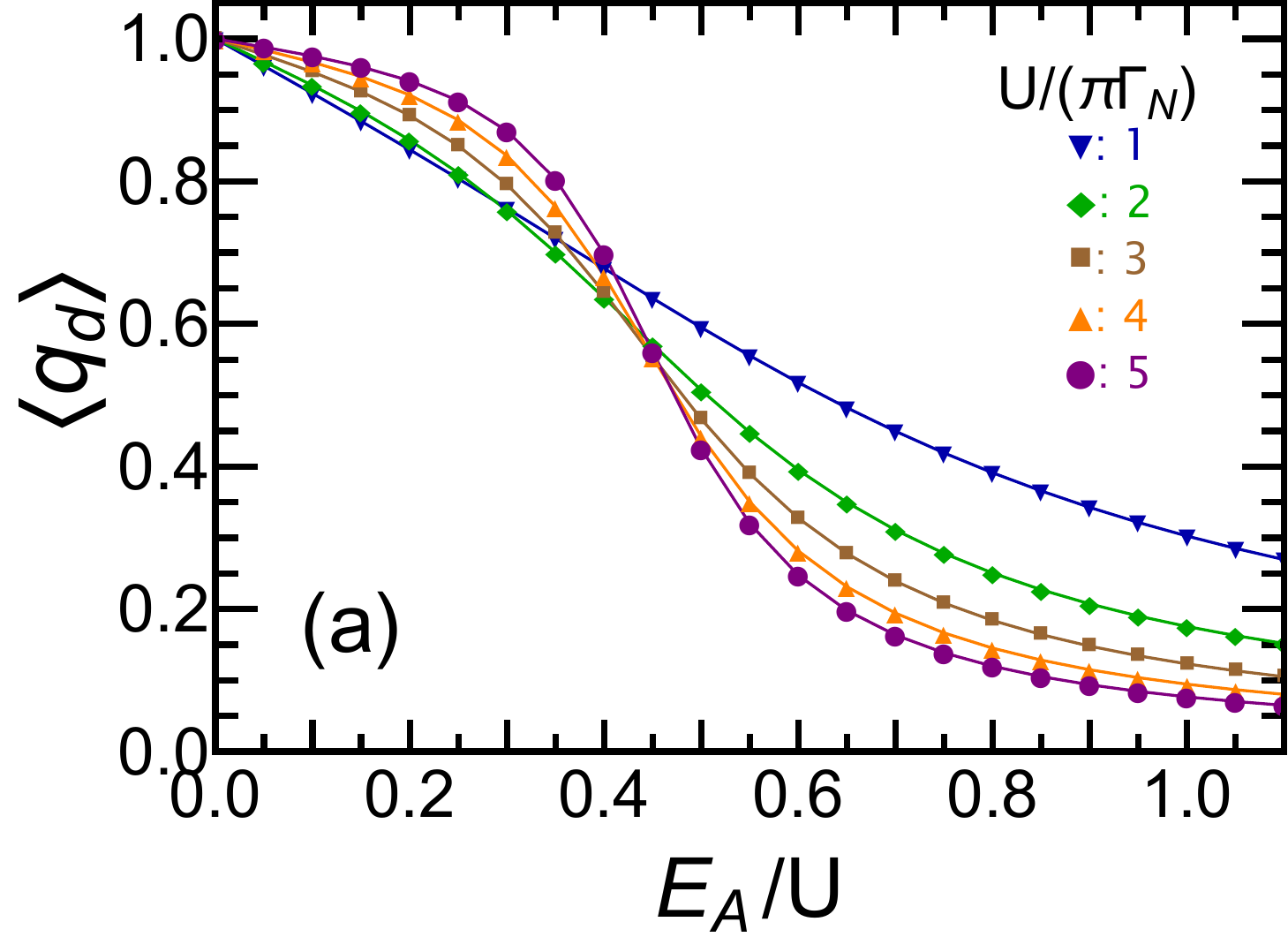}
\hspace{0.015\linewidth}
\includegraphics[width=0.465\linewidth]{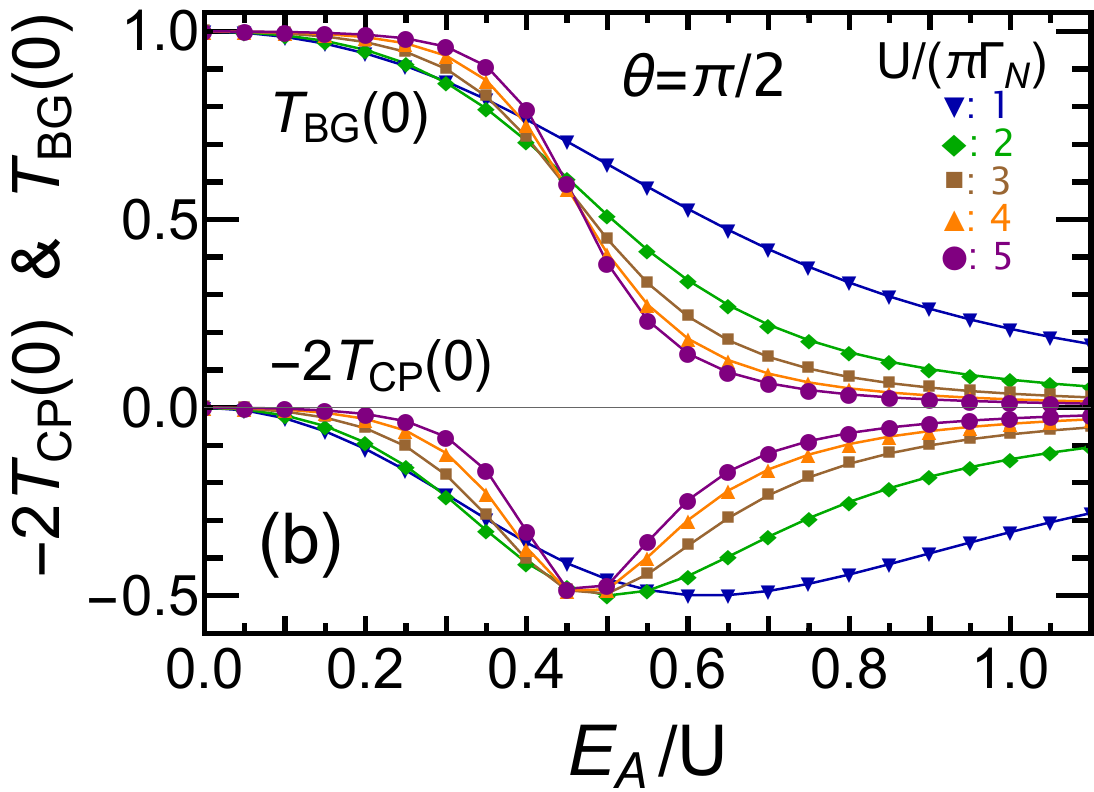}
 \\
\includegraphics[width=0.48\linewidth]{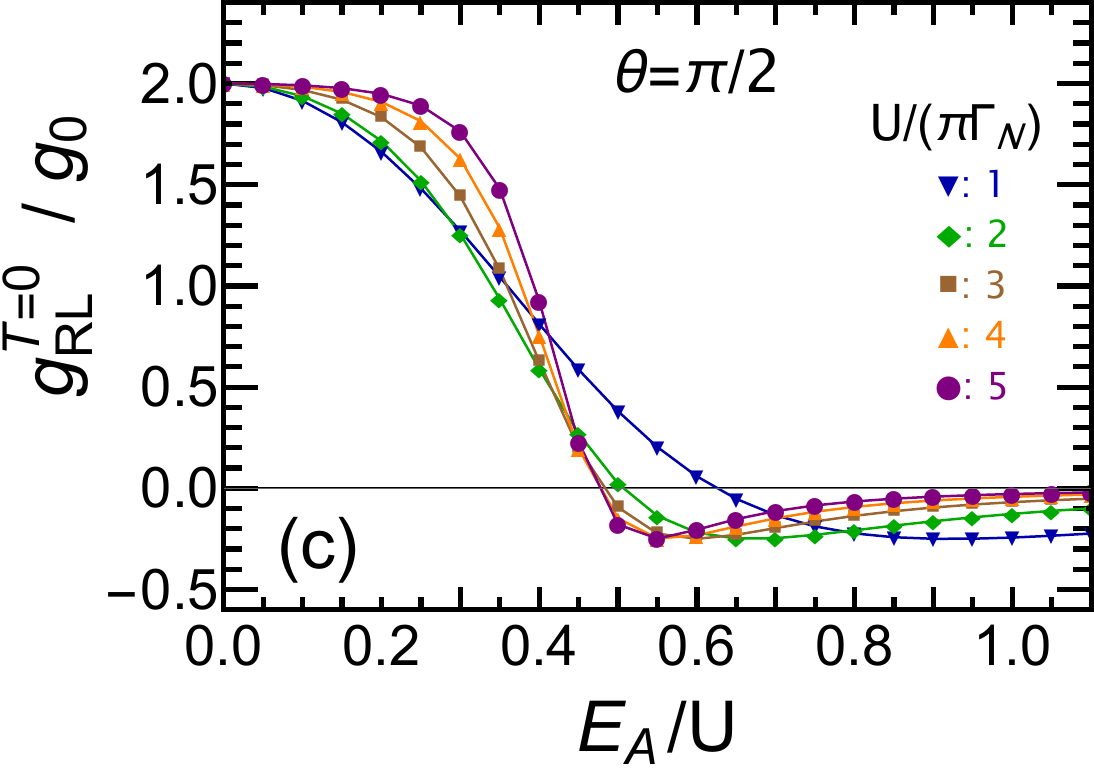}
\rule{0cm}{0.02\linewidth}
\includegraphics[width=0.48\linewidth]{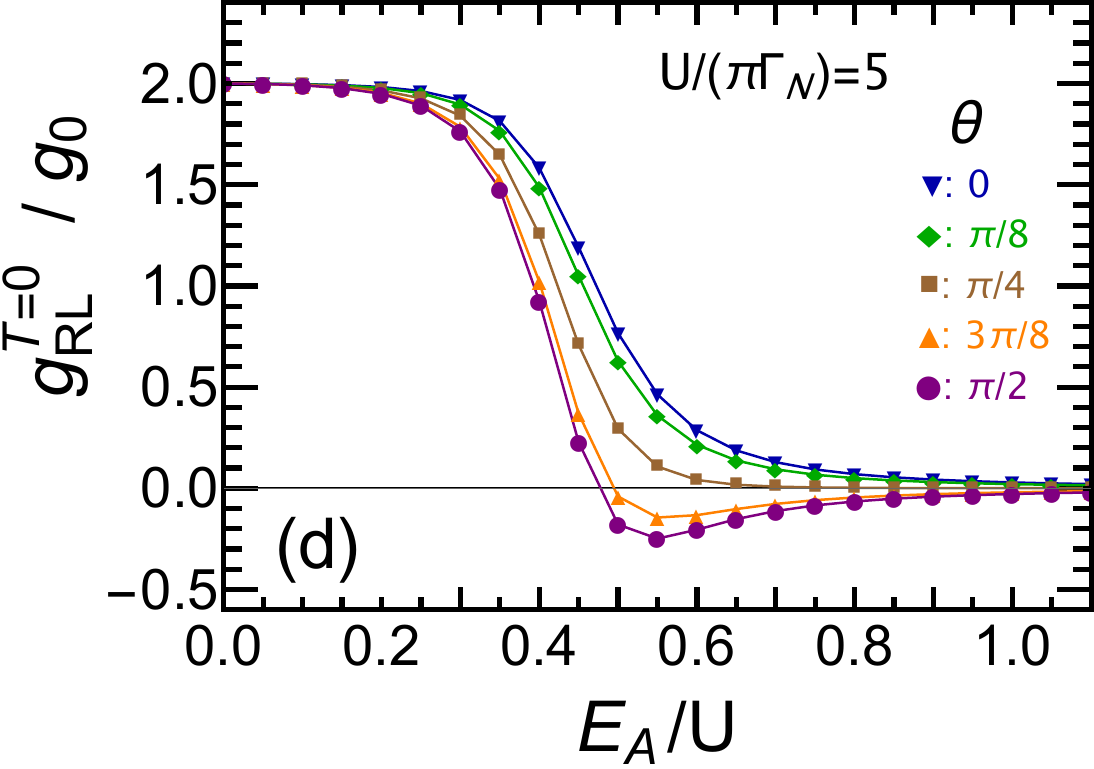}
 \\
\includegraphics[width=0.48\linewidth]{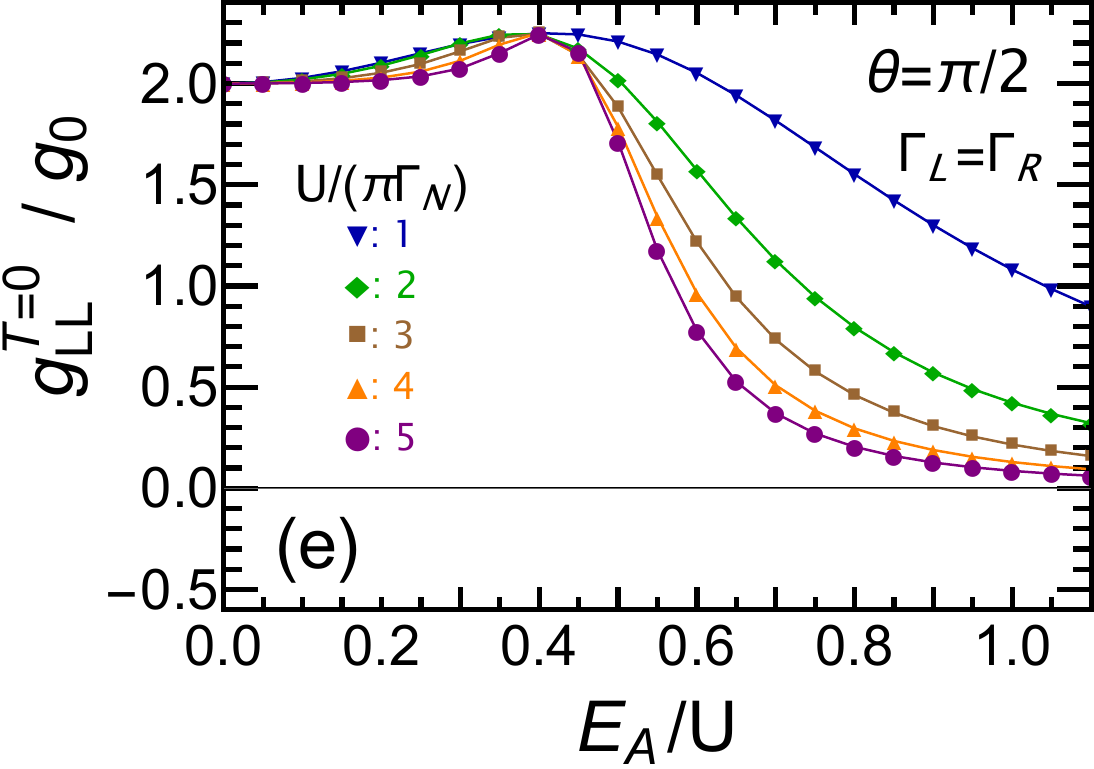}
\rule{0cm}{0.02\linewidth}
\includegraphics[width=0.48\linewidth]{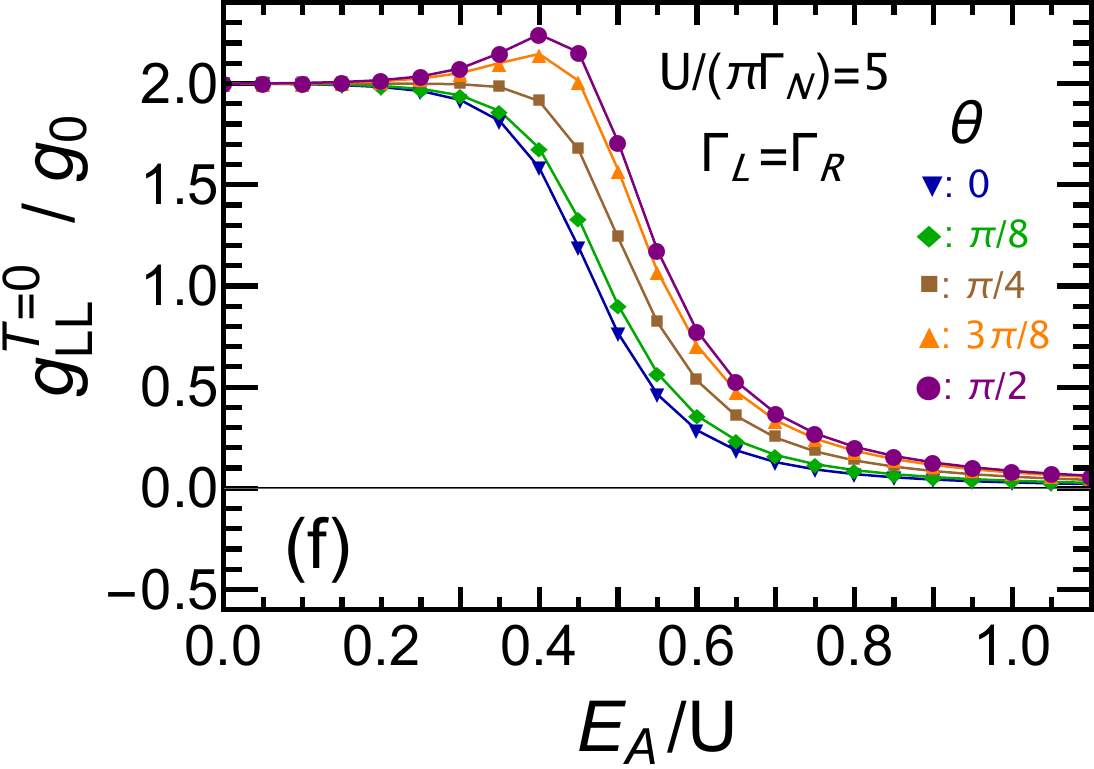}
\caption{Phase shift and transmission probabilities at $T=0$ plotted as functions 
of $E_A^{}/U$.
(a) Occupation number of Bogoliubov quasiparticles, $\langle q_{d}^{} \rangle 
=2\delta/\pi$.
 (b) Transmission probability of Bogoliubov quasiparticles, 
$\mathcal{T}_{\mathrm{BG}}(0)=\sin^2 \delta$ 
and the Cooper-pair contributions, 
$-2\mathcal{T}_{\mathrm{CP}}^{}(0)$ 
with $\mathcal{T}_{\mathrm{CP}}(0)=(1/4) \sin \theta\, \sin^2 2\delta$.  
 (c)(d) Nonlocal conductance, 
$
g_{RL}^{T=0} / (2 g_0^{})
= \mathcal{T}_{\mathrm{BG}}^{}(0) 
- 2\mathcal{T}_{\mathrm{CP}}^{}(0)
$. 
 (e)(f)  Local conductance, 
$
g_{LL}^{T=0} / (2 g_0^{})
= \mathcal{T}_{\mathrm{BG}}^{}(0) 
+ 2(\Gamma_L^{}/\Gamma_R^{}) \mathcal{T}_{\mathrm{CP}}^{}(0)
$ for  $\Gamma_R^{}=\Gamma_L^{}$.
Here, 
$g_0^{} = \frac{e^2}{h}{4\Gamma_R^{}\Gamma_L^{}}/{\Gamma_N^{2}}$.
The results plotted in (a)--(c), and (e) are obtained 
 for $U/(\pi \Gamma_{N})=1.0$, $2.0$, $3.0$, $4.0$, and $5.0$. 
Specifically, the Bogoliubov angle is set to $\theta = \pi/2$  
for $-2\mathcal{T}_{\mathrm{CP}}^{}(0)$,  $g_{RL}^{T=0}$, 
and $g_{LL}^{T=0}$.
For comparison,
 (d) and (f) show the results for 
$g_{RL}^{T=0}$ and $g_{LL}^{T=0}$
at $\theta =0$, $\pi/8$, $\pi/4$, $3\pi/8$, and  $\pi/2$,   
 with $U/(\pi \Gamma_{N})=5.0$. 
}
\label{fig:NRGOneDot_FLparametersxi0}
\end{figure}

\section{Behavior of Andreev transport at finite temperatures}
\label{CARZeroMag}

In this section, we demonstrate how the three-body correlations 
between the Bogoliubov quasiparticles contribute to 
the $T^2$ term of the Andreev transport at zero magnetic field, $b=0$.   
Note that the eigenstates and eigenvalues of the effective Hamiltonian 
 $\mathcal{H}_\mathrm{eff}^{}$, defined in Eq.\ \eqref{eq:SingleDot_new},  
evolve as $E_A^{}$ varies 
along the radial direction in the $\xi_d^{}$--$\Gamma_S^{}$ plane shown 
in Fig.\ \ref{fig:SingleDotPhase},
 while they do not depend on the angular coordinate $\theta$.

\subsection{Phase shift and ground-state properties}

We first of all discuss the behavior 
of the phase shift and the nonlocal conductance at zero temperature, $T=0$, 
to briefly demonstrate how the leading-order terms vary across the 
parameter space shown in Fig.\  \ref{fig:SingleDotPhase}  \cite{Hashimoto2024}.

The occupation number of Bogoliubov quasiparticles at the impurity site,  
 $\langle q_{d}^{} \rangle = 2\delta/\pi$,   
varies only along the radial direction in the $\xi_d^{}$--$\Gamma_{S}^{}$ plane 
 and is independent of the polar angle $\theta$.
In Fig.\ \ref{fig:NRGOneDot_FLparametersxi0}(a),    
the NRG results for $\langle q_{d}^{} \rangle$ are plotted  
as a function of $E_A^{}$ 
for $U/(\pi \Gamma_{N}) = 1.0$, $2.0$, $3.0$, $4.0$, and $5.0$.     
The occupation number $\left \langle q_{d\sigma}^{} \right\rangle$ 
decreases with increasing $E_A^{}$, 
particularly around $E_A^{} \simeq U/2$,  
where a crossover between the Kondo regime 
and the valence-fluctuation regime of Bogoliubov quasiparticles occurs 
 for strong interactions, $U/(\pi \Gamma_N^{}) \gtrsim 2.0$.

The local superconducting pair correlation 
$\bigl \langle d^\dagger_{\uparrow} d^\dagger_{\downarrow}
 +d_{\downarrow}^{} d_{\uparrow}^{} \bigr \rangle$ 
 can be deduced from $\langle q_{d}^{} \rangle$ 
via Eq.\ \eqref{eq:SingleDotK}.   
In particular, its magnitude is maximized at $\theta =\pi/2$ and is given by
 $|\bigl\langle d^\dagger_\uparrow d^\dagger_\downarrow 
+  d^{}_\downarrow d^{}_\uparrow \bigr\rangle| 
=1- \left \langle q_{d}^{}\right\rangle$. 
Therefore, as $E_A^{}$ increases, 
the magnitude of the superconducting correlation 
grows significantly around $E_A^{} \simeq U/2$, 
and then saturates at its upper bound, $1.0$, reflecting the behavior of 
$\langle q_{d}^{}\rangle$.
Note that at  $\theta =\pi/2$, corresponding to  $E_A^{} = \Gamma_S^{}$,    
the occupation number of impurity electrons is fixed 
at  $\langle n_d^{}\rangle =1$ 
as a direct consequence of the electron-hole symmetry  ($\xi_d^{}=0$) 
of the original Hamiltonian $H$.

The ground-state value of the transmission probability of Bogoliubov quasiparticles, 
$\mathcal{T}_{\mathrm{BG}}^{}(0) =\sin^2 \delta$,   
 is plotted as a function of $E_A^{}$ in Fig.\ \ref{fig:NRGOneDot_FLparametersxi0}(b). 
This leading-order term exhibits a clear Kondo ridge at $E_A^{} \lesssim U/2$ 
for large $U$, and then decreases for $E_A^{} \gtrsim U/2$, 
where $\langle q_{d}^{} \rangle$ rapidly decreases.
The corresponding leading-order term of the Cooper-pair transmission probability, 
 $\mathcal{T}_{\mathrm{CP}}^{}(0) 
= \frac{1}{4}\sin^2 \theta\, \sin^2 2\delta$, 
reaches its maximum value at $\delta ={\pi}/{4}$, 
where the local Bogoliubov quasiparticle state is at quarter filling
 $\left \langle q_{d\sigma}^{} \right\rangle ={1}/{4}$.
This Cooper-pair contribution is also plotted in 
 Fig.\ \ref{fig:NRGOneDot_FLparametersxi0}(b)  for $\theta=\pi/2$, 
multiplied by a factor of $-2$, which arises 
in the nonlocal conductance as 
\begin{align}
 g_{RL}^{T=0}
 \,  = & \  
2 g_0^{} \Bigl[\,
\mathcal{T}_{\mathrm{BG}}^{}(0)-2\mathcal{T}_{\mathrm{CP}}^{}(0)
\,\Bigr]\,,
\nonumber  
\\
   = & \  
2 g_0^{} 
\sin^2 \delta 
\,  \Bigl( 1
-  2 \sin^2 \theta\,\cos^2  \delta \Bigr)  ,  
\label{eq:gRL_b0_T0}
 \end{align}
where $g_0^{} = 
 \frac{e^2}{h}\, {4\Gamma_R^{}\Gamma_L^{}}/{\Gamma_N^{2}}$. 
The negative sign of this factor reflects the fact that  
crossed Andreev reflection induces a counterflow, 
with current flowing from the right lead toward the QD.  
The NRG results demonstrate that 
 a dip with a depth of $-2\mathcal{T}_{\mathrm{CP}}^{}(0) = -{1}/{2}$
appears in the crossover region $E_A^{} \simeq U/2$,   
 and that its width becomes of the order of $\Gamma_N^{}$.

Equation \eqref{eq:gRL_b0_T0} indicates that the 
nonlocal conductance becomes negative, $g_{RL}^{T=0} <0$,  
in the region in which Cooper-pair tunneling dominates, i.e., 
 $\pi/4<\theta< 3\pi/4$ in the angular direction, and 
$\cos^2 \delta >1/(2 \sin^2 \theta)$ in the radial direction.
The NRG results shown in Figs.\  
\ref{fig:NRGOneDot_FLparametersxi0}(c) and  
\ref{fig:NRGOneDot_FLparametersxi0}(d) clearly demonstrate  
that the nonlocal conductance decreases 
from its unitary-limit value,  $g_{RL}^{T=0}=2g_0^{}$, as $E_A^{}$ increases,  
 and becomes negative in the region in which the above conditions are satisfied.
In particular, for $\theta = {\pi}/{2}$,
 $g_{RL}^{T=0}$ vanishes at the point where 
the phase shift reaches $\delta = {\pi}/{4}$ 
and exhibits a minimum value of $g_{RL}^{T=0}/ g_0^{} = -1/4$  
at $\delta ={\pi}/{6}$.

For comparison, the local conductance $g_{LL}^{}$ at $T=0$,
\begin{align}
 g_{LL}^{T=0}
 \,  = & \  
2 g_0^{} \left[\,
\mathcal{T}_{\mathrm{BG}}^{}(0)+\frac{2\Gamma_L^{}}{\Gamma_R^{}} 
\,\mathcal{T}_{\mathrm{CP}}^{}(0)
\,\right]\,,
\nonumber \\
= & \ 
2 g_0^{} \sin^2 \delta 
\,  \left( 1+  
\frac{2\Gamma_L^{}}{\Gamma_R^{}}
\, \sin^2 \theta\,\cos^2  \delta \right)  ,  
\label{eq:gLL_b0_T0}
 \end{align}
is shown for $\Gamma_L^{} = \Gamma_R^{}$ 
 in Figs.\  \ref{fig:NRGOneDot_FLparametersxi0}(e) and  
\ref{fig:NRGOneDot_FLparametersxi0}(f).  
The Bogoliubov-quasiparticle and Cooper-pair contributions 
 add constructively to the local conductance $g_{LL}^{}$, 
whereas they interfere destructively 
in the nonlocal conductance $g_{RL}^{}$. 
In particular,  $g_{LL}^{T=0}$  is enhanced by the Cooper-pair contribution 
and exhibits a peak because of direct Andreev reflection   
near the crossover region, $E_A^{} \simeq U/2$. 
More specifically, for $\theta = {\pi}/{2}$, 
it reaches a maximum value of $g_{LL}^{T=0}/ g_0^{} = 9/4$  
at $\delta ={\pi}/{3}$.
Experimentally, the contributions 
from $\mathcal{T}_{\mathrm{BG}}^{}$ 
and $\mathcal{T}_{\mathrm{CP}}^{}$ 
can be extracted separately by measuring 
 both the local and the nonlocal conductances, 
$g_{LL}^{}$ and $g_{RL}^{}$.

\begin{figure}[t]
\includegraphics[width=0.48\linewidth]{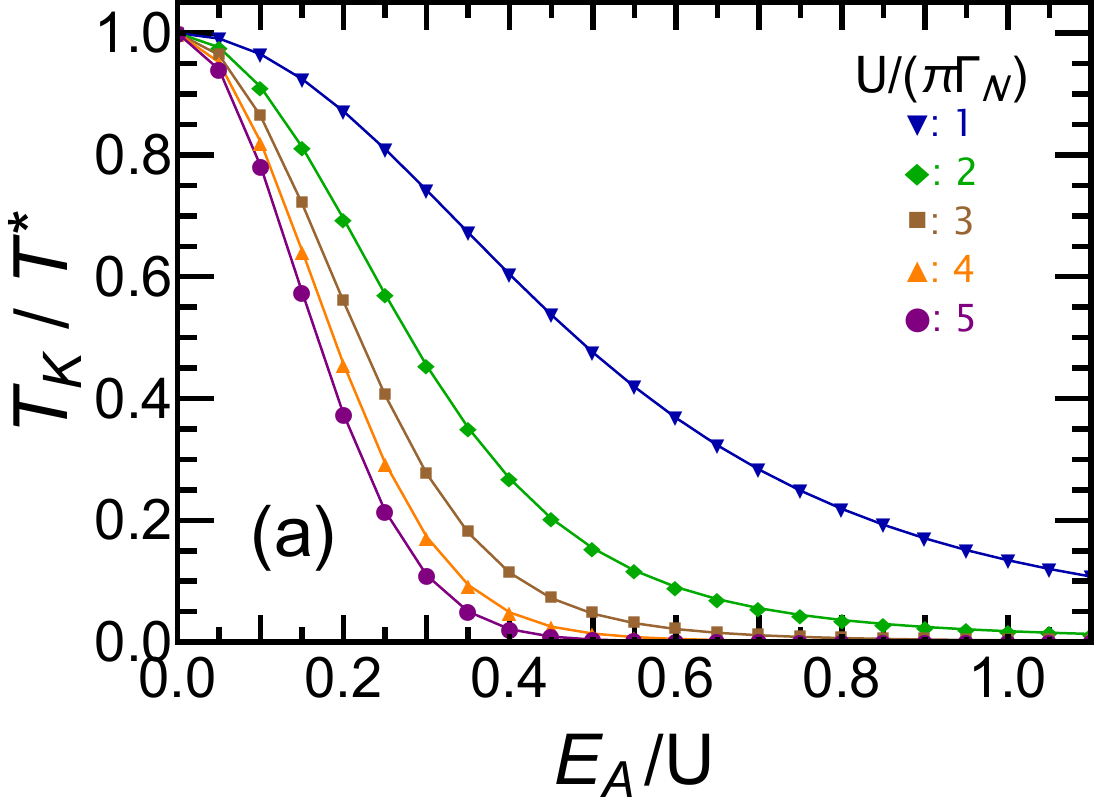}
\rule{0cm}{0.02\linewidth}
\includegraphics[width=0.48\linewidth]{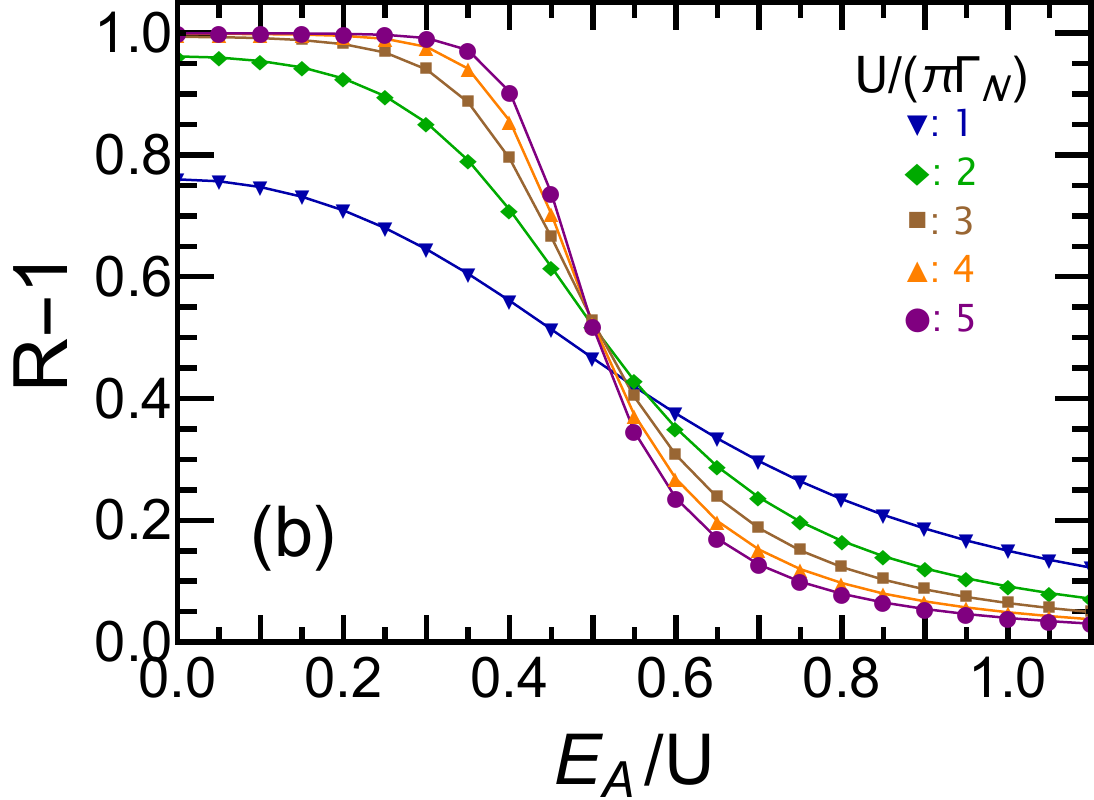}
\caption{
Inverse of the Fermi-liquid energy scale  $1/T^{*}$ and the Wilson ratio $R-1$, 
defined in Eqs.\ \eqref{eq:T*_def} and \eqref{eq:Wilson_ratio_def},
 plotted as functions of $E_A/U$ 
    for  $U/(\pi \Gamma_{N})= 1.0$, $2.0$, $3.0$, $4.0$, and $5.0$. 
Here, $T_{K}^{} \equiv T^{*}\big|_{E_A^{}\to 0}^{}$ corresponds to the 
Kondo temperature, which takes the values $T_{K}^{}/\Gamma_{N}^{} 
=0.494$, $0.188$, $0.063$, $0.020$, and $0.006$, 
respectively, for these five values of $U$.  
     }
\label{fig:NRGOneDotXi0}
\end{figure}

\subsection{Two-body correlations:  linear susceptibilities}
\label{subsec:2body_correction_NRG}

The Fermi-liquid energy scale and the Wilson ratio, defined by 
$T^*=1/(4\chi_{\uparrow\uparrow}^{})$ and   
$R-1=-\chi_{\uparrow\downarrow}^{}/\chi_{\uparrow\uparrow}^{}$, 
are quantities determined by the two-body correlation function 
$\chi_{\sigma\sigma'}^{}$. 
The NRG results for these quantities are shown in Fig.\ \ref{fig:NRGOneDotXi0}  
    for  $U/(\pi \Gamma_{N})= 1.0$, $2.0$, $3.0$, $4.0$, and  $5.0$.
The Kondo temperature, defined at $E_A^{}=0$ as  
 $T_{K}^{} \equiv T^{*}\big|_{E_A^{}\to 0}^{}$,  
 takes the values $T_{K}^{}/\Gamma_{N}^{} 
=0.494$, $0.188$, $0.063$, $0.020$, and $0.006$, 
respectively, 
which are much smaller than $\Gamma_N^{}$ for large $U$.

 Figure \ref{fig:NRGOneDotXi0}(a)  
shows the $E_A^{}$ dependence of $1/T^*$,  normalized by $T_{K}^{}$. 
The Fermi-liquid energy scale  $T^*$ 
increases rapidly as $E_A^{}$ moves away from the Fermi level ($E_F^{}=0$).
The Wilson ratio, plotted in Fig.\ \ref{fig:NRGOneDotXi0}(b), 
  exhibits a broad ridge, the height of which reaches 
the strong-coupling value 
$R-1 \simeq 1$ for $U \gtrsim 3.0 \pi \Delta $, 
over a wide Kondo-dominated region   $0\leq E_A^{} \lesssim U/2$ 
 inside the semicircle illustrated in Fig.\ \ref{fig:SingleDotPhase}. 
In contrast, for $E_A^{} \gtrsim U/2$, i.e.,  
 in the valence-fluctuation or empty-orbital regime of Bogoliubov quasiparticles,  
electron correlations become less important and the Wilson ratio 
approaches $R-1 \simeq 0$.

\begin{figure}
\leavevmode 
\begin{minipage}[b]{1\linewidth}
\hspace{0.2cm}
\includegraphics[width=0.75\linewidth]{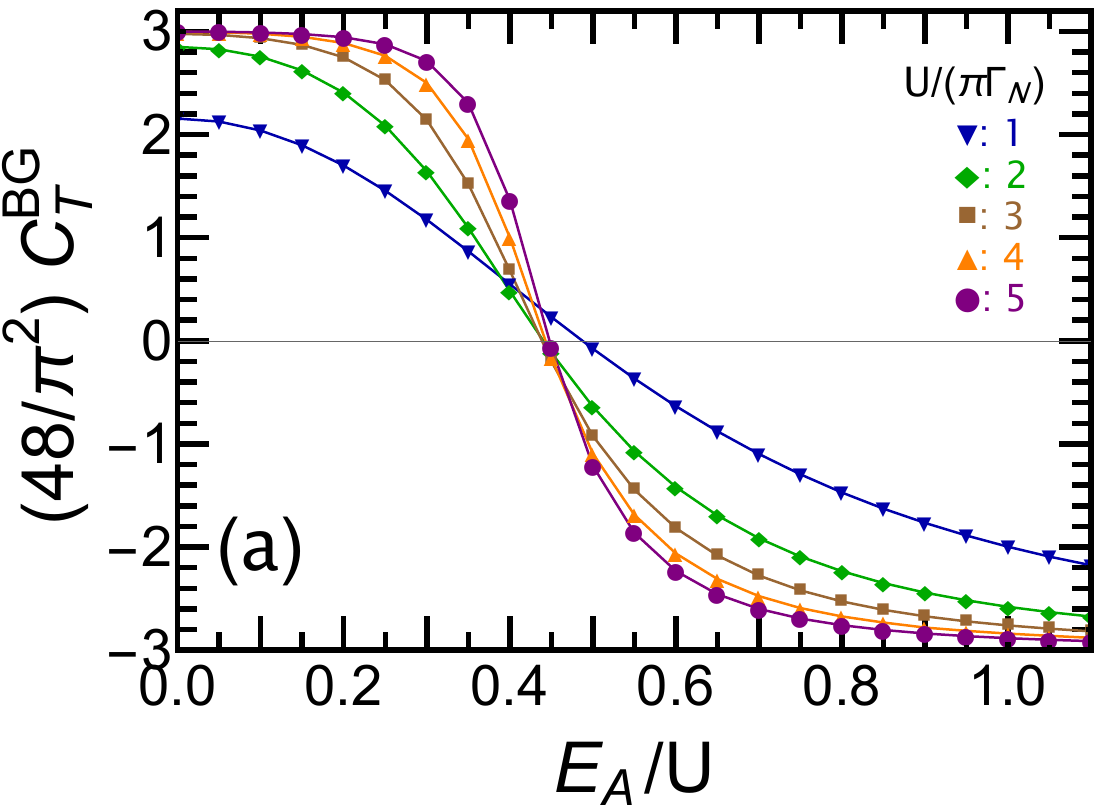}
\includegraphics[width=0.79\linewidth]{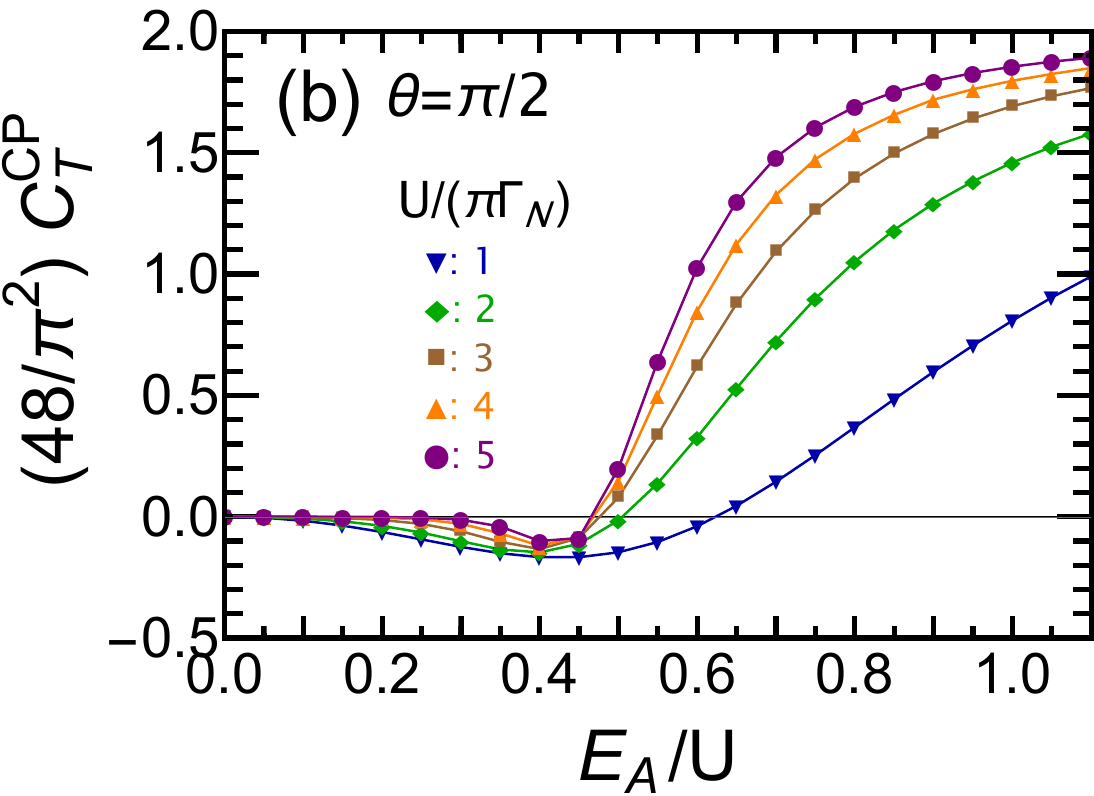}
\includegraphics[width=0.79\linewidth]{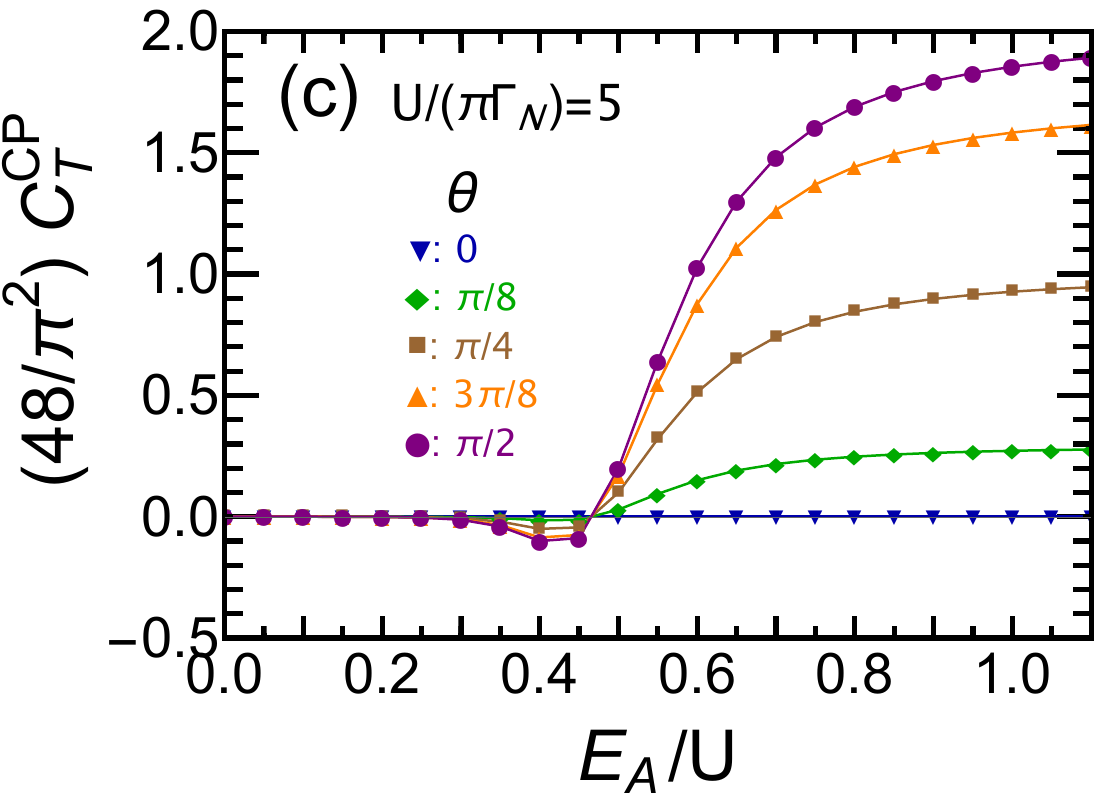}
\caption{
Coefficients of the $T^2$ term in the conductance, 
defined in Eqs.\ 
\eqref{eq:Transmission_BG_def_b=0}--\eqref{eq:Theta_CP_def_b=0}, 
are plotted as functions of $E_{A}^{}/U$ 
 for  $U/(\pi \Gamma_{N})= 1.0$, $2.0$, $3.0$, $4.0$, and $5.0$, 
with the Bogoliubov angle fixed at $\theta = \pi/2$. 
Panels (a) and (b) show $C_{T}^{\mathrm{BG}}$ and $C_{T}^{\mathrm{CP}}$,
which represent the contributions from Bogoliubov-quasiparticle tunneling, 
$\mathcal{T}_{\mathrm{BG}}^{}(T)$,
and Cooper-pair tunneling, $\mathcal{T}_{\mathrm{CP}}^{}(T)$, respectively. 
Panel (c) demonstrates the $\sin^2 \theta$ 
dependence of $C_{T}^{\mathrm{CP}}$, 
for  $\theta =0$, $\pi/8$, $\pi/4$, $3\pi/8$, and $\pi/2$,  
with $U/(\pi \Gamma_{N})=  5.0$. 
} 
\label{fig:3body_Andreev_partA}
\end{minipage}
\end{figure}

\begin{figure}
\leavevmode 
\begin{minipage}[b]{1\linewidth}
 \includegraphics[width=0.75\linewidth]{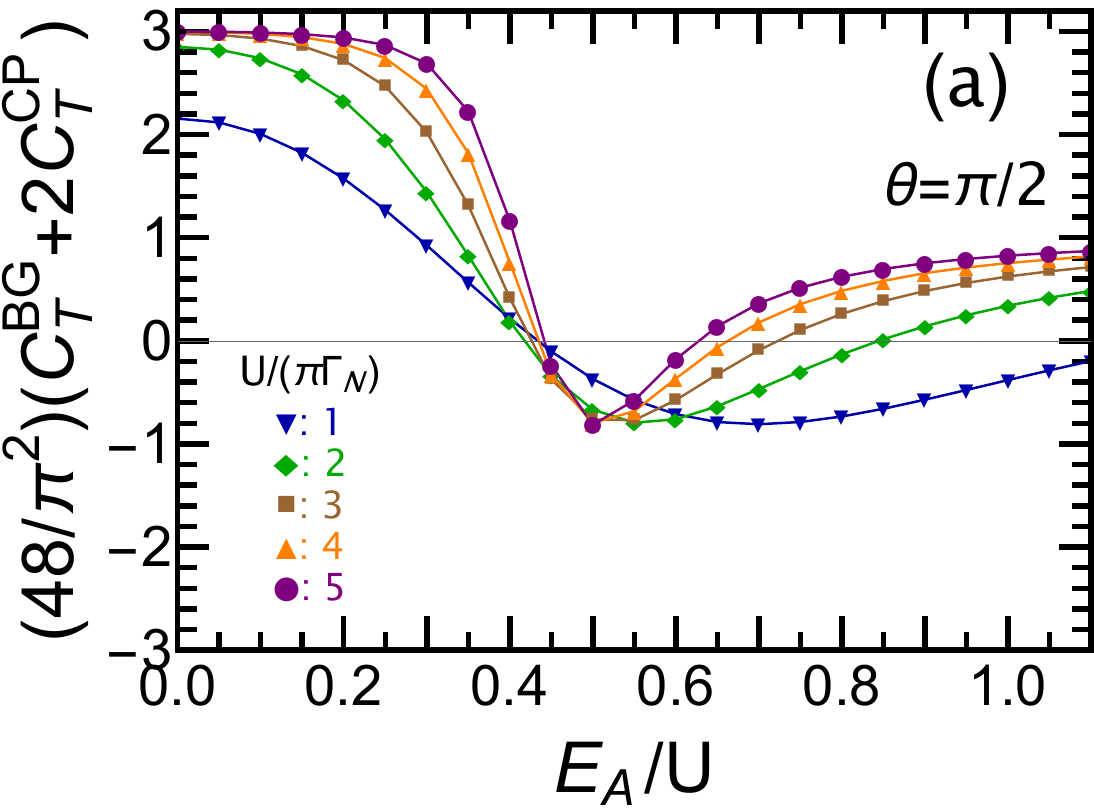}
 \includegraphics[width=0.75\linewidth]{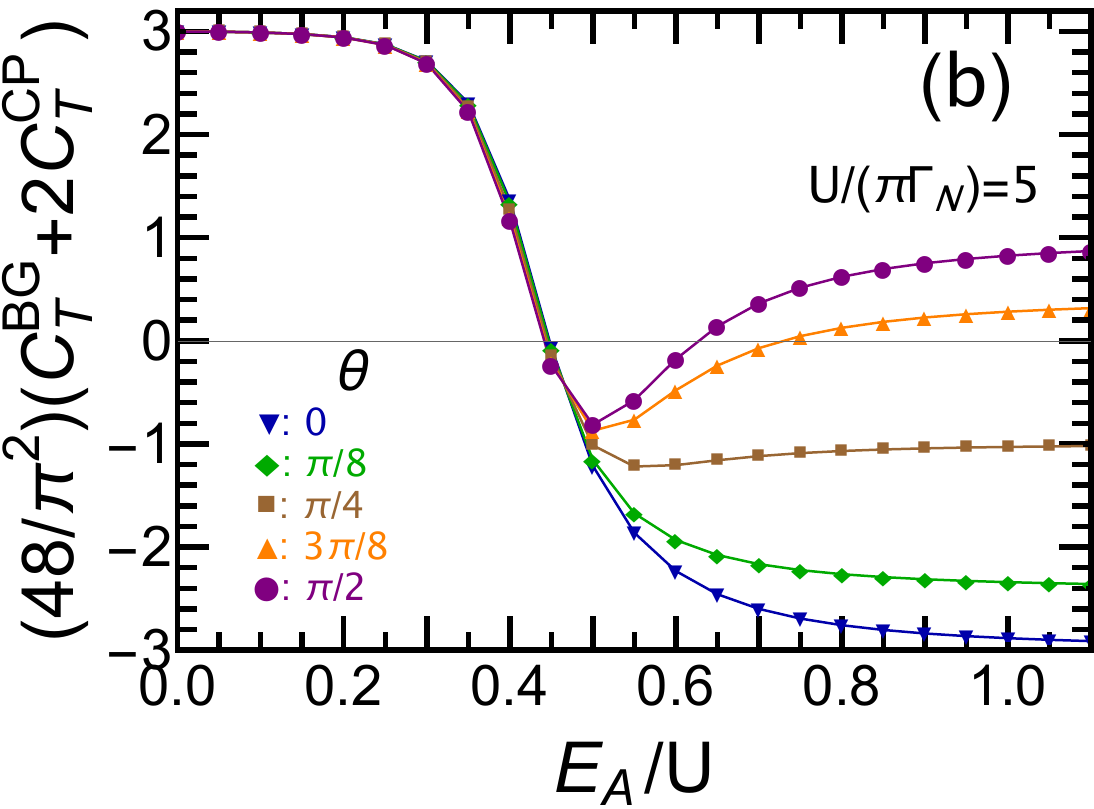}
\caption{
Coefficient of the $T^2$ term 
in the nonlocal conductance  $g_{RL}^{}$, 
defined in Eq.\ \eqref{eq:gRL},  is plotted as a function of $E_{A}^{}/U$. 
This coefficient, $C_{T}^{\mathrm{BG}}+2C_{T}^{\mathrm{CP}}$, 
arises from the combination 
$\mathcal{T}_{\mathrm{BG}}^{}(T)-2\mathcal{T}_{\mathrm{CP}}^{}(T)$, 
as described in Eqs.\ \eqref{eq:Transmission_BG_def_b=0} 
and \eqref{eq:Transmission_CP_def_b=0}.
 (a) Results for  $U/(\pi \Gamma_{N})= 1.0$, $2.0$, $3.0$, $4.0$, and $5.0$, 
with  $\theta = \pi/2$.  
 (b) Results for $\theta =  0$, $\pi/8$, $\pi/4$, $3\pi/8$, and $\pi/2$,  
with $U/(\pi \Gamma_{N})= 5.0$. 
}
\label{fig:3body_Andreev_partB}
\end{minipage}
\end{figure}

\begin{figure}
\leavevmode 
\begin{minipage}[b]{1\linewidth}
 \includegraphics[width=0.75\linewidth]{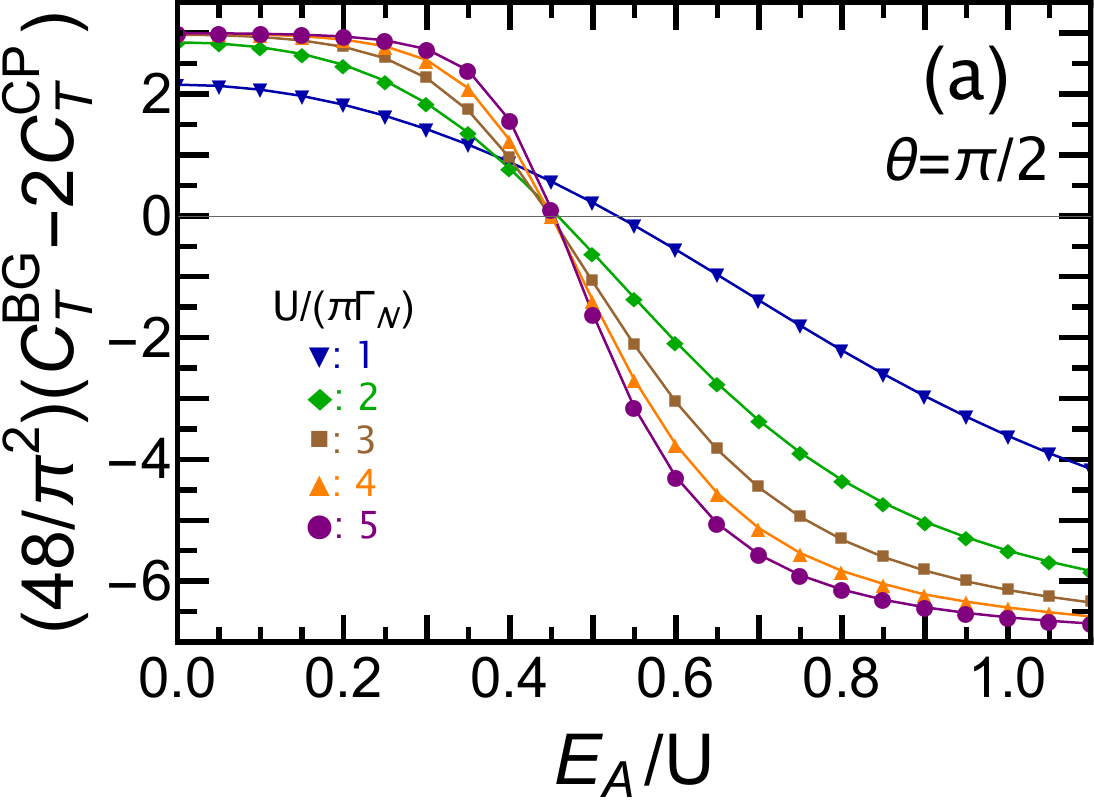}
 \includegraphics[width=0.75\linewidth]{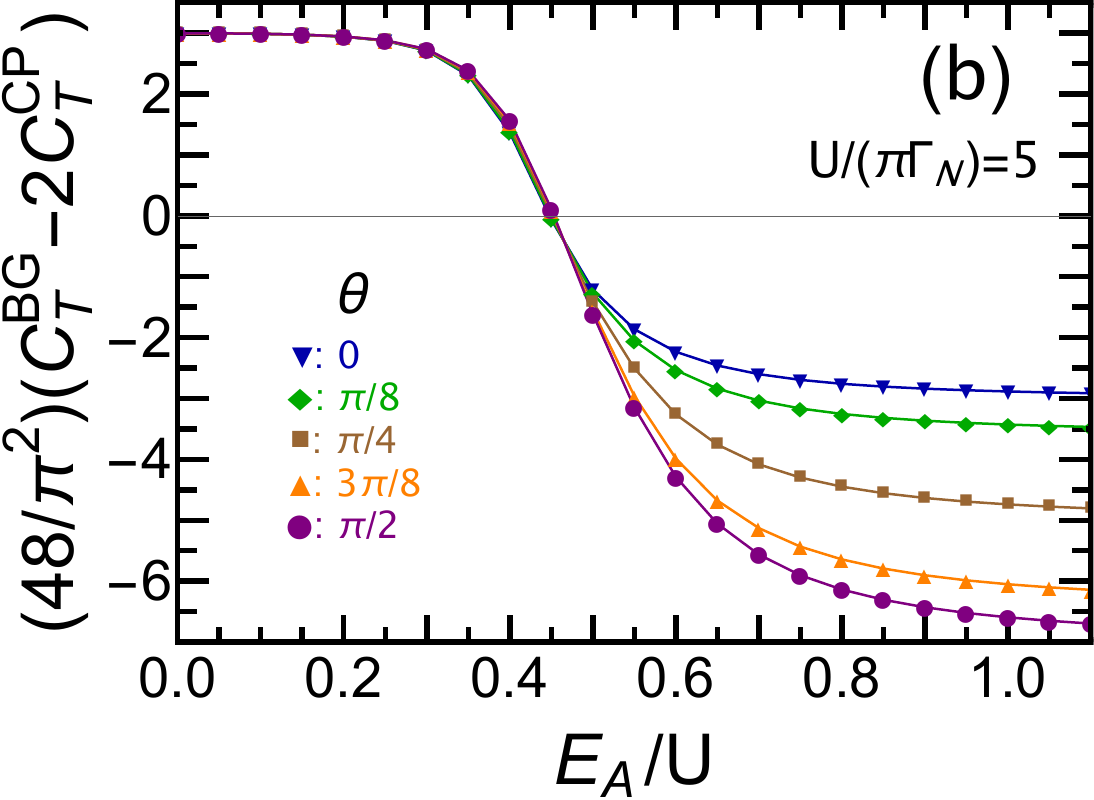}
\caption{
Coefficient of the $T^2$ term of the local conductance
$g_{LL}^{}$,  shown in Eq.\ \eqref{eq:gLL_b0_T}, 
 is plotted as a function of $E_{A}^{}/U$ 
for symmetric tunnel coupling $\Gamma_L^{}=\Gamma_R^{}$.
(a) Results for  $U/(\pi \Gamma_{N})= 1.0$, $2.0$, $3.0$, $4.0$, and $5.0$, 
with  $\theta = \pi/2$.  
(b) Results for $\theta =0$, $\pi/8$, $\pi/4$, $3\pi/8$, and $\pi/2$,  
with $U/(\pi \Gamma_{N})= 5.0$. 
}
\label{fig:3body_Andreev_partC}
\end{minipage}
\end{figure}

\subsection{Three-body FL corrections to CAR}
\label{subsec:3body_correction_NRG}

We next discuss the NRG results for 
the coefficients, $C_{T}^{\mathrm{BG}}$ and $C_{T}^{\mathrm{CP}}$,   
which characterize the $T ^2$ terms 
of the Bogoliubov-quasiparticle 
and Cooper-pair transmission probabilities, defined in Eqs.\ 
\eqref{eq:Transmission_BG_def_b=0}--\eqref{eq:Theta_CP_def_b=0}.

As shown in Fig.\  \ref{fig:3body_Andreev_partA} (a),
 $C_{T}^{\mathrm{BG}}$ exhibits a positive plateau 
for large $U$ over the region $E_A^{} \lesssim U/2$, 
inside the semicircle illustrated in Fig.\ \ref{fig:SingleDotPhase}. 
This behavior originates from the Kondo effect,  
in which the phase shift is almost locked at $\delta \simeq \pi/2$. 
In this region,  the three-body contribution 
almost vanishes,  $\Theta_T^{\mathrm{BG}}\simeq 0$,  
reflecting the particle-hole symmetry of Bogoliubov quasiparticles, 
and the plateau height is determined by the two-body correlations.  
In contrast, outside the semicircle, for $E_A^{} \gtrsim U/2$, 
 the three-body contribution becomes comparable to 
the two-body contribution and approaches its lower bound, 
$\Theta_{T}^{\mathrm{BG}} \xrightarrow{E_A^{}\to \infty} -2$,  
while $C_{T}^{\mathrm{BG}}$ also becomes negative and approaches 
$(48/\pi^2) C_{T}^{\mathrm{BG}} \xrightarrow{E_A^{}\to \infty} -3$.
Note that the expression for $C_{T}^{\mathrm{BG}}$, given in Eqs.\ 
\eqref{eq:cT_BG_def_b=0} and \eqref{eq:Theta_BG_def_b=0}, 
is identical to that for  
normal electrons tunneling through an N/QD/N junction 
\cite{MoraMocaVonDelftZarand,FMvDM2018,AO2017_I,AO2017_II,AO2017_III}.

Figures \ref{fig:3body_Andreev_partA}(b) and \ref{fig:3body_Andreev_partA}(c) 
show the coefficient $C_{T}^{\mathrm{CP}}$ of the $T^2$ term in the 
 Cooper-pair transmission probability. 
At zero magnetic field,  as mentioned above, the Cooper-pair contribution 
 is determined solely by the three-body correlations 
as shown in Eq.\ \eqref{eq:cT_CP_def_b=0}, 
namely,   
$(48/\pi^2) C_{T}^{\mathrm{CP}}= \Theta_{T}^{\mathrm{CP}} \sin^2 \theta$. 
Therefore,
this Cooper-pair contribution to the $T^2$ term almost vanishes,
  $C_{T}^{\mathrm{CP}}\simeq 0$, 
 over the region $E_A^{} \lesssim U/2$, 
where the Bogoliubov quasiparticles exhibit genuine ($E_A^{}=0$)
or emergent particle-hole symmetry. 
 In the crossover region $E_A^{} \simeq U/2$, 
the coefficient $C_{T}^{\mathrm{CP}}$ takes a negative minimum
 and changes sign at the quarter-filling point $ \langle q_{d}^{} \rangle = 1/2$, 
 reflecting the factor $\sin 4\delta$ inherent in $\Theta_{T}^{\mathrm{CP}}$. 
The coefficient $C_{T}^{\mathrm{CP}}$ increases for $E_A^{} \gtrsim U/2$, 
in the region in which the superconducting pair correlation,  
$|\bigl \langle d^\dagger_{\uparrow} d^\dagger_{\downarrow}
 +d_{\downarrow}^{} d_{\uparrow}^{} \bigr \rangle|=(1-\langle q_{d}^{}\rangle) 
\sin \theta$,  is enhanced [see Fig.\ \ref{fig:NRGOneDot_FLparametersxi0} (a) for 
the behavior of $\langle q_{d}^{}\rangle$]. 
It asymptotically approaches the upper bound,  
which is determined by 
 the coherence factor $\sin^2 \theta$ and by 
the three-body correlation $\Theta_{T}^{\mathrm{CP}}$ 
approaching its noninteracting value, 
\begin{align}
\Theta_{T}^{\mathrm{CP}} \,\xrightarrow{\,E_A^{}\to \infty \,}\, 2\,.
\label{eq:Theta_EA_inf} 
 \end{align}
The coefficient $C_{T}^{\mathrm{CP}}$ is maximized at  $\theta = \pi/2$, 
which corresponds to $\xi_d^{}=0$, and varies with the angle $\theta$,
as demonstrated in Fig.\ \ref{fig:3body_Andreev_partA}(c) 
for $U/(\pi\Gamma_N^{})=5.0$.

Figure \ref{fig:3body_Andreev_partB} 
shows the NRG results for 
the $T^2$ term of the nonlocal conductance $g_{RL}^{}$,
 defined in Eq.\ \eqref{eq:gRL}, which can also be expressed 
in the following form:  
\begin{align}
 g_{RL}^{}
 \,  = & \  \, 
 g_{RL}^{T=0}
\,- 2 g_0^{} \Bigl(
  C_{T}^{\mathrm{BG}}  + 2 C_{T}^{\mathrm{CP}} 
\Bigr)\!
\left(\frac{\pi T}{T^*}\right)^2 . 
\label{eq:gRL_b0_T} 
 \end{align}
The coefficient $C_{T}^{\mathrm{BG}} +2 C_{T}^{\mathrm{CP}}$ 
depends on the coherence factor,  $\sin^2 \theta$, 
which enters through the Cooper-pairing contribution $C_{T}^{\mathrm{CP}}$.
The Bogoliubov-quasiparticle contribution $C_{T}^{\mathrm{BG}}$ 
 dominates the $T^2$ term of  $g_{RL}^{}$
over the region $E_A^{} \lesssim U/2$ 
since $C_{T}^{\mathrm{CP}}$ almost vanishes or remains very small 
in this region, particularly for large $U$. 
In contrast,  for $E_A^{} \gtrsim U/2$, i.e., 
outside the semicircle illustrated in Fig.\ \ref{fig:SingleDotPhase}, 
the Cooper-pair contribution $C_{T}^{\mathrm{CP}}$ 
becomes comparable to $C_{T}^{\mathrm{BG}}$, particularly  
in the angular range $\pi/4 \lesssim \theta \lesssim 3\pi/4$.  
More specifically, 
the coefficient $C_{T}^{\mathrm{BG}} +2 C_{T}^{\mathrm{CP}}$ 
converges to the following value in the limit $E_A^{} \to \infty$:
 \begin{align}
\frac{48}{\pi^2}
 \Bigl( 
 C_{T}^{\mathrm{BG}}  + 2 C_{T}^{\mathrm{CP}} 
\Bigr)
 \,\xrightarrow{\,E_A^{}\to \infty \,}\, -3+4 \sin^2 \theta \,,
\label{eq:gRL_T2_EA_inf}
 \end{align}
and it remains positive for $\pi/3 < \theta < 2\pi/3$.
This coefficient also exhibits a negative minimum 
in the crossover region $E_A^{} \simeq U/2$, 
particularly for large $U$ and $\theta \simeq \pi/2$.

Similarly,  
the  $T^2$ term in the local conductance $g_{LL}^{}$, 
defined in Eq.\ \eqref{eq:gLL}, can be expressed 
in the following form for $b=0$: 
\begin{align}
 &\!\!\! 
g_{LL}^{}
 \,  = \,
 g_{LL}^{T=0}
\,- 2 g_0^{} \left(
  C_{T}^{\mathrm{BG}}  - \frac{2\Gamma_L^{}}{\Gamma_R^{}}\, 
C_{T}^{\mathrm{CP}} 
\right)\!
\left(\frac{\pi T}{T^*}\right)^2 . 
\label{eq:gLL_b0_T} 
 \end{align}
Figure \ref{fig:3body_Andreev_partC} shows 
the coefficient of the $T^2$ term of $g_{LL}^{}$ 
for a symmetric junction, $\Gamma_L^{}=\Gamma_R^{}$. 
In contrast to the $T^2$ term in the nonlocal 
conductance $g_{RL}^{}$ discussed above,  
the Bogoliubov-quasiparticle and 
Cooper-pair contributions add constructively  
to the local conductance $g_{LL}^{}$.  
Consequently, the coefficient  
$C_{T}^{\mathrm{BG}}  -2C_{T}^{\mathrm{CP}}$   
exhibits neither a peak nor a dip near the crossover region 
$E_A^{}\simeq U/2$, and decreases monotonically 
as $E_A^{}$ increases.  
However, its magnitude depends sensitively on the Bogoliubov angle 
$\theta=  \cot^{-1}(\xi_d^{}/\Gamma_S^{})$  
outside the semicircle, i.e., for $E_A^{}\gtrsim U/2$, 
and approaches the following value in the limit $E_A^{} \to \infty$:
\begin{align}
\!\!\!\!
\frac{48}{\pi^2}
 \Bigl( 
 C_{T}^{\mathrm{BG}}  -\frac{2\Gamma_L^{}}{\Gamma_R^{}} 
\,C_{T}^{\mathrm{CP}} 
\Bigr)
 \xrightarrow{\,E_A^{}\to \infty \,}\, -3-\frac{4\Gamma_L^{}}{\Gamma_R^{}} 
\,\sin^2 \theta .
\label{eq:gLL_T2_EA_inf}
\end{align}
In this regime, the $T^2$ term of the nonlocal conductance $g_{RL}^{}$,  
shown in Fig.\ \ref{fig:3body_Andreev_partB}(b),  
also depends significantly on the angle $\theta$.
This strong $\theta$ dependence reflects the behavior of the coherence factor 
$\sin^2 \theta$, which governs the magnitude of the Cooper-pair contribution 
$C_{T}^{\mathrm{CP}}$, particularly for $E_A^{}\gtrsim U/2$, 
as seen in Fig.\ \ref{fig:3body_Andreev_partA}(c).

These coefficients $C_{T}^{\mathrm{BG}}$ and $C_{T}^{\mathrm{CP}}$  
can be experimentally extracted by measuring 
the $T^2$ terms of both $g_{RL}^{}$ and $g_{LL}^{}$. 
Alternatively, these coefficients can also be obtained 
from measurements of a single conductance component, $g_{RL}^{}$ or $g_{LL}^{}$, 
by varying the Bogoliubov rotation angle $\theta$ 
through the gate voltage $\xi_d^{}$ or the tunnel coupling $\Gamma_S^{}$. 
From the observed coefficients 
$C_{T}^{\mathrm{BG}}$ and $C_{T}^{\mathrm{CP}}$, 
the three-body contributions can, in principle, be deduced using the theoretical formulas 
Eqs.\ \eqref{eq:cT_BG_def_b=0}--\eqref{eq:Theta_CP_def_b=0}.

\section{Microscopic Fermi-liquid theory for the $T^2$ Andreev transport}
\label{sec:3body_correction_derivation}

In this section, we provide an outline of the derivation of the conductance 
up to order $T^2$, 
leaving the additional details required for a complete proof 
to Appendix \ref{sec:current_vertex_sc}.

\subsection{Nambu--Keldysh formalism}

We use the spinor representation for the electron operators,  
$\bm{\psi}_{d}^{}$ and $\bm{\psi}_{j}^{}$ for $j=L, R$, 
defined with respect to the Nambu pseudo-spin space, 
\begin{align}
 \bm{\psi}_{d}^{} 
\,= &  
 \begin{pmatrix}
  \psi_{d,1}^{} \cr
  \psi_{d,2}^{} \rule{0cm}{0.5cm}\cr
 \end{pmatrix}
\,\equiv\,
 \begin{pmatrix}
  d_{\uparrow}^{} \cr
  d_{\downarrow}^{\dagger} \rule{0cm}{0.5cm}\cr
 \end{pmatrix}
,
 \\ 
 \bm{\psi}_{j}^{} 
\,= &   
 \begin{pmatrix}
  \psi_{j,1}^{} \cr
  \psi_{j,2}^{} \rule{0cm}{0.5cm}\cr
 \end{pmatrix}
\equiv 
\int_{-D}^{D}  \! d\epsilon \,\sqrt{\rho_c^{}}\,
 \begin{pmatrix}
  c_{\epsilon, j, \uparrow}^{} \cr
  -c_{-\epsilon, j, \downarrow}^{\dagger} 
\rule{0cm}{0.5cm}\cr
 \end{pmatrix}
\,.
\rule{0cm}{0.8cm}
\label{eq:normal_interface}
\end{align}
In the following, we assign  the label  $p$  ($=d,L,R$) 
to spatial configurations,  $\zeta$ ($=1,2$) 
to the Nambu spinor components,  and  $\nu$ ($=-,+$) to  
the Keldysh Green's functions \cite{Hershfield1,Caroli_1971,Keldysh},
  defined with respect to the nonequilibrium steady state 
under finite bias voltages $eV$,    
\begin{subequations}
\begin{align}
G^{--}_{pp';\zeta\zeta'} (t, t') \, \equiv&  
\, -i\, \langle \,\mbox{T}\,
\psi^{}_{p,\zeta}(t)\, \psi^{\dagger}_{p',\zeta'}(t')
\,\rangle_{V}^{}
\,,
\\
G^{++}_{pp';\zeta\zeta'} (t, t') \, \equiv&  
\,-i\, 
\langle \,\widetilde{\mbox{T}}\,
\psi^{}_{p,\zeta}(t)\, \psi^{\dagger}_{p',\zeta'}(t')
\,\rangle_{V}^{}
\,,
 \\
G^{+-}_{pp';\zeta\zeta'} (t, t') \, \equiv& 
\,-i\, 
\langle 
\psi^{}_{p,\zeta}(t)\, \psi^{\dagger}_{p',\zeta'}(t')
\,\rangle_{V}^{} 
\;,
\\
G^{-+}_{pp';\zeta\zeta'} (t, t') \equiv&  
\ \  i\, 
\langle 
\psi^{\dagger}_{p',\zeta'}(t') \, \psi^{}_{p,\zeta}(t) 
\,\rangle_{V}^{} 
\,.
\end{align}
\label{eq:Keldysh_Green's_function_def}  
\!\!\!\!\!\! 
\end{subequations}
Here,  
 $\mbox{T}$ and $\widetilde{\mbox{T}}$ are the time-ordering 
and anti-time-ordering operators, respectively.
These Keldysh Green's functions are linearly dependent and 
can be expressed in terms 
of three independent components, for instance,  
the retarded  $G_{}^{r}$,  
the advanced $G_{}^{a}$, 
and the symmetrized $G_{}^\mathrm{K}$ Green's functions, 
\begin{subequations}
\begin{align}
G_{}^{r}= & \ 
 G_{}^{--}-G_{}^{-+} ,
\label{eq:retarded_advanced_Keldysh}
\qquad \quad  
G_{}^{a}=  
 G_{}^{--}-G_{}^{+-},
\\
G_{}^\mathrm{K}\equiv  & \ 
 G_{}^{+-}+G_{}^{-+}
\,= \, 
 G_{}^{--}+G_{}^{++}\,.
\label{eq:linear_dependency_GK}
\rule{0cm}{0.3cm}
\end{align}
\end{subequations}
The Fourier transform of the Green's function, defined with respect to 
 the steady state, becomes a function of a single frequency $\omega$, 
\begin{align}
G_{pp';\zeta\zeta'}^{\nu\nu'}(\omega)= \int_{-\infty}^{\infty} 
\! dt \,e^{i\omega t} G_{pp';\zeta\zeta'}^{\nu\nu'}(t,0)\,.
\end{align}
We use $2\times 2$ matrices 
to treat the components in the Nambu space, 
\begin{align}
\bm{G}_{pp'}^{\nu\nu'}  \, = &  
 \begin{pmatrix}
 G_{pp';11}^{\nu\nu'} & \, G_{pp';12}^{\nu\nu'}   \cr
 G_{pp';21}^{\nu\nu'}  & \, G_{pp';22}^{\nu\nu'}  
\rule{0cm}{0.5cm}
\cr  
 \end{pmatrix}
 ,
\end{align}
and also $4 \times 4$ matrices including all Keldysh components, 
 \begin{align}
\widehat{\bm{G}}_{pp'}^{}  \, = &    
\left[ 
 \begin{matrix}
 \bm{G}_{pp'}^{--}  & \ \bm{G}_{pp'}^{-+}   \cr
 \bm{G}_{pp'}^{+-}  & \ \bm{G}_{pp'}^{++}  
\rule{0cm}{0.4cm}
\cr  
 \end{matrix}
 \right]  ,
\qquad
\widehat{\bm{\rho}}_3^{} 
 \, = 
\left[ 
 \begin{matrix}
 \bm{1} & \ \bm{0}   \cr
 \bm{0}  &  -\bm{1}  \rule{0cm}{0.4cm}\cr  
 \end{matrix}
 \right]  . 
 \end{align}
Here,   $\bm{1}$ and $\bm{0}$
are the $2 \times 2$ unit and zero matrices, respectively.  
We will also use the Pauli matrices, 
 \begin{align}
 \!\!\!
 \bm{\tau}_1 =
 \begin{pmatrix}
 0 &   1 \cr
 1 &  0 \cr  
 \end{pmatrix}
, 
 \quad
 \bm{\tau}_2 =
 \begin{pmatrix}
 0 &   -i \cr
 i &  \ 0  \cr  
 \end{pmatrix}
, 
 \quad
 \bm{\tau}_3 =
 \begin{pmatrix}
 1 &  \  0 \cr
 0 &  -1 \cr  
 \end{pmatrix}
 .
 \end{align}

\subsection{Green's function for the isolated normal leads}

For isolated normal leads disconnected from the dot,  
the retarded and advanced Green's functions 
of the conduction electrons $\bm{\psi}_{j}^{}$   
take the following form for small frequency $|\omega|\ll D$:  
 \begin{align}
\bm{g}_{j}^{r}(\omega)  \, = & \   
 - i\pi \rho_c^{} \bm{1}\,, 
\qquad 
\bm{g}_{j}^{a}(\omega)  \,=\,  i\pi \rho_c^{} \bm{1} 
\,. 
 \end{align}
Correspondingly,  the lesser $\bm{g}_{j}^{-+}$ 
and greater $\bm{g}_{j}^{+-}$ Green's functions are given by 
 \begin{align}
\bm{g}_{j}^{-+}(\omega)  \, = & \   
i 2 \pi \rho_c^{}
\bm{f}_j^{}(\omega)
\,, \\
%
\bm{g}_{j}^{+-}(\omega)  \, = &    
-i2 \pi \rho_c^{}
\Bigl[ \bm{1} - \bm{f}_j^{}(\omega)  \Bigr] \,. 
 \end{align}
 Here, $\bm{f}_j^{}(\omega)$ is a diagonal matrix 
describing the  distribution functions for electrons ($e$) and holes ($h$),
\begin{align}
\bm{f}_j^{}(\omega) \, \equiv & \  
 \begin{pmatrix}
  f_j^{(e)}(\omega) & \ 0 \cr
  0 \rule{0cm}{0.5cm}&\  f_j^{(h)}(\omega) \cr
 \end{pmatrix}
,
 \end{align}
where $f_j^{(e)}(\omega) \equiv f(\omega-\mu_j^{e})$ and      
$f_j^{(h)}(\omega) \equiv f(\omega-\mu_j^{h})$, 
with $f(\omega)=\bigl[e^{\omega/T} + 1 \bigr]^{-1}$. 
The bias voltage $V_{j}^{}$ applied to the normal lead on $j$ ($=L,R$)  
shifts the chemical potentials for electrons and holes such that  
\begin{align}
 \mu_j^{e} \,=\, -\mu_j^{h} \,=\, eV_j \,. 
\end{align}
Including the causal component 
$\bm{g}_{j}^{--} = \bm{g}_{j}^{-+}+\bm{g}_{j}^{r}$ 
and its counterpart 
$\bm{g}_{j}^{++} = \bm{g}_{j}^{+-}-\bm{g}_{j}^{r}$,
the Keldysh Green's functions for isolated leads can be expressed 
in the $4\times 4$ matrix form,      
\begin{align}
 \widehat{\bm{g}}_{j}^{}(\omega) \, =\, 
\left[ 
 \begin{matrix}
 \bm{g}_{j}^{--}(\omega)  & \ \bm{g}_{j}^{-+}(\omega)   \cr
 \bm{g}_{j}^{+-}(\omega)  & \ \bm{g}_{j}^{++}(\omega)  
\rule{0cm}{0.45cm}\cr  
 \end{matrix}
 \right] \,.
\end{align}

The tunnel coupling $v_j^{}$ gives rise not only to level broadening but 
also to a nonequilibrium distribution in the impurity level, 
which can be described by the following noninteracting Nambu--Keldysh self-energy: 
 \begin{align}
\widehat{\bm{\Sigma}}_{0}^{}(\omega)  
\,= &  
\left[ 
 \begin{matrix}
\bm{\Sigma}_{0}^{--}(\omega) &  
 \bm{\Sigma}_{0}^{-+} (\omega)   \cr
 \bm{\Sigma}_{0}^{+-} (\omega)  &  
 \bm{\Sigma}_{0}^{++} (\omega) 
\rule{0cm}{0.45cm} \cr  
 \end{matrix}
 \right] 
\,  \equiv  
\sum_{j=L,R} 
 v_j^2 \,
\widehat{\bm{\rho}}_3^{} \,
 \widehat{\bm{g}}_{j}^{}(\omega) 
\, \widehat{\bm{\rho}}_3^{} \,. 
\rule{0cm}{0.8cm} 
\label{eq:U0_self_Keldysh_Nambu}
 \end{align}
The lesser   $\bm{\Sigma}_{0}^{-+}$ and 
greater   $\bm{\Sigma}_{0}^{+-}$ 
components depend on the distribution function $\bm{f}_j^{}$, 
\begin{align}
  \bm{\Sigma}_{0}^{-+} (\omega) \,=& \ 
-2i \sum_{j=L,R} 
\Gamma_{j}^{} \, \bm{f}_j^{}(\omega) \,,
\label{eq:lesser_self_0}
\\  
  \bm{\Sigma}_{0}^{+-} (\omega) \,=& \ 
2i \sum_{j=L,R} 
\Gamma_{j}^{} 
\Bigl[ 
\bm{1} - \bm{f}_j^{}(\omega) 
 \Bigr] \,.
\label{eq:greater_self_0}
\end{align}
Thus,  the components 
$\bm{\Sigma}_{0}^{\mathrm{--}}
=   
\bm{\Sigma}_{0}^{r} - \bm{\Sigma}_{0}^{-+}$ 
and 
 $\bm{\Sigma}_{0}^{\mathrm{++}}
=   -\bm{\Sigma}_{0}^{r} - \bm{\Sigma}_{0}^{+-}$ 
also depend on the distribution function  $\bm{f}_j^{}$, 
while the retarded component does not,      
\begin{align}
\bm{\Sigma}_{0}^{r} = 
 -i \left( \Gamma_L^{} + \Gamma_R^{} \right) \bm{1} \,.
\label{eq:retarded_self_0}
\end{align}
Similarly, the symmetrized (Keldysh) component of the noninteracting 
self-energy, 
  $\bm{\Sigma}_{0}^{\mathrm{K}} \equiv 
  -\bm{\Sigma}_{0}^{-+}  -\bm{\Sigma}_{0}^{+-}$, 
is given by  
\begin{align}
  \bm{\Sigma}_{0}^{\mathrm{K}} (\omega) \,=& \ 
-2i \sum_{j=L,R} 
\Gamma_{j}^{} 
\Bigl[ 
\bm{1} - 2 \bm{f}_j^{}(\omega) 
 \Bigr] \,. 
\label{eq:Keldysh_self_0}
\end{align}

The weighted sum of  $\bm{f}_L^{}$ and $\bm{f}_R^{}$
plays an essential role in the transport properties, 
and its matrix elements represent 
 the nonequilibrium distribution functions 
for particles $f_\mathrm{eff}^{(e)}(\omega)$ 
and holes $f_\mathrm{eff}^{(h)}(\omega)$,   
\begin{align}
&
\!\!\!\!
\bm{f}_\mathrm{eff}^{}(\omega) \,\equiv  
\sum_{j=L,R}  
\frac{\Gamma_{j}^{}\, \bm{f}_j^{}(\omega)}{\Gamma_L^{} +\Gamma_R^{}} 
\  = 
 \begin{pmatrix}
  f_\mathrm{eff}^{(e)}(\omega) & 0 \cr
  0 \rule{0cm}{0.5cm}&  f_\mathrm{eff}^{(h)}(\omega) \cr
 \end{pmatrix}
, 
\label{eq:f_eff_matrix}
\\
&
\!\!\!\!
f_\mathrm{eff}^{(s)}(\omega) \,=\, 
\frac{ \Gamma_L^{}\, f_L^{(s)}(\omega) 
+ \Gamma_R^{} \,f_R^{(s)}(\omega) }{\Gamma_L^{} +\Gamma_R^{}} ,  
\qquad s=e, h \, .
\rule{0cm}{0.9cm}
\end{align}
Note that, at equilibrium $eV_L=eV_R=0$,  
the distribution-function matrices $\bm{f}_j^{}(\omega)$ and  
 the noninteracting 
self-energy  $\bm{\Sigma}_{0}^{\mathrm{\nu\nu'}}$ 
become proportional to the unit matrix, 
\begin{align}
\!\!\!\! 
\bm{f}_j^{}(\omega) 
\, \xrightarrow{\,eV_j^{}\to 0 \,}\,  f(\omega)  \,\bm{1} , 
\qquad 
\left. \bm{\Sigma}_{0}^{\nu\nu'} (\omega) \right|_{eV \to 0}^{}
 \propto  \bm{1} .  
\end{align}

 \subsection{Impurity Green's function at finite bias voltages}
\label{subsec:Impurity_Keldysh_GF}

We next consider the nonequilibrium Green's function 
for a quantum dot connected to normal and superconducting leads. 
Specifically, 
we set the chemical potential for the SC lead  $\mu_S^{}$ 
 at the Fermi level  ($\mu_S^{} \equiv 0$)  
and  
consider the large superconducting-gap limit  $\left| \Delta_{S}^{}\right| \to \infty$, 
as mentioned in Sec.\ \ref{subsec:Large_gap_limit}.
In this limit, the SC proximity effects penetrating into the impurity site 
can be described faithfully by an effective Hamiltonian 
$\mathcal{H}_\mathrm{eff}^{}$,   
the free-impurity part of which can be expressed in the following form: 
\begin{align}
 & \!\!\!\! \!\!\!\! \!\!\!\! \!\!\!\!
\xi_d^{}  \bigl( n_d^{} \! - \! 1 \bigr) 
-  b  \bigl(
n_{d\uparrow}^{} - \!  n_{d\downarrow}^{}
\bigr)  + \Gamma_S^{}\, 
\bigl(  
d^\dagger_{\uparrow} d^\dagger_{\downarrow} 
+ d^{}_{\downarrow} d^{}_{\uparrow}  
\bigr) 
\nonumber \\
= & \  \, 
\bm{\psi}_{d}^{\dagger}\, 
\bm{\mathcal{H}}_d^{0}  
\bm{\psi}_{d}^{} \,+ \, b  \,, 
\nonumber \\
\bm{\mathcal{H}}_d^{0} 
 \, \equiv  & \,   
 \begin{pmatrix}
 \xi_d^{} &  \Gamma_S^{} \cr
 \Gamma_S^{} & - \xi_d^{}
\rule{0cm}{0.35cm}
 \end{pmatrix}
-b \,\bm{1} \,. 
\rule{0cm}{0.7cm}
\end{align}

The Keldysh Green's function for the isolated impurity site,    
disconnected from the normal leads,  is given by  
\begin{align}
\Bigl \{ \widehat{\bm{g}}_{dd}^{} (\omega) \Bigr \}^{-1} 
=\,  
\left[ \,
 \begin{matrix}
\omega\, \bm{1} \, -\, \bm{\mathcal{H}}_{d}^{0} & \ \bm{0}   \cr
 \bm{0} & \ -\left(\, \omega\, \bm{1} \, -\, \bm{\mathcal{H}}_{d}^{0}\right) 
\rule{0cm}{0.4cm} \cr  
 \end{matrix}
\, \right] 
 \,.
\end{align}
The bias dependence enters through 
the tunnel couplings between the impurity and normal leads,  
which can be taken into account using the Dyson equation, 
\begin{align}
\Bigl \{ \widehat{\bm{G}}_{0}^{} (\omega) \Bigr \}^{-1} 
=\,  
\Bigl \{ \widehat{\bm{g}}_{dd}^{} (\omega) \Bigr \}^{-1} 
 -\,  \widehat{\bm{\Sigma}}_{0}^{}(\omega)  
\label{eq:G0_imp_Dyson}
\,.
\end{align}
Similarly, electron correlations arising from $H_d^{U}$, 
defined in Eq.\ \eqref{eq:H_dot_U},  enter through  
 the interacting self-energy  $\widehat{\bm{\Sigma}}_{U}^{}(\omega)$ 
in the Nambu--Keldysh formalism,  
\begin{align}
\Bigl \{ \widehat{\bm{G}}_{dd}^{} (\omega) \Bigr \}^{-1} 
= & \   
\Bigl \{ \widehat{\bm{G}}_{0}^{} (\omega) \Bigr \}^{-1} 
 -\,  \widehat{\bm{\Sigma}}_{U}^{}(\omega) \,, 
\label{eq:Dyson_Nambu_Keldysh_tot}
\\
\widehat{\bm{\Sigma}}_\mathrm{tot}^{}(\omega)  \, = & \ 
  \widehat{\bm{\Sigma}}_{0}^{}(\omega)  
 +  \widehat{\bm{\Sigma}}_{U}^{}(\omega).
\rule{0cm}{0.5cm}
\label{eq:self_energy_tot}
\end{align}
From Eq.\ \eqref{eq:Dyson_Nambu_Keldysh_tot}, 
it follows that the symmetrized Green's function 
$\bm{G}_{dd}^\mathrm{K} \equiv 
\bm{G}_{dd}^{+-} + \bm{G}_{dd}^{-+}$ 
can be factorized as 
\begin{align}
\bm{G}_{dd}^{\mathrm{K}}(\omega) 
\,=  & \, 
 \bm{G}_{dd}^{r}(\omega)  \,
 \bm{\Sigma}_\mathrm{tot}^{\mathrm{K}} (\omega) \,  
 \bm{G}_{dd}^{a}(\omega)  \,,    
\label{eq:Dyson_GK}
\end{align}
where  $\bm{\Sigma}_\mathrm{tot}^{\mathrm{K}} 
  =  
 -\bm{\Sigma}_\mathrm{tot}^{+-} 
 - \bm{\Sigma}_\mathrm{tot}^{-+}$. 
Similar relations also hold for the lesser and greater Green's functions, 
\begin{align}
\bm{G}_{dd}^{-+}(\omega)
\, = & \  
-\,\bm{G}_{dd}^{r}(\omega) \,\bm{\Sigma}_\mathrm{tot}^{-+}(\omega) 
\,\bm{G}_{dd}^{a}(\omega) 
\,, 
\\
\bm{G}_{dd}^{+-}(\omega)
\, = & \  
-\,\bm{G}_{dd}^{r}(\omega) \,\bm{\Sigma}_\mathrm{tot}^{+-}(\omega) 
\,\bm{G}_{dd}^{a}(\omega) 
\,.
\end{align}

\subsection{Non-local Green's functions}

The non-local Green's functions between the impurity site and 
the normal leads ($j=L,R$) satisfy the following relations:  
\begin{align}
\widehat{\bm{G}}_{jd}^{}(\omega) \, = & \ 
-v_j^{} \,
\widehat{\bm{g}}_{j}^{}(\omega)\,\widehat{\bm{\rho}}_{3}^{} \,
\widehat{\bm{G}}_{dd}^{}(\omega)\,, 
\label{eq:Keldysh_Nambu_recursion1}
\\
\widehat{\bm{G}}_{dj}^{}(\omega) \, =  & \ 
-v_j^{} \,
\widehat{\bm{G}}_{dd}^{}\,\widehat{\bm{\rho}}_{3}^{}\, 
\widehat{\bm{g}}_{j}^{}(\omega) \,. 
\label{eq:Keldysh_Nambu_recursion2}
\end{align}

The expectation values of the currents, 
$I_{j,\sigma}^{} \equiv \langle \widehat{I}_{j,\sigma} \rangle$, for $j=L\,,R$,  
 defined in Eqs.\ \eqref{eq:IR_def} and \eqref{eq:IL_def},  
can be expressed in terms of the non-local 
lesser and greater Green's functions, 
\begin{align}
 & \!\!
I_{j,\uparrow}^{} 
= \  
\frac{e}{h} \int_{-\infty}^{\infty} \!\!  d\omega
\,\eta_{j}^{} v_j^{} 
\left[\, \bm{G}_{dj}^{-+}(\omega) 
-  \bm{G}_{jd}^{-+}(\omega) \,\right]_{11}^{}, 
\label{eq:current_expectation_up}
\\
 & \!\! 
I_{j,\downarrow}^{} 
=   \,  
- \,\frac{e}{h} \int_{-\infty}^{\infty} \!\! d\omega
\, \eta_{j}^{} v_j^{} 
 \left[\, \bm{G}_{dj}^{+-}(\omega) 
-  \bm{G}_{jd}^{+-}(\omega) \,\right]_{22}^{} , 
\label{eq:current_expectation_down}
\end{align}
where  $\eta_R^{}=1$, and $\eta_L^{}=-1$. 
The (1,1) and  (2,2) components 
 of matrices on the right-hand sides of these equations 
represent  the currents carried by electrons 
with spin $\sigma=\uparrow$ and  $\downarrow$, 
respectively.   
These matrix parts can also be expressed 
 in terms of the impurity Green's function 
by using  Eqs.\ 
\eqref{eq:Dyson_GK}--\eqref{eq:Keldysh_Nambu_recursion2},
\begin{align}
& v_j^{}
\Bigl[\,
\bm{G}_{dj}^{-+}
-
\bm{G}_{jd}^{-+}
\,\Bigr]_{11}^{} 
\nonumber \\
&   \qquad \ \ 
=  \, 
-i\Gamma_j^{} \left[\,  
\bm{ G}_{dd}^\mathrm{K}
 -\left( 1-2f_j^{(e)}\right)\Bigl\{\bm{G}_{dd}^{r} -\bm{G}_{dd}^{a}\Bigr\}
\,\right]_{11}^{} 
\nonumber \\
&   \qquad \ \ =\, 
-i\Gamma_j^{} 
\Bigl[\,
\bm{G}_{dd}^{r}
\,\bm{\Pi}_\mathrm{tot}^{j;(e)}\,
\bm{G}_{dd}^{a} \,\Bigr]_{11}^{} ,
\label{eq:current_Ker_upspin}
\\
& v_j^{}
\Bigl[\,
\bm{G}_{dj}^{+-}
-
\bm{G}_{jd}^{+-}
\,\Bigr]_{22}^{} 
\rule{0cm}{0.6cm}
\nonumber \\ 
&  \qquad \ \ 
=\,  
-i\Gamma_j^{} \left[\, 
\bm{G}_{dd}^\mathrm{K}
 -\left( 1-2f_j^{(h)}\right)\Bigl\{\bm{G}_{dd}^{r} -\bm{G}_{dd}^{a}\Bigr\}
\,\right]_{22}^{} 
\nonumber \\
& \qquad \ \ 
=  \, 
-i\Gamma_j^{} 
\Bigl[\,
\bm{G}_{dd}^{r}
\,\bm{\Pi}_\mathrm{tot}^{j;(h)}\,
\bm{G}_{dd}^{a} \,\Bigr]_{22}^{} .
\label{eq:current_Ker_downspin}
\end{align}
Here, 
 $\bm{\Pi}_\mathrm{tot}^{j;(e)}$ and 
$\bm{\Pi}_\mathrm{tot}^{j;(h)}$ 
are given by linear combinations of the self-energy matrices 
which are purely imaginary   
and depend on $j$ ($=L,R$) through the Fermi distribution functions 
$f_j^{(e)}$ and  $f_j^{(h)}$ as follows: 
\begin{align}
\bm{\Pi}_\mathrm{tot}^{j;(e)}
\,\equiv & \  
\bm{\Sigma}_\mathrm{tot}^\mathrm{K} 
 -\left( 1-2 f_j^{(e)}\right)\Bigr(
\bm{\Sigma}_\mathrm{tot}^{-+} -\bm{\Sigma}_\mathrm{tot}^{+-} 
\Bigr) \,,
\label{eq:current_Self_upspin}
\\
\bm{\Pi}_\mathrm{tot}^{j;(h)}
\,\equiv & \  
\bm{\Sigma}_\mathrm{tot}^\mathrm{K} 
 -\left( 1-2f_j^{(h)}\right)\Bigr(
\bm{\Sigma}_\mathrm{tot}^{-+} -\bm{\Sigma}_\mathrm{tot}^{+-} 
\Bigr) \,.
\label{eq:current_Self_downspin}
\end{align}
Note that $
\bm{\Sigma}_\mathrm{tot}^{-+} 
- \bm{\Sigma}_\mathrm{tot}^{+-}
=
\bm{\Sigma}_\mathrm{tot}^{r} 
-  \bm{\Sigma}_\mathrm{tot}^{a} 
$.
Thus, the average currents,
Eqs.\ \eqref{eq:current_expectation_up}
and \eqref{eq:current_expectation_down},
 can also be expressed in the following forms: 
\begin{align}
 & \!\!
I_{j,\uparrow}^{} 
= \,  
\frac{e}{h} \int_{-\infty}^{\infty} \!\!  d\omega
\,\left(-i \eta_{j}^{} \Gamma_j^{}
\right) 
\Bigl[\,
\bm{G}_{dd}^{r}
\,\bm{\Pi}_\mathrm{tot}^{j;(e)}\,
\bm{G}_{dd}^{a} \,\Bigr]_{11}^{}  \,,
\label{eq:current_up_formula}
\\
 & \!\! 
I_{j,\downarrow}^{} 
=   \,  
- \,\frac{e}{h} \int_{-\infty}^{\infty} \!\! d\omega
\,\left(-i \eta_{j}^{} \Gamma_j^{}
\right) \Bigl[\,
\bm{G}_{dd}^{r}
\,\bm{\Pi}_\mathrm{tot}^{j;(h)}\,
\bm{G}_{dd}^{a} \,\Bigr]_{22}^{}.
\label{eq:current_down_formula}
\end{align}

At equilibrium $eV=0$, the symmetrized 
 Keldysh self-energy $\bm{\Sigma}_\mathrm{tot}^\mathrm{K}$ 
can be expressed in terms 
of the imaginary part of the retarded self-energy as
\begin{align}
\bm{\Sigma}_\mathrm{tot}^\mathrm{K}
\Big|_{eV=0}^{} 
\,=\,
 (1-2f)
\Bigr(
\bm{\Sigma}_\mathrm{tot}^{r} -\bm{\Sigma}_\mathrm{tot}^{a} 
\Bigr) 
\Big|_{eV=0}^{} 
\,.
%
\end{align}
Thus, the integrands in Eqs.\ 
\eqref{eq:current_up_formula} and \eqref{eq:current_down_formula}
 vanish at $eV=0$,  since 
the first and second terms of $\bm{\Pi}_\mathrm{tot}^{j;(s)}$,  
 shown on the right-hand sides of 
Eqs.\ \eqref{eq:current_Self_upspin} and \eqref{eq:current_Self_downspin},  
cancel each other out, 
\begin{align}
\bm{\Pi}_\mathrm{tot}^{j;(s)} \xrightarrow{\,eV\to 0 \,}\, 0\,,
\qquad \  s= e,h \,.
\end{align}
Therefore, the linear terms in $eV$ of the currents  
$I_{j, \uparrow}^{}$ and that of  $I_{j, \downarrow}^{}$   
arise from $\bm{\Pi}_\mathrm{tot}^{j;(e)}$ and
$\bm{\Pi}_\mathrm{tot}^{j;(h)}$, respectively.  
Consequently, in order to calculate the linear-response currents,
the Green's functions $\bm{G}_{dd}^{r}(\omega)$ 
and  $\bm{G}_{dd}^{a}(\omega)$ appearing in   
Eqs.\ \eqref{eq:current_up_formula} and \eqref{eq:current_down_formula}
 can be replaced by their equilibrium values evaluated at $eV=0$, 
the explicit forms of which are summarized in Appendix \ref{sec:microscopic_FL}.

The self-energy part $\bm{\Pi}_\mathrm{tot}^{j;(s)}$ 
 for $s= e, h$ can be decomposed into two parts,   
$\bm{\Pi}_0^{j;(s)}$ and $\bm{\Pi}_U^{j;(s)}$,   
which represent the contributions from 
the noninteracting self-energy 
and the interacting self-energy, respectively, as  
\begin{align}
\bm{\Pi}_\mathrm{tot}^{j;(s)} 
\, = & \  
\bm{\Pi}_0^{j;(s)}  + \bm{\Pi}_U^{j;(s)}  ,
\label{eq:Pi_0+U} 
\\
\bm{\Pi}_{0}^{j;(s)}
\,\equiv & \  
\bm{\Sigma}_{0}^\mathrm{K} 
 -\left( 1-2 f_j^{(s)}\right)\Bigr(
\bm{\Sigma}_{0}^{-+} -\bm{\Sigma}_{0}^{+-} 
\Bigr) \,,
\label{eq:current_Self_0}
\rule{0cm}{0.5cm}
\\
\bm{\Pi}_{U}^{j;(s)}
\,\equiv & \  
\bm{\Sigma}_{U}^\mathrm{K} 
 -\left( 1-2 f_j^{(s)}\right)\Bigr(
\bm{\Sigma}_{U}^{-+} -\bm{\Sigma}_{U}^{+-} 
\Bigr) \,.
\label{eq:current_Self_U}
\end{align}
Correspondingly, the average current $I_{j,\sigma}^{}$
can also be decomposed into two parts,   
$I_{j,\sigma}^{(0)}$ and $I_{j,\sigma}^{(U)}$, 
which represent the contributions from 
$\bm{\Pi}_0^{j;(s)}$  and  $\bm{\Pi}_U^{j;(s)}$,
 respectively, 
\begin{align}
I_{j,\sigma}^{} \,=\,
I_{j,\sigma}^{(0)} + I_{j,\sigma}^{(U)} \,.
\label{eq:I0_Iu_decompose} 
\end{align}

The linear-response contribution to $I_{j,\sigma}^{(0)}$
arises from the noninteracting part $\bm{\Pi}_0^{j;(s)}$, 
which can be expressed in the following form 
using Eqs.\ \eqref{eq:lesser_self_0}--\eqref{eq:Keldysh_self_0}:  
\begin{align}
& 
 \!\!\!\! \! 
\left. 
\frac{\partial \bm{\Pi}_0^{j';(s)}}{\partial eV_{j}^{}}\right|_{eV=0}^{} 
\! =   
4i \, \Bigl[  \,
\Gamma_{j}^{}  \, \bm{\tau}_3^{}  
-  \eta^{(s)}\Gamma_N^{}\, \delta_{jj'}   
 \bm{1}^{} 
\,\Bigr]\left(-\frac{\partial f(\omega)}{\partial \omega} \right) , 
\label{eq:Pi_0_linear}
\end{align} 
where $\eta^{(s)}$  is  a sign factor defined by  
$\eta^{(e)} = 1$ and $\eta^{(h)} = -1$,  
and  $\Gamma_N^{} \equiv 
\Gamma_L^{} +\Gamma_R^{}$.

The interacting counterpart $\bm{\Pi}_U^{j;(s)}$ 
 exhibits $\omega^2$ and $T^2$ dependences at low energies, 
reflecting the finite lifetime of the Bogoliubov quasiparticles. 
We will show more specifically 
in Sec.\ \ref{subsec:I^U} and Appendix \ref{sec:current_vertex_sc} 
that the first derivative of 
$\bm{\Pi}_U^{j;(s)}$ with respect to $eV$  
exhibits the following low-energy asymptotic form at $eV=0$, 
\begin{align}
\Bigl[\, \omega^2  + (\pi T)^2\, \Bigr] 
\left(-\frac{\partial f(\omega)}{\partial \omega} \right) \,. 
\end{align}
Therefore,  the linear-response term of $I_{j,\sigma}^{(U)}$, 
arising from the interacting part $\bm{\Pi}_U^{j;(s)}$,  
 is proportional to $T^2$ at low temperatures and vanishes at $T=0$.   
This $T^2$ dependence 
can be obtained by setting the frequency and temperature to $\omega=T=0$ 
in $\bm{G}_{dd}^{r}(\omega)$ and $\bm{G}_{dd}^{a}(\omega)$ 
appearing on the right-hand side of Eqs.\ \eqref{eq:current_up_formula} 
and \eqref{eq:current_down_formula}.
Consequently, the three-body contribution arises not from 
$I_{j,\sigma}^{(U)}$ but from $I_{j,\sigma}^{(0)}$,  
for which the $\omega^2$ and $T^2$ terms in     
 $\bm{G}_{dd}^{r}(\omega)$ and $\bm{G}_{dd}^{a}(\omega)$ 
generate the $T^2$ dependence of the conductance  
through Eqs.\ \eqref{eq:current_up_formula} and \eqref{eq:current_down_formula}.

\subsection{Linear-response current $I_{j,\sigma}^{(0)}$ up to order $T^2$}

In order to obtain the  $T^2$ contribution to the conductance arising from 
$I_{j,\sigma}^{(0)}$,  the Green's functions  
  $\bm{G}_{dd}^{r}(\omega)$ and $\bm{G}_{dd}^{a}(\omega)$, 
appearing on the right-hand sides of 
 Eqs.\ \eqref{eq:current_up_formula} and \eqref{eq:current_down_formula},   
need to be expanded up to terms of order $\omega^2$ and $T^2$, 
as mentioned above. 
More specifically, these terms arise from  
$G_{\gamma,\sigma}^{r}(\omega)$ 
defined in Eq.\ \eqref{eq:GreenBogoliubovTrans}, 
whose low-energy behavior has been investigated in detail 
in previous works \cite{FMvDM2018,AO2017_II}.   
The exact low-energy expansion of the corresponding self-energy 
$\Sigma_{\gamma,\sigma}^{r}(\omega)$  
 is given in Appendix \ref{sec:microscopic_FL}.  
In particular, the three-body contribution originates from its real part, 
presented in Eq.\ \eqref{eq:ReSelf}, 
and governs the next-to-leading-order transport in the Fermi-liquid regime.

We calculate $I_{j,\sigma}^{(0)}$  
by substituting Eq.\ \eqref{eq:Pi_0_linear}
into Eqs.\ \eqref{eq:current_up_formula} and \eqref{eq:current_down_formula},    
and then performing the Bogoliubov transformation,  
$\bm{\mathcal{U}}^{}  =  e^{-i \frac{\theta}{2} \bm{\tau}_2}$ 
 defined in Eq.\ \eqref{eq:Bogoliubov_rotation},  
which diagonalizes  $\bm{\mathcal{H}}_d^{0}$,
\begin{align}
&\!\! 
\bm{\mathcal{U}}^\dagger\,
\bm{\mathcal{H}}_d^{0} \  
\bm{\mathcal{U}}^{}  
\, =\, E^{}_A \bm{\tau}_3
-b \bm{1} \,.
 \label{eq:E_A_def}
\end{align}
Here, 
 $E_A^{}=\sqrt{\xi_d^2+\Gamma_S^2}$, 
$\sin \theta = \frac{\Gamma_S^{}}{E^{}_A}$, 
$\cos \theta = \frac{\xi_d^{}}{E^{}_A}$, and 
\begin{align}
 & 
\bm{\mathcal{U}}^\dagger
 \,\bm{\tau}_1 \,
\bm{\mathcal{U}}^{}
 \, = \,   
 \cos \theta  \, \bm{\tau}_1 + \sin  \theta  \, \bm{\tau}_3 
\,, 
\rule{0cm}{0.6cm}
\\
& 
\bm{\mathcal{U}}^\dagger
\,\bm{\tau}_3 \,
\bm{\mathcal{U}}^{}
\, = \,   
- \sin \theta \, \bm{\tau}_1 
+\cos \theta  \, \bm{\tau}_3 
\,. 
\end{align}
As shown in Eq.\ \eqref{eq:GreenBogoliubovTrans},  
$\widetilde{\bm{G}}^r_{}(\omega)
  \equiv \bm{\mathcal{U}}^\dagger 
\bm{G}^r_{dd}(\omega) \,\bm{\mathcal{U}}^{}$ 
corresponds to the propagators of Bogoliubov quasiparticles:  
$\widetilde{G}_{1}^{r}(\omega)
\,=\,
 G_{\gamma,\uparrow}^{r}(\omega)$, and
$\widetilde{G}_{2}^{r}(\omega)
\,=\,
 -G_{\gamma,\downarrow}^{a}(-\omega)$.
Using, in addition, the low-energy asymptotic form of the retarded self-energy 
$\Sigma_{\gamma,\sigma}^{r}(\omega)$ 
given in 
 Eqs.\ \eqref{eq:ImSelf} and \eqref{eq:ReSelf}: 
we obtain the $T^2$ contribution to conductance arising from $I_{j,\sigma}^{(0)}$,  
\begin{widetext}
\begin{align}
& 
\!\!\!\!\!\! 
\sum_{\sigma}
\left.\frac{\partial I_{j',\sigma}^{(0)}}{\partial V_j}
\right|_{eV=0}^{} =  
\frac{e^2}{h} \int_{-\infty}^{\infty} \! d\omega \,
\left(-i\eta_{j'}^{}\right) \Gamma_{j'}^{}  
\Biggl[\,
\left\{
\bm{G}_{dd}^{r}(\omega)\,
\left. 
\frac{\partial \bm{\Pi}_0^{j';(e)}}{\partial eV_{j}^{}}
\right|_{eV=0}^{} 
\bm{G}_{dd}^{a}(\omega)
\right\}_{11}^{}
 - 
\left\{
\bm{G}_{dd}^{r}(\omega)\,
\left. 
\frac{\partial \bm{\Pi}_0^{j';(h)}}{\partial eV_{j}^{}}
\right|_{eV=0}^{} 
\bm{G}_{dd}^{a}(\omega)
\right\}_{22}^{}
\,\Biggr]
\nonumber \\
& 
\qquad \qquad \qquad 
=  \, 
\frac{e^2}{h} \int_{-\infty}^{\infty} \! d\omega 
\left(-\frac{\partial f}{\partial \omega} \right)\, 
\frac{(-4) \eta_{j'}^{}\Gamma_{j'}^{}}{\Gamma_N^{}}
\Biggl\{\,
\delta_{jj'} \,\Gamma_N^{2} \,
\mathrm{Tr} \Bigl[ \bm{G}_{dd}^{r}(\omega)\,
\bm{G}_{dd}^{a}(\omega) \Bigr]
\, -  \,
\frac{\Gamma_{j}^{}}{\Gamma_N^{}}  \,
\Gamma_N^{2}
\mathrm{Tr}\Bigl[
\bm{G}_{dd}^{r}(\omega)\,
 \bm{\tau}_3^{}  
\,\bm{G}_{dd}^{a}(\omega)\, 
 \bm{\tau}_3^{}  
\Bigr]
\,\Biggr\}
\nonumber
\\
& 
\qquad \qquad \qquad 
=  \, 
\frac{e^2}{h} 
\frac{(-4) \eta_{j'}^{}\Gamma_{j'}^{}}{\Gamma_N^{}}
\Biggl[\,
\left(\delta_{jj'} \,-\,\frac{\Gamma_{j}^{}}{\Gamma_N^{}}
\right) 
\left\{ \sum_{\sigma}  \sin^2 \delta_{\sigma}^{}  
\,-\,a_\mathrm{I}^{(0)} \left(\pi T\right)^2
\right\}
+\frac{\Gamma_{j}^{}}{\Gamma_N^{}}
 \Bigl\{\,
\sin^2 ( \delta_{\uparrow}^{} + \delta_{\downarrow}^{} ) 
 \, + \, a_\mathrm{II}^{(0)} \left(\pi T\right)^2 
\,\Bigr\}
\, \sin^2 \theta
\,\Biggr], 
 \label{eq:Pi_0_result}
\rule{0cm}{0.7cm}
\end{align}
\end{widetext}
where  $\eta_R^{}=1$ and $\eta_L^{}=-1$. 
The coefficients $a_\mathrm{I}^{(0)}$ and $a_\mathrm{II}^{(0)}$ 
of the $T^2$ terms are given by
\begin{align}
& 
a_\mathrm{I}^{(0)} \equiv   \, 
- \frac{\pi^2}{3} 
\sum_{\sigma} 
\Biggl[\,
\cos 2 \delta_{\sigma}^{}
 \, \chi_{\sigma\sigma}^{2}
- 
4\sin^2 \delta_{\sigma}^{} 
\, \chi_{\uparrow\downarrow}^{2}
\,
\nonumber \\
& 
\qquad \qquad \qquad  \quad 
- \frac{\sin 2 \delta_{\sigma}^{}}{2\pi} 
\left(\,
\chi_{\sigma\sigma\sigma}^{[3]}
 \, +\,   
\chi_{\sigma \overline{\sigma}\overline{\sigma}}^{[3]}
\,\right)
\,\Biggr] , 
 \end{align}
\begin{align}
& \! 
a_\mathrm{II}^{(0)} \equiv    \frac{\pi^2}{3}
\Biggl[ 
\cos( 2\delta_{\uparrow}^{}+2\delta_{\downarrow}^{})
\bigl(
\chi_{\uparrow\uparrow}^{}
- \chi_{\downarrow\downarrow}^{}
\bigr )^2
 -  
8 \sin^2 (\delta_{\uparrow}^{}+\delta_{\downarrow}^{})\,
\chi_{\uparrow\downarrow}^{2}
\nonumber \\
& 
\qquad \quad \ \  
-
\frac{\sin (2 \delta_{\uparrow}^{} + 2 \delta_{\downarrow}^{}) 
 }{2\pi}
\left(\,
\chi_{\uparrow\uparrow\uparrow}^{[3]}\,  
 + 
\chi_{\uparrow\downarrow\downarrow}^{[3]}
+ 
\chi_{\downarrow\downarrow\downarrow}^{[3]}\,  
 +   
\chi_{\downarrow\uparrow\uparrow}^{[3]}
\,\right)
\Biggr] ,
 \end{align}
and $\overline{\sigma}$ represents the opposite spin component to $\sigma$.
To obtain the last line of Eq.\ \eqref{eq:Pi_0_result},
we have used the diagonal representation of 
$\bm{G}_{dd}^{r}$ and $\bm{G}_{dd}^{a}$,  
\begin{align}
\mathrm{Tr} \Bigl[ \bm{G}_{dd}^{r}(\omega)\,\bm{G}_{dd}^{a}(\omega) \Bigr] 
\, = & \  
\mathrm{Tr} \Bigl[\widetilde{\bm{G}}_{}^{r}(\omega)\,\widetilde{\bm{G}}_{}^{a}(\omega) 
\Bigr] 
\nonumber \\
 = & \  
\Big|\widetilde{G}_{1}^{r}(\omega)\Bigr|^2 
+ \,\Bigl|\widetilde{G}_{2}^{r}(\omega)\Bigr|^2 ,
\label{eq:I0_tr_GrGa}
\end{align}
and 
\begin{align}
& \ \ 
\mathrm{Tr} \Bigl[
\bm{G}_{dd}^{r}\,\bm{\tau}_3\,
\bm{G}_{dd}^{a}\,\bm{\tau}_3\,
\Bigr] 
\, = \, 
\mathrm{Tr} \Bigl[
\widetilde{\bm{G}}_{}^{r}\,
\bm{\mathcal{U}}^{\dagger} \,\bm{\tau}_3\,\bm{\mathcal{U}}^{} \,
\widetilde{\bm{G}}_{}^{a}\,
\bm{\mathcal{U}}^{\dagger} \,\bm{\tau}_3\,\bm{\mathcal{U}}^{} \,
\Bigr] 
\nonumber \\ 
& =  \ 
\mathrm{Tr} \Bigl[
\widetilde{\bm{G}}_{}^{r}  
\Bigl( \cos \theta  \, \bm{\tau}_3  -  \sin \theta \, \bm{\tau}_1 \Bigr)\, 
\widetilde{\bm{G}}_{}^{a}
\Bigl( \cos \theta  \, \bm{\tau}_3 - \sin \theta \, \bm{\tau}_1 \Bigr) 
\Bigr] 
\nonumber \\ 
 &=  \ 
\mathrm{Tr} \Bigl[
\widetilde{\bm{G}}_{}^{r}  
 \bm{\tau}_3 \,
\widetilde{\bm{G}}_{}^{a}  
  \bm{\tau}_3  \,
\Bigr]  \cos^2 \theta
+ 
\mathrm{Tr} \Bigl[
\widetilde{\bm{G}}_{}^{r} 
  \bm{\tau}_1  \,
\widetilde{\bm{G}}_{}^{a}  
  \bm{\tau}_1  \,
\Bigr]  \sin^2 \theta
\nonumber \\ 
& =  \ 
\left[\, \bigl|\widetilde{G}_{1}^{r}\bigr|^2 
+ \,\bigl|\widetilde{G}_{2}^{r} \bigr|^2 
\,\right] \,\cos^2 \theta
+
\left[\, \widetilde{G}_{1}^{r}\, \widetilde{G}_{2}^{a}
\,+\widetilde{G}_{2}^{r}\, \widetilde{G}_{1}^{a}
\,\right] \,\sin^2 \theta
. 
\label{eq:I0_tr_GrTau3GaTau3}
\end{align}

The contribution from the term 
 $\mathrm{Tr} \Bigl[ \bm{G}_{dd}^{r} \bm{G}_{dd}^{a}\Bigr]$  
 can be expressed in the following form, 
using Eqs.\ \eqref{eq:ImSelf} and \eqref{eq:ReSelf}: 
 \begin{align}
& 
\int^{\infty}_{-\infty} \! d \omega\,
\left(- \frac{\partial f(\omega)}{\partial \omega }\right) 
\Gamma_N^2 \,
\mathrm{Tr} \Bigl[ \bm{G}_{dd}^{r}(\omega)\,
\bm{G}_{dd}^{a}(\omega) \Bigr]
\nonumber \\
& = \int^{\infty}_{-\infty} \! d \omega\,
\left(- \frac{\partial f(\omega)}{\partial \omega }\right) 
\Gamma_N^2 
\sum_{\sigma}
\left|G_{\gamma,\sigma}^{r}(\omega)\right|^2 
\nonumber \\ 
& =   \     
\sum_{\sigma} 
 \sin^2 \delta_{\sigma}^{}  
\,-\,a_\mathrm{I}^{(0)}
\left(\pi T\right)^2 
\ + \ O(T^4)
\,.
\label{eq:I0_unit_part}
 \end{align}
The dependence on the Bogoliubov angle $\theta$ arises from 
the counterpart involving  
$\mathrm{Tr}\Bigl[
\bm{G}_{dd}^{r}\,
 \bm{\tau}_3^{}  
\,\bm{G}_{dd}^{a}\, 
 \bm{\tau}_3^{}  
\Bigr]$, 
\begin{align}
&
\int^{\infty}_{-\infty} \! d \omega\,
\left(- \frac{\partial f(\omega)}{\partial \omega }\right) \,
\Gamma_N^2 
\Biggl\{\,
\mathrm{Tr} \Bigl[ \bm{G}_{dd}^{r}(\omega)\,
\bm{G}_{dd}^{a}(\omega) \Bigr]
\nonumber \\
& 
\qquad\qquad\qquad\qquad\qquad\qquad
-\mathrm{Tr}\Bigl[
\bm{G}_{dd}^{r}(\omega)\,
 \bm{\tau}_3^{}  
\,\bm{G}_{dd}^{a}(\omega)\, 
 \bm{\tau}_3^{}  
\Bigr]
\,\Biggr\}
\nonumber \\
&
= \sin^2 \theta \, 
\int^{\infty}_{-\infty} \! d \omega\,
\left(- \frac{\partial f(\omega)}{\partial \omega }\right) \,
\Gamma_N^2
 \biggl[\,
 \sum_{\sigma} 
\left|G_{\gamma,\sigma}^{r}(\omega)\right|^2 
\nonumber \\
& 
\qquad\qquad\qquad
+G_{\gamma,\uparrow }^{r}(\omega)\,G_{\gamma, \downarrow}^{r}(-\omega)
+G_{\gamma,\uparrow}^{a}(\omega)\,G_{\gamma,\downarrow}^{a}(-\omega)
\,\biggr]
\nonumber\\
&
= \ 
 \Bigl[\,
\sin^2 ( \delta_{\uparrow}^{} + \delta_{\downarrow}^{} ) 
 \, + \, a_\mathrm{II}^{(0)} \left(\pi T\right)^2 
\,\Bigr]
\, \sin^2 \theta 
\ + \ O(T^4)
%
%
 \,.
\rule{0cm}{0.7cm}
\label{eq:I0_tau3_part}
 \end{align}
Note that the integrals over $\omega$ can be  
evaluated using the relations,
\begin{align}
& \int_{-\infty}^{\infty} \!\! d\omega   \,
\left(-\frac{\partial f(\omega)}{\partial \omega}\right)
\,= \, 1   \,, 
\\ 
 & 
\int_{-\infty}^{\infty} \!\! d\omega   \,
\omega^2 
\left(-\frac{\partial f(\omega)}{\partial \omega}\right)
\,= \,\frac{1}{3}(\pi T)^2  \,.  
\end{align}


\begin{figure}[b]
\leavevmode 
\begin{minipage}[t]{1\linewidth}
\includegraphics[width=0.6\linewidth]{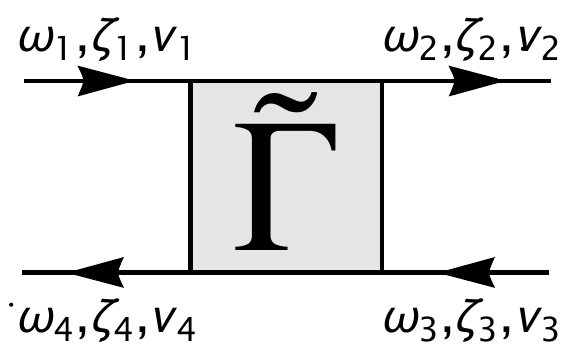}
\caption{Keldysh vertex correction 
$
\widetilde{\Gamma}_{\zeta_1\zeta_2;\zeta_3\zeta_4}^{\nu_1\nu_2;\nu_3\nu_4}
(\omega_1,\omega_2;\omega_3\omega_4)$ between 
Bogoliubov quasiparticles. Here, $\zeta_i$ ($=1,2$) labels the operators 
$\widetilde{\psi}^{}_{d,1} \equiv \gamma^{}_{d\uparrow}$ and 
$\widetilde{\psi}^{}_{d,2} \equiv \gamma^{\dagger}_{d\downarrow}$, 
defined in Eq.\ \eqref{eq:H_dot_U_Bogo} 
with respect to the principal axes in the Nambu pseudo-spin space. 
The superscript $\nu_i $ 
specifies the branches of Keldysh time-loop contour, 
for which $\nu=-$ and $+$ denote 
the forward and backward branches, respectively.
Frequency conservation requires 
 $\omega_1+\omega_3=\omega_2+\omega_4$. 
}
\label{fig:Keldysh_vertex}
\end{minipage}
\end{figure}


\subsection{Linear-response current $I_{j,\sigma}^{(U)}$ up to order $T^2$}
\label{subsec:I^U}

We next evaluate the $T^2$ term in the linear-response current
$I_{j,\sigma}^{(U)}$, arising from the interacting parts 
$\bm{\Pi}_{U}^{j;(e)}$ and $\bm{\Pi}_{U}^{j;(h)}$, 
which enter Eqs.\ \eqref{eq:current_up_formula} and
\eqref{eq:current_down_formula}, respectively, 
 through Eq.\ \eqref{eq:Pi_0+U}. 
Here, we outline the derivation,   
leaving the full details to Appendix \ref{sec:current_vertex_sc}.

To this end, we rewrite  
the Coulomb interaction part 
 $H_d^{U}$ of the Hamiltonian, defined in Eq.\ \eqref{eq:H_dot_U},  
into several equivalent forms: 
\begin{align}
 H_d^{U} \equiv & \  
\frac{U}{2}\bigl( n_d^{} -1 \bigr)^2 
= 
\frac{U}{2}\left(
  \widetilde{\psi}_{d,1}^{\dagger}  \widetilde{\psi}_{d,1}^{}
 \,- \, \widetilde{\psi}_{d,2}^{\dagger}  \widetilde{\psi}_{d,2}^{}
  \right)^2 
\nonumber 
\\
= & \ 
\frac{U}{2}\left(
\gamma^\dagger_{d\uparrow}\gamma^{}_{d\uparrow} 
 + \gamma^\dagger_{d\downarrow} \gamma^{}_{d\downarrow} -1 
\right)^2 . 
\rule{0cm}{0.6cm} 
\label{eq:H_dot_U_Bogo}  
\end{align}
In particular, at finite bias voltages, 
the Keldysh perturbation theory with respect to $H_d^{U}$ 
is most naturally formulated using the second representation in terms of     
$\widetilde{\psi}^{}_{d,1} \equiv \gamma^{}_{d\uparrow}$ and 
$\widetilde{\psi}^{}_{d,2} \equiv \gamma^{\dagger}_{d\downarrow}$.


\begin{figure}[t]
\leavevmode 
\begin{minipage}[t]{1\linewidth}
 \includegraphics[width=0.46\linewidth]{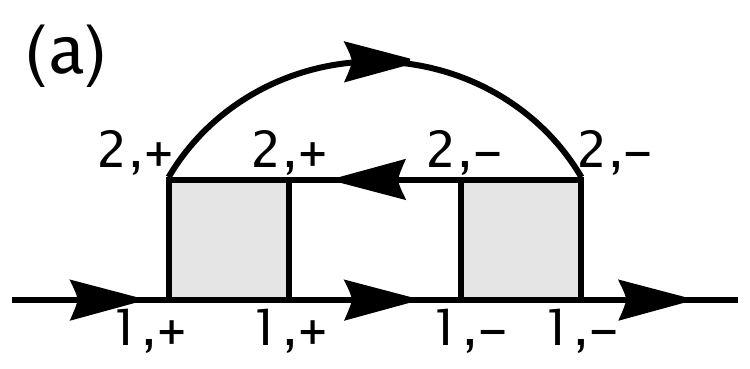}
\rule{0.08\linewidth}{0cm}
\includegraphics[width=0.42\linewidth]{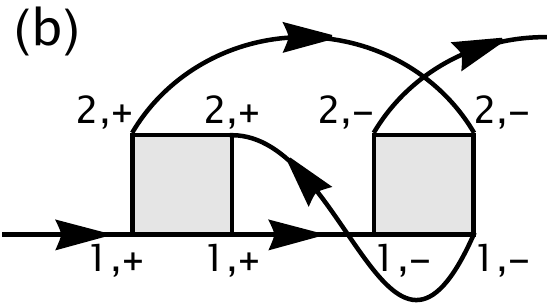}
\caption{
Feynman diagrams for 
(a) $\widetilde{\Sigma}_{U;11}^{-+}(\omega)$, and  
(b) $\widetilde{\Sigma}_{U;21}^{-+}(\omega)$ 
which yield the imaginary parts of order $\omega^2$ and $T^2$ 
contributing to the linear conductance through 
$\widetilde{\bm{\Pi}}_{U}^{j;(s)}$ with $s$ ($=e,h$).
The shaded squares represent 
 the scattering amplitudes of quasiparticles, 
given by the full vertex corrections  at $T=eV=0$: 
$\widetilde{\Gamma}_{12;21}^{--;--}(0, 0; 0, 0)
=- \widetilde{\Gamma}_{12;21}^{++;++}(0, 0; 0, 0)$. 
The off-diagonal components ($\zeta \neq \zeta'$)   
of the scattering amplitude are real and finite. 
}
\label{fig:Keldysh_self_K_renorm}
\end{minipage}
\end{figure}


The effects of multiple scattering processes caused by $H_d^{U}$ 
can be taken into account using the vertex function, 
as described in Fig.\ \ref{fig:Keldysh_vertex}. 
Specifically, the two processes 
 described in Figs.\ \ref{fig:Keldysh_self_K_renorm} (a)
 and \ref{fig:Keldysh_self_K_renorm} (b),  
 give rise to the $T^2$ term in the linear-response current $I_{j,\sigma}^{(U)}$.  
In the interacting self-energy part $\bm{\Pi}_{U}^{j;(s)}$, 
these two processes can be formulated in terms of 
the superconducting collision integrals of Bogoliubov quasiparticles 
 shown in Appendix \ref{sec:current_vertex_sc},  
 leading to an imaginary part 
of order $\omega^2$ and $T^2$, i.e., 
 the same order as the damping rate of quasiparticles in the Fermi-liquid regime.  
Consequently, the contributions from  
$\bm{\Pi}_{U}^{j;(s)}$ for $s=e,h$ 
can be expressed in the following form: 
\begin{align}
& 
\!\!\!\!
 \left\{\,
\bm{G}_{dd}^{r}(0)
\left.
\frac{\partial\,\bm{\Pi}_{U}^{j';(e)}}{\partial eV_j}\right|_{eV=0}^{}
\! \bm{G}_{dd}^{a}(0) 
\,\right\}_{11}^{}
\nonumber \\
&  
\!\!\!\!
= \, 
 -i\,  \frac{2\pi^2 \chi_{\uparrow\downarrow}^{2}}{\Gamma_N}\, 
\Bigl[\,  \omega^2 + (\pi T)^2\,\Bigr]
\left(-\frac{\partial f(\omega)}{\partial\omega} \right)
\rule{0cm}{0.7cm}  
\nonumber \\
& 
\!
 \times  
\left [\, 
 \left(\delta_{jj'}  -  \frac{\Gamma_{j}}{\Gamma_N^{}} \right)
 +  \frac{2\Gamma_{j}}{\Gamma_N^{}} 
\, \sin^2 (\delta_{\uparrow}+\delta_{\downarrow}) 
\, \sin^2 \theta\Bigr\} 
 \, \right ] ,
\rule{0cm}{0.7cm}  
\label{eq: Pi_U_result}
\end{align}
and
\begin{align}
& 
\!\!\!\!\!\!\!\!\! 
\!\!\!\!\!\!\!\!\!
\!\!\!\! 
-\left\{\,
\bm{G}_{dd}^{r}(0)
\left.
\frac{\partial\,\bm{\Pi}_{U}^{j';(h)}}{\partial eV_j}
\right|_{eV=0}^{}
\! \bm{G}_{dd}^{a}(0) \,\right\}_{22}^{} 
\rule{0cm}{0.7cm}  
\nonumber \\
& 
\!\!\!\!\!\!\!\!\! 
\!\!\!\!\!\!\!     
= \,  
  \left\{\,
\bm{G}_{dd}^{r}(0)
\left.
\frac{\partial\,\bm{\Pi}_{U}^{j';(e)}}{\partial eV_j}\right|_{eV=0}^{}
\! \bm{G}_{dd}^{a}(0) 
\,\right\}_{11}^{}\,.
\label{eq: Pi_U_result_add}
\rule{0cm}{0.7cm}  
\end{align}
Here, the off-diagonal susceptibility $\chi_{\uparrow\downarrow}^{}$  
appears through the Ward identity, 
$\chi_{\uparrow\downarrow}^{} =\,
- \Gamma_{\uparrow\downarrow;\downarrow\uparrow}^{--;--}(0,0;0,0) 
\,\rho_{d\uparrow}^{}\,\rho_{d\downarrow}^{}$, 
which relates the susceptibility to the vertex function  
 of the Bogoliubov quasiparticles, 
as explained in Appendix \ref{sec:microscopic_FL}.
 The $T^2$ term of the linear-response current $I_{j,\sigma}^{(U)}$ 
can be obtained by integrating Eq.\ \eqref{eq: Pi_U_result}  
over $\omega$,  while setting $\omega=T=0$  
 in the Green's functions $\bm{G}_{dd}^{r}$ and $\bm{G}_{dd}^{a}$ 
appearing in that equation:   
\begin{align}
& \left. \frac{\partial I_{j',\sigma}^{(U)}}{\partial V_j}
\right|_{eV=0}^{}
\nonumber \\ 
 & =    
\frac{e^2}{h}
  (-i\eta_{j'}^{})\Gamma_{j'}^{} \!\!  
\int_{-\infty}^{\infty} \!\! d\omega 
%
\left\{
\bm{G}_{dd}^{r}(0)
\left.
\frac{\partial\,\bm{\Pi}_{U}^{j';(e)}}{\partial eV_j}\right|_{eV=0}^{}
\!\!
\bm{G}_{dd}^{a}(0) 
\right\}_{11}^{} 
\nonumber \\ 
& = \, 
\frac{e^2}{h} \   
(-4\eta_{j'}^{})\, 
\frac{\Gamma_{j'}^{}}{\Gamma_N}\, 
\frac{2\pi^2}{3}\,
\chi_{\uparrow\downarrow}^{2}
\, \left(\pi T\right)^2\, 
\rule{0cm}{0.7cm}  
\nonumber \\ 
&  \   \times 
\left[ \, 
\left(\delta_{jj'} - \frac{\Gamma_{j}}{\Gamma_N^{}} \right)
 + \frac{2\Gamma_{j}}{\Gamma_N^{}}
\sin^2 (\delta_{\uparrow}+\delta_{\downarrow}) \, \sin^2 \theta 
 \, \right ]  ,
\rule{0cm}{0.6cm}  
\label{eq:I^U_result}
\end{align}
where  $\eta_R^{}=1$, and $\eta_L^{}=-1$. 
This result holds for both spin components, $\sigma=\uparrow$ and $\downarrow$. 
Consequently, 
the conductance formula presented in  
Sec.\ \ref{sec:multi_terminal_formulation} follows 
from $I_{j,\sigma}^{(U)}$ obtained here,   
conbined with the other contribution $I_{j,\sigma}^{(0)}$ 
given in Eq.\ \eqref{eq:Pi_0_result}.

\section{Summary}
\label{summary}

We have presented a complete microscopic Fermi-liquid description 
of crossed Andreev reflection in a quantum dot connected to multiple terminals 
consisting of superconducting and normal leads at finite temperatures,
in the limit of  a large superconducting gap $|\Delta_S^{}| \to \infty$, 
where single-electron tunneling 
(or tunneling of unpaired Bogoliubov quasiparticles) 
into the superconducting lead is suppressed 
and Cooper-pair tunneling is described by 
the effective Hamiltonian $\mathcal{H}_\mathrm{eff}^{}$ 
given in Eq.\ \eqref{eq:Heff_single}.
Specifically, we focus on the role of three-body correlations of Bogoliubov quasiparticles, 
which play an essential role in the $T^2$ term of conductance. 
We show that the nonlocal and local conductances, 
$g_{RL}^{}$ and $g_{LL}^{}$, are determined by 
the transmission probabilities of Cooper pairs 
 between the quantum dot and the superconducting lead,
$\mathcal{T}_{\mathrm{CP}}^{}(T)$, 
 and of Bogoliubov quasiparticles between the two normal leads via the quantum dot,  
$\mathcal{T}_{\mathrm{BG}}^{}(T)$.  
In particular, the Bogoliubov-quasiparticle and 
 Cooper-pair contributions interfere destructively 
for the nonlocal conductance $g_{RL}^{}$, 
whereas they add constructively for the local conductance $g_{LL}^{}$. 
Experimentally, these two contributions, $\mathcal{T}_{\mathrm{BG}}^{}(T)$ 
and $\mathcal{T}_{\mathrm{CP}}^{}(T)$,  
can be extracted separately by measuring both $g_{LL}^{}$ and $g_{RL}^{}$.

Our formulation is based on the Anderson impurity model 
and employs the Nambu-Keldysh Green's function approach.
We have shown that the $T^2$ term of the linear conductance  
can be expressed in terms of the three-body correlation functions 
$\chi^{[3]}_{\sigma_1\sigma_2\sigma_3}$,  
in addition to the phase shifts $\delta_{\sigma}^{}$, 
the linear susceptibilities $\chi_{\sigma_1\sigma_2}^{}$, and
 the Bogoliubov rotation angle $\theta$.
The resulting conductance formula is given 
in Sec.\ \ref{subsec:General_formula_result}, 
with its derivation presented in Sec.\ \ref{sec:3body_correction_derivation}.  
The derivation is based on the exact low-energy asymptotic forms of the self-energy 
and vertex corrections of the Bogoliubov quasiparticles 
up to terms of order $\omega^2$ and $T^2$, 
as detailed in the Appendixes \ref{sec:microscopic_FL} and \ref{sec:current_vertex_sc},
 where the superconducting collision integral is also formulated.

We have also examined the behavior of 
the nonlocal and local conductances 
over a wide parameter range at zero magnetic field 
 using the numerical renormalization group approach (NRG).
The Kondo effect dominates the low-temperature transport 
for large $U$ in the region $E_A^{} \lesssim U/2$, 
corresponding to the area 
inside the semicircle illustrated in Fig.\ \ref{fig:SingleDotPhase},  
where $E_A^{} = \sqrt{\xi_d^2 + \Gamma_S^2}$ is the Andreev level. 
In contrast,  for $E_A^{} \gtrsim U/2$, i.e., 
outside the semicircle, the Cooper-pair contribution 
$\mathcal{T}_{\mathrm{CP}}^{}(T)$
becomes comparable to the Bogoliubov-quasiparticle contribution 
$\mathcal{T}_{\mathrm{BG}}^{}(T)$. 
In this regime, 
the $T^2$ coefficient $C_{T}^{\mathrm{CP}}$  
for $\mathcal{T}_{\mathrm{CP}}^{}(T)$ depends significantly 
on the Bogoliubov angle $\theta$. 
Consequently, the $T^2$ coefficient 
$C_{T}^{\mathrm{BG}} +2 C_{T}^{\mathrm{CP}}$ 
for the nonlocal conductance $g_{RL}^{}$ 
 exhibits a negative minimum in the crossover region $E_A^{} \simeq U/2$ 
 at $\theta \simeq \pi/2$, as a result of the competition between the two contributions.
In particular, for the angular range $\pi/3 < \theta < 2\pi/3$,    
this coefficient changes sign and becomes positive for $E_A^{} \gtrsim U/2$. 
In contrast, the $T^2$ coefficient 
$C_{T}^{\mathrm{BG}} -(2\Gamma_L^{}/\Gamma_R^{}) 
\,C_{T}^{\mathrm{CP}}$ 
for the local conductance $g_{LL}^{}$ 
exhibits a monotonic dependence on $E_A^{}$, 
reflecting the constructive interplay between Bogoliubov quasiparticles and Cooper pairs.

 The coefficients $C_{T}^{\mathrm{BG}}$ and $C_{T}^{\mathrm{CP}}$, 
 which are determined by three-body correlations, 
can also be experimentally extracted 
by measuring both $g_{RL}^{}$ and $g_{LL}^{}$. 
The $T^2$ terms of the local and nonlocal conductances exhibit 
a strong $\theta$ dependence,  particularly for $E_A^{}\gtrsim U/2$, 
through the coherence factor $\sin^2 \theta$, 
which governs the magnitude of the Cooper-pair contribution $C_{T}^{\mathrm{CP}}$.
Therefore, the $T^2$ coefficients can also be obtained 
from measurements of a single conductance component, $g_{RL}^{}$ or $g_{LL}^{}$, 
by varying the Bogoliubov rotation angle 
$\theta=  \cot^{-1}(\xi_d^{}/\Gamma_S^{})$ 
through the gate voltage $\xi_d^{}$ or the tunnel coupling $\Gamma_S^{}$ 
between the dot and the superconducting lead. 
From the observed coefficients 
$C_{T}^{\mathrm{BG}}$ and $C_{T}^{\mathrm{CP}}$, 
the three-body contributions can, in principle, be deduced using the theoretical formulas 
 Eqs.\ \eqref{eq:cT_BG_def_b=0}--\eqref{eq:Theta_CP_def_b=0}.
While the present analysis focuses on the $T^2$ term of the linear conductance,  
the three-body correlations also play a crucial role in the 
nonlinear current of order $(eV)^3$, which will be addressed in future work.

\begin{acknowledgments}
We would like to thank Yoichi Tanaka and Yasuhiro Yamada for valuable discussions. 
This work was supported by JSPS KAKENHI Grant No.\ JP23K03284. 
\end{acknowledgments}

\appendix

\section{Fermi liquid theory for equilibrium Bogoliubov quasiparticles}
\label{sec:microscopic_FL}

Here, we provide a brief overview of the recent developments in the 
Fermi liquid theory for the Anderson impurity model, 
which we apply in this paper to interacting Bogoliubov quasiparticles.

The Fermi liquid properties of Bogoliubov quasiparticles, 
described by the effective Hamiltonian  
 $\mathcal{H}_\mathrm{eff}^{}$ of Eq.\ \eqref{eq:SingleDot_new}, 
reflect the low-energy asymptotic form of the retarded Green's function, 
\begin{align}
G_{\gamma,\sigma}^{r}(\omega)\,=& 
\ \frac{1}{\omega-E_{A,\sigma}^{}+i\Gamma_N^{}
-\Sigma_{\gamma,\sigma}^{r}(\omega)}\,.
\label{eq:Green_Dyson}
\end{align}
Fermi-liquid corrections arising from low-lying energy excitations 
can be systematically obtained by expanding the self-energy 
$\Sigma_{\gamma,\sigma}^{r}(\omega)$   
around the Fermi energy $\omega=0$.
The phase shift is given by 
$\delta_\sigma^{} 
= \cot^{-1} \frac{E_{A,\sigma}^{}
+\Sigma_{\gamma,\sigma}^{r}(0)}{\Gamma_N^{}}$ 
at $T=eV=0$.
It plays a primary role in the ground-state properties 
through the Friedel sum rule, 
 $\langle q_{d\sigma}^{} \rangle \xrightarrow{T \to 0} 
\delta_\sigma /\pi$  \cite{ShibaKorringa}.   
The spectral weight  $\rho_{d\sigma}^{}$
of impurity levels at the Fermi level can also be expressed in the form, 
 \begin{align}
 \rho_{d\sigma}^{} 
\,\equiv\, -\frac{1}{\pi}
 \mathrm{Im}\,G_{\gamma,\sigma}^{r}(0)\Big|_{T=eV=0}^{} 
\,= \,  \frac{\sin^2\delta_\sigma^{}}{\pi\Gamma_N^{}}  
\,.
 \label{eq:rho_def} 
\end{align}

The leading-order Fermi-liquid corrections are characterized by    
the wave function renormalization factor $z_\sigma^{}$ 
and  the linear susceptibilities $\chi_{\sigma\sigma'}^{}$,  
defined at $T=eV=0$ with respect to the equilibrium ground state, 
\begin{align}
\frac{1}{z_\sigma^{}}\, \equiv & \ 
1-\left.\frac{\partial \Sigma_{\gamma,\sigma}^r(\omega)}
{\partial \omega}
\right|_{\omega=0}^{} 
\,, \\
\label{eq:Zdef}
\chi_{\sigma\sigma'}^{}\, \equiv & \ 
- \frac{\partial \bigl\langle q_{d\sigma}^{}\bigr\rangle}
{\partial E_{A,\sigma'}^{}}
\ =  \  
\rho_{d\sigma}^{} 
\widetilde{\chi}_{\sigma\sigma'}^{}\,, 
\\
\widetilde{\chi}_{\sigma\sigma'}^{} 
\,\equiv& \ 
\delta_{\sigma\sigma'}^{}+
 \frac{\partial \Sigma_{\gamma,\sigma}^r(0)}
{\partial E_{A,\sigma'}^{}} 
\,. 
\label{eq:YamadaYsidaRelationAppendix}
\rule{0cm}{0.6cm}
\end{align}
Yamada and Yosida showed that 
the renormalization factor  $z_\sigma^{}$ and 
the diagonal component of $\widetilde{\chi}_{\sigma\sigma}^{}$ 
are related to each other \cite{YamadaYosida2},  as 
\begin{align}
\frac{1}{z_\sigma^{}}
\, = \, 
\widetilde{\chi}_{\sigma\sigma}^{}
\,. 
\label{eq:YamadaYsidaRelationAppendixAdd}
\end{align}
Similarly, the off-diagonal susceptibility is related to 
the causal vertex function $\Gamma_{\sigma\sigma';\sigma'\sigma}^{--;--}
(\omega,\omega';\omega',\omega)$ for Bogoliubov quasiparticles
defined with respect to  $\mathcal{H}_\mathrm{eff}^{}$ 
in Eq.\ \eqref{eq:SingleDot_new} at $T=eV=0$,   
\begin{align}
& \!\!\!\!
\chi_{\sigma\sigma'}^{} =\,
- \Gamma_{\sigma\sigma';\sigma'\sigma}^{--;--}(0,0;0,0) 
\,\rho_{d\sigma}^{}\,\rho_{d\sigma'}^{} , 
\quad  \ \ \sigma\neq \sigma' . 
\label{eq:YamadaYsidaRelation_vertex}
\end{align}

These relations were derived in the early studies of the Kondo physics 
from the Ward identity  
\cite{YamadaYosida2,YamadaYosida4,ShibaKorringa,Yoshimori},  
which can be expressed in the following form at $T=eV=0$:  
\begin{align}
\delta_{\sigma\sigma'} \frac{\partial \Sigma_{\gamma,\sigma}^{--}(\omega) }{\partial \omega} 
+ \frac{\partial \Sigma_{\gamma,\sigma}^{--}(\omega) }{\partial E_{A,\sigma'}^{}} 
 = \, - \Gamma_{\sigma\sigma';\sigma'\sigma}^{--;--}(\omega,0;0,\omega) 
\,\rho_{d\sigma'}^{}\,.
\label{eq:YYY_T0_causal} 
\end{align}
The causal self-energy $\Sigma_{\gamma,\sigma}^{--}$
and the retarded self-energy are related at $T=eV=0$ through 
\begin{align}
\Sigma_{\gamma,\sigma}^{--}(\omega) 
\, =\, 
\mathrm{Re}\,\Sigma_{\gamma,\sigma}^{r}(\omega)\,+\, i\,
\mathrm{Im}\,\Sigma_{\gamma,\sigma}^{r}(\omega)
\,\mathrm{sgn} \, \omega\,.
\label{eq:T=0selfenergy}
\end{align}
Physically, Eq.\ \eqref{eq:YYY_T0_causal} reflects 
current conservation around the impurity site \cite{Oguri2022}. 
Equations \eqref{eq:YamadaYsidaRelationAppendixAdd}
and \eqref{eq:YamadaYsidaRelation_vertex} follow from 
Eq.\ \eqref{eq:YYY_T0_causal} and from  
the property 
that the diagonal vertex component ($\sigma=\sigma'$)
vanishes  at zero frequencies  ($\omega=\omega'=0$),   
 \begin{align}
 &\Gamma_{\sigma\sigma;\sigma\sigma}^{--;--}(0,0;0,0) \,=\, 0 \,. 
\label{eq:self_w1_N}
\end{align}

It has recently been clarified 
that the vertex function for $\sigma=\sigma'$ 
also satisfies the additional  property 
\cite{AO2017_I,AO2017_II},  
\begin{align}
& \left. \frac{\partial }{\partial \omega}
\mathrm{Re}\, 
\Gamma_{\sigma\sigma;\sigma\sigma}^{--;--}(\omega,0;0,\omega) 
\right|_{\omega\to 0} =\,0  
\;.
\label{eq:self_w2_N}
\end{align}
This implies that the real part of  
$\Gamma_{\sigma\sigma;\sigma\sigma}^{--;--}(\omega,0;0,\omega)$ 
contains no linear term in $\omega$. 
As a consequence,  
  the $\omega^2$ term of the real part of  the self-energy 
can be expressed in terms of the derivative of the susceptibility, 
or equivalently, the three-body correlation function, at $T=eV=0$ 
\cite{FMvDM2018,AO2017_I,AO2017_II,AO2017_III}, 
\begin{align}
\!\!\! 
\left.
\frac{\partial^2}{\partial \omega^2}
\mathrm{Re}\,\Sigma_{\gamma,\sigma}^{r}(\omega)
\right|_{\omega \to 0}^{} 
\,=\  \frac{\partial^2 \Sigma_{\gamma,\sigma}^{r}(0)}
{\partial E_{A,\sigma}^{2}}
\ = \ 
\frac{\partial \widetilde{\chi}_{\sigma\sigma}^{}}{\partial E_{A,\sigma}^{}} \,.
\label{eq:self_w2}
\end{align}
Similarly, the coefficient of the  $T^2$ term 
in $\mathrm{Re}\,\Sigma_{\gamma,\sigma}^r(0)$ 
can also be derived from these properties  
of the vertex function \cite{AO2017_II}.

 Consequently,  
the low-energy asymptotic form of 
$\Sigma_{\gamma,\sigma}^r(\omega)$ 
 can be exactly determined exactly 
up to order $\omega^2$ and  $T^2$:
\begin{align}
& \mathrm{Im} \, \Sigma_{\gamma, \sigma}^r (\omega)\,=  \,   
-\frac{\pi}{2\rho_{d\sigma}^{}}
\chi_{\uparrow \downarrow}^2 
 \Bigl[ \,
\omega^2 
 +  (\pi T)^2 \,\Bigr] ,
\label{eq:ImSelf}
\\
&E_{A,\sigma}^{} + 
\mathrm{Re} \, \Sigma_{\gamma,\sigma}^r(\omega) \, =   \,    
\Gamma_N^{} \cot \delta_\sigma^{}
+ 
\left(1- \frac{\chi_{\sigma \sigma}^{}}{\rho_{d\sigma}^{}}\right)\, \omega
\nonumber \\
&  \quad  \ 
+\frac{1}{2}
\left(
 \chi_{\sigma\sigma\sigma}^{[3]}\,
+ 2\pi \cot \delta_\sigma^{}\,  \chi_{\sigma\sigma}^{2}\,
\right)
\,\omega^2 
+ \frac{1}{6\rho_{d\sigma}^{}}
\, \chi_{\sigma \overline{\sigma} \overline{\sigma}}^{[3]}\,
(\pi T)^2 .
\label{eq:ReSelf}
\end{align}

\begin{table*}[t]
\caption{
$\widetilde{\bm{G}}_{0}^{-+}(\omega)$:
Noninteracting lesser Green's function. 
}  
\begin{tabular}{l} 
\hline \hline
$
 \widetilde{G}_{0;11}^{-+}(\omega) 
\,= \, 
i \Gamma_N \,
\widetilde{G}_{0;1}^{r}(\omega)\,
\widetilde{G}_{0;1}^{a}(\omega)\, 
\Bigl[\,
\left(1+\cos \theta \right)
 f_\mathrm{eff}^{(e)}(\omega)  
+  \left(1-\cos \theta \right) f_\mathrm{eff}^{(h)}(\omega) 
\,\Bigr],
$
\rule{0cm}{0.45cm}
\\
$
\widetilde{G}_{0;22}^{-+}(\omega) 
\,= \, 
i \Gamma_N \,
\widetilde{G}_{0;2}^{r}(\omega)\,
\widetilde{G}_{0;2}^{a}(\omega)\,
\Bigl[\,
\left(1-\cos \theta \right)  f_\mathrm{eff}^{(e)}(\omega)  
+  \left(1+\cos \theta \right) f_\mathrm{eff}^{(h)}(\omega) 
\,\Bigr] ,
$
\rule{0cm}{0.5cm}
\\
$
\widetilde{G}_{0;21}^{-+}(\omega) 
\,= \, 
-i \Gamma_N \,
\widetilde{G}_{0;2}^{r}(\omega)\,
\widetilde{G}_{0;1}^{a}(\omega)
\,\sin \theta
\, \Bigl[\,
\, f_\mathrm{eff}^{(e)}(\omega)  - f_\mathrm{eff}^{(h)}(\omega)\,
\,\Bigr] ,
$
\rule{0cm}{0.5cm}
\\
$
\widetilde{G}_{0;12}^{-+}(\omega) 
\, = \,  
-i \Gamma_N \,
\widetilde{G}_{0;1}^{r}(\omega)\,
\widetilde{G}_{0;2}^{a}(\omega)\,
\sin \theta
\, \Bigl[\,
\, f_\mathrm{eff}^{(e)}(\omega)  - f_\mathrm{eff}^{(h)}(\omega)\,
\,\Bigr] .
$
\rule{0cm}{0.5cm}
\\
\hline
\hline
\end{tabular}
\label{tab:G0_lesser}
\end{table*}

\begin{table*}[t]
\caption{
$\widetilde{\bm{G}}_{0}^{+-}(\omega)$: 
Noninteracting greater Green's function. 
}  
\begin{tabular}{l} 
\hline \hline
$
 \widetilde{G}_{0;11}^{+-}(\omega) 
\,= \, 
-i \Gamma_N \,
\widetilde{G}_{0;1}^{r}(\omega)\,
\widetilde{G}_{0;1}^{a}(\omega)\, 
\Bigl[\,
\left(1+\cos \theta \right)
\left\{ 1-f_\mathrm{eff}^{(e)}(\omega) \right\}  
+\left(1- \cos \theta \right)
\left\{1-  f_\mathrm{eff}^{(h)}(\omega) \right\}  
\,\Bigr],
$
\rule{0cm}{0.5cm}
\\
$
\widetilde{G}_{0;22}^{+-}(\omega) 
\,= \, 
-i \Gamma_N \,
\widetilde{G}_{0;2}^{r}(\omega)\,
\widetilde{G}_{0;2}^{a}(\omega)\,
\Bigl[\,
\left(1- \cos \theta \right)
\left\{1-  f_\mathrm{eff}^{(e)}(\omega) \right\} 
+ \left(1+ \cos \theta \right)
\left\{ 1- f_\mathrm{eff}^{(h)}(\omega)  \right\}
\,\Bigr] ,
$
\rule{0cm}{0.5cm}
\\
$
\widetilde{G}_{0;21}^{+-}(\omega) 
\,= \, 
i \Gamma_N \,
\widetilde{G}_{0;2}^{r}(\omega)\,
\widetilde{G}_{0;1}^{a}(\omega)
\,\sin \theta
\, \Bigl[
\,
 \left\{1- f_\mathrm{eff}^{(e)}(\omega) \right\} \,-\, 
 \left\{1 - f_\mathrm{eff}^{(h)}(\omega) \right\}
\,\Bigr] ,
$
\rule{0cm}{0.5cm}
\\
$
\widetilde{G}_{0;12}^{+-}(\omega) 
\, = \,  
i \Gamma_N \,
\widetilde{G}_{0;1}^{r}(\omega)\,
\widetilde{G}_{0;2}^{a}(\omega)\,
\sin \theta
\, \Bigl[\,
 \left\{1- f_\mathrm{eff}^{(e)}(\omega) \right\} \,-\, 
 \left\{1 - f_\mathrm{eff}^{(h)}(\omega) \right\}
\,\Bigr] .
$
\rule{0cm}{0.5cm}
\\
\hline
\hline
\end{tabular}
\label{tab:G0_greater}
\end{table*}


\begin{table*}[t]
\caption{
Order-$U^2$ self-energy 
$\widetilde{\bm{\Sigma}}_{U}^{(a)\nu\nu'}(\omega)$ 
corresponding to Fig.\ \ref{fig:2nd_order_self-energy_rev} (a). 
}  
\begin{tabular}{l} 
\hline \hline
$
\widetilde{\Sigma}^{(a) -+}_{U;11}(\omega)
= 
-U^2\!\! 
\int\!\! \frac{d\varepsilon_1}{2\pi}
\!\!\int\!\!\frac{d\varepsilon_2}{2\pi}
\,\widetilde{G}_{0;22}^{-+}(\varepsilon_2) 
\,\widetilde{G}_{0;22}^{+-}(\varepsilon_1+\varepsilon_2-\omega) 
\,\widetilde{G}_{0;11}^{-+}(\varepsilon_1),
$
\rule{0cm}{0.5cm}
\\
$
\widetilde{\Sigma}^{(a) +-}_{U;11}(\omega)
=  
-U^2\!\!
\int\!\!\frac{d\varepsilon_1}{2\pi}
\!\!\int\!\!\frac{d\varepsilon_2}{2\pi}
\,\widetilde{G}_{0;22}^{+-}(\varepsilon_2) 
\,\widetilde{G}_{0;22}^{-+}(\varepsilon_1+\varepsilon_2-\omega) 
\,\widetilde{G}_{0;11}^{+-}(\varepsilon_1) , 
$
\rule{0cm}{0.5cm}
\\
$
\widetilde{\Sigma}^{(a) -+}_{U;21}(\omega)
=  
-U^2\!\!
\int\!\!\frac{d\varepsilon_1}{2\pi}
\!\!\int\!\!\frac{d\varepsilon_2}{2\pi}
\,\widetilde{G}_{0;12}^{-+}(\varepsilon_2) 
\,\widetilde{G}_{0;21}^{+-}(\varepsilon_1+\varepsilon_2-\omega) 
\,\widetilde{G}_{0;21}^{-+}(\varepsilon_1) 
\, \xrightarrow{\,eV\to 0\,} \, O\left((eV)^3\right)
,
$
\rule{0cm}{0.5cm}
\\
$
\widetilde{\Sigma}^{(a) +-}_{U;21}(\omega)
= 
-U^2
\!\!\int\!\!\frac{d\varepsilon_1}{2\pi}
\!\!\int\!\!\frac{d\varepsilon_2}{2\pi}
\,\widetilde{G}_{0;12}^{+-}(\varepsilon_2) 
\,\widetilde{G}_{0;21}^{-+}(\varepsilon_1+\varepsilon_2-\omega) 
\,\widetilde{G}_{0;21}^{+-}(\varepsilon_1)
\,\xrightarrow{\,eV\to 0\,} \, O\left((eV)^3\right).
$
\rule{0cm}{0.45cm}
\\
\hline
\hline
\end{tabular}
\label{tab:Self_U2_a}
\end{table*}


\begin{table*}[t]
\caption{
Order-$U^2$ self-energy 
$\widetilde{\bm{\Sigma}}_{U}^{(b)\nu\nu'}(\omega)$ 
corresponding to Fig.\ \ref{fig:2nd_order_self-energy_rev} (b). 
}  
\begin{tabular}{l} 
\hline \hline
$
\widetilde{\Sigma}^{(b) -+}_{U;21}(\omega)
=  
\,U^2\!\!
\int\!\frac{d\varepsilon_1}{2\pi}
\!\int\!\frac{d\varepsilon_2}{2\pi}
\,\widetilde{G}_{0;22}^{-+}(\varepsilon_2) 
\,\widetilde{G}_{0;21}^{+-}(\varepsilon_1+\varepsilon_2-\omega) 
\,\widetilde{G}_{0;11}^{-+}(\varepsilon_1)
\,\xrightarrow{\,eV\to 0\,} \ O\left(eV\right), 
$
\rule{0cm}{0.5cm}
\\
$
\widetilde{\Sigma}^{(b) +-}_{U;21}(\omega)
=   \, 
U^2 \!\! 
\int\!\frac{d\varepsilon_1}{2\pi}
\!\int\!\frac{d\varepsilon_2}{2\pi}
\,\widetilde{G}_{0;22}^{+-}(\varepsilon_2) 
\,\widetilde{G}_{0;21}^{-+}(\varepsilon_1+\varepsilon_2-\omega) 
\,\widetilde{G}_{0;11}^{+-}(\varepsilon_1)
\, \xrightarrow{\,eV\to 0\,} \ O\left(eV\right)
, 
$
\rule{0cm}{0.45cm}
\\
$
\widetilde{\Sigma}^{(b) -+}_{U;11}(\omega)
=  
\, U^2\!\!
\int\!\frac{d\varepsilon_1}{2\pi}
\!\int\!\frac{d\varepsilon_2}{2\pi}
\,\widetilde{G}_{0;12}^{-+}(\varepsilon_2) 
\,\widetilde{G}_{0;22}^{+-}(\varepsilon_1+\varepsilon_2-\omega)
\,\widetilde{G}_{0;21}^{-+}(\varepsilon_1)
\, \xrightarrow{\,eV\to 0\,} \ O\left((eV)^2\right) , 
$
\rule{0cm}{0.5cm}
\\
$
\widetilde{\Sigma}^{(b) +-}_{U;11}(\omega)
=  \, 
U^2 \!\!
\int\!\frac{d\varepsilon_1}{2\pi}
\!\int\!\frac{d\varepsilon_2}{2\pi}
\,\widetilde{G}_{0;12}^{+-}(\varepsilon_2) 
\,\widetilde{G}_{0;22}^{-+}(\varepsilon_1+\varepsilon_2-\omega)
\,\widetilde{G}_{0;21}^{+-}(\varepsilon_1)
\, \xrightarrow{\,eV\to 0\,} \ O\left((eV)^2\right) . 
$
\rule{0cm}{0.5cm}
\\
\hline
\hline
\end{tabular}
\label{tab:Self_U2_b}
\end{table*}

\begin{table*}[t]
\caption{
Low-energy asymptotic form of the order-$U^2$ lesser self-energy
 $\widetilde{\bm{\Sigma}}^{-+}_{U}(\omega)$ 
contributing to the linear conductance. 
The corresponding greater self-energy  $\widetilde{\bm{\Sigma}}^{+-}_{U}(\omega)$ 
is obtained from the lesser component by replacing 
$\mathcal{I}^{-+}_{s's'';s'''}(\omega)$ with 
$\mathcal{I}^{+-}_{s's'';s'''}(\omega)$ 
and multiplying by an overall factor of $-1$.
}  
\begin{tabular}{l} 
\hline \hline
$
 \widetilde{\Sigma}^{(a)-+}_{U;11}(\omega)
$
\rule{0cm}{0.45cm}
\\
$
\simeq \ 
-i\, \Gamma_N^3 \,
|\widetilde{G}_{0;1}^{r}(0)|^2\,
|\widetilde{G}_{0;2}^{r}(0)|^4 \, 
\displaystyle \frac{U^2}{(2\pi)^2} 
%
\biggl[ \, 
\left(1+ \cos \theta \right)^3
\mathcal{I}^{-+}_{eh;h}(\omega) 
+\left(1- \cos \theta \right)^3
\mathcal{I}^{-+}_{he;e}(\omega) 
$
\\
$
\quad +\left(1+ \cos \theta \right)\left(1- \cos \theta \right)^2
\left\{
\mathcal{I}^{-+}_{ee;e}(\omega) 
+\mathcal{I}^{-+}_{he;h}(\omega) 
+\mathcal{I}^{-+}_{hh;e}(\omega) 
\right\}
+\left(1+ \cos \theta \right)^2\left(1- \cos \theta \right)
\left\{
 \mathcal{I}^{-+}_{hh;h}(\omega) 
+\mathcal{I}^{-+}_{eh;e}(\omega) 
+\mathcal{I}^{-+}_{ee;h}(\omega) 
\right\}
\,\biggr], 
$
\\
\hline
\rule{0cm}{0.6cm}
$
 \widetilde{\Sigma}^{(a)-+}_{U;22}(\omega)
$
\\
$
\simeq \ 
-i\, \Gamma_N^3 \,
|\widetilde{G}_{0;1}^{r}(0)|^4\,
|\widetilde{G}_{0;2}^{r}(0)|^2 \, 
\displaystyle \frac{U^2}{(2\pi)^2} 
%
\biggl[ \,
\left(1+ \cos \theta \right)^3
\mathcal{I}^{-+}_{he;e}(\omega) 
+\left(1- \cos \theta \right)^3
\mathcal{I}^{-+}_{eh;h}(\omega) 
$
\\
$
\quad 
+\left(1+ \cos \theta \right)\left(1- \cos \theta \right)^2
\left\{
\mathcal{I}^{-+}_{hh;h}(\omega) 
+\mathcal{I}^{-+}_{eh;e}(\omega) 
+ \mathcal{I}^{-+}_{ee;h}(\omega) 
\right\}
+\left(1+ \cos \theta \right)^2\left(1- \cos \theta \right)
\left\{
 \mathcal{I}^{-+}_{ee;e}(\omega) 
+\mathcal{I}^{-+}_{he;h}(\omega) 
+\mathcal{I}^{-+}_{hh;e}(\omega) 
\right\}
\,\biggr],  
$
\\
\hline
$
\widetilde{\Sigma}^{(b)-+}_{U;21}(\omega)
$
\rule{0cm}{0.6cm}
\\
$
\simeq\,
-i\, \Gamma_N^3 \,
|\widetilde{G}_{0;1}^{r}(0)|^2\,
|\widetilde{G}_{0;2}^{r}(0)|^2 \,\widetilde{G}_{0;2}^{r}(0)\,\widetilde{G}_{0;1}^{a}(0)\,
 \displaystyle \frac{U^2}{(2\pi)^2} \, \sin \theta\,
\biggl[\, 
\sin^2 \theta
\left\{\,
\mathcal{I}^{-+}_{ee;e}(\omega) 
-\mathcal{I}^{-+}_{hh;h}(\omega) 
+\mathcal{I}^{-+}_{hh;e}(\omega) 
-\mathcal{I}^{-+}_{ee;h}(\omega) 
\right\}
$
\\
$ \rule{6cm}{0cm}  
+\left(1+ \cos \theta \right)^2
\left\{\,\mathcal{I}^{-+}_{eh;e}(\omega) 
-
\mathcal{I}^{-+}_{eh;h}(\omega) 
\right\}
\,+\,\left(1- \cos \theta \right)^2
\left\{\,\mathcal{I}^{-+}_{he;e}(\omega) 
-\mathcal{I}^{-+}_{he;h}(\omega) \,\right\}
\,\biggr] 
. 
$
\\
\hline
$
\widetilde{\Sigma}^{(b)-+}_{U;12}(\omega)
$ 
\rule{0cm}{0.6cm}
\\
$
\simeq \,  
-i\, \Gamma_N^3 \,
|\widetilde{G}_{0;1}^{r}(0)|^2\,
|\widetilde{G}_{0;2}^{r}(0)|^2 \,\widetilde{G}_{0;1}^{r}(0)\,\widetilde{G}_{0;2}^{a}(0)\,  
\displaystyle  \frac{U^2}{(2\pi)^2} \, \sin \theta\,
\biggl[\, 
\sin^2 \theta
\left\{\, 
\mathcal{I}^{-+}_{ee;e}(\omega) 
-\mathcal{I}^{-+}_{hh;h}(\omega) 
+\mathcal{I}^{-+}_{hh;e}(\omega) 
-\mathcal{I}^{-+}_{ee;h}(\omega) 
\right\}
$
\\
$
\rule{6cm}{0cm} 
 +\left(1+ \cos \theta \right)^2
\left\{\,\mathcal{I}^{-+}_{eh;e}(\omega) 
-\mathcal{I}^{-+}_{eh;h}(\omega) \right\}
+
\left(1- \cos \theta \right)^2
\left\{\, \mathcal{I}^{-+}_{he;e}(\omega) 
- \mathcal{I}^{-+}_{he;h}(\omega) \right\}
\,\biggr] . 
$
\\
\hline
\hline
\end{tabular}
\label{tab:lesser_U2_low_energy}
\end{table*}


\begin{table*}[t]
\caption{
Low-energy asymptotic form of the order-$U^2$ self-energy   
 $\widetilde{\bm{\Pi}}_{U}^{ j;(s)}(\omega)$  
 with $s=e,h$ contributing to the linear conductance. 
The prefactor is defined as 
 $\mathcal{A} \equiv \Gamma_N^3 \,
|\widetilde{G}_{0;1}^{r}(0)|^2\,
|\widetilde{G}_{0;2}^{r}(0)|^2  
\displaystyle \frac{U^2}{(2\pi)^2}$. 
}  
\begin{tabular}{l} 
\hline \hline
$
\widetilde{\Pi}_{U;11}^{(a) j;(s)}(\omega) 
\, \simeq \,  
-i \,\displaystyle \frac{\mathcal{A}}{2}\, 
\bigl|\widetilde{G}_{0;2}^{r}(0)\bigr|^2 \, 
%
\biggl[ \, 
\left(1+ \cos \theta \right)^3
\mathcal{I}^{j;(s)}_{ee;e}(\omega) 
+\left(1- \cos \theta \right)^3
\mathcal{I}^{j;(s)}_{hh;h}(\omega) 
$
\rule{0cm}{0.5cm}
\\
$
\rule{2.5cm}{0cm} 
+\left(1+ \cos \theta \right)\left(1- \cos \theta \right)^2
\left\{\,
2\,\mathcal{I}^{j;(s)}_{ee;e}(\omega) 
+\mathcal{I}^{j;(s)}_{hh;e}(\omega) 
\right\}
+\left(1+ \cos \theta \right)^2\left(1- \cos \theta \right)
\left\{\, 
 2\,\mathcal{I}^{j;(s)}_{hh;h}(\omega) 
+\mathcal{I}^{j;(s)}_{ee;h}(\omega) 
\right\}
\,\biggr], 
$
\rule{0cm}{0.5cm}
\\
\hline
\rule{0cm}{0.6cm}
$
\widetilde{\Pi}_{U;22}^{(a) j;(s)}(\omega)
\,\simeq \, 
-i \,\displaystyle \frac{\mathcal{A}}{2}\, 
\bigl|\widetilde{G}_{0;1}^{r}(0)\bigr|^2 \, 
%
\biggl[ \,
\left(1+ \cos \theta \right)^3
\mathcal{I}^{j;(s)}_{hh;h}(\omega) 
+\left(1- \cos \theta \right)^3
\mathcal{I}^{j;(s)}_{ee;e}(\omega) 
$
\rule{0cm}{0.5cm}
\\
$
\rule{2.5cm}{0cm} 
+\left(1+ \cos \theta \right)\left(1- \cos \theta \right)^2
\left\{\,
2\,\mathcal{I}^{j;(s)}_{hh;h}(\omega) 
+\mathcal{I}^{j;(s)}_{ee;h}(\omega) 
\right\}
+\left(1+ \cos \theta \right)^2\left(1- \cos \theta \right)
\left\{\, 
 2\,\mathcal{I}^{j;(s)}_{ee;e}(\omega) 
+\mathcal{I}^{j;(s)}_{hh;e}(\omega) 
\right\}
\,\biggr],  
$
\rule{0cm}{0.5cm}
\\
\hline
$
\widetilde{\Pi}_{U;21}^{(b) j;(s)}(\omega)
\, \simeq\,
-i \,\displaystyle \frac{\mathcal{A}}{2}\, 
 \widetilde{G}_{0;2}^{r}(0)\,\widetilde{G}_{0;1}^{a}(0)\,
 \sin \theta\,
\biggl[\, 
\sin^2 \theta \, 
\left\{
\mathcal{I}^{j;(s)}_{ee;e}(\omega) 
-\mathcal{I}^{j;(s)}_{hh;h}(\omega) 
+\mathcal{I}^{j;(s)}_{hh;e}(\omega) 
-\mathcal{I}^{j;(s)}_{ee;h}(\omega) 
\right\}
$
\rule{0cm}{0.5cm}
\\
$ \rule{6cm}{0cm}  
+\left(1+ \cos \theta \right)^2
\left\{\mathcal{I}^{j;(s)}_{hh;h}(\omega) 
-\mathcal{I}^{j;(s)}_{ee;e}(\omega) 
\right\}
+\left(1- \cos \theta \right)^2
\left\{\mathcal{I}^{j;(s)}_{hh;h}(\omega) 
-\mathcal{I}^{j;(s)}_{ee;e}(\omega) \right\}
\,\biggr] ,
$
\rule{0cm}{0.5cm}
\\
\hline
$
\widetilde{\Pi}_{U;12}^{(b) j;(s)}(\omega)
\, \simeq\,
-i \,\displaystyle \frac{\mathcal{A}}{2}\, 
 \widetilde{G}_{0;1}^{r}(0)\,\widetilde{G}_{0;2}^{a}(0)\,
 \sin \theta\,
\biggl[\, 
\sin^2 \theta \, 
\left\{
\mathcal{I}^{j;(s)}_{ee;e}(\omega) 
-\mathcal{I}^{j;(s)}_{hh;h}(\omega) 
+\mathcal{I}^{j;(s)}_{hh;e}(\omega) 
-\mathcal{I}^{j;(s)}_{ee;h}(\omega) 
\right\}
$
\rule{0cm}{0.5cm}
\\
$ \rule{6cm}{0cm}  
+\left(1+ \cos \theta \right)^2
\left\{\mathcal{I}^{j;(s)}_{hh;h}(\omega) 
-\mathcal{I}^{j;(s)}_{ee;e}(\omega) 
\right\}
+\left(1- \cos \theta \right)^2
\left\{\mathcal{I}^{j;(s)}_{hh;h}(\omega) 
-\mathcal{I}^{j;(s)}_{ee;e}(\omega) \right\}
\,\biggr] .
$
\rule{0cm}{0.5cm}
\\
\hline
\hline
\end{tabular}
\label{tab:Pi_for_current_U2_low_energy}
\end{table*}


\section{Low-energy behavior of  $\bm{\Pi}_{U}^{j;(s)}$: 
derivation of Eq.\ (\ref{eq: Pi_U_result}) }

\label{sec:current_vertex_sc}

In Sec.\ \ref{subsec:I^U}, 
 we discussed the result for the order-$T^2$ contribution to 
 the linear-response current  $I_{j,\sigma}^{(U)}$ for $j=L,R$,  
given by Eq.\ \eqref{eq:I^U_result},
which arises from the first derivative of the interacting self-energy part  
$\bm{\Pi}_{U}^{j;(s)}(\omega)$ with respect to $eV$.  
This appendix is devoted to a detailed derivation of 
Eq.\ \eqref{eq: Pi_U_result}.

The function $\bm{\Pi}_{U}^{j;(s)}$, 
defined in Eq.\ \eqref{eq:current_Self_U}, 
is purely imaginary and 
consists of the lesser $\bm{\Sigma}^{-+}_{U}(\omega)$  and 
the greater $\bm{\Sigma}^{+-}_{U}(\omega)$ self-energies,  
together with the distribution functions $f_j^{(s)}$ 
for electrons ($s=e$)  and holes ($s=h$).
We start with second-order perturbation 
theory with respect to $H_d^{U}$ and  
show how 
these self-energy components capture 
the $\omega^2$ and $T^2$ terms  
 through the superconducting collision integrals of Bogoliubov quasiparticles.  
We then take into account the Fermi-liquid corrections arising from 
multiple scattering processes of the Bogoliubov quasiparticles.

\subsection{Lesser and greater self-energies for nonequibrium Bogoliubov quasiparticles}

The unperturbed retarded Green's function 
$\bm{G}_{0}^{r}$ for the impurity site, 
derived from Eq.\ \eqref{eq:G0_imp_Dyson}, 
can be diagonalized via the Bogoliubov transformation as follows: 
\begin{align}
\widetilde{\bm{G}}_{0}^{r}(\omega)  
\equiv & \   
\bm{\mathcal{U}}^{\dagger} \,
\bm{G}_{0}^{r}(\omega)   \,
\bm{\mathcal{U}}^{} 
\, =    
 \begin{pmatrix}
 \widetilde{G}_{0;1}^{r}(\omega)  &   0   \cr
 0  &  \widetilde{G}_{0;2}^{r}(\omega) 
\rule{0cm}{0.4cm} \cr  
 \end{pmatrix}
, 
\\ 
\widetilde{G}_{0;1}^{r}(\omega) \,= &  \   
\frac{1}{\omega \, -\, E_{A,\uparrow}^{} + i \Gamma_N^{}}
\,, 
\rule{0cm}{0.8cm}
\\
 \widetilde{G}_{0;2}^{r}(\omega) \,= &  \    
\frac{1}{\omega \, +\, E_{A,\downarrow}^{} + i \Gamma_N^{}}
\,.
\end{align}
The complex conjugate 
$\widetilde{G}_{0;\zeta}^{a}(\omega)  
= \bigl\{\widetilde{G}_{0;\zeta}^{r}(\omega)\bigr\} ^*$ 
for $\zeta =1,2$   
corresponds to the advanced Green's function.  

In contrast, the Keldysh Green's functions still  
retain nonzero off-diagonal elements 
at finite bias voltages $eV \neq 0$,  
even after carrying out the Bogoliubov transformation, 
\begin{align}
\widetilde{\bm{G}}_{0}^{\nu\nu'}  
 \equiv 
\bm{\mathcal{U}}^{\dagger} \,
\bm{G}_{0}^{\nu\nu'} \,
\bm{\mathcal{U}}^{}  
\ =   
 \begin{pmatrix}
\widetilde{G}_{0;11}^{\nu\nu'} & 
\widetilde{G}_{0;12}^{\nu\nu'}    \cr
\widetilde{G}_{0;21}^{\nu\nu'}  & 
\widetilde{G}_{0;22}^{\nu\nu'}    
\rule{0cm}{0.5cm}
\cr  
 \end{pmatrix} .
\end{align}
Specifically, the unperturbed lesser $\widetilde{\bm{G}}_{0}^{-+}$  
and the greater $\widetilde{\bm{G}}_{0}^{+-}$ components  
 can be expressed in the following forms,  
using Eqs.\ \eqref{eq:lesser_self_0} and \eqref{eq:greater_self_0} 
together with Eq.\ \eqref{eq:f_eff_matrix}:   
\begin{align}
\widetilde{\bm{G}}_{0}^{-+}(\omega) 
\equiv & \,  
 -
\bm{\mathcal{U}}^{\dagger}\,
\bm{G}_{0}^{r}(\omega)   \, 
 \bm{\Sigma}_{0}^{-+}(\omega)   \,
 \bm{G}_{0}^{a}(\omega)   \     
\bm{\mathcal{U}}^{}
 \nonumber \\
= &  \ 
2i \Gamma_N \,
 \widetilde{\bm{G}}_{0}^{r}(\omega) \,    
\widetilde{\bm{f}}_\mathrm{eff}^{}(\omega)   \,
 \widetilde{\bm{G}}_{0}^{a}(\omega) 
\,, 
\\
\widetilde{\bm{G}}_{0}^{+-}(\omega) 
\equiv & \,    
 -
\bm{\mathcal{U}}^{\dagger} \, 
\bm{G}_{0}^{r}(\omega)   \, 
 \bm{\Sigma}_{0}^{+-}(\omega)   \,
 \bm{G}_{0}^{a}(\omega)   \   
\bm{\mathcal{U}}^{}
\rule{0cm}{0.6cm}
\nonumber \\
= &  
-2i \Gamma_N \,
 \widetilde{\bm{G}}_{0}^{r}(\omega)  \,     
\Bigl[\,\bm{1} - \widetilde{\bm{f}}_\mathrm{eff}^{}(\omega)   \,
\,\Bigr]
 \widetilde{\bm{G}}_{0}^{a}(\omega)   \,.
\end{align}
Here, 
$\widetilde{\bm{f}}_\mathrm{eff}^{}(\omega)  
 \equiv  \bm{\mathcal{U}}^{\dagger}
 \bm{f}_\mathrm{eff}^{}(\omega)   \,\bm{\mathcal{U}}^{}$ 
represents the effective distribution function, which takes the following form: 
\begin{align}
\widetilde{\bm{f}}_\mathrm{eff}^{} \, 
&  = \, 
 \frac{f_\mathrm{eff}^{(e)}  +   f_\mathrm{eff}^{(h)}}{2} \,\bm{1} 
+ \frac{f_\mathrm{eff}^{(e)}  -   f_\mathrm{eff}^{(h)}}{2} 
 \Bigl( \cos \theta  \, \bm{\tau}_3 - \sin \theta \, \bm{\tau}_1 \Bigr) 
\nonumber \\
 & \xrightarrow{\,eV\to 0\,} \,  f(\omega)\, \bm{1}\,.
\end{align}
The unperturbed 
lesser and greater Green's functions  
depend on the temperature and bias voltage  
through the distribution functions 
$f_\mathrm{eff}^{(e)}(\omega)$ and   
$f_\mathrm{eff}^{(h)}(\omega)$,   
as explicitly listed in Tables \ref{tab:G0_lesser} and \ref{tab:G0_greater}.

The off-diagonal elements of  
 both $\widetilde{\bm{G}}_{0}^{-+}(\omega)$ and 
 $\widetilde{\bm{G}}_{0}^{+-}(\omega)$ 
vanish at $eV=0$ and 
exhibit linear dependence on $eV$,  
with the coefficient given by  
\begin{align}
\!\!\!\! 
\left.\frac{\partial}{\partial eV_j}
\Bigl[
 f_\mathrm{eff}^{(e)}(\omega)  - f_\mathrm{eff}^{(h)}(\omega)
\Bigr]\right|_{eV =0}^{}  
= \frac{2\Gamma_j}{\Gamma_N} 
\left(-\frac{\partial f (\omega)}{\partial \omega}\right) .
\end{align}
Furthermore, the off-diagonal elements of 
 the lesser and greater Green's functions 
 vanish in the limit $\sin \theta =0$, i.e.,  
when the superconducting proximity effect 
induced by $\Gamma_S^{}$ is absent.


\begin{figure}[b]
\leavevmode 
\begin{minipage}[t]{1\linewidth}
\includegraphics[width=0.96\linewidth]{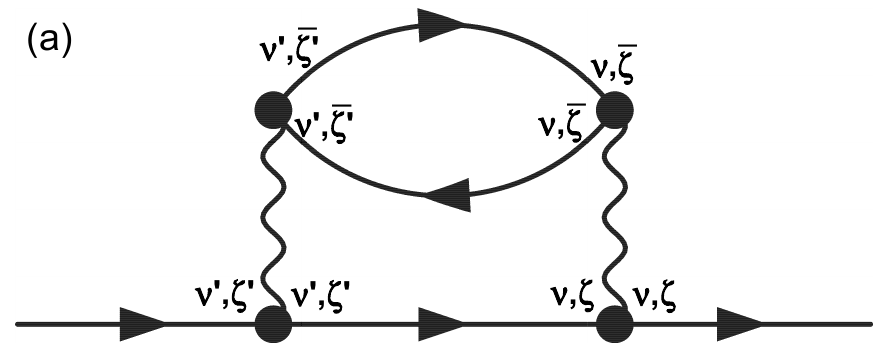} 

 \rule{0\linewidth}{1cm}

 \includegraphics[width=0.8\linewidth]{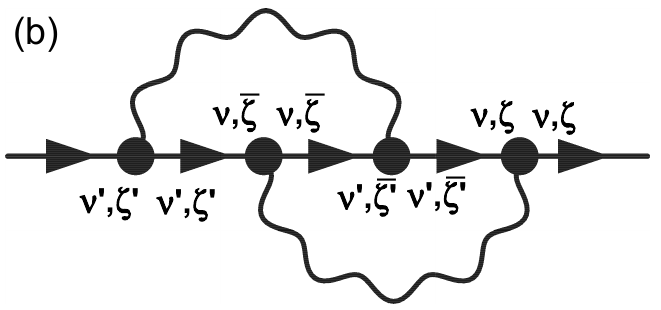}
\caption{Feynman diagrams for the order-$U^2$ self-energy 
$\widetilde{\Sigma}^{\nu\nu'}_{U;\zeta\zeta'}$ of Bogoliubov quasiparticles. 
The label  $\zeta$ and its counterpart  $\overline{\zeta}$ denote  
the matrix components in the Nambu pseudo-spin space 
 ($\overline{1} = 2$ and $\overline{2}=1$), while $\nu$ ($=-,+$)
 labels the Keldysh components.
}
\label{fig:2nd_order_self-energy_rev}
\end{minipage}
\end{figure}


Next, in order to clarify the low-energy behavior   
of $\bm{\Pi}_U^{j;(s)}$,    
we calculate the interacting lesser 
$\bm{\Sigma}_{U}^{-+}$ 
and the greater $\bm{\Sigma}_{U}^{+-}$ self-energies 
by carrying out the Bogoliubov transformation,   
\begin{align}
  \widetilde{\bm{\Sigma}}_{U}^{\nu_4\nu_1}
 \equiv  \,  \bm{\mathcal{U}}^{\dagger}\,
\bm{\Sigma}_{U}^{\nu_4\nu_1}\, \bm{\mathcal{U}}^{} ,
\qquad
\widetilde{\bm{\Pi}}_{U}^{j;(s)} 
 \equiv   \,  
 \bm{\mathcal{U}}^{\dagger}
\bm{\Pi}_U^{j;(s)} \,
 \bm{\mathcal{U}}^{} \,.
\label{eq:current_Pi_tilde}
\end{align}
The Keldysh perturbation theory  
with respect to the interacting Hamiltonian $H_d^{U}$ is 
naturally formulated at finite bias voltages 
by using the representation in terms of 
$\widetilde{\psi}^{}_{d,\zeta}$, defined as       
$\widetilde{\psi}^{}_{d,1} \equiv \gamma^{}_{d\uparrow}$ and 
$\widetilde{\psi}^{}_{d,2} \equiv \gamma^{\dagger}_{d\downarrow}$, 
 introduced in Eq.\ \eqref{eq:H_dot_U_Bogo}.
In this basis, the interaction Hamiltonian reads  
\begin{align}
 H_d^{U} \equiv & \  
\frac{U}{2}\left(
  \widetilde{\psi}_{d,1}^{\dagger}  \widetilde{\psi}_{d,1}^{}
 \,- \, \widetilde{\psi}_{d,2}^{\dagger}  \widetilde{\psi}_{d,2}^{}
  \right)^2 \, .
\label{eq:H_dot_U_Bogo_beta}
\end{align}
Figure \ref{fig:2nd_order_self-energy_rev} 
illustrates the Feynman diagrams for the  self-energy  
 $\widetilde{\Sigma}^{\nu\nu'}_{U;\zeta\zeta'}$ at order $U^2$.

Expressions for the lesser $\widetilde{\bm{\Sigma}}^{(a) -+}_{U}$ 
and the greater $\widetilde{\bm{\Sigma}}^{(a) +-}_{U}$ 
self-energies corresponding 
to Fig.\ \ref{fig:2nd_order_self-energy_rev}(a) 
are listed in Table \ref{tab:Self_U2_a} 
 for the (1,1) and (2,1) components.
The remaining components, namely, 
$\widetilde{\Sigma}^{(a) -+}_{U;22}$,
$\widetilde{\Sigma}^{(a) +-}_{U;22}$, 
$\widetilde{\Sigma}^{(a) -+}_{U;12}$, 
and 
$\widetilde{\Sigma}^{(a) +-}_{U;12}$,  
can be obtained from those listed in the table    
by interchanging the indices in the subscripts,  
 $1 \Leftrightarrow 2$. 
For the self-energies $\widetilde{\bm{\Sigma}}^{(a) -+}_{U}$ and 
$\widetilde{\bm{\Sigma}}^{(a) +-}_{U}$, 
the terms linear in  $eV$ arise   
only from diagonal components, namely $\widetilde{\Sigma}^{(a) -+}_{U;11}$,   
$\widetilde{\Sigma}^{(a) +-}_{U;11}$, $\widetilde{\Sigma}^{(a) -+}_{U;22}$, 
and $\widetilde{\Sigma}^{(a) +-}_{U;22}$. 
In contrast, the off-diagonal components, i.e.,  
  $\widetilde{\Sigma}^{(a) -+}_{U;21}$, 
$\widetilde{\Sigma}^{(a) +-}_{U;21}$, $\widetilde{\Sigma}^{(a) -+}_{U;12}$, 
and $\widetilde{\Sigma}^{(a) +-}_{U;12}$,   
exhibit a $(eV)^3$ dependence at low bias voltages. 
This behavior originates from 
the linear dependence on $eV$ of  
the off-diagonal components of the lesser and 
greater Green's functions, 
 $\widetilde{G}_{0;12}^{-+}(\omega)$,  
 $\widetilde{G}_{0;12}^{+-}(\omega)$,  
 $\widetilde{G}_{0;21}^{-+}(\omega)$,  and 
 $\widetilde{G}_{0;21}^{+-}(\omega)$,  
 as discussed above. 
   
Figure \ref{fig:2nd_order_self-energy_rev}(b)
depicts a second class of order-$U^2$ scattering processes,
which become operative only in the presence of the superconducting proximity effect,
i.e., for $\sin\theta \neq 0$.
The resulting lesser 
$\widetilde{\bm{\Sigma}}^{(b) -+}_{U}$ and 
greater $\widetilde{\bm{\Sigma}}^{(b) +-}_{U}$ self-energies 
are summarized in 
Table \ref{tab:Self_U2_b} for the $(1,1)$ and $(2,1)$ components.
All other components, namely,
$\widetilde{\Sigma}^{(b),-+}_{U;12}$,
$\widetilde{\Sigma}^{(b),+-}_{U;12}$,
$\widetilde{\Sigma}^{(b),-+}_{U;22}$, and
$\widetilde{\Sigma}^{(b),+-}_{U;22}$, 
follow straightforwardly from those listed by interchanging the indices 
$1 \Leftrightarrow 2$. 
At low bias voltages, the off-diagonal (1,2) and (2,1) elements of 
$\widetilde{\bm{\Sigma}}^{(b) -+}_{U}$ 
and  $\widetilde{\bm{\Sigma}}^{(b) +-}_{U}$ exhibit 
a linear dependence on $eV$,
 whereas the diagonal (1,1) and (2,2) elements 
show a  $(eV)^2$ dependence.
This behavior originates from the linear dependence on $eV$ of the 
off-diagonal components of 
 $\widetilde{\bm{G}}_{0}^{-+}(\omega)$  and 
 $\widetilde{\bm{G}}_{0}^{+-}(\omega)$.

\subsection{Collision integrals for Andreev Scatterings}
\label{subsec:Collision_intergrals}

We next carry out the integrations over $\omega$ 
for the lesser and greater self-energies listed in Tables  
\ref{tab:Self_U2_a} and \ref{tab:Self_U2_b} 
in order to clarify the low-energy behavior of 
 $\widetilde{\bm{\Pi}}_{U}^{j;(s)}$.    
The low-energy asymptotic forms of these self-energy components 
are determined by the distribution functions 
$f_\mathrm{eff}^{(e)}$ and $f_\mathrm{eff}^{(h)}$,  
which enter the integrands through the lesser 
$\widetilde{G}_{0;\zeta\zeta'}^{-+}$ and the 
greater $\widetilde{G}_{0;\zeta\zeta'}^{+-}$  Green's functions. 
The contributions arising from these distribution functions 
can be decomposed into a sum of 
the superconducting collision integrals 
$\mathcal{I}^{-+}_{s' s'';s'''} $ and 
$\mathcal{I}^{+-}_{s' s'';s'''}$ for $s= e, h$, defined as  
 \begin{align}
 &\mathcal{I}^{-+}_{s' s'';s'''}(\omega) 
 \, \equiv  
 \int_{-\infty}^{\infty}\! d\varepsilon_1\! \int_{-\infty}^{\infty}\! d\varepsilon_2
  \nonumber \\
 & 
\rule{1.8cm}{0cm}
 \times
 \,f_\mathrm{eff}^{(s')}(\varepsilon_1)\,f_\mathrm{eff}^{(s'')}(\varepsilon_2)
 \Bigl[ 1- f_\mathrm{eff}^{(s''')}(\varepsilon_1+\varepsilon_2-\omega)
 \Bigr] ,
\label{eq:collision_def_-+}
\\
& \mathcal{I}^{+-}_{s' s'';s'''}(\omega) \,  \equiv     
\int_{-\infty}^{\infty}\! d\varepsilon_1 \!\int_{-\infty}^{\infty}\! d\varepsilon_2
 \nonumber \\
 & \rule{0.8cm}{0cm}
 \times
\, \Bigl[1-f_\mathrm{eff}^{(s')}(\varepsilon_1)\Bigr]
 \Bigl[1-f_\mathrm{eff}^{(s'')}(\varepsilon_2)\Bigr]
 \,f_\mathrm{eff}^{(s''')}(\varepsilon_1+\varepsilon_2-\omega) .
\label{eq:collision_def_+-}
\end{align}
Note that 
the retarded $\widetilde{G}_{0;\zeta}^{r}(\varepsilon)$ 
and the advanced $\widetilde{G}_{0;\zeta}^{a}(\varepsilon)$ Green's functions  
 also appear in the integrands together with the distribution functions. 
However, their frequency dependence does not affect 
the leading low-energy behavior and can therefore be evaluated at $\varepsilon=0$.

The lesser collision integral 
 $\mathcal{I}^{-+}_{s' s'';s'''}$ 
has four independent components, since 
some of the components take identical values 
owing to the relation $\mu_j^h=-\mu_j^e$, 
\begin{align}
&\mathcal{I}^{-+}_{ee;h}\,, \qquad 
\mathcal{I}^{-+}_{ee;e}=\mathcal{I}^{-+}_{eh;h}
=\mathcal{I}^{-+}_{he;h}\,, 
\nonumber \\
&\mathcal{I}^{-+}_{hh;e}\,,
\qquad 
\mathcal{I}^{-+}_{hh;h}=\mathcal{I}^{-+}_{he;e}
=\mathcal{I}^{-+}_{eh;e}\,. 
\end{align}
The same relations also hold 
for the greater collision integral $\mathcal{I}^{+-}_{s' s'';s'''}$. 
 These collision integrals play an essential role  
 in the transport properties of Fermi liquids
\cite{MorelAnderson1962,Aligia2012,Aligia2014,Oguri2022}.  
They can be evaluated analytically  
and expressed in the following symmetrized forms:   
\begin{align}
& \mathcal{I}^\mathrm{diff}_{s' s'';s'''}(\omega)  
\, \equiv \,   
2\,\Bigl[\,
\mathcal{I}_{s' s'';s'''}^{+-}(\omega) +  
\mathcal{I}_{s' s'';s'''}^{-+}(\omega) \,\Bigr]  
\nonumber \\
 &  
\rule{0cm}{0.8cm}
=   
\sum_{j,k,\ell\atop = L,R} 
\frac{\Gamma_j\Gamma_k\Gamma_\ell}{(\Gamma_L+\Gamma_R)^{3}}\,
\left[ \left(\omega-\mu_j^{s'}-\mu_k^{s''}+\mu_\ell^{s'''}\right)^2 
+\bigl(\pi T\bigr)^2 
\right] , 
\label{eq:collision_integral_diff}
\end{align}
\begin{align}
& \mathcal{I}_{s' s'';s'''}^\mathrm{K} (\omega) 
  \,  \equiv \,        
2\,\Bigl[\,\mathcal{I}_{s' s'';s'''}^{+-}(\omega)\, -\,  
\mathcal{I}_{s' s'';s'''}^{-+}(\omega)\, \Bigr]  
\nonumber \\
& 
\rule{0cm}{0.8cm}
 = \sum_{j,k,\ell\atop = L,R} 
\frac{\Gamma_j\Gamma_k\Gamma_\ell}{(\Gamma_L+\Gamma_R)^{3}}\,
\left[ \left(\omega-\mu_j^{s'}-\mu_k^{s''}+\mu_\ell^{s'''}\right)^2 
+\bigl(\pi T\bigr)^2 
\right]
\nonumber \\
& \qquad \qquad \qquad  \qquad 
\times 
\biggl[\,1-2f\!\left(\omega-\mu_j^{s'}-\mu_k^{s''}+\mu_\ell^{s'''}\right) 
\,\biggr] . 
\label{eq:collision_integral_K}
\end{align}

Table \ref{tab:lesser_U2_low_energy} 
summarizes the low-energy asymptotic forms 
of the order-$U^2$ lesser self-energy components 
that contribute to the linear conductance.  
The corresponding greater self-energy components 
can be obtained from the results listed in this table  
by replacing $\mathcal{I}^{-+}_{s's'';s'''}(\omega)$ with 
$\mathcal{I}^{+-}_{s's'';s'''}(\omega)$ and multiplying by a factor of $-1$.
These  lesser and greater self-energies 
contribute to the partial current $I_{j,\sigma}^{(U)}$ 
via Eqs.\ \eqref{eq:current_Self_0} and \eqref{eq:current_Self_U}.
Table \ref{tab:Pi_for_current_U2_low_energy} 
summarizes 
the order $U^2$-results 
for  $\widetilde{\bm{\Pi}}_U^{j;(s)}$ 
 in terms of the collision integrals 
 $\mathcal{I}^{j;(s)}_{s' s'';s'''}(\omega)$,  
which are rearranged for the current expectation values as
\begin{align}
& 
\!\!\!\!
 \mathcal{I}^{j;(s)}_{s' s'';s'''}(\omega) 
\,\equiv  \   
\mathcal{I}^\mathrm{K}_{s' s'';s'''}(\omega) - 
\Bigl[ 1-2 f_j^{(s)}(\omega) \Bigr] 
\,\mathcal{I}^\mathrm{diff}_{s' s'';s'''}(\omega).
\end{align}
In particular, the part linear in $eV$ of these collision integrals   
exhibits the $\omega^2$ and $T^2$ dependences of the form 
\begin{align}
\mathcal{F}(\omega) \, \equiv \,  2 \Bigl[\,  \omega^2 + (\pi T)^2\,\Bigr]
 \left(-\frac{\partial f(\omega)}{\partial\omega} \right), 
\end{align}
and, more precisely,  
\begin{align}
&\left.
\frac{\partial}{\partial eV_j} \mathcal{I}^{j'(s)}_{ee;e}(\omega) 
\right|_{eV=0}^{} 
 =   \,
\left(\eta^{(s)}\, \delta_{jj'} - \frac{\Gamma_{j}}{\Gamma_N^{}} \right)
 \mathcal{F}(\omega) \,,
\label{eq:collision_linear_1}
\\
&\left.
\frac{\partial}{\partial eV_j} \mathcal{I}^{j'(s)}_{hh;h}(\omega) 
\right|_{eV=0}^{} 
 =   \,
\left(\eta^{(s)}\, \delta_{jj'} + \frac{\Gamma_{j}}{\Gamma_N^{}} \right)
\mathcal{F}(\omega) 
\,, 
\label{eq:collision_linear_2}
\\
&\left.
\frac{\partial}{\partial eV_j} \mathcal{I}^{j'(s)}_{ee;h}(\omega) 
\right|_{eV=0}^{} 
 =   \,
\left( \eta^{(s)} \,\delta_{jj'} - \frac{3\Gamma_{j}}{\Gamma_N^{}} \right)
\mathcal{F}(\omega)
\,, 
\label{eq:collision_linear_3}
\\
&\left.
\frac{\partial}{\partial eV_j} \mathcal{I}^{j'(s)}_{hh;e}(\omega) 
\right|_{eV=0}^{} 
 =   \,
\left(\eta^{(s)} \,\delta_{jj'} + \frac{3\Gamma_{j}}{\Gamma_N^{}} \right)
\mathcal{F}(\omega)\,,
\label{eq:collision_linear_4}
\end{align}
where $\eta^{(e)} = 1$ and $\eta^{(h)} = -1$.



\subsection{Renormalization for $\widetilde{\bm{\Pi}}_{U}^{j;(s)}$}

The low-energy asymptotic form of the lesser self-energy 
$\widetilde{\bm{\Sigma}}^{-+}_{U}$  
is determined by the many-body multiple-scattering processes 
shown in Fig.\  \ref{fig:Keldysh_self_K_renorm}.
In these diagrams, 
the solid lines represent matrix elements of the full interacting Green's function
of the Bogoliubov quasiparticles, $\widetilde{\bm{G}}_{}^{\nu\nu'}  
 \equiv 
\bm{\mathcal{U}}^{\dagger} \,
\bm{G}_{dd}^{\nu\nu'} \,
\bm{\mathcal{U}}^{}$.  
The shaded squares represent the full vertex corrections  
with respect to $H_d^{U}$ in Eq.\ \eqref{eq:H_dot_U_Bogo_beta}, 
defined at $\omega=T=eV=0$.  
 Specifically, the vertex labeled $(--;--)$ denotes the causal component  
and the one labeled  $(++;++)$  is its anti-causal counterpart.
These vertices satisfy the following relations: 
$\widetilde{\Gamma}_{21;12}^{--;--}(0, 0; 0, 0)
=\widetilde{\Gamma}_{12;21}^{--;--}(0, 0; 0, 0)$ and 
 $\widetilde{\Gamma}_{12;21}^{++;++}(0,0;0,0)
=-\widetilde{\Gamma}_{12;21}^{--;--}(0,0;0,0)$. 
The (1,1) and (2,1) components of $\widetilde{\bm{\Sigma}}^{-+}_{U}$ 
can be calculated exactly, up to terms of order $\omega^2$ and $T^2$,  
in the linear-response regime with respect to $eV$,  
\begin{widetext}
\begin{align}
& 
\widetilde{\Sigma}^{-+}_{U;11}(\omega)
= 
\int\!\frac{d\varepsilon_1}{2\pi}
\!\int\!\frac{d\varepsilon_2}{2\pi}\,
\widetilde{\Gamma}_{21;12}^{--;--}(0,0; 0,0) \ 
\widetilde{\Gamma}_{12;21}^{++;++}(0,0; 0,0)  
\,\widetilde{G}_{22}^{-+}(\varepsilon_2) 
\,\widetilde{G}_{22}^{+-}(\varepsilon_1+\varepsilon_2-\omega) 
\,\widetilde{G}_{11}^{-+}(\varepsilon_1)
+ \cdots 
\nonumber \\
&=  \,   
-i \,\frac{2\Gamma_N^{3}}{\pi^2}
\left|\widetilde{\Gamma}_{12;21}^{--;--}(0,0; 0,0) \right|^2 
\left|\widetilde{G}_{1}^{r}(0)\right|^2
\left|\widetilde{G}_{2}^{r}(0)\right|^4
\int\!\! d\varepsilon_1 \!\int\!\! d\varepsilon_2 
\left\{\widetilde{\bm{f}}_\mathrm{eff}^{}(\varepsilon_1) \right\}_{11}^{}\, 
\left\{\widetilde{\bm{f}}_\mathrm{eff}^{}(\varepsilon_2) \right\}_{22}^{}\, 
\left\{
\bm{1}-\widetilde{\bm{f}}_\mathrm{eff}^{}(\varepsilon_1+\varepsilon_2-\omega) 
\right\}_{22}^{} ,
\rule{0cm}{0.6cm}
 \label{eq:lesser_asymptotic_sc_11}
\\
& 
\widetilde{\Sigma}^{-+}_{U;21}(\omega)
= 
- \int\!\frac{d\varepsilon_1}{2\pi}
\!\int\!\frac{d\varepsilon_2}{2\pi}\,
\widetilde{\Gamma}_{21;12}^{--;--}(0,0; 0,0) \ 
\widetilde{\Gamma}_{12;21}^{++;++}(0,0; 0,0)  
\,\widetilde{G}_{22}^{-+}(\varepsilon_2) 
\,\widetilde{G}_{21}^{+-}(\varepsilon_1+\varepsilon_2-\omega) 
\,\widetilde{G}_{11}^{-+}(\varepsilon_1)
+ \cdots 
\nonumber \\
&=  \,   
i \,\frac{2\Gamma_N^{3}}{\pi^2}
\left|\widetilde{\Gamma}_{12;21}^{--;--}(0,0; 0,0) \right|^2 
\left|\widetilde{G}_{1}^{r}(0)\right|^2
\left|\widetilde{G}_{2}^{r}(0)\right|^2
\widetilde{G}_{2}^{r}(0)\,\widetilde{G}_{1}^{a}(0)
\int\!\! d\varepsilon_1 \!\int\!\! d\varepsilon_2 
\left\{\widetilde{\bm{f}}_\mathrm{eff}^{}(\varepsilon_1) \right\}_{11}^{}\, 
\left\{\widetilde{\bm{f}}_\mathrm{eff}^{}(\varepsilon_2) \right\}_{22}^{}\, 
\left\{
\bm{1}-\widetilde{\bm{f}}_\mathrm{eff}^{}(\varepsilon_1+\varepsilon_2-\omega) 
\right\}_{21}^{} .
\rule{0cm}{0.6cm}
 \label{eq:lesser_asymptotic_sc_21}
\end{align}
\end{widetext}
In order to determine the low-energy expansion up to order $\omega^2$ and $T^2$, 
the lesser and greater Green's functions appearing in the integrands   
of the first lines of  Eqs.\ \eqref{eq:lesser_asymptotic_sc_11} and 
 \eqref{eq:lesser_asymptotic_sc_21},  
have been replaced by 
\begin{align}
\widetilde{\bm{G}}_{}^{-+}(\omega) 
\equiv & \,  
 -
\widetilde{\bm{G}}_{}^{r}(\omega)   \, 
\left[
 \widetilde{\bm{\Sigma}}_{0}^{-+}(\omega)   +
 \widetilde{\bm{\Sigma}}_{U}^{-+}(\omega)   
\right]
 \widetilde{\bm{G}}_{}^{a}(\omega)   \     
 \nonumber \\
=  &  \ 
2i \Gamma_N \,
 \widetilde{\bm{G}}_{}^{r}(0) \,    
\widetilde{\bm{f}}_\mathrm{eff}^{}(\omega)   \,
 \widetilde{\bm{G}}_{}^{a}(0)  
\  + \  \cdots 
\,, 
 \label{eq:lesser_asymptotic_internal}
\end{align}
\begin{align}
\widetilde{\bm{G}}_{}^{+-}(\omega) 
\equiv & \,    
 -
\widetilde{\bm{G}}_{}^{r}(\omega)   \, 
\left[
 \widetilde{\bm{\Sigma}}_{0}^{+-}(\omega)   
 + \widetilde{\bm{\Sigma}}_{U}^{+-}(\omega)   
\right]
 \widetilde{\bm{G}}_{}^{a}(\omega)   \   
\nonumber \\
= &  
-2i \Gamma_N \,
 \widetilde{\bm{G}}_{}^{r}(0)  \,     
\Bigl[\,\bm{1} - \widetilde{\bm{f}}_\mathrm{eff}^{}(\omega)   \,
\,\Bigr]
 \widetilde{\bm{G}}_{}^{a}(0)  
\ + \  \cdots 
 \label{eq:greater_asymptotic_internal}
 \,.
\end{align}
This is because the contributions from 
 $\widetilde{\bm{\Sigma}}^{-+}_{U}$ 
and 
 $\widetilde{\bm{\Sigma}}^{+-}_{U}$ terms, 
which enter through the internal lines of these Feynman diagrams,   
ultimately generate terms beyond order $T^2$ in the current. 
The integrals involving the distribution functions 
$\widetilde{\bm{f}}_\mathrm{eff}^{}$ in the last lines of Eqs.\ 
\eqref{eq:lesser_asymptotic_sc_11} and 
 \eqref{eq:lesser_asymptotic_sc_21} can therefore be 
 expressed in terms of the collision integrals $\mathcal{I}^{-+}_{s' s'';s'''}$ 
defined in Appendix \ref{subsec:Collision_intergrals}.

These low-energy asymptotic forms 
of $\widetilde{\Sigma}^{-+}_{U;11}(\omega)$ and 
 $\widetilde{\Sigma}^{-+}_{U;21}(\omega)$ 
can be compared with the corresponding perturbative results   
shown in Table \ref{tab:lesser_U2_low_energy}.
The perturbative results capture the essential functional structure 
of the asymptotic forms,    
and they become identical to the exact ones upon 
 replacing the bare interaction strength $U^2$  
with 
$\left|\widetilde{\Gamma}_{12;21}^{--;--}(0,0;0,0)\right|^2$,
the full scattering strength determined by the vertex function, 
and further substituting  the noninteracting Green's functions  
$\widetilde{G}_{0;\zeta}^{r}(0)$ and 
$\widetilde{G}_{0;\zeta}^{a}(0)$ with  
the corresponding full Green's functions 
$\widetilde{G}_{\zeta}^{r}(0)$
and $\widetilde{G}_{\zeta}^{a}(0)$, respectively. 
The other components of the lesser self-energy,  
  $\widetilde{\Sigma}^{-+}_{U;22}$ and $\widetilde{\Sigma}^{-+}_{U;12}$,  
 can also be obtained from Eqs.\ \eqref{eq:lesser_asymptotic_sc_11} and  
\eqref{eq:lesser_asymptotic_sc_21}
by interchanging the indices $1 \Leftrightarrow 2$.  
Furthermore, the low-energy asymptotic form of 
greater self-energy $\widetilde{\bm{\Sigma}}^{+-}_{U}$ can be obtained 
in a similar manner by exchanging $- \Leftrightarrow +$ 
and then using Eqs.\  \eqref{eq:lesser_asymptotic_internal} and 
 \eqref{eq:greater_asymptotic_internal}.
The low-energy asymptotic forms of all these lesser and greater self-energies  
 exhibit the same functional structure as the perturbative results,  
differing only in the overall coefficients renormalized by multiple scattering processes.

Similarly, the exact low-energy asymptotic 
form of  $\widetilde{\bm{\Pi}}_{U}^{j';(s)}$ 
can be derived from  
the corresponding perturbative results summarized 
in Table \ref{tab:Pi_for_current_U2_low_energy}, 
by replacing the bare parameters with 
 their renormalized counterparts,  
 $U^2  \Rightarrow 
 \left|\widetilde{\Gamma}_{12;21}^{--;--}(0,0;0,0)\right|^2$, 
 and $\widetilde{G}_{0;\zeta}^{r/a}(0)  \Rightarrow 
  \widetilde{G}_{\zeta}^{r/a}(0)$.  
Thus,  
 the first derivative of $\widetilde{\bm{\Pi}}_{U}^{j';(s)}$ with respect to $eV$ 
 can be expressed in the following form, 
 using also Eqs.\ \eqref{eq:collision_linear_1}--\eqref{eq:collision_linear_4}: 
\begin{widetext}
\begin{align}
& 
\left.
\frac{\partial\,\widetilde{\bm{\Pi}}_{U}^{j';(s)}}{\partial eV_j}\right|_{eV=0}^{}
= \    
 -2i\, 
\frac{\Gamma_N^3}{\pi^2} \, 
|\widetilde{G}_{1}^{r}(0)|^2\,
|\widetilde{G}_{2}^{r}(0)|^2 \, 
\left|\widetilde{\Gamma}_{12;21}^{--;--}(0,0;0,0)\right|^2
\ \Bigl[\,  \omega^2 + (\pi T)^2\,\Bigr]
\left(-\frac{\partial f(\omega)}{\partial\omega} \right)
\nonumber \\
& \times \,
\left \{ \, 
\eta^{(s)} \delta_{jj'}
\begin{pmatrix}
|\widetilde{G}_{2}^{r}(0)|^2 & 
0   \cr
0 & 
|\widetilde{G}_{1}^{r}(0) |^2   
\rule{0cm}{0.5cm}
\cr  
 \end{pmatrix}
 \, - 
\frac{\Gamma_{j}}{\Gamma_N^{}} \,\cos \theta  
\begin{pmatrix}
|\widetilde{G}_{2}^{r}(0)|^2 & 
0   \cr
0 & 
-|\widetilde{G}_{1}^{r}(0)|^2   
\rule{0cm}{0.5cm}
\cr  
 \end{pmatrix}
\, + 
\frac{\Gamma_{j}}{\Gamma_N^{}} \,  \sin \theta
\begin{pmatrix}
0 & 
\widetilde{G}_{1}^{r}(0)\,\widetilde{G}_{2}^{a}(0)  \cr
\widetilde{G}_{2}^{r}(0)\,\widetilde{G}_{1}^{a}(0) 
& 0
\rule{0cm}{0.7cm}
\cr  
 \end{pmatrix}
\, \right \}
,
\rule{0cm}{1.2cm}
\label{eq:tilde_Pi_U_results}
\end{align}
\end{widetext}
where $\eta^{(e)} = 1$ and $\eta^{(h)} = -1$.   

In order to derive the linear-response contribution to $I_{j,\sigma}^{(U)}$ 
using Eqs.\ \eqref{eq:current_up_formula} and \eqref{eq:current_down_formula}, 
we next evaluate the diagonal elements of the matrix product,   
$\bm{G}_{dd}^{r}(0)
\frac{\partial \bm{\Pi}_{U}^{j';(s)}}{\partial eV_j} 
\bm{G}_{dd}^{a}(0)$.
Carrying out the inverse Bogoliubov transformation 
for the three matrices appearing on the right-hand side of   
Eq.\ \eqref{eq:tilde_Pi_U_results}, 
the matrices in the first and second terms 
can be expressed in the following forms:  
\begin{align}
& 
\!\!\!\!
\bm{G}_{dd}^{r}(0)\ \bm{\mathcal{U}}^{}
\begin{pmatrix}
|\widetilde{G}_{2}^{r}(0)|^2 & 
0   \cr
0 & 
|\widetilde{G}_{1}^{r}(0) |^2   
\rule{0cm}{0.5cm}
\cr  
 \end{pmatrix}
\,\bm{\mathcal{U}}^{\dagger}\,\bm{G}_{dd}^{a}(0)
\nonumber \\
 & 
=  \,   
|\widetilde{G}_{1}^{r}(0)|^2 |\widetilde{G}_{2}^{r}(0)|^2  \, \bm{1} 
\,,
\rule{0cm}{0.7cm}
\label{eq:Pi_U_matrix_0}
\end{align}
and 
\begin{align}
& 
\!\!\!\!
\bm{G}_{dd}^{r}(0)\ \bm{\mathcal{U}}^{}
\begin{pmatrix}
|\widetilde{G}_{2}^{r}(0)|^2 & 
0   \cr
0 & 
- |\widetilde{G}_{1}^{r}(0) |^2   
\rule{0cm}{0.5cm}
\cr  
 \end{pmatrix}
\,\bm{\mathcal{U}}^{\dagger}\,\bm{G}_{dd}^{a}(0)
 \nonumber \\
& 
=    \,     
|\widetilde{G}_{1}^{r}(0)|^2 |\widetilde{G}_{2}^{r}(0)|^2 
\,\Bigl( \cos \theta  \ \bm{\tau}_3 \,+\, \sin \theta \  \bm{\tau}_1 \Bigr) \,,
\rule{0cm}{0.7cm}
\label{eq:Pi_U_matrix_3}
\end{align}
The third term is transformed into the following form: 
\begin{align}
& 
\!\!
\bm{G}_{dd}^{r}(0)\ \bm{\mathcal{U}}^{}
\begin{pmatrix}
0 & 
\widetilde{G}_{1}^{r}(0)\,\widetilde{G}_{2}^{a}(0)  \cr
\widetilde{G}_{2}^{r}(0)\,\widetilde{G}_{1}^{a}(0) 
& 0
\rule{0cm}{0.7cm}
\cr  
 \end{pmatrix}
\,\bm{\mathcal{U}}^{\dagger}\,\bm{G}_{dd}^{a}(0)
\nonumber \\ 
& 
= \ 
\bm{\mathcal{U}}^{}
\begin{pmatrix}
0 & 
\left\{
\widetilde{G}_{1}^{r}(0)\,\widetilde{G}_{2}^{a}(0)
\right\}^2  \cr
\left\{
\widetilde{G}_{2}^{r}(0)\,\widetilde{G}_{1}^{a}(0) 
\right\}^2 
& 0
\rule{0cm}{0.7cm}
\cr  
 \end{pmatrix}
\,\bm{\mathcal{U}}^{\dagger}
\rule{0cm}{1.1cm}
\nonumber \\
 & 
= \,   
-\  \frac{\left\{\widetilde{G}_{1}^{r}(0) \widetilde{G}_{2}^{a}(0) \right\}^2 
 + \left\{\widetilde{G}_{2}^{r}(0) \widetilde{G}_{1}^{a}(0) \right\}^2}{2} 
 \,\sin \theta \ \bm{\tau}_3 
\rule{0cm}{1.2cm}
\nonumber \\
& \quad   + \  (\mbox{off-diagonal matrix components}) 
 \rule{0cm}{0.5cm}
\,. 
\label{eq:Pi_U_matrix_off}
\end{align}
Thus, the diagonal elements of $\bm{G}_{dd}^{r}(0)
\frac{\partial \bm{\Pi}_{U}^{j';(s)}}{\partial eV_j} 
\bm{G}_{dd}^{a}(0)$ 
consist of a linear combination of  
 $\eta^{(s)}\bm{1}$ and $\bm{\tau}_3$.
Consequently, the (1,1) element of the matrix for $s=e$ 
and the (2,2) element of one for $s=h$ 
are equal in magnitude but opposite in sign, 
\begin{align}
& - \ \left\{\,
\bm{G}_{dd}^{r}(0)
\,
\left.
\frac{\partial\,\bm{\Pi}_{U}^{j';(h)}}{\partial eV_j}
\right|_{eV=0}^{}
\bm{G}_{dd}^{a}(0) \,\right\}_{22}^{} 
\nonumber \\
&  = \ \  
 \left\{\,
\bm{G}_{dd}^{r}(0)
\,
\left.
\frac{\partial\,\bm{\Pi}_{U}^{j';(e)}}{\partial eV_j}\right|_{eV=0}^{}
\bm{G}_{dd}^{a}(0) 
\,\right\}_{11}^{} \,.
\end{align}
This implies that $I_{j,\uparrow}^{(U)}=I_{j,\downarrow}^{(U)}$ 
in the linear-response regime, 
whereas $I_{j,\sigma}^{(0)}$ generally depends on the spin component $\sigma$.

The above discussion mainly employs 
the Keldysh Green's functions $\widetilde{G}_{\zeta\zeta'}^{\nu\nu'}$, 
defined in terms of 
the operators $\widetilde{\psi}_{d,\zeta}^{}$ with $\zeta=1,2$. 
The coefficients appearing in Eq.\ \eqref{eq:tilde_Pi_U_results} 
can alternatively be expressed 
 in terms of the Fermi-liquid parameters defined with respect 
to the operators 
$\gamma^{}_{d\uparrow}=\widetilde{\psi}^{}_{d,1}$ and 
$\gamma^{}_{d\downarrow}=\widetilde{\psi}^{\dagger}_{d,2}$.    
The correspondence between the vertex functions in these two representations 
is given by 
$
\widetilde{\Gamma}_{12;21}^{--;--}(0,0;0,0) = -
\Gamma_{\uparrow\downarrow;\downarrow\uparrow}^{--;--}(0,0;0,0)
$.  
For the retarded Green's functions,  we have 
$\widetilde{G}_{1}^{r}(\omega)
=
 G_{\gamma,\uparrow}^{r}(\omega)$ and
$\widetilde{G}_{2}^{r}(\omega)
=
 -G_{\gamma,\downarrow}^{a}(-\omega)$.  
In particular,  at  $\omega=T=eV=0$, 
these Green's functions can be expressed 
in terms of the phase shifts as follows:    
\begin{align}
& G_{\gamma,\sigma}^{r}(0) 
\ = \ -\, \frac{e^{i\delta_{\sigma}^{}}
\,\sin\delta_{\sigma}^{}}{\Gamma_N^{}}\,, 
 \qquad
\rho_{d\sigma}^{}\, \equiv  \,  
\frac{\sin^2 \delta_{\sigma}^{}}{\pi \Gamma_N}  \,.
\end{align}
%
Therefore, 
\begin{align}
& 
\!\!\!\!\!\!  \!\!\!\!\!\!  
\frac{\left\{\widetilde{G}_{1}^{r}(0) \widetilde{G}_{2}^{a}(0) \right\}^2 
 + \left\{\widetilde{G}_{2}^{r}(0) \widetilde{G}_{1}^{a}(0) \right\}^2}{2} 
\nonumber \\
& =\, 
\frac{
\left\{G_{\gamma,\uparrow}^{r}(0) G_{\gamma,\downarrow}^{r}(0) \right\}^2 
 + 
\left\{G_{\gamma,\uparrow}^{a}(0) G_{\gamma,\downarrow}^{a}(0) \right\}^2 
}{2} 
\nonumber \\
& =\, 
|G_{\gamma,\uparrow}^{r}(0)|^2\, 
|G_{\gamma,\downarrow}^{r}(0)|^2 
\,\cos 2(\delta_{\uparrow}+\delta_{\downarrow}) .
\end{align}
Moreover,  the vertex function can be expressed in terms of 
$\chi_{\uparrow\downarrow}^{}$
defined in Eq.\ \eqref{eq:YamadaYsidaRelation_vertex} as 
 \begin{align}
&
\!\!\!\!\!\!  \!\!\!\!\!\!  
\frac{\Gamma_N^2}{\pi^2} 
|\widetilde{G}_{1}^{r}(0)|^2\,
|\widetilde{G}_{2}^{r}(0)|^2 \, 
\widetilde{\Gamma}_{12;21}^{--;--}(0,0;0,0)
\nonumber \\ 
& =\, 
- \Gamma_{\uparrow\downarrow;\downarrow\uparrow}^{--;--}(0,0;0,0) 
\,\rho_{d\uparrow}^{}\,\rho_{d\downarrow}^{} 
\ = \  \chi_{\uparrow\downarrow}^{} \,.
\end{align}
Using these relations together with  
Eqs.\ \eqref{eq:Pi_U_matrix_0}--\eqref{eq:Pi_U_matrix_off},   
Eq.\ \eqref{eq:tilde_Pi_U_results} can be rewritten in  
the following form, which is also presented in the main text 
as Eq.\ \eqref{eq: Pi_U_result},   
\begin{widetext}
\begin{align}
&
\!\!\!\!\!\!\!\!\!\!\!\!\!\!\! 
\left\{\,
\bm{G}_{dd}^{r}(0)
\,
\left.
\frac{\partial\,\bm{\Pi}_{U}^{j';(e)}}{\partial eV_j}\right|_{eV=0}^{}
\bm{G}_{dd}^{a}(0) 
\,\right\}_{11}^{} 
\nonumber \\
= & \  
 -2i\ \frac{\Gamma_N^3}{\pi^2} \,
|G_{\gamma,\uparrow}^{r}(0)|^4\,
|G_{\gamma,\downarrow}^{r}(0)|^4 \, 
\left|\Gamma_{\uparrow\downarrow;\downarrow\uparrow}^{--;--}(0,0;0,0) 
\right|^2
\, \Bigl[\,  \omega^2 + (\pi T)^2\,\Bigr]
\left(-\frac{\partial f(\omega)}{\partial\omega} \right)
\rule{0cm}{0.7cm}
\nonumber \\
&\times 
\left[\,
\left(\,\delta_{jj'} -\frac{\Gamma_{j}^{}}{\Gamma_N^{}} 
\cos^2 \theta \,\right)\, 
 - 
\frac{\Gamma_{j}}{\Gamma_N^{}} \, 
\cos 2(\delta_{\uparrow}+\delta_{\downarrow}) \, \sin^2 \theta 
\,\right]
\rule{0cm}{0.7cm}
\nonumber\\
=& \ 
 -i\,  \frac{2\pi^2\,\chi_{\uparrow\downarrow}^{2}}{\Gamma_N}\, 
\ \Bigl[\,  \omega^2 + (\pi T)^2\,\Bigr]
\left(-\frac{\partial f(\omega)}{\partial\omega} \right)
\,\left [\, 
 \left(\delta_{jj'}
 \, - \,  \frac{\Gamma_{j}}{\Gamma_N^{}} \right)
\, + \,  \frac{\Gamma_{j}}{\Gamma_N^{}} 
\  2 \sin^2 (\delta_{\uparrow}+\delta_{\downarrow}) \,
 \sin^2 \theta
 \, \right ] .
\rule{0cm}{0.8cm}
\end{align}

\end{widetext}

%



\begin{thebibliography}{107}%
\makeatletter
\providecommand \@ifxundefined [1]{%
 \@ifx{#1\undefined}
}%
\providecommand \@ifnum [1]{%
 \ifnum #1\expandafter \@firstoftwo
 \else \expandafter \@secondoftwo
 \fi
}%
\providecommand \@ifx [1]{%
 \ifx #1\expandafter \@firstoftwo
 \else \expandafter \@secondoftwo
 \fi
}%
\providecommand \natexlab [1]{#1}%
\providecommand \enquote  [1]{``#1''}%
\providecommand \bibnamefont  [1]{#1}%
\providecommand \bibfnamefont [1]{#1}%
\providecommand \citenamefont [1]{#1}%
\providecommand \href@noop [0]{\@secondoftwo}%
\providecommand \href [0]{\begingroup \@sanitize@url \@href}%
\providecommand \@href[1]{\@@startlink{#1}\@@href}%
\providecommand \@@href[1]{\endgroup#1\@@endlink}%
\providecommand \@sanitize@url [0]{\catcode `\\12\catcode `\$12\catcode
  `\&12\catcode `\#12\catcode `\^12\catcode `\_12\catcode `\%12\relax}%
\providecommand \@@startlink[1]{}%
\providecommand \@@endlink[0]{}%
\providecommand \url  [0]{\begingroup\@sanitize@url \@url }%
\providecommand \@url [1]{\endgroup\@href {#1}{\urlprefix }}%
\providecommand \urlprefix  [0]{URL }%
\providecommand \Eprint [0]{\href }%
\providecommand \doibase [0]{https://doi.org/}%
\providecommand \selectlanguage [0]{\@gobble}%
\providecommand \bibinfo  [0]{\@secondoftwo}%
\providecommand \bibfield  [0]{\@secondoftwo}%
\providecommand \translation [1]{[#1]}%
\providecommand \BibitemOpen [0]{}%
\providecommand \bibitemStop [0]{}%
\providecommand \bibitemNoStop [0]{.\EOS\space}%
\providecommand \EOS [0]{\spacefactor3000\relax}%
\providecommand \BibitemShut  [1]{\csname bibitem#1\endcsname}%
\let\auto@bib@innerbib\@empty
\bibitem [{\citenamefont {Schindele}\ \emph {et~al.}(2014)\citenamefont
  {Schindele}, \citenamefont {Baumgartner}, \citenamefont {Maurand},
  \citenamefont {Weiss},\ and\ \citenamefont
  {Sch\"onenberger}}]{PhysRevB.89.045422}%
  \BibitemOpen
  \bibfield  {author} {\bibinfo {author} {\bibfnamefont {J.}~\bibnamefont
  {Schindele}}, \bibinfo {author} {\bibfnamefont {A.}~\bibnamefont
  {Baumgartner}}, \bibinfo {author} {\bibfnamefont {R.}~\bibnamefont
  {Maurand}}, \bibinfo {author} {\bibfnamefont {M.}~\bibnamefont {Weiss}},\
  and\ \bibinfo {author} {\bibfnamefont {C.}~\bibnamefont {Sch\"onenberger}},\
  }\bibfield  {title} {\bibinfo {title} {{Nonlocal spectroscopy of Andreev
  bound states}},\ }\href {https://doi.org/10.1103/PhysRevB.89.045422}
  {\bibfield  {journal} {\bibinfo  {journal} {Phys. Rev. B}\ }\textbf {\bibinfo
  {volume} {89}},\ \bibinfo {pages} {045422} (\bibinfo {year}
  {2014})}\BibitemShut {NoStop}%
\bibitem [{\citenamefont {Lee}\ \emph {et~al.}(2014)\citenamefont {Lee},
  \citenamefont {Jiang}, \citenamefont {Houzet}, \citenamefont {Aguado},
  \citenamefont {Lieber},\ and\ \citenamefont {De~Franceschi}}]{HLee2013}%
  \BibitemOpen
  \bibfield  {author} {\bibinfo {author} {\bibfnamefont {E.~J.~H.}\
  \bibnamefont {Lee}}, \bibinfo {author} {\bibfnamefont {X.}~\bibnamefont
  {Jiang}}, \bibinfo {author} {\bibfnamefont {M.}~\bibnamefont {Houzet}},
  \bibinfo {author} {\bibfnamefont {R.}~\bibnamefont {Aguado}}, \bibinfo
  {author} {\bibfnamefont {C.~M.}\ \bibnamefont {Lieber}},\ and\ \bibinfo
  {author} {\bibfnamefont {S.}~\bibnamefont {De~Franceschi}},\ }\bibfield
  {title} {\bibinfo {title} {{Spin-resolved Andreev levels and parity crossings
  in hybrid superconductor--semiconductor nanostructures}},\ }\href
  {https://doi.org/10.1038/nnano.2013.267} {\bibfield  {journal} {\bibinfo
  {journal} {Nature Nanotechnology}\ }\textbf {\bibinfo {volume} {9}},\
  \bibinfo {pages} {79} (\bibinfo {year} {2014})}\BibitemShut {NoStop}%
\bibitem [{\citenamefont {Bordoloi}\ \emph {et~al.}(2022)\citenamefont
  {Bordoloi}, \citenamefont {Zannier}, \citenamefont {Sorba}, \citenamefont
  {Sch{\"o}nenberger},\ and\ \citenamefont {Baumgartner}}]{Bordoloi2022}%
  \BibitemOpen
  \bibfield  {author} {\bibinfo {author} {\bibfnamefont {A.}~\bibnamefont
  {Bordoloi}}, \bibinfo {author} {\bibfnamefont {V.}~\bibnamefont {Zannier}},
  \bibinfo {author} {\bibfnamefont {L.}~\bibnamefont {Sorba}}, \bibinfo
  {author} {\bibfnamefont {C.}~\bibnamefont {Sch{\"o}nenberger}},\ and\
  \bibinfo {author} {\bibfnamefont {A.}~\bibnamefont {Baumgartner}},\
  }\bibfield  {title} {\bibinfo {title} {{Spin cross-correlation experiments in
  an electron entangler}},\ }\href {https://doi.org/10.1038/s41586-022-05436-z}
  {\bibfield  {journal} {\bibinfo  {journal} {Nature}\ }\textbf {\bibinfo
  {volume} {612}},\ \bibinfo {pages} {454} (\bibinfo {year}
  {2022})}\BibitemShut {NoStop}%
\bibitem [{\citenamefont {Golubev}\ and\ \citenamefont
  {Zaikin}(2007)}]{Golubev_2007}%
  \BibitemOpen
  \bibfield  {author} {\bibinfo {author} {\bibfnamefont {D.~S.}\ \bibnamefont
  {Golubev}}\ and\ \bibinfo {author} {\bibfnamefont {A.~D.}\ \bibnamefont
  {Zaikin}},\ }\bibfield  {title} {\bibinfo {title} {{Non-local Andreev
  reflection in superconducting quantum dots}},\ }\href
  {https://doi.org/10.1103/PhysRevB.76.184510} {\bibfield  {journal} {\bibinfo
  {journal} {Phys. Rev. B}\ }\textbf {\bibinfo {volume} {76}},\ \bibinfo
  {pages} {184510} (\bibinfo {year} {2007})}\BibitemShut {NoStop}%
\bibitem [{\citenamefont {Hofstetter}\ \emph {et~al.}(2009)\citenamefont
  {Hofstetter}, \citenamefont {Csonka}, \citenamefont {Nyg{\aa}rd},\ and\
  \citenamefont {Sch{\"o}nenberger}}]{Hofstetter_2009}%
  \BibitemOpen
  \bibfield  {author} {\bibinfo {author} {\bibfnamefont {L.}~\bibnamefont
  {Hofstetter}}, \bibinfo {author} {\bibfnamefont {S.}~\bibnamefont {Csonka}},
  \bibinfo {author} {\bibfnamefont {J.}~\bibnamefont {Nyg{\aa}rd}},\ and\
  \bibinfo {author} {\bibfnamefont {C.}~\bibnamefont {Sch{\"o}nenberger}},\
  }\bibfield  {title} {\bibinfo {title} {{Cooper pair splitter realized in a
  two-quantum-dot Y-junction}},\ }\href {https://doi.org/10.1038/nature08432}
  {\bibfield  {journal} {\bibinfo  {journal} {Nature}\ }\textbf {\bibinfo
  {volume} {461}},\ \bibinfo {pages} {960} (\bibinfo {year}
  {2009})}\BibitemShut {NoStop}%
\bibitem [{\citenamefont {Schindele}\ \emph {et~al.}(2012)\citenamefont
  {Schindele}, \citenamefont {Baumgartner},\ and\ \citenamefont
  {Sch\"onenberger}}]{Schindele_2012}%
  \BibitemOpen
  \bibfield  {author} {\bibinfo {author} {\bibfnamefont {J.}~\bibnamefont
  {Schindele}}, \bibinfo {author} {\bibfnamefont {A.}~\bibnamefont
  {Baumgartner}},\ and\ \bibinfo {author} {\bibfnamefont {C.}~\bibnamefont
  {Sch\"onenberger}},\ }\bibfield  {title} {\bibinfo {title} {{Near-Unity
  Cooper Pair Splitting Efficiency}},\ }\href
  {https://doi.org/10.1103/PhysRevLett.109.157002} {\bibfield  {journal}
  {\bibinfo  {journal} {Phys. Rev. Lett.}\ }\textbf {\bibinfo {volume} {109}},\
  \bibinfo {pages} {157002} (\bibinfo {year} {2012})}\BibitemShut {NoStop}%
\bibitem [{\citenamefont {Das}\ \emph {et~al.}(2012)\citenamefont {Das},
  \citenamefont {Ronen}, \citenamefont {Heiblum}, \citenamefont {Mahalu},
  \citenamefont {Kretinin},\ and\ \citenamefont {Shtrikman}}]{Das_2012}%
  \BibitemOpen
  \bibfield  {author} {\bibinfo {author} {\bibfnamefont {A.}~\bibnamefont
  {Das}}, \bibinfo {author} {\bibfnamefont {Y.}~\bibnamefont {Ronen}}, \bibinfo
  {author} {\bibfnamefont {M.}~\bibnamefont {Heiblum}}, \bibinfo {author}
  {\bibfnamefont {D.}~\bibnamefont {Mahalu}}, \bibinfo {author} {\bibfnamefont
  {A.~V.}\ \bibnamefont {Kretinin}},\ and\ \bibinfo {author} {\bibfnamefont
  {H.}~\bibnamefont {Shtrikman}},\ }\bibfield  {title} {\bibinfo {title}
  {{High-efficiency Cooper pair splitting demonstrated by two-particle
  conductance resonance and positive noise cross-correlation}},\ }\href
  {https://doi.org/10.1038/ncomms2169} {\bibfield  {journal} {\bibinfo
  {journal} {Nature Communications}\ }\textbf {\bibinfo {volume} {3}},\
  \bibinfo {pages} {1165} (\bibinfo {year} {2012})}\BibitemShut {NoStop}%
\bibitem [{\citenamefont {F\"ul\"op}\ \emph {et~al.}(2014)\citenamefont
  {F\"ul\"op}, \citenamefont {d'Hollosy}, \citenamefont {Baumgartner},
  \citenamefont {Makk}, \citenamefont {Guzenko}, \citenamefont {Madsen},
  \citenamefont {Nyg\aa{}rd}, \citenamefont {Sch\"onenberger},\ and\
  \citenamefont {Csonka}}]{PhysRevB.90.235412}%
  \BibitemOpen
  \bibfield  {author} {\bibinfo {author} {\bibfnamefont {G.}~\bibnamefont
  {F\"ul\"op}}, \bibinfo {author} {\bibfnamefont {S.}~\bibnamefont
  {d'Hollosy}}, \bibinfo {author} {\bibfnamefont {A.}~\bibnamefont
  {Baumgartner}}, \bibinfo {author} {\bibfnamefont {P.}~\bibnamefont {Makk}},
  \bibinfo {author} {\bibfnamefont {V.~A.}\ \bibnamefont {Guzenko}}, \bibinfo
  {author} {\bibfnamefont {M.~H.}\ \bibnamefont {Madsen}}, \bibinfo {author}
  {\bibfnamefont {J.}~\bibnamefont {Nyg\aa{}rd}}, \bibinfo {author}
  {\bibfnamefont {C.}~\bibnamefont {Sch\"onenberger}},\ and\ \bibinfo {author}
  {\bibfnamefont {S.}~\bibnamefont {Csonka}},\ }\bibfield  {title} {\bibinfo
  {title} {{Local electrical tuning of the nonlocal signals in a Cooper pair
  splitter}},\ }\href {https://doi.org/10.1103/PhysRevB.90.235412} {\bibfield
  {journal} {\bibinfo  {journal} {Phys. Rev. B}\ }\textbf {\bibinfo {volume}
  {90}},\ \bibinfo {pages} {235412} (\bibinfo {year} {2014})}\BibitemShut
  {NoStop}%
\bibitem [{\citenamefont {Tan}\ \emph {et~al.}(2015)\citenamefont {Tan},
  \citenamefont {Cox}, \citenamefont {Nieminen}, \citenamefont
  {L\"ahteenm\"aki}, \citenamefont {Golubev}, \citenamefont {Lesovik},\ and\
  \citenamefont {Hakonen}}]{PhysRevLett.114.096602}%
  \BibitemOpen
  \bibfield  {author} {\bibinfo {author} {\bibfnamefont {Z.~B.}\ \bibnamefont
  {Tan}}, \bibinfo {author} {\bibfnamefont {D.}~\bibnamefont {Cox}}, \bibinfo
  {author} {\bibfnamefont {T.}~\bibnamefont {Nieminen}}, \bibinfo {author}
  {\bibfnamefont {P.}~\bibnamefont {L\"ahteenm\"aki}}, \bibinfo {author}
  {\bibfnamefont {D.}~\bibnamefont {Golubev}}, \bibinfo {author} {\bibfnamefont
  {G.~B.}\ \bibnamefont {Lesovik}},\ and\ \bibinfo {author} {\bibfnamefont
  {P.~J.}\ \bibnamefont {Hakonen}},\ }\bibfield  {title} {\bibinfo {title}
  {{Cooper Pair Splitting by Means of Graphene Quantum Dots}},\ }\href
  {https://doi.org/10.1103/PhysRevLett.114.096602} {\bibfield  {journal}
  {\bibinfo  {journal} {Phys. Rev. Lett.}\ }\textbf {\bibinfo {volume} {114}},\
  \bibinfo {pages} {096602} (\bibinfo {year} {2015})}\BibitemShut {NoStop}%
\bibitem [{\citenamefont {Borzenets}\ \emph {et~al.}(2016)\citenamefont
  {Borzenets}, \citenamefont {Shimazaki}, \citenamefont {Jones}, \citenamefont
  {Craciun}, \citenamefont {Russo}, \citenamefont {Yamamoto},\ and\
  \citenamefont {Tarucha}}]{Borzenets2016}%
  \BibitemOpen
  \bibfield  {author} {\bibinfo {author} {\bibfnamefont {I.~V.}\ \bibnamefont
  {Borzenets}}, \bibinfo {author} {\bibfnamefont {Y.}~\bibnamefont
  {Shimazaki}}, \bibinfo {author} {\bibfnamefont {G.~F.}\ \bibnamefont
  {Jones}}, \bibinfo {author} {\bibfnamefont {M.~F.}\ \bibnamefont {Craciun}},
  \bibinfo {author} {\bibfnamefont {S.}~\bibnamefont {Russo}}, \bibinfo
  {author} {\bibfnamefont {M.}~\bibnamefont {Yamamoto}},\ and\ \bibinfo
  {author} {\bibfnamefont {S.}~\bibnamefont {Tarucha}},\ }\bibfield  {title}
  {\bibinfo {title} {{High Efficiency CVD Graphene-lead (Pb) Cooper Pair
  Splitter}},\ }\href {https://doi.org/10.1038/srep23051} {\bibfield  {journal}
  {\bibinfo  {journal} {Scientific Reports}\ }\textbf {\bibinfo {volume} {6}},\
  \bibinfo {pages} {23051} (\bibinfo {year} {2016})}\BibitemShut {NoStop}%
\bibitem [{\citenamefont {Tan}\ \emph {et~al.}(2021)\citenamefont {Tan},
  \citenamefont {Laitinen}, \citenamefont {Kirsanov}, \citenamefont {Galda},
  \citenamefont {Vinokur}, \citenamefont {Haque}, \citenamefont {Savin},
  \citenamefont {Golubev}, \citenamefont {Lesovik},\ and\ \citenamefont
  {Hakonen}}]{Tan2020}%
  \BibitemOpen
  \bibfield  {author} {\bibinfo {author} {\bibfnamefont {Z.~B.}\ \bibnamefont
  {Tan}}, \bibinfo {author} {\bibfnamefont {A.}~\bibnamefont {Laitinen}},
  \bibinfo {author} {\bibfnamefont {N.~S.}\ \bibnamefont {Kirsanov}}, \bibinfo
  {author} {\bibfnamefont {A.}~\bibnamefont {Galda}}, \bibinfo {author}
  {\bibfnamefont {V.~M.}\ \bibnamefont {Vinokur}}, \bibinfo {author}
  {\bibfnamefont {M.}~\bibnamefont {Haque}}, \bibinfo {author} {\bibfnamefont
  {A.}~\bibnamefont {Savin}}, \bibinfo {author} {\bibfnamefont {D.~S.}\
  \bibnamefont {Golubev}}, \bibinfo {author} {\bibfnamefont {G.~B.}\
  \bibnamefont {Lesovik}},\ and\ \bibinfo {author} {\bibfnamefont {P.~J.}\
  \bibnamefont {Hakonen}},\ }\bibfield  {title} {\bibinfo {title}
  {{Thermoelectric current in a graphene Cooper pair splitter}},\ }\href
  {https://doi.org/10.1038/s41467-020-20476-7} {\bibfield  {journal} {\bibinfo
  {journal} {Nature Communications}\ }\textbf {\bibinfo {volume} {12}},\
  \bibinfo {pages} {138} (\bibinfo {year} {2021})}\BibitemShut {NoStop}%
\bibitem [{\citenamefont {Rech}\ \emph {et~al.}(2012)\citenamefont {Rech},
  \citenamefont {Chevallier}, \citenamefont {Jonckheere},\ and\ \citenamefont
  {Martin}}]{PhysRevB.85.035419}%
  \BibitemOpen
  \bibfield  {author} {\bibinfo {author} {\bibfnamefont {J.}~\bibnamefont
  {Rech}}, \bibinfo {author} {\bibfnamefont {D.}~\bibnamefont {Chevallier}},
  \bibinfo {author} {\bibfnamefont {T.}~\bibnamefont {Jonckheere}},\ and\
  \bibinfo {author} {\bibfnamefont {T.}~\bibnamefont {Martin}},\ }\bibfield
  {title} {\bibinfo {title} {{Current correlations in an interacting
  Cooper-pair beam splitter}},\ }\href
  {https://doi.org/10.1103/PhysRevB.85.035419} {\bibfield  {journal} {\bibinfo
  {journal} {Phys. Rev. B}\ }\textbf {\bibinfo {volume} {85}},\ \bibinfo
  {pages} {035419} (\bibinfo {year} {2012})}\BibitemShut {NoStop}%
\bibitem [{\citenamefont {S\'anchez}\ \emph {et~al.}(2018)\citenamefont
  {S\'anchez}, \citenamefont {Burset},\ and\ \citenamefont
  {Yeyati}}]{PhysRevB.98.241414}%
  \BibitemOpen
  \bibfield  {author} {\bibinfo {author} {\bibfnamefont {R.}~\bibnamefont
  {S\'anchez}}, \bibinfo {author} {\bibfnamefont {P.}~\bibnamefont {Burset}},\
  and\ \bibinfo {author} {\bibfnamefont {A.~L.}\ \bibnamefont {Yeyati}},\
  }\bibfield  {title} {\bibinfo {title} {{Cooling by Cooper pair splitting}},\
  }\href {https://doi.org/10.1103/PhysRevB.98.241414} {\bibfield  {journal}
  {\bibinfo  {journal} {Phys. Rev. B}\ }\textbf {\bibinfo {volume} {98}},\
  \bibinfo {pages} {241414} (\bibinfo {year} {2018})}\BibitemShut {NoStop}%
\bibitem [{\citenamefont {Kirsanov}\ \emph {et~al.}(2019)\citenamefont
  {Kirsanov}, \citenamefont {Tan}, \citenamefont {Golubev}, \citenamefont
  {Hakonen},\ and\ \citenamefont {Lesovik}}]{PhysRevB.99.115127}%
  \BibitemOpen
  \bibfield  {author} {\bibinfo {author} {\bibfnamefont {N.~S.}\ \bibnamefont
  {Kirsanov}}, \bibinfo {author} {\bibfnamefont {Z.~B.}\ \bibnamefont {Tan}},
  \bibinfo {author} {\bibfnamefont {D.~S.}\ \bibnamefont {Golubev}}, \bibinfo
  {author} {\bibfnamefont {P.~J.}\ \bibnamefont {Hakonen}},\ and\ \bibinfo
  {author} {\bibfnamefont {G.~B.}\ \bibnamefont {Lesovik}},\ }\bibfield
  {title} {\bibinfo {title} {{Heat switch and thermoelectric effects based on
  Cooper-pair splitting and elastic cotunneling}},\ }\href
  {https://doi.org/10.1103/PhysRevB.99.115127} {\bibfield  {journal} {\bibinfo
  {journal} {Phys. Rev. B}\ }\textbf {\bibinfo {volume} {99}},\ \bibinfo
  {pages} {115127} (\bibinfo {year} {2019})}\BibitemShut {NoStop}%
\bibitem [{\citenamefont {Walldorf}\ \emph {et~al.}(2018)\citenamefont
  {Walldorf}, \citenamefont {Padurariu}, \citenamefont {Jauho},\ and\
  \citenamefont {Flindt}}]{PhysRevLett.120.087701}%
  \BibitemOpen
  \bibfield  {author} {\bibinfo {author} {\bibfnamefont {N.}~\bibnamefont
  {Walldorf}}, \bibinfo {author} {\bibfnamefont {C.}~\bibnamefont {Padurariu}},
  \bibinfo {author} {\bibfnamefont {A.-P.}\ \bibnamefont {Jauho}},\ and\
  \bibinfo {author} {\bibfnamefont {C.}~\bibnamefont {Flindt}},\ }\bibfield
  {title} {\bibinfo {title} {{Electron Waiting Times of a Cooper Pair
  Splitter}},\ }\href {https://doi.org/10.1103/PhysRevLett.120.087701}
  {\bibfield  {journal} {\bibinfo  {journal} {Phys. Rev. Lett.}\ }\textbf
  {\bibinfo {volume} {120}},\ \bibinfo {pages} {087701} (\bibinfo {year}
  {2018})}\BibitemShut {NoStop}%
\bibitem [{\citenamefont {Hussein}\ \emph {et~al.}(2019)\citenamefont
  {Hussein}, \citenamefont {Governale}, \citenamefont {Kohler}, \citenamefont
  {Belzig}, \citenamefont {Giazotto},\ and\ \citenamefont
  {Braggio}}]{PhysRevB.99.075429}%
  \BibitemOpen
  \bibfield  {author} {\bibinfo {author} {\bibfnamefont {R.}~\bibnamefont
  {Hussein}}, \bibinfo {author} {\bibfnamefont {M.}~\bibnamefont {Governale}},
  \bibinfo {author} {\bibfnamefont {S.}~\bibnamefont {Kohler}}, \bibinfo
  {author} {\bibfnamefont {W.}~\bibnamefont {Belzig}}, \bibinfo {author}
  {\bibfnamefont {F.}~\bibnamefont {Giazotto}},\ and\ \bibinfo {author}
  {\bibfnamefont {A.}~\bibnamefont {Braggio}},\ }\bibfield  {title} {\bibinfo
  {title} {{Nonlocal thermoelectricity in a Cooper-pair splitter}},\ }\href
  {https://doi.org/10.1103/PhysRevB.99.075429} {\bibfield  {journal} {\bibinfo
  {journal} {Phys. Rev. B}\ }\textbf {\bibinfo {volume} {99}},\ \bibinfo
  {pages} {075429} (\bibinfo {year} {2019})}\BibitemShut {NoStop}%
\bibitem [{\citenamefont {Ranni}\ \emph {et~al.}(2021)\citenamefont {Ranni},
  \citenamefont {Brange}, \citenamefont {Mannila}, \citenamefont {Flindt},\
  and\ \citenamefont {Maisi}}]{Ranni_2021}%
  \BibitemOpen
  \bibfield  {author} {\bibinfo {author} {\bibfnamefont {A.}~\bibnamefont
  {Ranni}}, \bibinfo {author} {\bibfnamefont {F.}~\bibnamefont {Brange}},
  \bibinfo {author} {\bibfnamefont {E.~T.}\ \bibnamefont {Mannila}}, \bibinfo
  {author} {\bibfnamefont {C.}~\bibnamefont {Flindt}},\ and\ \bibinfo {author}
  {\bibfnamefont {V.~F.}\ \bibnamefont {Maisi}},\ }\bibfield  {title} {\bibinfo
  {title} {{Real-time observation of Cooper pair splitting showing strong
  non-local correlations}},\ }\href
  {https://doi.org/10.1038/s41467-021-26627-8} {\bibfield  {journal} {\bibinfo
  {journal} {Nature Communications}\ }\textbf {\bibinfo {volume} {12}},\
  \bibinfo {pages} {6358} (\bibinfo {year} {2021})}\BibitemShut {NoStop}%
\bibitem [{\citenamefont {Rozhkov}\ and\ \citenamefont
  {Arovas}(1999)}]{RozhkovArovas1999}%
  \BibitemOpen
  \bibfield  {author} {\bibinfo {author} {\bibfnamefont {A.~V.}\ \bibnamefont
  {Rozhkov}}\ and\ \bibinfo {author} {\bibfnamefont {D.~P.}\ \bibnamefont
  {Arovas}},\ }\bibfield  {title} {\bibinfo {title} {{Josephson Coupling
  through a Magnetic Impurity}},\ }\href
  {https://doi.org/10.1103/PhysRevLett.82.2788} {\bibfield  {journal} {\bibinfo
   {journal} {Phys. Rev. Lett.}\ }\textbf {\bibinfo {volume} {82}},\ \bibinfo
  {pages} {2788} (\bibinfo {year} {1999})}\BibitemShut {NoStop}%
\bibitem [{\citenamefont {Clerk}\ and\ \citenamefont
  {Ambegaokar}(2000)}]{ClerkAmbegaokar2000}%
  \BibitemOpen
  \bibfield  {author} {\bibinfo {author} {\bibfnamefont {A.~A.}\ \bibnamefont
  {Clerk}}\ and\ \bibinfo {author} {\bibfnamefont {V.}~\bibnamefont
  {Ambegaokar}},\ }\bibfield  {title} {\bibinfo {title} {{Loss of
  $\ensuremath{\pi}$-junction behavior in an interacting impurity Josephson
  junction}},\ }\href {https://doi.org/10.1103/PhysRevB.61.9109} {\bibfield
  {journal} {\bibinfo  {journal} {Phys. Rev. B}\ }\textbf {\bibinfo {volume}
  {61}},\ \bibinfo {pages} {9109} (\bibinfo {year} {2000})}\BibitemShut
  {NoStop}%
\bibitem [{\citenamefont {Vecino}\ \emph {et~al.}(2003)\citenamefont {Vecino},
  \citenamefont {Mart\'{\i}n-Rodero},\ and\ \citenamefont
  {Yeyati}}]{Vecino_2003}%
  \BibitemOpen
  \bibfield  {author} {\bibinfo {author} {\bibfnamefont {E.}~\bibnamefont
  {Vecino}}, \bibinfo {author} {\bibfnamefont {A.}~\bibnamefont
  {Mart\'{\i}n-Rodero}},\ and\ \bibinfo {author} {\bibfnamefont {A.~L.}\
  \bibnamefont {Yeyati}},\ }\bibfield  {title} {\bibinfo {title} {{Josephson
  current through a correlated quantum level: Andreev states and
  $\ensuremath{\pi}$ junction behavior}},\ }\href
  {https://doi.org/10.1103/PhysRevB.68.035105} {\bibfield  {journal} {\bibinfo
  {journal} {Phys. Rev. B}\ }\textbf {\bibinfo {volume} {68}},\ \bibinfo
  {pages} {035105} (\bibinfo {year} {2003})}\BibitemShut {NoStop}%
\bibitem [{\citenamefont {Yoshioka}\ and\ \citenamefont
  {Ohashi}(2000)}]{Yoshioka_2000}%
  \BibitemOpen
  \bibfield  {author} {\bibinfo {author} {\bibfnamefont {T.}~\bibnamefont
  {Yoshioka}}\ and\ \bibinfo {author} {\bibfnamefont {Y.}~\bibnamefont
  {Ohashi}},\ }\bibfield  {title} {\bibinfo {title} {{Numerical Renormalization
  Group Studies on Single Impurity Anderson Model in Superconductivity: A
  Unified Treatment of Magnetic, Nonmagnetic Impurities, and Resonance
  Scattering}},\ }\href {https://doi.org/10.1143/JPSJ.69.1812} {\bibfield
  {journal} {\bibinfo  {journal} {J. Phys. Soc. Jpn.}\ }\textbf {\bibinfo
  {volume} {69}},\ \bibinfo {pages} {1812} (\bibinfo {year}
  {2000})}\BibitemShut {NoStop}%
\bibitem [{\citenamefont {Choi}\ \emph {et~al.}(2004)\citenamefont {Choi},
  \citenamefont {Lee}, \citenamefont {Kang},\ and\ \citenamefont
  {Belzig}}]{Choi2004}%
  \BibitemOpen
  \bibfield  {author} {\bibinfo {author} {\bibfnamefont {M.-S.}\ \bibnamefont
  {Choi}}, \bibinfo {author} {\bibfnamefont {M.}~\bibnamefont {Lee}}, \bibinfo
  {author} {\bibfnamefont {K.}~\bibnamefont {Kang}},\ and\ \bibinfo {author}
  {\bibfnamefont {W.}~\bibnamefont {Belzig}},\ }\bibfield  {title} {\bibinfo
  {title} {{Kondo effect and Josephson current through a quantum dot between
  two superconductors}},\ }\href {https://doi.org/10.1103/PhysRevB.70.020502}
  {\bibfield  {journal} {\bibinfo  {journal} {Phys. Rev. B}\ }\textbf {\bibinfo
  {volume} {70}},\ \bibinfo {pages} {020502} (\bibinfo {year}
  {2004})}\BibitemShut {NoStop}%
\bibitem [{\citenamefont {Siano}\ and\ \citenamefont
  {Egger}(2004)}]{SianoEgger2004}%
  \BibitemOpen
  \bibfield  {author} {\bibinfo {author} {\bibfnamefont {F.}~\bibnamefont
  {Siano}}\ and\ \bibinfo {author} {\bibfnamefont {R.}~\bibnamefont {Egger}},\
  }\bibfield  {title} {\bibinfo {title} {{Josephson Current through a Nanoscale
  Magnetic Quantum Dot}},\ }\href
  {https://doi.org/10.1103/PhysRevLett.93.047002} {\bibfield  {journal}
  {\bibinfo  {journal} {Phys. Rev. Lett.}\ }\textbf {\bibinfo {volume} {93}},\
  \bibinfo {pages} {047002} (\bibinfo {year} {2004})}\BibitemShut {NoStop}%
\bibitem [{\citenamefont {Oguri}\ \emph {et~al.}(2004)\citenamefont {Oguri},
  \citenamefont {Tanaka},\ and\ \citenamefont {C.~Hewson}}]{Oguri_2004}%
  \BibitemOpen
  \bibfield  {author} {\bibinfo {author} {\bibfnamefont {A.}~\bibnamefont
  {Oguri}}, \bibinfo {author} {\bibfnamefont {Y.}~\bibnamefont {Tanaka}},\ and\
  \bibinfo {author} {\bibfnamefont {A.}~\bibnamefont {C.~Hewson}},\ }\bibfield
  {title} {\bibinfo {title} {{Quantum Phase Transition in a Minimal Model for
  the Kondo Effect in a Josephson Junction}},\ }\href
  {https://doi.org/10.1143/JPSJ.73.2494} {\bibfield  {journal} {\bibinfo
  {journal} {J. Phys. Soc. Jpn.}\ }\textbf {\bibinfo {volume} {73}},\ \bibinfo
  {pages} {2494} (\bibinfo {year} {2004})}\BibitemShut {NoStop}%
\bibitem [{\citenamefont {Bauer}\ \emph {et~al.}(2007)\citenamefont {Bauer},
  \citenamefont {Oguri},\ and\ \citenamefont {Hewson}}]{Bauer_2007}%
  \BibitemOpen
  \bibfield  {author} {\bibinfo {author} {\bibfnamefont {J.}~\bibnamefont
  {Bauer}}, \bibinfo {author} {\bibfnamefont {A.}~\bibnamefont {Oguri}},\ and\
  \bibinfo {author} {\bibfnamefont {A.~C.}\ \bibnamefont {Hewson}},\ }\bibfield
   {title} {\bibinfo {title} {{Spectral properties of locally correlated
  electrons in a Bardeen-Cooper-Schrieffer superconductor}},\ }\href
  {https://doi.org/10.1088/0953-8984/19/48/486211} {\bibfield  {journal}
  {\bibinfo  {journal} {Journal of Physics: Condensed Matter}\ }\textbf
  {\bibinfo {volume} {19}},\ \bibinfo {pages} {486211} (\bibinfo {year}
  {2007})}\BibitemShut {NoStop}%
\bibitem [{\citenamefont {Tanaka}\ \emph
  {et~al.}(2007{\natexlab{a}})\citenamefont {Tanaka}, \citenamefont {Oguri},\
  and\ \citenamefont {Hewson}}]{YoshihideTanaka_2007}%
  \BibitemOpen
  \bibfield  {author} {\bibinfo {author} {\bibfnamefont {Y.}~\bibnamefont
  {Tanaka}}, \bibinfo {author} {\bibfnamefont {A.}~\bibnamefont {Oguri}},\ and\
  \bibinfo {author} {\bibfnamefont {A.~C.}\ \bibnamefont {Hewson}},\ }\bibfield
   {title} {\bibinfo {title} {{Kondo effect in asymmetric Josephson couplings
  through a quantum dot}},\ }\href {https://doi.org/10.1088/1367-2630/9/5/115}
  {\bibfield  {journal} {\bibinfo  {journal} {New J. Phys.}\ }\textbf {\bibinfo
  {volume} {9}},\ \bibinfo {pages} {115} (\bibinfo {year}
  {2007}{\natexlab{a}})}\BibitemShut {NoStop}%
\bibitem [{\citenamefont {Karrasch}\ \emph {et~al.}(2008)\citenamefont
  {Karrasch}, \citenamefont {Oguri},\ and\ \citenamefont
  {Meden}}]{Karrasch2008}%
  \BibitemOpen
  \bibfield  {author} {\bibinfo {author} {\bibfnamefont {C.}~\bibnamefont
  {Karrasch}}, \bibinfo {author} {\bibfnamefont {A.}~\bibnamefont {Oguri}},\
  and\ \bibinfo {author} {\bibfnamefont {V.}~\bibnamefont {Meden}},\ }\bibfield
   {title} {\bibinfo {title} {{Josephson current through a single Anderson
  impurity coupled to BCS leads}},\ }\href
  {https://doi.org/10.1103/PhysRevB.77.024517} {\bibfield  {journal} {\bibinfo
  {journal} {Phys. Rev. B}\ }\textbf {\bibinfo {volume} {77}},\ \bibinfo
  {pages} {024517} (\bibinfo {year} {2008})}\BibitemShut {NoStop}%
\bibitem [{\citenamefont {Meng}\ \emph {et~al.}(2009)\citenamefont {Meng},
  \citenamefont {Florens},\ and\ \citenamefont {Simon}}]{Florens2009}%
  \BibitemOpen
  \bibfield  {author} {\bibinfo {author} {\bibfnamefont {T.}~\bibnamefont
  {Meng}}, \bibinfo {author} {\bibfnamefont {S.}~\bibnamefont {Florens}},\ and\
  \bibinfo {author} {\bibfnamefont {P.}~\bibnamefont {Simon}},\ }\bibfield
  {title} {\bibinfo {title} {Self-consistent description of andreev bound
  states in josephson quantum dot devices},\ }\href
  {https://doi.org/10.1103/PhysRevB.79.224521} {\bibfield  {journal} {\bibinfo
  {journal} {Phys. Rev. B}\ }\textbf {\bibinfo {volume} {79}},\ \bibinfo
  {pages} {224521} (\bibinfo {year} {2009})}\BibitemShut {NoStop}%
\bibitem [{\citenamefont {Hecht}\ \emph {et~al.}(2008)\citenamefont {Hecht},
  \citenamefont {Weichselbaum}, \citenamefont {von Delft},\ and\ \citenamefont
  {Bulla}}]{Hecht_2008}%
  \BibitemOpen
  \bibfield  {author} {\bibinfo {author} {\bibfnamefont {T.}~\bibnamefont
  {Hecht}}, \bibinfo {author} {\bibfnamefont {A.}~\bibnamefont {Weichselbaum}},
  \bibinfo {author} {\bibfnamefont {J.}~\bibnamefont {von Delft}},\ and\
  \bibinfo {author} {\bibfnamefont {R.}~\bibnamefont {Bulla}},\ }\bibfield
  {title} {\bibinfo {title} {{Numerical renormalization group calculation of
  near-gap peaks in spectral functions of the Anderson model with
  superconducting leads}},\ }\href
  {https://doi.org/10.1088/0953-8984/20/27/275213} {\bibfield  {journal}
  {\bibinfo  {journal} {Journal of Physics: Condensed Matter}\ }\textbf
  {\bibinfo {volume} {20}},\ \bibinfo {pages} {275213} (\bibinfo {year}
  {2008})}\BibitemShut {NoStop}%
\bibitem [{\citenamefont {Tanaka}\ \emph
  {et~al.}(2007{\natexlab{b}})\citenamefont {Tanaka}, \citenamefont
  {Kawakami},\ and\ \citenamefont {Oguri}}]{YoichTanaka_2007}%
  \BibitemOpen
  \bibfield  {author} {\bibinfo {author} {\bibfnamefont {Y.}~\bibnamefont
  {Tanaka}}, \bibinfo {author} {\bibfnamefont {N.}~\bibnamefont {Kawakami}},\
  and\ \bibinfo {author} {\bibfnamefont {A.}~\bibnamefont {Oguri}},\ }\bibfield
   {title} {\bibinfo {title} {{Numerical Renormalization Group Approach to a
  Quantum Dot Coupled to Normal and Superconducting Leads}},\ }\href
  {https://doi.org/10.1143/JPSJ.76.074701} {\bibfield  {journal} {\bibinfo
  {journal} {J. Phys. Soc. Jpn.}\ }\textbf {\bibinfo {volume} {76}},\ \bibinfo
  {pages} {074701} (\bibinfo {year} {2007}{\natexlab{b}})}\BibitemShut
  {NoStop}%
\bibitem [{\citenamefont {Yamada}\ \emph {et~al.}(2011)\citenamefont {Yamada},
  \citenamefont {Tanaka},\ and\ \citenamefont {Kawakami}}]{Yamada_2011}%
  \BibitemOpen
  \bibfield  {author} {\bibinfo {author} {\bibfnamefont {Y.}~\bibnamefont
  {Yamada}}, \bibinfo {author} {\bibfnamefont {Y.}~\bibnamefont {Tanaka}},\
  and\ \bibinfo {author} {\bibfnamefont {N.}~\bibnamefont {Kawakami}},\
  }\bibfield  {title} {\bibinfo {title} {{Interplay of Kondo and
  superconducting correlations in the nonequilibrium Andreev transport through
  a quantum dot}},\ }\href {https://doi.org/10.1103/PhysRevB.84.075484}
  {\bibfield  {journal} {\bibinfo  {journal} {Phys. Rev. B}\ }\textbf {\bibinfo
  {volume} {84}},\ \bibinfo {pages} {075484} (\bibinfo {year}
  {2011})}\BibitemShut {NoStop}%
\bibitem [{\citenamefont {Buizert}\ \emph {et~al.}(2007)\citenamefont
  {Buizert}, \citenamefont {Oiwa}, \citenamefont {Shibata}, \citenamefont
  {Hirakawa},\ and\ \citenamefont {Tarucha}}]{Buizert_2007}%
  \BibitemOpen
  \bibfield  {author} {\bibinfo {author} {\bibfnamefont {C.}~\bibnamefont
  {Buizert}}, \bibinfo {author} {\bibfnamefont {A.}~\bibnamefont {Oiwa}},
  \bibinfo {author} {\bibfnamefont {K.}~\bibnamefont {Shibata}}, \bibinfo
  {author} {\bibfnamefont {K.}~\bibnamefont {Hirakawa}},\ and\ \bibinfo
  {author} {\bibfnamefont {S.}~\bibnamefont {Tarucha}},\ }\bibfield  {title}
  {\bibinfo {title} {{Kondo Universal Scaling for a Quantum Dot Coupled to
  Superconducting Leads}},\ }\href
  {https://doi.org/10.1103/PhysRevLett.99.136806} {\bibfield  {journal}
  {\bibinfo  {journal} {Phys. Rev. Lett.}\ }\textbf {\bibinfo {volume} {99}},\
  \bibinfo {pages} {136806} (\bibinfo {year} {2007})}\BibitemShut {NoStop}%
\bibitem [{\citenamefont {Deacon}\ \emph
  {et~al.}(2010{\natexlab{a}})\citenamefont {Deacon}, \citenamefont {Tanaka},
  \citenamefont {Oiwa}, \citenamefont {Sakano}, \citenamefont {Yoshida},
  \citenamefont {Shibata}, \citenamefont {Hirakawa},\ and\ \citenamefont
  {Tarucha}}]{Deacon_2010}%
  \BibitemOpen
  \bibfield  {author} {\bibinfo {author} {\bibfnamefont {R.~S.}\ \bibnamefont
  {Deacon}}, \bibinfo {author} {\bibfnamefont {Y.}~\bibnamefont {Tanaka}},
  \bibinfo {author} {\bibfnamefont {A.}~\bibnamefont {Oiwa}}, \bibinfo {author}
  {\bibfnamefont {R.}~\bibnamefont {Sakano}}, \bibinfo {author} {\bibfnamefont
  {K.}~\bibnamefont {Yoshida}}, \bibinfo {author} {\bibfnamefont
  {K.}~\bibnamefont {Shibata}}, \bibinfo {author} {\bibfnamefont
  {K.}~\bibnamefont {Hirakawa}},\ and\ \bibinfo {author} {\bibfnamefont
  {S.}~\bibnamefont {Tarucha}},\ }\bibfield  {title} {\bibinfo {title}
  {{Tunneling Spectroscopy of Andreev Energy Levels in a Quantum Dot Coupled to
  a Superconductor}},\ }\href {https://doi.org/10.1103/PhysRevLett.104.076805}
  {\bibfield  {journal} {\bibinfo  {journal} {Phys. Rev. Lett.}\ }\textbf
  {\bibinfo {volume} {104}},\ \bibinfo {pages} {076805} (\bibinfo {year}
  {2010}{\natexlab{a}})}\BibitemShut {NoStop}%
\bibitem [{\citenamefont {Deacon}\ \emph
  {et~al.}(2010{\natexlab{b}})\citenamefont {Deacon}, \citenamefont {Tanaka},
  \citenamefont {Oiwa}, \citenamefont {Sakano}, \citenamefont {Yoshida},
  \citenamefont {Shibata}, \citenamefont {Hirakawa},\ and\ \citenamefont
  {Tarucha}}]{Deacon_2010Rapid}%
  \BibitemOpen
  \bibfield  {author} {\bibinfo {author} {\bibfnamefont {R.~S.}\ \bibnamefont
  {Deacon}}, \bibinfo {author} {\bibfnamefont {Y.}~\bibnamefont {Tanaka}},
  \bibinfo {author} {\bibfnamefont {A.}~\bibnamefont {Oiwa}}, \bibinfo {author}
  {\bibfnamefont {R.}~\bibnamefont {Sakano}}, \bibinfo {author} {\bibfnamefont
  {K.}~\bibnamefont {Yoshida}}, \bibinfo {author} {\bibfnamefont
  {K.}~\bibnamefont {Shibata}}, \bibinfo {author} {\bibfnamefont
  {K.}~\bibnamefont {Hirakawa}},\ and\ \bibinfo {author} {\bibfnamefont
  {S.}~\bibnamefont {Tarucha}},\ }\bibfield  {title} {\bibinfo {title}
  {{Kondo-enhanced Andreev transport in single self-assembled InAs quantum dots
  contacted with normal and superconducting leads}},\ }\href
  {https://doi.org/10.1103/PhysRevB.81.121308} {\bibfield  {journal} {\bibinfo
  {journal} {Phys. Rev. B}\ }\textbf {\bibinfo {volume} {81}},\ \bibinfo
  {pages} {121308} (\bibinfo {year} {2010}{\natexlab{b}})}\BibitemShut
  {NoStop}%
\bibitem [{\citenamefont {Governale}\ \emph {et~al.}(2008)\citenamefont
  {Governale}, \citenamefont {Pala},\ and\ \citenamefont
  {K\"onig}}]{Governale_2008}%
  \BibitemOpen
  \bibfield  {author} {\bibinfo {author} {\bibfnamefont {M.}~\bibnamefont
  {Governale}}, \bibinfo {author} {\bibfnamefont {M.~G.}\ \bibnamefont
  {Pala}},\ and\ \bibinfo {author} {\bibfnamefont {J.}~\bibnamefont
  {K\"onig}},\ }\bibfield  {title} {\bibinfo {title} {{Real-time diagrammatic
  approach to transport through interacting quantum dots with normal and
  superconducting leads}},\ }\href {https://doi.org/10.1103/PhysRevB.77.134513}
  {\bibfield  {journal} {\bibinfo  {journal} {Phys. Rev. B}\ }\textbf {\bibinfo
  {volume} {77}},\ \bibinfo {pages} {134513} (\bibinfo {year}
  {2008})}\BibitemShut {NoStop}%
\bibitem [{\citenamefont {Futterer}\ \emph {et~al.}(2009)\citenamefont
  {Futterer}, \citenamefont {Governale}, \citenamefont {Pala},\ and\
  \citenamefont {K\"onig}}]{Futterer2009}%
  \BibitemOpen
  \bibfield  {author} {\bibinfo {author} {\bibfnamefont {D.}~\bibnamefont
  {Futterer}}, \bibinfo {author} {\bibfnamefont {M.}~\bibnamefont {Governale}},
  \bibinfo {author} {\bibfnamefont {M.~G.}\ \bibnamefont {Pala}},\ and\
  \bibinfo {author} {\bibfnamefont {J.}~\bibnamefont {K\"onig}},\ }\bibfield
  {title} {\bibinfo {title} {{Nonlocal Andreev transport through an interacting
  quantum dot}},\ }\href {https://doi.org/10.1103/PhysRevB.79.054505}
  {\bibfield  {journal} {\bibinfo  {journal} {Phys. Rev. B}\ }\textbf {\bibinfo
  {volume} {79}},\ \bibinfo {pages} {054505} (\bibinfo {year}
  {2009})}\BibitemShut {NoStop}%
\bibitem [{\citenamefont {Eldridge}\ \emph {et~al.}(2010)\citenamefont
  {Eldridge}, \citenamefont {Pala}, \citenamefont {Governale},\ and\
  \citenamefont {K\"onig}}]{Eldridge_2010}%
  \BibitemOpen
  \bibfield  {author} {\bibinfo {author} {\bibfnamefont {J.}~\bibnamefont
  {Eldridge}}, \bibinfo {author} {\bibfnamefont {M.~G.}\ \bibnamefont {Pala}},
  \bibinfo {author} {\bibfnamefont {M.}~\bibnamefont {Governale}},\ and\
  \bibinfo {author} {\bibfnamefont {J.}~\bibnamefont {K\"onig}},\ }\bibfield
  {title} {\bibinfo {title} {{Superconducting proximity effect in interacting
  double-dot systems}},\ }\href {https://doi.org/10.1103/PhysRevB.82.184507}
  {\bibfield  {journal} {\bibinfo  {journal} {Phys. Rev. B}\ }\textbf {\bibinfo
  {volume} {82}},\ \bibinfo {pages} {184507} (\bibinfo {year}
  {2010})}\BibitemShut {NoStop}%
\bibitem [{\citenamefont {Micha\l{}ek}\ \emph {et~al.}(2013)\citenamefont
  {Micha\l{}ek}, \citenamefont {Bu\l{}ka}, \citenamefont
  {Doma\ifmmode~\acute{n}\else \'{n}\fi{}ski},\ and\ \citenamefont
  {Wysoki\ifmmode~\acute{n}\else \'{n}\fi{}ski}}]{Michalek_2013}%
  \BibitemOpen
  \bibfield  {author} {\bibinfo {author} {\bibfnamefont {G.}~\bibnamefont
  {Micha\l{}ek}}, \bibinfo {author} {\bibfnamefont {B.~R.}\ \bibnamefont
  {Bu\l{}ka}}, \bibinfo {author} {\bibfnamefont {T.}~\bibnamefont
  {Doma\ifmmode~\acute{n}\else \'{n}\fi{}ski}},\ and\ \bibinfo {author}
  {\bibfnamefont {K.~I.}\ \bibnamefont {Wysoki\ifmmode~\acute{n}\else
  \'{n}\fi{}ski}},\ }\bibfield  {title} {\bibinfo {title} {{Interplay between
  direct and crossed Andreev reflections in hybrid nanostructures}},\ }\href
  {https://doi.org/10.1103/PhysRevB.88.155425} {\bibfield  {journal} {\bibinfo
  {journal} {Phys. Rev. B}\ }\textbf {\bibinfo {volume} {88}},\ \bibinfo
  {pages} {155425} (\bibinfo {year} {2013})}\BibitemShut {NoStop}%
\bibitem [{\citenamefont {Micha{\l}ek}\ \emph {et~al.}(2015)\citenamefont
  {Micha{\l}ek}, \citenamefont {Doma{\'n}ski}, \citenamefont {Bu{\l}ka},\ and\
  \citenamefont {Wysoki{\'n}ski}}]{Michalek_2015}%
  \BibitemOpen
  \bibfield  {author} {\bibinfo {author} {\bibfnamefont {G.}~\bibnamefont
  {Micha{\l}ek}}, \bibinfo {author} {\bibfnamefont {T.}~\bibnamefont
  {Doma{\'n}ski}}, \bibinfo {author} {\bibfnamefont {B.~R.}\ \bibnamefont
  {Bu{\l}ka}},\ and\ \bibinfo {author} {\bibfnamefont {K.~I.}\ \bibnamefont
  {Wysoki{\'n}ski}},\ }\bibfield  {title} {\bibinfo {title} {{Novel non-local
  effects in three-terminal hybrid devices with quantum dot}},\ }\href
  {https://doi.org/10.1038/srep14572} {\bibfield  {journal} {\bibinfo
  {journal} {Scientific Reports}\ }\textbf {\bibinfo {volume} {5}},\ \bibinfo
  {pages} {14572} (\bibinfo {year} {2015})}\BibitemShut {NoStop}%
\bibitem [{\citenamefont {Weymann}\ and\ \citenamefont
  {W\'ojcik}(2015)}]{Weymann2015}%
  \BibitemOpen
  \bibfield  {author} {\bibinfo {author} {\bibfnamefont {I.}~\bibnamefont
  {Weymann}}\ and\ \bibinfo {author} {\bibfnamefont {K.~P.}\ \bibnamefont
  {W\'ojcik}},\ }\bibfield  {title} {\bibinfo {title} {{Andreev transport in a
  correlated ferromagnet-quantum-dot-superconductor device}},\ }\href
  {https://doi.org/10.1103/PhysRevB.92.245307} {\bibfield  {journal} {\bibinfo
  {journal} {Phys. Rev. B}\ }\textbf {\bibinfo {volume} {92}},\ \bibinfo
  {pages} {245307} (\bibinfo {year} {2015})}\BibitemShut {NoStop}%
\bibitem [{\citenamefont {Oguri}\ \emph {et~al.}(2013)\citenamefont {Oguri},
  \citenamefont {Tanaka},\ and\ \citenamefont {Bauer}}]{Oguri_2013}%
  \BibitemOpen
  \bibfield  {author} {\bibinfo {author} {\bibfnamefont {A.}~\bibnamefont
  {Oguri}}, \bibinfo {author} {\bibfnamefont {Y.}~\bibnamefont {Tanaka}},\ and\
  \bibinfo {author} {\bibfnamefont {J.}~\bibnamefont {Bauer}},\ }\bibfield
  {title} {\bibinfo {title} {{Interplay between Kondo and Andreev-Josephson
  effects in a quantum dot coupled to one normal and two superconducting
  leads}},\ }\href {https://doi.org/10.1103/PhysRevB.87.075432} {\bibfield
  {journal} {\bibinfo  {journal} {Phys. Rev. B}\ }\textbf {\bibinfo {volume}
  {87}},\ \bibinfo {pages} {075432} (\bibinfo {year} {2013})}\BibitemShut
  {NoStop}%
\bibitem [{\citenamefont {Koga}(2013)}]{PhysRevB.87.115409}%
  \BibitemOpen
  \bibfield  {author} {\bibinfo {author} {\bibfnamefont {A.}~\bibnamefont
  {Koga}},\ }\bibfield  {title} {\bibinfo {title} {{Quantum Monte Carlo study
  of nonequilibrium transport through a quantum dot coupled to normal and
  superconducting leads}},\ }\href {https://doi.org/10.1103/PhysRevB.87.115409}
  {\bibfield  {journal} {\bibinfo  {journal} {Phys. Rev. B}\ }\textbf {\bibinfo
  {volume} {87}},\ \bibinfo {pages} {115409} (\bibinfo {year}
  {2013})}\BibitemShut {NoStop}%
\bibitem [{\citenamefont {Doma\ifmmode~\acute{n}\else \'{n}\fi{}ski}\ \emph
  {et~al.}(2017)\citenamefont {Doma\ifmmode~\acute{n}\else \'{n}\fi{}ski},
  \citenamefont {\ifmmode~\check{Z}\else \v{Z}\fi{}onda}, \citenamefont
  {Pokorn\'y}, \citenamefont {G\'orski}, \citenamefont
  {Jani\ifmmode~\check{s}\else \v{s}\fi{}},\ and\ \citenamefont
  {Novotn\'y}}]{Domanski_2017}%
  \BibitemOpen
  \bibfield  {author} {\bibinfo {author} {\bibfnamefont {T.}~\bibnamefont
  {Doma\ifmmode~\acute{n}\else \'{n}\fi{}ski}}, \bibinfo {author}
  {\bibfnamefont {M.}~\bibnamefont {\ifmmode~\check{Z}\else \v{Z}\fi{}onda}},
  \bibinfo {author} {\bibfnamefont {V.}~\bibnamefont {Pokorn\'y}}, \bibinfo
  {author} {\bibfnamefont {G.}~\bibnamefont {G\'orski}}, \bibinfo {author}
  {\bibfnamefont {V.}~\bibnamefont {Jani\ifmmode~\check{s}\else \v{s}\fi{}}},\
  and\ \bibinfo {author} {\bibfnamefont {T.}~\bibnamefont {Novotn\'y}},\
  }\bibfield  {title} {\bibinfo {title} {{Josephson-phase-controlled interplay
  between correlation effects and electron pairing in a three-terminal
  nanostructure}},\ }\href {https://doi.org/10.1103/PhysRevB.95.045104}
  {\bibfield  {journal} {\bibinfo  {journal} {Phys. Rev. B}\ }\textbf {\bibinfo
  {volume} {95}},\ \bibinfo {pages} {045104} (\bibinfo {year}
  {2017})}\BibitemShut {NoStop}%
\bibitem [{\citenamefont {Wrze\ifmmode~\acute{s}\else \'{s}\fi{}niewski}\ and\
  \citenamefont {Weymann}(2017)}]{Wrzeinfmmode_2017}%
  \BibitemOpen
  \bibfield  {author} {\bibinfo {author} {\bibfnamefont {K.}~\bibnamefont
  {Wrze\ifmmode~\acute{s}\else \'{s}\fi{}niewski}}\ and\ \bibinfo {author}
  {\bibfnamefont {I.}~\bibnamefont {Weymann}},\ }\bibfield  {title} {\bibinfo
  {title} {{Kondo physics in double quantum dot based Cooper pair splitters}},\
  }\href {https://doi.org/10.1103/PhysRevB.96.195409} {\bibfield  {journal}
  {\bibinfo  {journal} {Phys. Rev. B}\ }\textbf {\bibinfo {volume} {96}},\
  \bibinfo {pages} {195409} (\bibinfo {year} {2017})}\BibitemShut {NoStop}%
\bibitem [{\citenamefont {Lee}\ \emph {et~al.}(2022)\citenamefont {Lee},
  \citenamefont {L\'opez}, \citenamefont {Xu},\ and\ \citenamefont
  {Platero}}]{PhysRevLett.129.207701}%
  \BibitemOpen
  \bibfield  {author} {\bibinfo {author} {\bibfnamefont {M.}~\bibnamefont
  {Lee}}, \bibinfo {author} {\bibfnamefont {R.}~\bibnamefont {L\'opez}},
  \bibinfo {author} {\bibfnamefont {H.~Q.}\ \bibnamefont {Xu}},\ and\ \bibinfo
  {author} {\bibfnamefont {G.}~\bibnamefont {Platero}},\ }\bibfield  {title}
  {\bibinfo {title} {{Proposal for Detection of the ${0}^{\ensuremath{'}}$ and
  ${\ensuremath{\pi}}^{\ensuremath{'}}$ Phases in Quantum-Dot Josephson
  Junctions}},\ }\href {https://doi.org/10.1103/PhysRevLett.129.207701}
  {\bibfield  {journal} {\bibinfo  {journal} {Phys. Rev. Lett.}\ }\textbf
  {\bibinfo {volume} {129}},\ \bibinfo {pages} {207701} (\bibinfo {year}
  {2022})}\BibitemShut {NoStop}%
\bibitem [{\citenamefont {Pillet}\ \emph {et~al.}(2013)\citenamefont {Pillet},
  \citenamefont {Joyez}, \citenamefont {\ifmmode~\check{Z}\else
  \v{Z}\fi{}itko},\ and\ \citenamefont {Goffman}}]{Zitko2013}%
  \BibitemOpen
  \bibfield  {author} {\bibinfo {author} {\bibfnamefont {J.-D.}\ \bibnamefont
  {Pillet}}, \bibinfo {author} {\bibfnamefont {P.}~\bibnamefont {Joyez}},
  \bibinfo {author} {\bibfnamefont {R.}~\bibnamefont {\ifmmode~\check{Z}\else
  \v{Z}\fi{}itko}},\ and\ \bibinfo {author} {\bibfnamefont {M.~F.}\
  \bibnamefont {Goffman}},\ }\bibfield  {title} {\bibinfo {title} {{Tunneling
  spectroscopy of a single quantum dot coupled to a superconductor: From Kondo
  ridge to Andreev bound states}},\ }\href
  {https://doi.org/10.1103/PhysRevB.88.045101} {\bibfield  {journal} {\bibinfo
  {journal} {Phys. Rev. B}\ }\textbf {\bibinfo {volume} {88}},\ \bibinfo
  {pages} {045101} (\bibinfo {year} {2013})}\BibitemShut {NoStop}%
\bibitem [{\citenamefont {Kadlecov\'a}\ \emph {et~al.}(2017)\citenamefont
  {Kadlecov\'a}, \citenamefont {\ifmmode~\check{Z}\else \v{Z}\fi{}onda},\ and\
  \citenamefont {Novotn\'y}}]{Zonda2017}%
  \BibitemOpen
  \bibfield  {author} {\bibinfo {author} {\bibfnamefont {A.}~\bibnamefont
  {Kadlecov\'a}}, \bibinfo {author} {\bibfnamefont {M.}~\bibnamefont
  {\ifmmode~\check{Z}\else \v{Z}\fi{}onda}},\ and\ \bibinfo {author}
  {\bibfnamefont {T.}~\bibnamefont {Novotn\'y}},\ }\bibfield  {title} {\bibinfo
  {title} {{Quantum dot attached to superconducting leads: Relation between
  symmetric and asymmetric coupling}},\ }\href
  {https://doi.org/10.1103/PhysRevB.95.195114} {\bibfield  {journal} {\bibinfo
  {journal} {Phys. Rev. B}\ }\textbf {\bibinfo {volume} {95}},\ \bibinfo
  {pages} {195114} (\bibinfo {year} {2017})}\BibitemShut {NoStop}%
\bibitem [{\citenamefont {Pokorn\'y}\ and\ \citenamefont
  {\ifmmode~\check{Z}\else \v{Z}\fi{}onda}(2023)}]{Zonda2023}%
  \BibitemOpen
  \bibfield  {author} {\bibinfo {author} {\bibfnamefont {V.}~\bibnamefont
  {Pokorn\'y}}\ and\ \bibinfo {author} {\bibfnamefont {M.}~\bibnamefont
  {\ifmmode~\check{Z}\else \v{Z}\fi{}onda}},\ }\bibfield  {title} {\bibinfo
  {title} {{Effective low-energy models for superconducting impurity
  systems}},\ }\href {https://doi.org/10.1103/PhysRevB.107.155111} {\bibfield
  {journal} {\bibinfo  {journal} {Phys. Rev. B}\ }\textbf {\bibinfo {volume}
  {107}},\ \bibinfo {pages} {155111} (\bibinfo {year} {2023})}\BibitemShut
  {NoStop}%
\bibitem [{\citenamefont {W\'ojcik}\ and\ \citenamefont
  {Weymann}(2019)}]{PhysRevB.99.045120}%
  \BibitemOpen
  \bibfield  {author} {\bibinfo {author} {\bibfnamefont {K.~P.}\ \bibnamefont
  {W\'ojcik}}\ and\ \bibinfo {author} {\bibfnamefont {I.}~\bibnamefont
  {Weymann}},\ }\bibfield  {title} {\bibinfo {title} {{Nonlocal pairing as a
  source of spin exchange and Kondo screening}},\ }\href
  {https://doi.org/10.1103/PhysRevB.99.045120} {\bibfield  {journal} {\bibinfo
  {journal} {Phys. Rev. B}\ }\textbf {\bibinfo {volume} {99}},\ \bibinfo
  {pages} {045120} (\bibinfo {year} {2019})}\BibitemShut {NoStop}%
\bibitem [{\citenamefont {Walldorf}\ \emph {et~al.}(2020)\citenamefont
  {Walldorf}, \citenamefont {Brange}, \citenamefont {Padurariu},\ and\
  \citenamefont {Flindt}}]{PhysRevB.101.205422}%
  \BibitemOpen
  \bibfield  {author} {\bibinfo {author} {\bibfnamefont {N.}~\bibnamefont
  {Walldorf}}, \bibinfo {author} {\bibfnamefont {F.}~\bibnamefont {Brange}},
  \bibinfo {author} {\bibfnamefont {C.}~\bibnamefont {Padurariu}},\ and\
  \bibinfo {author} {\bibfnamefont {C.}~\bibnamefont {Flindt}},\ }\bibfield
  {title} {\bibinfo {title} {{Noise and full counting statistics of a Cooper
  pair splitter}},\ }\href {https://doi.org/10.1103/PhysRevB.101.205422}
  {\bibfield  {journal} {\bibinfo  {journal} {Phys. Rev. B}\ }\textbf {\bibinfo
  {volume} {101}},\ \bibinfo {pages} {205422} (\bibinfo {year}
  {2020})}\BibitemShut {NoStop}%
\bibitem [{\citenamefont {Hashimoto}\ \emph {et~al.}(2024)\citenamefont
  {Hashimoto}, \citenamefont {Yamada}, \citenamefont {Tanaka}, \citenamefont
  {Teratani}, \citenamefont {Kemi}, \citenamefont {Kawakami},\ and\
  \citenamefont {Oguri}}]{Hashimoto2024}%
  \BibitemOpen
  \bibfield  {author} {\bibinfo {author} {\bibfnamefont {M.}~\bibnamefont
  {Hashimoto}}, \bibinfo {author} {\bibfnamefont {Y.}~\bibnamefont {Yamada}},
  \bibinfo {author} {\bibfnamefont {Y.}~\bibnamefont {Tanaka}}, \bibinfo
  {author} {\bibfnamefont {Y.}~\bibnamefont {Teratani}}, \bibinfo {author}
  {\bibfnamefont {T.}~\bibnamefont {Kemi}}, \bibinfo {author} {\bibfnamefont
  {N.}~\bibnamefont {Kawakami}},\ and\ \bibinfo {author} {\bibfnamefont
  {A.}~\bibnamefont {Oguri}},\ }\bibfield  {title} {\bibinfo {title} {{Nonlocal
  Andreev transport through a quantum dot in a magnetic field: Interplay
  between Kondo, Zeeman, and Cooper-pair correlations}},\ }\href
  {https://doi.org/10.1103/PhysRevB.109.035404} {\bibfield  {journal} {\bibinfo
   {journal} {Phys. Rev. B}\ }\textbf {\bibinfo {volume} {109}},\ \bibinfo
  {pages} {035404} (\bibinfo {year} {2024})}\BibitemShut {NoStop}%
\bibitem [{\citenamefont {Tanaka}\ \emph {et~al.}(2008)\citenamefont {Tanaka},
  \citenamefont {Kawakami},\ and\ \citenamefont {Oguri}}]{Tanaka_2008}%
  \BibitemOpen
  \bibfield  {author} {\bibinfo {author} {\bibfnamefont {Y.}~\bibnamefont
  {Tanaka}}, \bibinfo {author} {\bibfnamefont {N.}~\bibnamefont {Kawakami}},\
  and\ \bibinfo {author} {\bibfnamefont {A.}~\bibnamefont {Oguri}},\ }\bibfield
   {title} {\bibinfo {title} {{Andreev transport through side-coupled double
  quantum dots}},\ }\href {https://doi.org/10.1103/PhysRevB.78.035444}
  {\bibfield  {journal} {\bibinfo  {journal} {Phys. Rev. B}\ }\textbf {\bibinfo
  {volume} {78}},\ \bibinfo {pages} {035444} (\bibinfo {year}
  {2008})}\BibitemShut {NoStop}%
\bibitem [{\citenamefont {Goldhaber-Gordon}\ \emph
  {et~al.}(1998{\natexlab{a}})\citenamefont {Goldhaber-Gordon}, \citenamefont
  {Shtrikman}, \citenamefont {Mahalu}, \citenamefont {Abusch-Magder},
  \citenamefont {Meirav},\ and\ \citenamefont
  {Kastner}}]{Goldhaber-Gordon1998nature}%
  \BibitemOpen
  \bibfield  {author} {\bibinfo {author} {\bibfnamefont {D.}~\bibnamefont
  {Goldhaber-Gordon}}, \bibinfo {author} {\bibfnamefont {H.}~\bibnamefont
  {Shtrikman}}, \bibinfo {author} {\bibfnamefont {D.}~\bibnamefont {Mahalu}},
  \bibinfo {author} {\bibfnamefont {D.}~\bibnamefont {Abusch-Magder}}, \bibinfo
  {author} {\bibfnamefont {U.}~\bibnamefont {Meirav}},\ and\ \bibinfo {author}
  {\bibfnamefont {M.~A.}\ \bibnamefont {Kastner}},\ }\bibfield  {title}
  {\bibinfo {title} {{Kondo effect in a single-electron transistor}},\ }\href
  {https://doi.org/10.1038/34373} {\bibfield  {journal} {\bibinfo  {journal}
  {Nature}\ }\textbf {\bibinfo {volume} {391}},\ \bibinfo {pages} {156}
  (\bibinfo {year} {1998}{\natexlab{a}})}\BibitemShut {NoStop}%
\bibitem [{\citenamefont {Goldhaber-Gordon}\ \emph
  {et~al.}(1998{\natexlab{b}})\citenamefont {Goldhaber-Gordon}, \citenamefont
  {G\"ores}, \citenamefont {Kastner}, \citenamefont {Shtrikman}, \citenamefont
  {Mahalu},\ and\ \citenamefont {Meirav}}]{Goldhaber-Goldon1998PRL}%
  \BibitemOpen
  \bibfield  {author} {\bibinfo {author} {\bibfnamefont {D.}~\bibnamefont
  {Goldhaber-Gordon}}, \bibinfo {author} {\bibfnamefont {J.}~\bibnamefont
  {G\"ores}}, \bibinfo {author} {\bibfnamefont {M.~A.}\ \bibnamefont
  {Kastner}}, \bibinfo {author} {\bibfnamefont {H.}~\bibnamefont {Shtrikman}},
  \bibinfo {author} {\bibfnamefont {D.}~\bibnamefont {Mahalu}},\ and\ \bibinfo
  {author} {\bibfnamefont {U.}~\bibnamefont {Meirav}},\ }\bibfield  {title}
  {\bibinfo {title} {{From the Kondo Regime to the Mixed-Valence Regime in a
  Single-Electron Transistor}},\ }\href
  {https://doi.org/10.1103/PhysRevLett.81.5225} {\bibfield  {journal} {\bibinfo
   {journal} {Phys. Rev. Lett.}\ }\textbf {\bibinfo {volume} {81}},\ \bibinfo
  {pages} {5225} (\bibinfo {year} {1998}{\natexlab{b}})}\BibitemShut {NoStop}%
\bibitem [{\citenamefont {Cronenwett}\ \emph {et~al.}(1998)\citenamefont
  {Cronenwett}, \citenamefont {Oosterkamp},\ and\ \citenamefont
  {Kouwenhoven}}]{Cronenwett-Oosterkamp-Kouwenhoven1998}%
  \BibitemOpen
  \bibfield  {author} {\bibinfo {author} {\bibfnamefont {S.~M.}\ \bibnamefont
  {Cronenwett}}, \bibinfo {author} {\bibfnamefont {T.~H.}\ \bibnamefont
  {Oosterkamp}},\ and\ \bibinfo {author} {\bibfnamefont {L.~P.}\ \bibnamefont
  {Kouwenhoven}},\ }\bibfield  {title} {\bibinfo {title} {{A Tunable Kondo
  Effect in Quantum Dots}},\ }\href
  {https://doi.org/10.1126/science.281.5376.540} {\bibfield  {journal}
  {\bibinfo  {journal} {Science}\ }\textbf {\bibinfo {volume} {281}},\ \bibinfo
  {pages} {540} (\bibinfo {year} {1998})}\BibitemShut {NoStop}%
\bibitem [{\citenamefont {van~der Wiel}\ \emph {et~al.}(2000)\citenamefont
  {van~der Wiel}, \citenamefont {Franceschi}, \citenamefont {Fujisawa},
  \citenamefont {Elzerman}, \citenamefont {Tarucha},\ and\ \citenamefont
  {Kouwenhoven}}]{vanderWiel2000}%
  \BibitemOpen
  \bibfield  {author} {\bibinfo {author} {\bibfnamefont {W.~G.}\ \bibnamefont
  {van~der Wiel}}, \bibinfo {author} {\bibfnamefont {S.~D.}\ \bibnamefont
  {Franceschi}}, \bibinfo {author} {\bibfnamefont {T.}~\bibnamefont
  {Fujisawa}}, \bibinfo {author} {\bibfnamefont {J.~M.}\ \bibnamefont
  {Elzerman}}, \bibinfo {author} {\bibfnamefont {S.}~\bibnamefont {Tarucha}},\
  and\ \bibinfo {author} {\bibfnamefont {L.~P.}\ \bibnamefont {Kouwenhoven}},\
  }\bibfield  {title} {\bibinfo {title} {{The Kondo Effect in the Unitary
  Limit}},\ }\href {https://doi.org/10.1126/science.289.5487.2105} {\bibfield
  {journal} {\bibinfo  {journal} {Science}\ }\textbf {\bibinfo {volume}
  {289}},\ \bibinfo {pages} {2105} (\bibinfo {year} {2000})}\BibitemShut
  {NoStop}%
\bibitem [{\citenamefont {V.~Borzenets}\ \emph {et~al.}(2020)\citenamefont
  {V.~Borzenets}, \citenamefont {Shim}, \citenamefont {Chen}, \citenamefont
  {Ludwig}, \citenamefont {Wieck}, \citenamefont {Tarucha}, \citenamefont
  {Sim},\ and\ \citenamefont {Yamamoto}}]{KondoCloud2020}%
  \BibitemOpen
  \bibfield  {author} {\bibinfo {author} {\bibfnamefont {I.}~\bibnamefont
  {V.~Borzenets}}, \bibinfo {author} {\bibfnamefont {J.}~\bibnamefont {Shim}},
  \bibinfo {author} {\bibfnamefont {J.~C.~H.}\ \bibnamefont {Chen}}, \bibinfo
  {author} {\bibfnamefont {A.}~\bibnamefont {Ludwig}}, \bibinfo {author}
  {\bibfnamefont {A.~D.}\ \bibnamefont {Wieck}}, \bibinfo {author}
  {\bibfnamefont {S.}~\bibnamefont {Tarucha}}, \bibinfo {author} {\bibfnamefont
  {H.-S.}\ \bibnamefont {Sim}},\ and\ \bibinfo {author} {\bibfnamefont
  {M.}~\bibnamefont {Yamamoto}},\ }\bibfield  {title} {\bibinfo {title}
  {{Observation of the Kondo screening cloud}},\ }\href
  {https://doi.org/10.1038/s41586-020-2058-6} {\bibfield  {journal} {\bibinfo
  {journal} {Nature}\ }\textbf {\bibinfo {volume} {579}},\ \bibinfo {pages}
  {210} (\bibinfo {year} {2020})}\BibitemShut {NoStop}%
\bibitem [{\citenamefont {Grobis}\ \emph {et~al.}(2008)\citenamefont {Grobis},
  \citenamefont {Rau}, \citenamefont {Potok}, \citenamefont {Shtrikman},\ and\
  \citenamefont {Goldhaber-Gordon}}]{GrobisGoldhaber-Gordon}%
  \BibitemOpen
  \bibfield  {author} {\bibinfo {author} {\bibfnamefont {M.}~\bibnamefont
  {Grobis}}, \bibinfo {author} {\bibfnamefont {I.~G.}\ \bibnamefont {Rau}},
  \bibinfo {author} {\bibfnamefont {R.~M.}\ \bibnamefont {Potok}}, \bibinfo
  {author} {\bibfnamefont {H.}~\bibnamefont {Shtrikman}},\ and\ \bibinfo
  {author} {\bibfnamefont {D.}~\bibnamefont {Goldhaber-Gordon}},\ }\bibfield
  {title} {\bibinfo {title} {{Universal Scaling in Nonequilibrium Transport
  through a Single Channel Kondo Dot}},\ }\href
  {https://doi.org/10.1103/PhysRevLett.100.246601} {\bibfield  {journal}
  {\bibinfo  {journal} {Phys. Rev. Lett.}\ }\textbf {\bibinfo {volume} {100}},\
  \bibinfo {pages} {246601} (\bibinfo {year} {2008})}\BibitemShut {NoStop}%
\bibitem [{\citenamefont {Scott}\ \emph {et~al.}(2009)\citenamefont {Scott},
  \citenamefont {Keane}, \citenamefont {Ciszek}, \citenamefont {Tour},\ and\
  \citenamefont {Natelson}}]{ScottNatelson}%
  \BibitemOpen
  \bibfield  {author} {\bibinfo {author} {\bibfnamefont {G.~D.}\ \bibnamefont
  {Scott}}, \bibinfo {author} {\bibfnamefont {Z.~K.}\ \bibnamefont {Keane}},
  \bibinfo {author} {\bibfnamefont {J.~W.}\ \bibnamefont {Ciszek}}, \bibinfo
  {author} {\bibfnamefont {J.~M.}\ \bibnamefont {Tour}},\ and\ \bibinfo
  {author} {\bibfnamefont {D.}~\bibnamefont {Natelson}},\ }\bibfield  {title}
  {\bibinfo {title} {{Universal scaling of nonequilibrium transport in the
  Kondo regime of single molecule devices}},\ }\href
  {https://doi.org/10.1103/PhysRevB.79.165413} {\bibfield  {journal} {\bibinfo
  {journal} {Phys. Rev. B}\ }\textbf {\bibinfo {volume} {79}},\ \bibinfo
  {pages} {165413} (\bibinfo {year} {2009})}\BibitemShut {NoStop}%
\bibitem [{\citenamefont {Zarchin}\ \emph {et~al.}(2008)\citenamefont
  {Zarchin}, \citenamefont {Zaffalon}, \citenamefont {Heiblum}, \citenamefont
  {Mahalu},\ and\ \citenamefont {Umansky}}]{Heiblum}%
  \BibitemOpen
  \bibfield  {author} {\bibinfo {author} {\bibfnamefont {O.}~\bibnamefont
  {Zarchin}}, \bibinfo {author} {\bibfnamefont {M.}~\bibnamefont {Zaffalon}},
  \bibinfo {author} {\bibfnamefont {M.}~\bibnamefont {Heiblum}}, \bibinfo
  {author} {\bibfnamefont {D.}~\bibnamefont {Mahalu}},\ and\ \bibinfo {author}
  {\bibfnamefont {V.}~\bibnamefont {Umansky}},\ }\bibfield  {title} {\bibinfo
  {title} {{Two-electron bunching in transport through a quantum dot induced by
  Kondo correlations}},\ }\href {https://doi.org/10.1103/PhysRevB.77.241303}
  {\bibfield  {journal} {\bibinfo  {journal} {Phys. Rev. B}\ }\textbf {\bibinfo
  {volume} {77}},\ \bibinfo {pages} {241303(R)} (\bibinfo {year}
  {2008})}\BibitemShut {NoStop}%
\bibitem [{\citenamefont {Delattre}\ \emph {et~al.}(2009)\citenamefont
  {Delattre}, \citenamefont {Feuillet-Palma}, \citenamefont {Herrmann},
  \citenamefont {Morfin}, \citenamefont {Berroir}, \citenamefont {F{\`e}ve},
  \citenamefont {Pla{\c c}ais}, \citenamefont {Glattli}, \citenamefont {Choi},
  \citenamefont {Mora},\ and\ \citenamefont {Kontos}}]{Delattre2009}%
  \BibitemOpen
  \bibfield  {author} {\bibinfo {author} {\bibfnamefont {T.}~\bibnamefont
  {Delattre}}, \bibinfo {author} {\bibfnamefont {C.}~\bibnamefont
  {Feuillet-Palma}}, \bibinfo {author} {\bibfnamefont {L.~G.}\ \bibnamefont
  {Herrmann}}, \bibinfo {author} {\bibfnamefont {P.}~\bibnamefont {Morfin}},
  \bibinfo {author} {\bibfnamefont {J.-M.}\ \bibnamefont {Berroir}}, \bibinfo
  {author} {\bibfnamefont {G.}~\bibnamefont {F{\`e}ve}}, \bibinfo {author}
  {\bibfnamefont {B.}~\bibnamefont {Pla{\c c}ais}}, \bibinfo {author}
  {\bibfnamefont {D.~C.}\ \bibnamefont {Glattli}}, \bibinfo {author}
  {\bibfnamefont {M.-S.}\ \bibnamefont {Choi}}, \bibinfo {author}
  {\bibfnamefont {C.}~\bibnamefont {Mora}},\ and\ \bibinfo {author}
  {\bibfnamefont {T.}~\bibnamefont {Kontos}},\ }\bibfield  {title} {\bibinfo
  {title} {{Noisy Kondo impurities}},\ }\href
  {https://doi.org/10.1038/nphys1186} {\bibfield  {journal} {\bibinfo
  {journal} {Nat. Phys.}\ }\textbf {\bibinfo {volume} {5}},\ \bibinfo {pages}
  {208} (\bibinfo {year} {2009})}\BibitemShut {NoStop}%
\bibitem [{\citenamefont {Yamauchi}\ \emph {et~al.}(2011)\citenamefont
  {Yamauchi}, \citenamefont {Sekiguchi}, \citenamefont {Chida}, \citenamefont
  {Arakawa}, \citenamefont {Nakamura}, \citenamefont {Kobayashi}, \citenamefont
  {Ono}, \citenamefont {Fujii},\ and\ \citenamefont
  {Sakano}}]{KobayashiKondoShot}%
  \BibitemOpen
  \bibfield  {author} {\bibinfo {author} {\bibfnamefont {Y.}~\bibnamefont
  {Yamauchi}}, \bibinfo {author} {\bibfnamefont {K.}~\bibnamefont {Sekiguchi}},
  \bibinfo {author} {\bibfnamefont {K.}~\bibnamefont {Chida}}, \bibinfo
  {author} {\bibfnamefont {T.}~\bibnamefont {Arakawa}}, \bibinfo {author}
  {\bibfnamefont {S.}~\bibnamefont {Nakamura}}, \bibinfo {author}
  {\bibfnamefont {K.}~\bibnamefont {Kobayashi}}, \bibinfo {author}
  {\bibfnamefont {T.}~\bibnamefont {Ono}}, \bibinfo {author} {\bibfnamefont
  {T.}~\bibnamefont {Fujii}},\ and\ \bibinfo {author} {\bibfnamefont
  {R.}~\bibnamefont {Sakano}},\ }\bibfield  {title} {\bibinfo {title}
  {{Evolution of the Kondo Effect in a Quantum Dot Probed by Shot Noise}},\
  }\href {https://doi.org/10.1103/PhysRevLett.106.176601} {\bibfield  {journal}
  {\bibinfo  {journal} {Phys. Rev. Lett.}\ }\textbf {\bibinfo {volume} {106}},\
  \bibinfo {pages} {176601} (\bibinfo {year} {2011})}\BibitemShut {NoStop}%
\bibitem [{\citenamefont {Ferrier}\ \emph {et~al.}(2016)\citenamefont
  {Ferrier}, \citenamefont {Arakawa}, \citenamefont {Hata}, \citenamefont
  {Fujiwara}, \citenamefont {Delagrange}, \citenamefont {Weil}, \citenamefont
  {Deblock}, \citenamefont {Sakano}, \citenamefont {Oguri},\ and\ \citenamefont
  {Kobayashi}}]{Ferrier2016}%
  \BibitemOpen
  \bibfield  {author} {\bibinfo {author} {\bibfnamefont {M.}~\bibnamefont
  {Ferrier}}, \bibinfo {author} {\bibfnamefont {T.}~\bibnamefont {Arakawa}},
  \bibinfo {author} {\bibfnamefont {T.}~\bibnamefont {Hata}}, \bibinfo {author}
  {\bibfnamefont {R.}~\bibnamefont {Fujiwara}}, \bibinfo {author}
  {\bibfnamefont {R.}~\bibnamefont {Delagrange}}, \bibinfo {author}
  {\bibfnamefont {R.}~\bibnamefont {Weil}}, \bibinfo {author} {\bibfnamefont
  {R.}~\bibnamefont {Deblock}}, \bibinfo {author} {\bibfnamefont
  {R.}~\bibnamefont {Sakano}}, \bibinfo {author} {\bibfnamefont
  {A.}~\bibnamefont {Oguri}},\ and\ \bibinfo {author} {\bibfnamefont
  {K.}~\bibnamefont {Kobayashi}},\ }\bibfield  {title} {\bibinfo {title}
  {{Universality of non-equilibrium fluctuations in strongly correlated quantum
  liquids}},\ }\href {https://doi.org/10.1038/nphys3556} {\bibfield  {journal}
  {\bibinfo  {journal} {Nat. Phys.}\ }\textbf {\bibinfo {volume} {12}},\
  \bibinfo {pages} {230} (\bibinfo {year} {2016})}\BibitemShut {NoStop}%
\bibitem [{\citenamefont {Hata}\ \emph {et~al.}(2021)\citenamefont {Hata},
  \citenamefont {Teratani}, \citenamefont {Arakawa}, \citenamefont {Lee},
  \citenamefont {Ferrier}, \citenamefont {Deblock}, \citenamefont {Sakano},
  \citenamefont {Oguri},\ and\ \citenamefont {Kobayashi}}]{Hata2021}%
  \BibitemOpen
  \bibfield  {author} {\bibinfo {author} {\bibfnamefont {T.}~\bibnamefont
  {Hata}}, \bibinfo {author} {\bibfnamefont {Y.}~\bibnamefont {Teratani}},
  \bibinfo {author} {\bibfnamefont {T.}~\bibnamefont {Arakawa}}, \bibinfo
  {author} {\bibfnamefont {S.}~\bibnamefont {Lee}}, \bibinfo {author}
  {\bibfnamefont {M.}~\bibnamefont {Ferrier}}, \bibinfo {author} {\bibfnamefont
  {R.}~\bibnamefont {Deblock}}, \bibinfo {author} {\bibfnamefont
  {R.}~\bibnamefont {Sakano}}, \bibinfo {author} {\bibfnamefont
  {A.}~\bibnamefont {Oguri}},\ and\ \bibinfo {author} {\bibfnamefont
  {K.}~\bibnamefont {Kobayashi}},\ }\bibfield  {title} {\bibinfo {title}
  {{Three-body correlations in nonlinear response of correlated quantum
  liquid}},\ }\href {https://doi.org/10.1038/s41467-021-23467-4} {\bibfield
  {journal} {\bibinfo  {journal} {Nature Communications}\ }\textbf {\bibinfo
  {volume} {12}},\ \bibinfo {pages} {3233} (\bibinfo {year}
  {2021})}\BibitemShut {NoStop}%
\bibitem [{\citenamefont {Hsu}\ \emph {et~al.}(2022)\citenamefont {Hsu},
  \citenamefont {Costi}, \citenamefont {Vogel}, \citenamefont {Wegeberg},
  \citenamefont {Mayor}, \citenamefont {van~der Zant},\ and\ \citenamefont
  {Gehring}}]{Costi2022}%
  \BibitemOpen
  \bibfield  {author} {\bibinfo {author} {\bibfnamefont {C.}~\bibnamefont
  {Hsu}}, \bibinfo {author} {\bibfnamefont {T.~A.}\ \bibnamefont {Costi}},
  \bibinfo {author} {\bibfnamefont {D.}~\bibnamefont {Vogel}}, \bibinfo
  {author} {\bibfnamefont {C.}~\bibnamefont {Wegeberg}}, \bibinfo {author}
  {\bibfnamefont {M.}~\bibnamefont {Mayor}}, \bibinfo {author} {\bibfnamefont
  {H.~S.~J.}\ \bibnamefont {van~der Zant}},\ and\ \bibinfo {author}
  {\bibfnamefont {P.}~\bibnamefont {Gehring}},\ }\bibfield  {title} {\bibinfo
  {title} {{Magnetic-Field Universality of the Kondo Effect Revealed by
  Thermocurrent Spectroscopy}},\ }\href
  {https://doi.org/10.1103/PhysRevLett.128.147701} {\bibfield  {journal}
  {\bibinfo  {journal} {Phys. Rev. Lett.}\ }\textbf {\bibinfo {volume} {128}},\
  \bibinfo {pages} {147701} (\bibinfo {year} {2022})}\BibitemShut {NoStop}%
\bibitem [{\citenamefont {Svilans}\ \emph {et~al.}(2018)\citenamefont
  {Svilans}, \citenamefont {Josefsson}, \citenamefont {Burke}, \citenamefont
  {Fahlvik}, \citenamefont {Thelander}, \citenamefont {Linke},\ and\
  \citenamefont {Leijnse}}]{Svilansexperimentthermopower}%
  \BibitemOpen
  \bibfield  {author} {\bibinfo {author} {\bibfnamefont {A.}~\bibnamefont
  {Svilans}}, \bibinfo {author} {\bibfnamefont {M.}~\bibnamefont {Josefsson}},
  \bibinfo {author} {\bibfnamefont {A.~M.}\ \bibnamefont {Burke}}, \bibinfo
  {author} {\bibfnamefont {S.}~\bibnamefont {Fahlvik}}, \bibinfo {author}
  {\bibfnamefont {C.}~\bibnamefont {Thelander}}, \bibinfo {author}
  {\bibfnamefont {H.}~\bibnamefont {Linke}},\ and\ \bibinfo {author}
  {\bibfnamefont {M.}~\bibnamefont {Leijnse}},\ }\bibfield  {title} {\bibinfo
  {title} {{Thermoelectric Characterization of the Kondo Resonance in Nanowire
  Quantum Dots}},\ }\href {https://doi.org/10.1103/PhysRevLett.121.206801}
  {\bibfield  {journal} {\bibinfo  {journal} {Phys. Rev. Lett.}\ }\textbf
  {\bibinfo {volume} {121}},\ \bibinfo {pages} {206801} (\bibinfo {year}
  {2018})}\BibitemShut {NoStop}%
\bibitem [{\citenamefont {Hershfield}\ \emph {et~al.}(1992)\citenamefont
  {Hershfield}, \citenamefont {Davies},\ and\ \citenamefont
  {Wilkins}}]{Hershfield1}%
  \BibitemOpen
  \bibfield  {author} {\bibinfo {author} {\bibfnamefont {S.}~\bibnamefont
  {Hershfield}}, \bibinfo {author} {\bibfnamefont {J.~H.}\ \bibnamefont
  {Davies}},\ and\ \bibinfo {author} {\bibfnamefont {J.~W.}\ \bibnamefont
  {Wilkins}},\ }\bibfield  {title} {\bibinfo {title} {{Resonant tunneling
  through an Anderson impurity. I. Current in the symmetric model}},\ }\href
  {https://doi.org/10.1103/PhysRevB.46.7046} {\bibfield  {journal} {\bibinfo
  {journal} {Phys. Rev. B}\ }\textbf {\bibinfo {volume} {46}},\ \bibinfo
  {pages} {7046} (\bibinfo {year} {1992})}\BibitemShut {NoStop}%
\bibitem [{\citenamefont {Izumida}\ \emph {et~al.}(2001)\citenamefont
  {Izumida}, \citenamefont {Sakai},\ and\ \citenamefont
  {Suzuki}}]{Izumida2001}%
  \BibitemOpen
  \bibfield  {author} {\bibinfo {author} {\bibfnamefont {W.}~\bibnamefont
  {Izumida}}, \bibinfo {author} {\bibfnamefont {O.}~\bibnamefont {Sakai}},\
  and\ \bibinfo {author} {\bibfnamefont {S.}~\bibnamefont {Suzuki}},\
  }\bibfield  {title} {\bibinfo {title} {{Kondo Effect in Tunneling through a
  Quantum Dot}},\ }\href {https://doi.org/10.1143/JPSJ.70.1045} {\bibfield
  {journal} {\bibinfo  {journal} {J. Phys. Soc. Japan}\ }\textbf {\bibinfo
  {volume} {70}},\ \bibinfo {pages} {1045} (\bibinfo {year}
  {2001})}\BibitemShut {NoStop}%
\bibitem [{\citenamefont {Oguri}(2001)}]{AO2001}%
  \BibitemOpen
  \bibfield  {author} {\bibinfo {author} {\bibfnamefont {A.}~\bibnamefont
  {Oguri}},\ }\bibfield  {title} {\bibinfo {title} {{Fermi-liquid theory for
  the Anderson model out of equilibrium}},\ }\href
  {https://doi.org/10.1103/PhysRevB.64.153305} {\bibfield  {journal} {\bibinfo
  {journal} {Phys. Rev. B}\ }\textbf {\bibinfo {volume} {64}},\ \bibinfo
  {pages} {153305} (\bibinfo {year} {2001})}\BibitemShut {NoStop}%
\bibitem [{\citenamefont {Sela}\ and\ \citenamefont
  {Malecki}(2009)}]{Sela-Malecki2009}%
  \BibitemOpen
  \bibfield  {author} {\bibinfo {author} {\bibfnamefont {E.}~\bibnamefont
  {Sela}}\ and\ \bibinfo {author} {\bibfnamefont {J.}~\bibnamefont {Malecki}},\
  }\bibfield  {title} {\bibinfo {title} {{Nonequilibrium conductance of
  asymmetric nanodevices in the Kondo regime}},\ }\href
  {https://doi.org/10.1103/PhysRevB.80.233103} {\bibfield  {journal} {\bibinfo
  {journal} {Phys. Rev. B}\ }\textbf {\bibinfo {volume} {80}},\ \bibinfo
  {pages} {233103} (\bibinfo {year} {2009})}\BibitemShut {NoStop}%
\bibitem [{\citenamefont {Aligia}(2012)}]{Aligia2012}%
  \BibitemOpen
  \bibfield  {author} {\bibinfo {author} {\bibfnamefont {A.~A.}\ \bibnamefont
  {Aligia}},\ }\bibfield  {title} {\bibinfo {title} {{Nonequilibrium
  conductance of a nanodevice for small bias voltage}},\ }\href
  {https://doi.org/10.1088/0953-8984/24/1/015306} {\bibfield  {journal}
  {\bibinfo  {journal} {J. Phys.: Condens. Matter}\ }\textbf {\bibinfo {volume}
  {24}},\ \bibinfo {pages} {015306} (\bibinfo {year} {2012})}\BibitemShut
  {NoStop}%
\bibitem [{\citenamefont {Aligia}(2014)}]{Aligia2014}%
  \BibitemOpen
  \bibfield  {author} {\bibinfo {author} {\bibfnamefont {A.~A.}\ \bibnamefont
  {Aligia}},\ }\bibfield  {title} {\bibinfo {title} {{Nonequilibrium
  self-energies, Ng approach, and heat current of a nanodevice for small bias
  voltage and temperature}},\ }\href
  {https://doi.org/10.1103/PhysRevB.89.125405} {\bibfield  {journal} {\bibinfo
  {journal} {Phys. Rev. B}\ }\textbf {\bibinfo {volume} {89}},\ \bibinfo
  {pages} {125405} (\bibinfo {year} {2014})}\BibitemShut {NoStop}%
\bibitem [{\citenamefont {Hershfield}(1992)}]{Hershfield2}%
  \BibitemOpen
  \bibfield  {author} {\bibinfo {author} {\bibfnamefont {S.}~\bibnamefont
  {Hershfield}},\ }\bibfield  {title} {\bibinfo {title} {{Resonant tunneling
  through an Anderson impurity. II. Noise in the Hartree approximation}},\
  }\href {https://doi.org/10.1103/PhysRevB.46.7061} {\bibfield  {journal}
  {\bibinfo  {journal} {Phys. Rev. B}\ }\textbf {\bibinfo {volume} {46}},\
  \bibinfo {pages} {7061} (\bibinfo {year} {1992})}\BibitemShut {NoStop}%
\bibitem [{\citenamefont {Gogolin}\ and\ \citenamefont
  {Komnik}(2006)}]{GogolinKomnikPRL}%
  \BibitemOpen
  \bibfield  {author} {\bibinfo {author} {\bibfnamefont {A.~O.}\ \bibnamefont
  {Gogolin}}\ and\ \bibinfo {author} {\bibfnamefont {A.}~\bibnamefont
  {Komnik}},\ }\bibfield  {title} {\bibinfo {title} {{Full Counting Statistics
  for the Kondo Dot in the Unitary Limit}},\ }\href
  {https://doi.org/10.1103/PhysRevLett.97.016602} {\bibfield  {journal}
  {\bibinfo  {journal} {Phys. Rev. Lett.}\ }\textbf {\bibinfo {volume} {97}},\
  \bibinfo {pages} {016602} (\bibinfo {year} {2006})}\BibitemShut {NoStop}%
\bibitem [{\citenamefont {Sela}\ \emph {et~al.}(2006)\citenamefont {Sela},
  \citenamefont {Oreg}, \citenamefont {von Oppen},\ and\ \citenamefont
  {Koch}}]{Sela2006}%
  \BibitemOpen
  \bibfield  {author} {\bibinfo {author} {\bibfnamefont {E.}~\bibnamefont
  {Sela}}, \bibinfo {author} {\bibfnamefont {Y.}~\bibnamefont {Oreg}}, \bibinfo
  {author} {\bibfnamefont {F.}~\bibnamefont {von Oppen}},\ and\ \bibinfo
  {author} {\bibfnamefont {J.}~\bibnamefont {Koch}},\ }\bibfield  {title}
  {\bibinfo {title} {{Fractional Shot Noise in the Kondo Regime}},\ }\href
  {https://doi.org/10.1103/PhysRevLett.97.086601} {\bibfield  {journal}
  {\bibinfo  {journal} {Phys. Rev. Lett.}\ }\textbf {\bibinfo {volume} {97}},\
  \bibinfo {pages} {086601} (\bibinfo {year} {2006})}\BibitemShut {NoStop}%
\bibitem [{\citenamefont {Golub}(2006)}]{Golub}%
  \BibitemOpen
  \bibfield  {author} {\bibinfo {author} {\bibfnamefont {A.}~\bibnamefont
  {Golub}},\ }\bibfield  {title} {\bibinfo {title} {{Shot noise near the
  unitary limit of a Kondo quantum dot}},\ }\href
  {https://doi.org/10.1103/PhysRevB.73.233310} {\bibfield  {journal} {\bibinfo
  {journal} {Phys. Rev. B}\ }\textbf {\bibinfo {volume} {73}},\ \bibinfo
  {pages} {233310} (\bibinfo {year} {2006})}\BibitemShut {NoStop}%
\bibitem [{\citenamefont {Oguri}\ \emph {et~al.}(2011)\citenamefont {Oguri},
  \citenamefont {Sakano},\ and\ \citenamefont {Fujii}}]{OguriSakanoFujii2011}%
  \BibitemOpen
  \bibfield  {author} {\bibinfo {author} {\bibfnamefont {A.}~\bibnamefont
  {Oguri}}, \bibinfo {author} {\bibfnamefont {R.}~\bibnamefont {Sakano}},\ and\
  \bibinfo {author} {\bibfnamefont {T.}~\bibnamefont {Fujii}},\ }\bibfield
  {title} {\bibinfo {title} {{$1/(N\ensuremath{-}1)$ expansion based on a
  perturbation theory in $U$ for the Anderson model with $N$-fold
  degeneracy}},\ }\href {https://doi.org/10.1103/PhysRevB.84.113301} {\bibfield
   {journal} {\bibinfo  {journal} {Phys. Rev. B}\ }\textbf {\bibinfo {volume}
  {84}},\ \bibinfo {pages} {113301} (\bibinfo {year} {2011})}\BibitemShut
  {NoStop}%
\bibitem [{\citenamefont {Costi}\ and\ \citenamefont
  {Zlati\ifmmode~\acute{c}\else \'{c}\fi{}}(2010)}]{CostiZlatic2010}%
  \BibitemOpen
  \bibfield  {author} {\bibinfo {author} {\bibfnamefont {T.~A.}\ \bibnamefont
  {Costi}}\ and\ \bibinfo {author} {\bibfnamefont {V.}~\bibnamefont
  {Zlati\ifmmode~\acute{c}\else \'{c}\fi{}}},\ }\bibfield  {title} {\bibinfo
  {title} {{Thermoelectric transport through strongly correlated quantum
  dots}},\ }\href {https://doi.org/10.1103/PhysRevB.81.235127} {\bibfield
  {journal} {\bibinfo  {journal} {Phys. Rev. B}\ }\textbf {\bibinfo {volume}
  {81}},\ \bibinfo {pages} {235127} (\bibinfo {year} {2010})}\BibitemShut
  {NoStop}%
\bibitem [{\citenamefont {Costi}(2019)}]{Costimagthermopower}%
  \BibitemOpen
  \bibfield  {author} {\bibinfo {author} {\bibfnamefont {T.~A.}\ \bibnamefont
  {Costi}},\ }\bibfield  {title} {\bibinfo {title} {{Magnetic field dependence
  of the thermopower of Kondo-correlated quantum dots: Comparison with
  experiment}},\ }\href {https://doi.org/10.1103/PhysRevB.100.155126}
  {\bibfield  {journal} {\bibinfo  {journal} {Phys. Rev. B}\ }\textbf {\bibinfo
  {volume} {100}},\ \bibinfo {pages} {155126} (\bibinfo {year}
  {2019})}\BibitemShut {NoStop}%
\bibitem [{\citenamefont {Nozi{\`e}res}(1974)}]{NozieresFermiLiquid}%
  \BibitemOpen
  \bibfield  {author} {\bibinfo {author} {\bibfnamefont {P.}~\bibnamefont
  {Nozi{\`e}res}},\ }\bibfield  {title} {\bibinfo {title} {{A Fermi-liquid
  description of the Kondo problem at low temperatures}},\ }\href
  {https://doi.org/10.1007/BF00654541} {\bibfield  {journal} {\bibinfo
  {journal} {J. Low Temp. Phys.}\ }\textbf {\bibinfo {volume} {17}},\ \bibinfo
  {pages} {31} (\bibinfo {year} {1974})}\BibitemShut {NoStop}%
\bibitem [{\citenamefont {Yamada}(1975{\natexlab{a}})}]{YamadaYosida2}%
  \BibitemOpen
  \bibfield  {author} {\bibinfo {author} {\bibfnamefont {K.}~\bibnamefont
  {Yamada}},\ }\bibfield  {title} {\bibinfo {title} {{Perturbation Expansion
  for the Anderson Hamiltonian. II}},\ }\href
  {https://doi.org/10.1143/PTP.53.970} {\bibfield  {journal} {\bibinfo
  {journal} {Prog. Theor. Phys.}\ }\textbf {\bibinfo {volume} {53}},\ \bibinfo
  {pages} {970} (\bibinfo {year} {1975}{\natexlab{a}})}\BibitemShut {NoStop}%
\bibitem [{\citenamefont {Yamada}(1975{\natexlab{b}})}]{YamadaYosida4}%
  \BibitemOpen
  \bibfield  {author} {\bibinfo {author} {\bibfnamefont {K.}~\bibnamefont
  {Yamada}},\ }\bibfield  {title} {\bibinfo {title} {{Perturbation Expansion
  for the Anderson Hamiltonian. IV}},\ }\href
  {https://doi.org/10.1143/PTP.54.316} {\bibfield  {journal} {\bibinfo
  {journal} {Prog. Theor. Phys.}\ }\textbf {\bibinfo {volume} {54}},\ \bibinfo
  {pages} {316} (\bibinfo {year} {1975}{\natexlab{b}})}\BibitemShut {NoStop}%
\bibitem [{\citenamefont {Shiba}(1975)}]{ShibaKorringa}%
  \BibitemOpen
  \bibfield  {author} {\bibinfo {author} {\bibfnamefont {H.}~\bibnamefont
  {Shiba}},\ }\bibfield  {title} {\bibinfo {title} {{The Korringa Relation for
  the Impurity Nuclear Spin-Lattice Relaxation in Dilute Kondo Alloys}},\
  }\href {https://doi.org/10.1143/PTP.54.967} {\bibfield  {journal} {\bibinfo
  {journal} {Prog. Theor. Phys.}\ }\textbf {\bibinfo {volume} {54}},\ \bibinfo
  {pages} {967} (\bibinfo {year} {1975})}\BibitemShut {NoStop}%
\bibitem [{\citenamefont {Yoshimori}(1976)}]{Yoshimori}%
  \BibitemOpen
  \bibfield  {author} {\bibinfo {author} {\bibfnamefont {A.}~\bibnamefont
  {Yoshimori}},\ }\bibfield  {title} {\bibinfo {title} {{Perturbation Analysis
  on Orbital-Degenerate Anderson Model}},\ }\href
  {https://doi.org/10.1143/PTP.55.67} {\bibfield  {journal} {\bibinfo
  {journal} {Prog. Theor. Phys.}\ }\textbf {\bibinfo {volume} {55}},\ \bibinfo
  {pages} {67} (\bibinfo {year} {1976})}\BibitemShut {NoStop}%
\bibitem [{\citenamefont {Mora}\ \emph {et~al.}(2009)\citenamefont {Mora},
  \citenamefont {Vitushinsky}, \citenamefont {Leyronas}, \citenamefont
  {Clerk},\ and\ \citenamefont {Le~Hur}}]{Mora_etal_2009}%
  \BibitemOpen
  \bibfield  {author} {\bibinfo {author} {\bibfnamefont {C.}~\bibnamefont
  {Mora}}, \bibinfo {author} {\bibfnamefont {P.}~\bibnamefont {Vitushinsky}},
  \bibinfo {author} {\bibfnamefont {X.}~\bibnamefont {Leyronas}}, \bibinfo
  {author} {\bibfnamefont {A.~A.}\ \bibnamefont {Clerk}},\ and\ \bibinfo
  {author} {\bibfnamefont {K.}~\bibnamefont {Le~Hur}},\ }\bibfield  {title}
  {\bibinfo {title} {{Theory of nonequilibrium transport in the $\text{SU}(N)$
  Kondo regime}},\ }\href {https://doi.org/10.1103/PhysRevB.80.155322}
  {\bibfield  {journal} {\bibinfo  {journal} {Phys. Rev. B}\ }\textbf {\bibinfo
  {volume} {80}},\ \bibinfo {pages} {155322} (\bibinfo {year}
  {2009})}\BibitemShut {NoStop}%
\bibitem [{\citenamefont {Mora}(2009)}]{Mora2009}%
  \BibitemOpen
  \bibfield  {author} {\bibinfo {author} {\bibfnamefont {C.}~\bibnamefont
  {Mora}},\ }\bibfield  {title} {\bibinfo {title} {{Fermi-liquid theory for
  $\text{SU}(N)$ Kondo model}},\ }\href
  {https://doi.org/10.1103/PhysRevB.80.125304} {\bibfield  {journal} {\bibinfo
  {journal} {Phys. Rev. B}\ }\textbf {\bibinfo {volume} {80}},\ \bibinfo
  {pages} {125304} (\bibinfo {year} {2009})}\BibitemShut {NoStop}%
\bibitem [{\citenamefont {Mora}\ \emph {et~al.}(2015)\citenamefont {Mora},
  \citenamefont {Moca}, \citenamefont {von Delft},\ and\ \citenamefont
  {Zar\'{a}nd}}]{MoraMocaVonDelftZarand}%
  \BibitemOpen
  \bibfield  {author} {\bibinfo {author} {\bibfnamefont {C.}~\bibnamefont
  {Mora}}, \bibinfo {author} {\bibfnamefont {C.~P.}\ \bibnamefont {Moca}},
  \bibinfo {author} {\bibfnamefont {J.}~\bibnamefont {von Delft}},\ and\
  \bibinfo {author} {\bibfnamefont {G.}~\bibnamefont {Zar\'{a}nd}},\ }\bibfield
   {title} {\bibinfo {title} {{Fermi-liquid theory for the single-impurity
  Anderson model}},\ }\href {https://doi.org/10.1103/PhysRevB.92.075120}
  {\bibfield  {journal} {\bibinfo  {journal} {Phys. Rev. B}\ }\textbf {\bibinfo
  {volume} {92}},\ \bibinfo {pages} {075120} (\bibinfo {year}
  {2015})}\BibitemShut {NoStop}%
\bibitem [{\citenamefont {Filippone}\ \emph {et~al.}(2018)\citenamefont
  {Filippone}, \citenamefont {Moca}, \citenamefont {Weichselbaum},
  \citenamefont {von Delft},\ and\ \citenamefont {Mora}}]{FMvDM2018}%
  \BibitemOpen
  \bibfield  {author} {\bibinfo {author} {\bibfnamefont {M.}~\bibnamefont
  {Filippone}}, \bibinfo {author} {\bibfnamefont {C.~P.}\ \bibnamefont {Moca}},
  \bibinfo {author} {\bibfnamefont {A.}~\bibnamefont {Weichselbaum}}, \bibinfo
  {author} {\bibfnamefont {J.}~\bibnamefont {von Delft}},\ and\ \bibinfo
  {author} {\bibfnamefont {C.}~\bibnamefont {Mora}},\ }\bibfield  {title}
  {\bibinfo {title} {{At which magnetic field, exactly, does the Kondo
  resonance begin to split? A Fermi liquid description of the low-energy
  properties of the Anderson model}},\ }\href
  {https://doi.org/10.1103/PhysRevB.98.075404} {\bibfield  {journal} {\bibinfo
  {journal} {Phys. Rev. B}\ }\textbf {\bibinfo {volume} {98}},\ \bibinfo
  {pages} {075404} (\bibinfo {year} {2018})}\BibitemShut {NoStop}%
\bibitem [{\citenamefont {Oguri}\ and\ \citenamefont
  {Hewson}(2018{\natexlab{a}})}]{AO2017_I}%
  \BibitemOpen
  \bibfield  {author} {\bibinfo {author} {\bibfnamefont {A.}~\bibnamefont
  {Oguri}}\ and\ \bibinfo {author} {\bibfnamefont {A.~C.}\ \bibnamefont
  {Hewson}},\ }\bibfield  {title} {\bibinfo {title} {{Higher-Order Fermi-Liquid
  Corrections for an Anderson Impurity Away from Half Filling}},\ }\href
  {https://doi.org/10.1103/PhysRevLett.120.126802} {\bibfield  {journal}
  {\bibinfo  {journal} {Phys. Rev. Lett.}\ }\textbf {\bibinfo {volume} {120}},\
  \bibinfo {pages} {126802} (\bibinfo {year} {2018}{\natexlab{a}})}\BibitemShut
  {NoStop}%
\bibitem [{\citenamefont {Oguri}\ and\ \citenamefont
  {Hewson}(2018{\natexlab{b}})}]{AO2017_II}%
  \BibitemOpen
  \bibfield  {author} {\bibinfo {author} {\bibfnamefont {A.}~\bibnamefont
  {Oguri}}\ and\ \bibinfo {author} {\bibfnamefont {A.~C.}\ \bibnamefont
  {Hewson}},\ }\bibfield  {title} {\bibinfo {title} {{Higher-order Fermi-liquid
  corrections for an Anderson impurity away from half filling : Equilibrium
  properties}},\ }\href {https://doi.org/10.1103/PhysRevB.97.045406} {\bibfield
   {journal} {\bibinfo  {journal} {Phys. Rev. B}\ }\textbf {\bibinfo {volume}
  {97}},\ \bibinfo {pages} {045406} (\bibinfo {year}
  {2018}{\natexlab{b}})}\BibitemShut {NoStop}%
\bibitem [{\citenamefont {Oguri}\ and\ \citenamefont
  {Hewson}(2018{\natexlab{c}})}]{AO2017_III}%
  \BibitemOpen
  \bibfield  {author} {\bibinfo {author} {\bibfnamefont {A.}~\bibnamefont
  {Oguri}}\ and\ \bibinfo {author} {\bibfnamefont {A.~C.}\ \bibnamefont
  {Hewson}},\ }\bibfield  {title} {\bibinfo {title} {{Higher-order Fermi-liquid
  corrections for an Anderson impurity away from half filling: Nonequilibrium
  transport}},\ }\href {https://doi.org/10.1103/PhysRevB.97.035435} {\bibfield
  {journal} {\bibinfo  {journal} {Phys. Rev. B}\ }\textbf {\bibinfo {volume}
  {97}},\ \bibinfo {pages} {035435} (\bibinfo {year}
  {2018}{\natexlab{c}})}\BibitemShut {NoStop}%
\bibitem [{\citenamefont {Karki}\ \emph {et~al.}(2018)\citenamefont {Karki},
  \citenamefont {Mora}, \citenamefont {von Delft},\ and\ \citenamefont
  {Kiselev}}]{KarkiMora2018}%
  \BibitemOpen
  \bibfield  {author} {\bibinfo {author} {\bibfnamefont {D.~B.}\ \bibnamefont
  {Karki}}, \bibinfo {author} {\bibfnamefont {C.}~\bibnamefont {Mora}},
  \bibinfo {author} {\bibfnamefont {J.}~\bibnamefont {von Delft}},\ and\
  \bibinfo {author} {\bibfnamefont {M.~N.}\ \bibnamefont {Kiselev}},\
  }\bibfield  {title} {\bibinfo {title} {{Two-color Fermi-liquid theory for
  transport through a multilevel Kondo impurity}},\ }\href
  {https://doi.org/10.1103/PhysRevB.97.195403} {\bibfield  {journal} {\bibinfo
  {journal} {Phys. Rev. B}\ }\textbf {\bibinfo {volume} {97}},\ \bibinfo
  {pages} {195403} (\bibinfo {year} {2018})}\BibitemShut {NoStop}%
\bibitem [{\citenamefont {Karki}\ and\ \citenamefont
  {Kiselev}(2017)}]{KarkiKiselev}%
  \BibitemOpen
  \bibfield  {author} {\bibinfo {author} {\bibfnamefont {D.~B.}\ \bibnamefont
  {Karki}}\ and\ \bibinfo {author} {\bibfnamefont {M.~N.}\ \bibnamefont
  {Kiselev}},\ }\bibfield  {title} {\bibinfo {title} {{Thermoelectric transport
  through a $\text{SU}(N)$ Kondo impurity}},\ }\href
  {https://doi.org/10.1103/PhysRevB.96.121403} {\bibfield  {journal} {\bibinfo
  {journal} {Phys. Rev. B}\ }\textbf {\bibinfo {volume} {96}},\ \bibinfo
  {pages} {121403} (\bibinfo {year} {2017})}\BibitemShut {NoStop}%
\bibitem [{\citenamefont {Moca}\ \emph
  {et~al.}(2018{\natexlab{a}})\citenamefont {Moca}, \citenamefont {Mora},
  \citenamefont {Weymann},\ and\ \citenamefont {Zar\'and}}]{MocaMora2018}%
  \BibitemOpen
  \bibfield  {author} {\bibinfo {author} {\bibfnamefont {C.~P.}\ \bibnamefont
  {Moca}}, \bibinfo {author} {\bibfnamefont {C.}~\bibnamefont {Mora}}, \bibinfo
  {author} {\bibfnamefont {I.}~\bibnamefont {Weymann}},\ and\ \bibinfo {author}
  {\bibfnamefont {G.}~\bibnamefont {Zar\'and}},\ }\bibfield  {title} {\bibinfo
  {title} {{Noise of a Chargeless Fermi Liquid}},\ }\href
  {https://doi.org/10.1103/PhysRevLett.120.016803} {\bibfield  {journal}
  {\bibinfo  {journal} {Phys. Rev. Lett.}\ }\textbf {\bibinfo {volume} {120}},\
  \bibinfo {pages} {016803} (\bibinfo {year} {2018}{\natexlab{a}})}\BibitemShut
  {NoStop}%
\bibitem [{\citenamefont {Teratani}\ \emph {et~al.}(2020)\citenamefont
  {Teratani}, \citenamefont {Sakano},\ and\ \citenamefont
  {Oguri}}]{Teratani2020PRL}%
  \BibitemOpen
  \bibfield  {author} {\bibinfo {author} {\bibfnamefont {Y.}~\bibnamefont
  {Teratani}}, \bibinfo {author} {\bibfnamefont {R.}~\bibnamefont {Sakano}},\
  and\ \bibinfo {author} {\bibfnamefont {A.}~\bibnamefont {Oguri}},\ }\bibfield
   {title} {\bibinfo {title} {{Fermi Liquid Theory for Nonlinear Transport
  through a Multilevel Anderson Impurity}},\ }\href
  {https://doi.org/10.1103/PhysRevLett.125.216801} {\bibfield  {journal}
  {\bibinfo  {journal} {Phys. Rev. Lett.}\ }\textbf {\bibinfo {volume} {125}},\
  \bibinfo {pages} {216801} (\bibinfo {year} {2020})}\BibitemShut {NoStop}%
\bibitem [{\citenamefont {Oguri}\ \emph {et~al.}(2022)\citenamefont {Oguri},
  \citenamefont {Teratani}, \citenamefont {Tsutsumi},\ and\ \citenamefont
  {Sakano}}]{Oguri2022}%
  \BibitemOpen
  \bibfield  {author} {\bibinfo {author} {\bibfnamefont {A.}~\bibnamefont
  {Oguri}}, \bibinfo {author} {\bibfnamefont {Y.}~\bibnamefont {Teratani}},
  \bibinfo {author} {\bibfnamefont {K.}~\bibnamefont {Tsutsumi}},\ and\
  \bibinfo {author} {\bibfnamefont {R.}~\bibnamefont {Sakano}},\ }\bibfield
  {title} {\bibinfo {title} {{Current noise and Keldysh vertex function of an
  Anderson impurity in the Fermi-liquid regime}},\ }\href
  {https://doi.org/10.1103/PhysRevB.105.115409} {\bibfield  {journal} {\bibinfo
   {journal} {Phys. Rev. B}\ }\textbf {\bibinfo {volume} {105}},\ \bibinfo
  {pages} {115409} (\bibinfo {year} {2022})}\BibitemShut {NoStop}%
\bibitem [{\citenamefont {Tsutsumi}\ \emph {et~al.}(2021)\citenamefont
  {Tsutsumi}, \citenamefont {Teratani}, \citenamefont {Sakano},\ and\
  \citenamefont {Oguri}}]{Tsutsumi2021}%
  \BibitemOpen
  \bibfield  {author} {\bibinfo {author} {\bibfnamefont {K.}~\bibnamefont
  {Tsutsumi}}, \bibinfo {author} {\bibfnamefont {Y.}~\bibnamefont {Teratani}},
  \bibinfo {author} {\bibfnamefont {R.}~\bibnamefont {Sakano}},\ and\ \bibinfo
  {author} {\bibfnamefont {A.}~\bibnamefont {Oguri}},\ }\bibfield  {title}
  {\bibinfo {title} {{Nonlinear Fermi liquid transport through a quantum dot in
  asymmetric tunnel junctions}},\ }\href
  {https://doi.org/10.1103/PhysRevB.104.235147} {\bibfield  {journal} {\bibinfo
   {journal} {Phys. Rev. B}\ }\textbf {\bibinfo {volume} {104}},\ \bibinfo
  {pages} {235147} (\bibinfo {year} {2021})}\BibitemShut {NoStop}%
\bibitem [{\citenamefont {Tsutsumi}\ \emph {et~al.}(2023)\citenamefont
  {Tsutsumi}, \citenamefont {Teratani}, \citenamefont {Motoyama}, \citenamefont
  {Sakano},\ and\ \citenamefont {Oguri}}]{Tsutsumi2023}%
  \BibitemOpen
  \bibfield  {author} {\bibinfo {author} {\bibfnamefont {K.}~\bibnamefont
  {Tsutsumi}}, \bibinfo {author} {\bibfnamefont {Y.}~\bibnamefont {Teratani}},
  \bibinfo {author} {\bibfnamefont {K.}~\bibnamefont {Motoyama}}, \bibinfo
  {author} {\bibfnamefont {R.}~\bibnamefont {Sakano}},\ and\ \bibinfo {author}
  {\bibfnamefont {A.}~\bibnamefont {Oguri}},\ }\bibfield  {title} {\bibinfo
  {title} {{Role of bias and tunneling asymmetries in nonlinear Fermi-liquid
  transport through an $\mathrm{SU}(N)$ quantum dot}},\ }\href
  {https://doi.org/10.1103/PhysRevB.108.045109} {\bibfield  {journal} {\bibinfo
   {journal} {Phys. Rev. B}\ }\textbf {\bibinfo {volume} {108}},\ \bibinfo
  {pages} {045109} (\bibinfo {year} {2023})}\BibitemShut {NoStop}%
\bibitem [{\citenamefont {Teratani}\ \emph {et~al.}(2024)\citenamefont
  {Teratani}, \citenamefont {Tsutsumi}, \citenamefont {Motoyama}, \citenamefont
  {Sakano},\ and\ \citenamefont {Oguri}}]{teratani2024thermoelectric}%
  \BibitemOpen
  \bibfield  {author} {\bibinfo {author} {\bibfnamefont {Y.}~\bibnamefont
  {Teratani}}, \bibinfo {author} {\bibfnamefont {K.}~\bibnamefont {Tsutsumi}},
  \bibinfo {author} {\bibfnamefont {K.}~\bibnamefont {Motoyama}}, \bibinfo
  {author} {\bibfnamefont {R.}~\bibnamefont {Sakano}},\ and\ \bibinfo {author}
  {\bibfnamefont {A.}~\bibnamefont {Oguri}},\ }\bibfield  {title} {\bibinfo
  {title} {{Thermoelectric transport and current noise through a multilevel
  Anderson impurity: Three-body Fermi liquid corrections in quantum dots and
  magnetic alloys}},\ }\href {https://doi.org/10.1103/PhysRevB.110.035308}
  {\bibfield  {journal} {\bibinfo  {journal} {Phys. Rev. B}\ }\textbf {\bibinfo
  {volume} {110}},\ \bibinfo {pages} {035308} (\bibinfo {year}
  {2024})}\BibitemShut {NoStop}%
\bibitem [{\citenamefont {Motoyama}\ \emph {et~al.}(2025)\citenamefont
  {Motoyama}, \citenamefont {Teratani}, \citenamefont {Tsutsumi}, \citenamefont
  {Wake}, \citenamefont {Kobayashi}, \citenamefont {Sakano},\ and\
  \citenamefont {Oguri}}]{MotoyamaUinf2025}%
  \BibitemOpen
  \bibfield  {author} {\bibinfo {author} {\bibfnamefont {K.}~\bibnamefont
  {Motoyama}}, \bibinfo {author} {\bibfnamefont {Y.}~\bibnamefont {Teratani}},
  \bibinfo {author} {\bibfnamefont {K.}~\bibnamefont {Tsutsumi}}, \bibinfo
  {author} {\bibfnamefont {K.}~\bibnamefont {Wake}}, \bibinfo {author}
  {\bibfnamefont {R.}~\bibnamefont {Kobayashi}}, \bibinfo {author}
  {\bibfnamefont {R.}~\bibnamefont {Sakano}},\ and\ \bibinfo {author}
  {\bibfnamefont {A.}~\bibnamefont {Oguri}},\ }\bibfield  {title} {\bibinfo
  {title} {{Three-body Fermi liquid corrections for an infinite-$U$ SU($N$)
  Anderson impurity model}},\ }\href
  {https://doi.org/10.1103/PhysRevB.111.235301} {\bibfield  {journal} {\bibinfo
   {journal} {Phys. Rev. B}\ }\textbf {\bibinfo {volume} {111}},\ \bibinfo
  {pages} {235301} (\bibinfo {year} {2025})}\BibitemShut {NoStop}%
\bibitem [{\citenamefont {Caroli}\ \emph {et~al.}(1971)\citenamefont {Caroli},
  \citenamefont {Combescot}, \citenamefont {Nozieres},\ and\ \citenamefont
  {Saint-James}}]{Caroli_1971}%
  \BibitemOpen
  \bibfield  {author} {\bibinfo {author} {\bibfnamefont {C.}~\bibnamefont
  {Caroli}}, \bibinfo {author} {\bibfnamefont {R.}~\bibnamefont {Combescot}},
  \bibinfo {author} {\bibfnamefont {P.}~\bibnamefont {Nozieres}},\ and\
  \bibinfo {author} {\bibfnamefont {D.}~\bibnamefont {Saint-James}},\
  }\bibfield  {title} {\bibinfo {title} {{Direct calculation of the tunneling
  current}},\ }\href {https://doi.org/10.1088/0022-3719/4/8/018} {\bibfield
  {journal} {\bibinfo  {journal} {Journal of Physics C: Solid State Physics}\
  }\textbf {\bibinfo {volume} {4}},\ \bibinfo {pages} {916} (\bibinfo {year}
  {1971})}\BibitemShut {NoStop}%
\bibitem [{\citenamefont {Keldysh}(1965)}]{Keldysh}%
  \BibitemOpen
  \bibfield  {author} {\bibinfo {author} {\bibfnamefont {L.~V.}\ \bibnamefont
  {Keldysh}},\ }\bibfield  {title} {\bibinfo {title} {Diagram technique for
  nonequilibrium processes},\ }\href@noop {} {\bibfield  {journal} {\bibinfo
  {journal} {Sov.\ Phys.\ JETP}\ }\textbf {\bibinfo {volume} {20}},\ \bibinfo
  {pages} {1018} (\bibinfo {year} {1965})}\BibitemShut {NoStop}%
\bibitem [{\citenamefont {Krishna-murthy}\ \emph
  {et~al.}(1980{\natexlab{a}})\citenamefont {Krishna-murthy}, \citenamefont
  {Wilkins},\ and\ \citenamefont {Wilson}}]{KWW1}%
  \BibitemOpen
  \bibfield  {author} {\bibinfo {author} {\bibfnamefont {H.~R.}\ \bibnamefont
  {Krishna-murthy}}, \bibinfo {author} {\bibfnamefont {J.~W.}\ \bibnamefont
  {Wilkins}},\ and\ \bibinfo {author} {\bibfnamefont {K.~G.}\ \bibnamefont
  {Wilson}},\ }\bibfield  {title} {\bibinfo {title} {{Renormalization-group
  approach to the Anderson model of dilute magnetic alloys. I. Static
  properties for the symmetric case}},\ }\href
  {https://doi.org/10.1103/PhysRevB.21.1003} {\bibfield  {journal} {\bibinfo
  {journal} {Phys. Rev. B}\ }\textbf {\bibinfo {volume} {21}},\ \bibinfo
  {pages} {1003} (\bibinfo {year} {1980}{\natexlab{a}})}\BibitemShut {NoStop}%
\bibitem [{\citenamefont {Krishna-murthy}\ \emph
  {et~al.}(1980{\natexlab{b}})\citenamefont {Krishna-murthy}, \citenamefont
  {Wilkins},\ and\ \citenamefont {Wilson}}]{KWW2}%
  \BibitemOpen
  \bibfield  {author} {\bibinfo {author} {\bibfnamefont {H.~R.}\ \bibnamefont
  {Krishna-murthy}}, \bibinfo {author} {\bibfnamefont {J.~W.}\ \bibnamefont
  {Wilkins}},\ and\ \bibinfo {author} {\bibfnamefont {K.~G.}\ \bibnamefont
  {Wilson}},\ }\bibfield  {title} {\bibinfo {title} {{Renormalization-group
  approach to the Anderson model of dilute magnetic alloys. II. Static
  properties for the asymmetric case}},\ }\href
  {https://doi.org/10.1103/PhysRevB.21.1044} {\bibfield  {journal} {\bibinfo
  {journal} {Phys. Rev. B}\ }\textbf {\bibinfo {volume} {21}},\ \bibinfo
  {pages} {1044} (\bibinfo {year} {1980}{\natexlab{b}})}\BibitemShut {NoStop}%
\bibitem [{\citenamefont {Hewson}\ \emph {et~al.}(2004)\citenamefont {Hewson},
  \citenamefont {Oguri},\ and\ \citenamefont {Meyer}}]{HewsonOguriMeyer}%
  \BibitemOpen
  \bibfield  {author} {\bibinfo {author} {\bibfnamefont {A.~C.}\ \bibnamefont
  {Hewson}}, \bibinfo {author} {\bibfnamefont {A.}~\bibnamefont {Oguri}},\ and\
  \bibinfo {author} {\bibfnamefont {D.}~\bibnamefont {Meyer}},\ }\bibfield
  {title} {\bibinfo {title} {{Renormalized parameters for impurity models }},\
  }\href {https://doi.org/10.1140/epjb/e2004-00256-0} {\bibfield  {journal}
  {\bibinfo  {journal} {Eur. Phys. J. B}\ }\textbf {\bibinfo {volume} {40}},\
  \bibinfo {pages} {177} (\bibinfo {year} {2004})}\BibitemShut {NoStop}%
\bibitem [{\citenamefont {Morel}\ and\ \citenamefont
  {Nozi\`eres}(1962)}]{MorelAnderson1962}%
  \BibitemOpen
  \bibfield  {author} {\bibinfo {author} {\bibfnamefont {P.}~\bibnamefont
  {Morel}}\ and\ \bibinfo {author} {\bibfnamefont {P.}~\bibnamefont
  {Nozi\`eres}},\ }\bibfield  {title} {\bibinfo {title} {{Lifetime Effects in
  Condensed Helium-3}},\ }\href {https://doi.org/10.1103/PhysRev.126.1909}
  {\bibfield  {journal} {\bibinfo  {journal} {Phys. Rev.}\ }\textbf {\bibinfo
  {volume} {126}},\ \bibinfo {pages} {1909} (\bibinfo {year}
  {1962})}\BibitemShut {NoStop}%
\end{thebibliography}

\end{document}